\documentclass[]{IEEEtran}
\usepackage{graphicx}
\usepackage{subfigure}
\usepackage{cite}
\usepackage{amsmath}
\usepackage[colorlinks=true,
            linkcolor=blue,
            citecolor=blue,
            urlcolor=blue]{hyperref}
\usepackage{setspace}
\usepackage{array}
\usepackage{lineno}
\usepackage{bm}
\usepackage[noend]{algpseudocode}
\usepackage{algorithmicx, algorithm}
\usepackage{color}
\usepackage{xcolor}
\usepackage{multirow}
\usepackage{ulem}
\usepackage{lineno}
\usepackage{hhline}  
\usepackage{booktabs}
\usepackage{amssymb}
\usepackage{algpseudocode}
\usepackage{subcaption}

\begin{document}
\title{Multi-Condition Guided Diffusion Model for Controllable Elastic Parameter Synthesis }

\author{Hongling Chen, Qi Pang, Chuangji Meng, Shian Shen, Jinghuai Gao
\thanks{Hongling Chen, Qi Pang, Chuangji Meng, and Jinghuai Gao are with the School of Information and Communications Engineering, Faculty of Electronic and Information Engineering, Xi$'$an
Jiaotong University, Xi$'$an 710049, China. Shian Shen is with CCTEG Xi’an Research Institute (Group) Co., Ltd., Xi’an 710077, China. (\textit{Corresponding author: Hongling Chen. E-mail: 859311743$@$qq.com}).}}

\maketitle

\begin{abstract}
Prestack elastic parameter inversion is important for reservoir characterization and quantitative seismic interpretation. Most existing deep-learning-based methods have achieved promising results, but they generally require sufficient labeled training data and have limited flexibility in integrating multi-source conditioning information. To address this issue, we propose a multi-condition guided diffusion model for controllable elastic parameter synthesis. Elastic parameter training datasets are first constructed based on well log statistics and geological characteristics of the target area and are used to train the diffusion model. A unified multi-condition guided diffusion framework is then developed to incorporate both implicit and explicit conditioning information. Specifically, iterative latent variable refinement, Adapter-based conditioning, and a diffusion posterior sampling (DPS)-projection guidance strategy are introduced for implicit model-domain constraints, implicit structural constraints, and explicit conditioning-operator constraints, respectively. Synthetic examples demonstrate that the proposed method can generate elastic parameter samples that are consistent with the prescribed conditions under both single-condition and multi-condition guidance. When seismic data are used as conditioning information, the framework can be further adapted to seismic elastic parameter inversion. Experiments show that the proposed method improves the prediction of representative elastic parameters, including P-wave velocity, S-wave velocity, and density, compared with baseline methods. The synthesized samples can also support downstream deep-learning-based inversion under limited labeled data, achieving competitive performance. \textcolor{blue}{\textbf{Code will be released soon.}}
\end{abstract}


\IEEEpeerreviewmaketitle

\section{Introduction}
\IEEEPARstart{E}lastic parameters provide key quantitative information for describing subsurface lithology, pore-fluid properties, and reservoir heterogeneity \cite{avseth2010quantitative}. Prestack seismic inversion aims to estimate these parameters from angle-dependent seismic responses and has therefore become an important tool in reservoir characterization. However, prestack inversion is usually an ill-posed problem because of the band-limited nature of seismic data, noise contamination, nonlinear forward modeling, and strong coupling among multiple elastic parameters. As a result, obtaining high-accuracy estimates of elastic parameters remains challenging. Conventional seismic inversion methods alleviate the ill-posedness by introducing prior information through regularization terms or Bayesian frameworks \cite{buland2003bayesian, wang2019three, huang2021slope, lin2024improved}. Although these methods have achieved wide applications, the prescribed priors are often insufficient to fully characterize the complex solution space, and their effectiveness depends on specific assumptions about model smoothness, sparsity, statistical distribution, or geological structure. To overcome these limitations, increasing attention has been paid to deep-learning-based seismic inversion methods, which aim to learn nonlinear mappings or model priors directly from data. Following this direction, this study focuses on deep-learning-based seismic inversion for elastic parameter prediction.

Deep learning methods generally have stronger nonlinear representation capability and higher inference efficiency than conventional inversion methods. However, the performance of deep-learning-based prestack inversion is affected by several key factors, including the construction of training datasets, the design of neural network architectures, and the training strategy. Among these factors, the availability and representativeness of labeled training data are particularly important. In seismic inversion, well logs are commonly used as labels for seismic data near well locations. However, well logs are usually sparse and limited in number, making it difficult to construct sufficient training samples directly from field data. To alleviate this problem, some studies have constructed 1D synthetic labeled datasets by augmenting limited well log information through stochastic simulation, rock-physics modeling, and prestack seismic forward modeling \cite{das2020petrophysical}.
Other studies have further generated spatially organized 2D or 3D synthetic models by incorporating geological structures and lateral variations \cite{cao2022multi,deng2022openfwi,Synthoseis}. These efforts indicate that training dataset construction has gradually evolved from trace-wise 1D samples to spatially organized multidimensional datasets.

With the increasing use of spatially organized datasets, network architectures have also evolved from trace-wise 1D models to multidimensional models. Early studies commonly employed 1D deep neural networks to learn the nonlinear mapping between prestack angle gathers and elastic parameters \cite{biswas2019prestack,cao2022multi,wang2023avo}. These trace-wise networks are computationally efficient and can capture the local relationship between angle-dependent amplitudes and elastic parameters. However, because lateral spatial information is not explicitly incorporated, 1D networks may suffer from insufficient lateral continuity. To address this limitation, multidimensional neural networks have been developed to incorporate spatial contextual information and improve the lateral consistency of inversion results. When only sparse well log labels are available, mask-weighted losses are often introduced during training to impose supervision at well locations \cite{wu2021deep,chen2024multidimensional,pang2025iterative,li2025intelligent}. In addition, 2D or 3D synthetic training datasets can be used to train multidimensional neural networks, allowing the models to learn spatially coherent relationships between seismic data and subsurface properties \cite{tao2025porosity}.

In addition to training datasets and network architectures, the training strategy also significantly affects the performance of deep-learning-based inversion. Based on the use of labeled samples, existing methods can be broadly divided into supervised, semi-supervised, and unsupervised learning approaches. When sufficient labeled training datasets are available, supervised learning can be used to optimize network parameters, and the mapping relationship between seismic data and model parameters is mainly learned from paired samples \cite{zheng2019applications,zhang2022deep}. However, the generalization ability of supervised models is often limited by the distribution and diversity of the training datasets. To reduce the dependence on dense labeled samples, semi-supervised and unsupervised learning strategies have been introduced into seismic inversion. These strategies either combine limited well log supervision with unlabeled seismic data, or use physical forward-operator-based data-consistency losses to train networks. \cite{feng2020unsupervised,chen2021seismic,chen2024multidimensional,zhang2025joint}. Nevertheless, the performance of these methods still depends on the quality of well logs, low-frequency models, or other prior information. Moreover, most existing methods still have limited flexibility in incorporating multi-source conditioning information.

To address these limitations, diffusion models have recently attracted increasing attention as generative models because of their strong capability to characterize complex data distributions. They have been widely explored for inverse problems \cite{daras2024survey} and have also been introduced into geophysical applications \cite{wang2024controllable,mengchuangji, chen2025unsupervised, chen2025unsupervisedEAGE, zhang2025diffusionvel,taufik2025wavenumber,
Zhoulin2026,pangqiEAGE,pang2026geovoldiff}.   
In seismic inversion, diffusion models can serve as learned generative priors to constrain the solution space of ill-posed seismic inversion. For example, Chen et al. \cite{chen2025unsupervised} developed an unsupervised seismic impedance inversion method based on a generative diffusion model, thereby reducing the dependence on low-frequency models and eliminating the need for labeled data. Considering the high computational cost of pixel-space diffusion models, they further developed a seismic inversion method based on a conditional latent diffusion model \cite{chen2025unsupervisedEAGE}. Additionally, diffusion models can provide a flexible framework for incorporating multi-source information. For example, Wang et al. \cite{wang2024controllable} incorporated multi-source prior information, including velocity classes, well log information, reflectivity images, and their combinations, into a conditional diffusion model to achieve controllable seismic velocity synthesis.

However, most existing diffusion-model-based studies in seismic inversion focus on single-parameter prediction, such as velocity or impedance. Although Taufik and Alkhalifah \cite{taufik2025wavenumber} introduced diffusion models into multi-parameter elastic full-waveform inversion, their method is formulated within an elastic wave-equation-based inversion framework. While their study demonstrates the potential of diffusion models for elastic parameter inversion, it remains focused on solving a specific full-waveform inversion problem. By contrast, elastic parameter synthesis can be formulated as a broader conditional generation task. Within this formulation, seismic inversion can be viewed as a special case in which seismic observations are incorporated as one type of conditioning information. Motivated by this perspective, we propose a multi-condition guided diffusion model for controllable elastic parameter synthesis, aiming to generate condition-consistent elastic parameter models under different geological and geophysical constraints. The main contributions of this work are summarized as follows:

(1) We propose a unified multi-condition guided diffusion framework based on explicit and implicit conditioning strategies. The proposed framework enables flexible incorporation of different types of conditioning information, including seismic data, low-frequency models, well logs, interpolated well-log models, and structural information such as horizons and faults.

(2) For implicit conditioning, we introduce iterative latent variable refinement (ILVR) and an Adapter module to incorporate conditions without explicit conditioning operators. ILVR imposes model-domain reference constraints during sampling, whereas the Adapter module injects non-model-domain structural information into the diffusion model.

(3) For explicit conditioning, we develop a DPS-projection hybrid guidance strategy by introducing an additional projection step into diffusion posterior sampling (DPS) \cite{chung2023diffusion}. This strategy enhances the enforcement of conditions with explicit conditioning operators and improves the consistency between the generated elastic parameters and the prescribed conditioning information.

(4) We develop a multi-elastic parameter dataset construction strategy by combining a Bayesian Gaussian mixture model (BGMM), vertical shearing fields, and displacement vector fields. The constructed datasets incorporate well-log statistics and representative structural characteristics, providing sufficient training data for learning the diffusion prior.

(5) We conduct extensive experiments to validate the effectiveness of the proposed method under both single-condition and multi-condition guidance. Seismic inversion experiments are further performed to demonstrate the advantages of diffusion models as learned generative priors and to evaluate the usefulness of synthesized samples for downstream deep-learning-based inversion.

\section{Background}
\subsection{Diffusion Models}
\label{sec:2A}
Diffusion models consist of a forward noising process and a reverse denoising process. Following Song et al. (2021) \cite{song2020score}, the continuous-time formulation of the Denoising Diffusion Probabilistic Model (DDPM) forward noising process $\mathbf{m}_t, t\in[0,T]$, can be described by the following variance-preserving stochastic differential equation (VP-SDE)
\begin{equation} 
\label{eq:1}
d\mathbf{m}_t = -\frac{\beta(t)}{2}\mathbf{m}_t dt + \sqrt{\beta(t)}d\mathbf{w}_t,
\end{equation}
where $\beta(t)$ is the noise schedule, and $\mathbf{w}_t$ is the standard Brownian motion. The forward process gradually perturbs the data distribution $\mathbf{m}_0 \backsim p_{data}$ to a standard Gaussian distribution $\mathbf{m}_T \backsim \mathcal{N}(\mathbf{0}, \mathbf{I})$. The corresponding reverse process is given by the following reverse SDE
\begin{equation} 
\label{eq:2}
d\mathbf{m}_t = [-\frac{\beta(t)}{2}\mathbf{m}_t-\beta(t) \nabla_{\mathbf{m}_t} \log p_t(\mathbf{m}_t)]dt + \sqrt{\beta(t)} d\overline{\mathbf{w}}_t,
\end{equation}
where $\overline{\mathbf{w}}_t$ denotes a standard Brownian motion in the reverse-time process from $T$ to $0$, $p_t(\mathbf{m}_t)$ is the probability density of $\mathbf{m}_t$, and $\nabla_{\mathbf{m}_t} \log p_t(\mathbf{m}_t)$ is the time-dependent score function, which is approximated by a UNet-based score network  $s_\theta(\mathbf{m}_t, t)$ trained with denosing score matching 
\begin{equation}
\label{eq:3}
\min_\theta\mathbb{E}_{t, \mathbf{m}_0, \mathbf{m}_t} \left[ || s_\theta(\mathbf{m}_t, t) -  \nabla_{\mathbf{m}_t} \log p_{0t}(\mathbf{m}_t | \mathbf{m}_0) ||_2^2 \right].
\end{equation}
Although the training objective involves the conditional score of the transition kernel $\nabla_{\mathbf{m}_t} \log p_{0t}(\mathbf{m}_t | \mathbf{m}_0)$, denoising score matching enables the network to approximate the score function $\nabla_{\mathbf{m}_t} \log p_t(\mathbf{m}_t)$, which is required in the reverse SDE. 
\subsection{Inversion Framework Based on Diffusion Models}
Inversion methods aim to recover unknown model parameters $\mathbf{m}$ from measurements $\mathbf{y}$. Generally, the forward problem is defined as 
\begin{equation}
\label{eq:4}
\mathbf{y} = f(\mathbf{m}) + \mathbf{n},
\end{equation}
where $\mathbf{n}$ denotes the noise, and $f$ is the forward operator which can be either linear or nonlinear. Generally, the corresponding inverse problem is ill-posed, making the incorporation of prior information essential. Diffusion models can capture complex distributions from training datasets and thus provide a learned prior for model parameters. Given conditioning information, posterior samples can then be obtained by solving the following conditional reverse SDE
\begin{equation}
\label{eq:5}
d\mathbf{m}_t = [-\frac{\beta(t)}{2}\mathbf{m}_t-\beta(t) \nabla_{\mathbf{m}_t} \log p_t(\mathbf{m}_t|\mathbf{y})]dt + \sqrt{\beta(t)} d\overline{\mathbf{w}}_t, 
\end{equation} 
where $\nabla_{\mathbf{m}_t} \log p_t(\mathbf{m}_t|\mathbf{y})$ is the conditional score function. According to Bayes’ rule, the conditional score function can be decomposed as
\begin{equation}
\label{eq:6}
\nabla_{\mathbf{m}_t} \log p_t(\mathbf{m}_t|\mathbf{y})=\nabla_{\mathbf{m}_t} \log p_t(\mathbf{m}_t)+ \nabla_{\mathbf{m}_t} \log p_t(\mathbf{y}|\mathbf{m}_t). 
\end{equation}
The pre-trained score network $s_\theta(\mathbf{m}_t, t)$, introduced in Section~\ref{sec:2A}, is used to approximate the prior score function $\nabla_{\mathbf{m}_t} \log p_t(\mathbf{m}_t)$. However, the likelihood score $\nabla_{\mathbf{m}_t} \log p_t(\mathbf{y}|\mathbf{m}_t)$ is intractable to compute because the measurements $\mathbf{y}$ are related to the clean data $\mathbf{m}_0$, rather than directly to the noisy variable $\mathbf{m}_t$. Various approaches have been proposed to address this problem. Among them, DPS \cite{chung2023diffusion} is one of the most widely used methods, as it can handle both linear and nonlinear inverse problems. Under the assumption of Gaussian measurement noise, DPS provides a practical approximation to this likelihood score as follows 
\begin{equation}
\label{eq:7}
\begin{aligned}
\nabla_{\mathbf{m}_t} \log p_t(\mathbf{y}|\mathbf{m}_t) &\approx -\lambda \nabla_{\mathbf{m}_t} ||\mathbf{y}-f(\widehat{\mathbf{m}}_0(\mathbf{m}_t)) ||_2^2 \\
& = -\lambda \nabla_{\mathbf{m}_t} \varepsilon (\mathbf{y}, \widehat{\mathbf{m}}_0(\mathbf{m}_t)),
\end{aligned}
\end{equation} 
where $\lambda$ is a likelihood guidance weight related to the measurement noise level, $\widehat{\mathbf{m}}_0 = \mathbb{E}[\mathbf{m}_0|\mathbf{m}_t] = \frac{1}{\sqrt{\overline{\alpha}_{t}}}(\mathbf{m}_t + (1- \overline{\alpha}_{t})\nabla_{\mathbf{m}_t}\log p_t(\mathbf{m}_t) )$, and $\varepsilon$ is the data-misfit energy function. The notation $\overline{\alpha}_{t}$ denotes a time-dependent parameter associated with the variance schedule.

\section{Method}
\subsection{Unified Multi-Condition Guided Diffusion Framework for Elastic Parameter Synthesis}
A natural strategy for condition-guided synthesis is to perform posterior sampling under different conditions by incorporating them into the conditional score function in \eqref{eq:5}. In this way, elastic parameter samples can be synthesized from the learned diffusion prior while remaining consistent with the prescribed conditioning information. However, the likelihood guidance in \eqref{eq:7} requires an explicit conditioning operator $f$, which may not be available for some types of conditions. To address this limitation, we divide the conditions $\mathbf{y}$ into two groups: $\mathbf{y}_{f}$, for which an explicit conditioning operator is available, and $\mathbf{y}_{uf}$, for which such an operator is unavailable. The former can be incorporated through likelihood guidance, while the latter is used to construct a conditional prior score, denoted as $\nabla_{\mathbf{m}_t}\log p_t(\mathbf{m}_t|\mathbf{y}_{uf})$, which guides the synthesis toward samples consistent with operator-free conditions. Accordingly, the conditional score function in \eqref{eq:5} is approximately formulated as
\begin{equation}
\label{eq:8}
\begin{aligned}
 \nabla_{\mathbf{m}_t} \log p_t(\mathbf{m}_t|\mathbf{y}_{uf}, \mathbf{y}_f)\approx \nabla_{\mathbf{m}_t} \log p_t(\mathbf{m}_t|\mathbf{y}_{uf}) \\ +  \nabla_{\mathbf{m}_t} \log p_t(\mathbf{y}_f|\mathbf{m}_t),
\end{aligned}
\end{equation}  
where we assume that $\mathbf{y}_{uf}$ is conditionally independent of $\mathbf{y}_f$ given $\mathbf{m}_t$. Note that $\mathbf{m}$ denotes multiple elastic parameters, such as P-wave velocity, S-wave velocity, and density, rather than a single elastic parameter. The computation of the conditional prior score $\nabla_{\mathbf{m}_t} \log p_t(\mathbf{m}_t|\mathbf{y}_{uf})$ and the likelihood score $\nabla_{\mathbf{m}_t} \log p_t(\mathbf{y}_f|\mathbf{m}_t)$ is detailed in the following sections. The relevant pseudocode is presented in Algorithm \ref{alg:cond_sampling}. 

\subsubsection{Computation of $\nabla_{\mathbf{m}_t} \log p_t(\mathbf{m}_t|\mathbf{y}_{uf})$} In general, the conditions $\mathbf{y}_{uf}$ consist of heterogeneous auxiliary information. Some of them are available in the model domain, including previously obtained inversion results and interpolated well-log models. Others, such as horizons, faults, and facies, are not directly defined in the model domain, but still provide valuable geological or structural constraints.  For clarity, we denote the model-domain conditions as $\mathbf{y}_{uf}^{m}$ and the non-model-domain conditions as $\mathbf{y}_{uf}^{nm}$. 
Accordingly, we approximate $\nabla_{\mathbf{m}_t}\log p_t(\mathbf{m}_t|\mathbf{y}_{uf})$ using two complementary strategies tailored to $\mathbf{y}_{uf}^{m}$ and $\mathbf{y}_{uf}^{nm}$, respectively.

\textit{ILVR for Model-Domain Reference Constraints:} When the conditions are defined in the model domain, we introduce ILVR \cite{choi2021ilvr} to 
guide the reverse sampling process without additional training. These model-domain conditions are represented in the same model space as the target model and can therefore be directly used as reference data. Instead of explicitly estimating the score $\nabla_{\mathbf{m}_t} \log p_t(\mathbf{m}_t|\mathbf{y}_{uf})$, this strategy modifies the reverse trajectory by enforcing low-frequency consistency between the generated samples and the corrupted reference condition. In this way, the reference condition is implicitly incorporated into the sampling process while preserving the high-frequency components generated by the pretrained unconditional diffusion model. The implementation is expressed as
\begin{align}
\label{eq:9}
\mathbf{m}'_{t-1} &\sim p_\theta (\mathbf{m}'_{t-1}|\mathbf{m}_{t}), 
\\
[\mathbf{y}_{uf}^{m}]_{t-1}&\sim q([\mathbf{y}_{uf}^{m}]_{t-1}|\mathbf{y}_{uf}^{m}),
\\
\mathbf{m}_{t-1} &=  \mathbf{m}'_{t-1}-\phi_N(\mathbf{m}'_{t-1}) + \phi_N([\mathbf{y}_{uf}^{m}]_{t-1}), 
\end{align} 
where $p_\theta (\mathbf{m}'_{t-1}|\mathbf{m}_{t})$ denotes the one-step transition kernel of the unconditional reverse process, which is obtained from a numerical discretization of the reverse-time SDE in \eqref{eq:2}, and $q([\mathbf{y}_{uf}^{m}]_{t-1}|\mathbf{y}_{uf}^{m})= \mathcal{N}\left([\mathbf{y}_{uf}^{m}]_{t-1}; \sqrt{\bar{\alpha}_t}\mathbf{y}_{uf}^{m}, (1-\bar{\alpha}_t)\mathbf{I}\right)$. Notation $\phi_N$ denotes the linear low-pass filtering operation implemented by a sequence of downsampling and upsampling operations with a factor of $N$. A larger $N$ enforces consistency only at coarser scales, allowing more diverse samples, whereas a smaller $N$ imposes stronger similarity to the reference condition.

\textit{Adapter-Based Conditioning for Non-Model-Domain Structural Constraints:} When the conditions are not defined in the model domain, it is difficult to incorporate them using the ILVR strategy. This is because such conditions do not have the same physical meaning as the target model and therefore can not be directly used to replace the low-frequency components of generated samples. To incorporate these non-model-domain conditions, a mapping from the external conditions to the feature space of the score network is required. A straightforward solution is to train a conditional diffusion model, in which the score network estimates a condition-dependent score $s_\theta(\mathbf{m}_t,t, \mathbf{y}_{uf}^{nm})$ to approximate $\nabla_{\mathbf{m}_t} \log p_t(\mathbf{m}_t|\mathbf{y}_{uf})$. However, retraining the entire diffusion model is computationally expensive and lacks flexibility when different types of conditions are considered. 

Therefore, we introduce a lightweight Adapter module \cite{mou2024t2i} to extract condition-aware features from $\mathbf{y}_{uf}^{nm}$ and inject them into the pretrained score network. The employed Adapter consists of four feature extraction blocks and three downsample blocks, producing multi-scale conditional features $\mathbf{H}=\lbrace \mathbf{H}^1, \mathbf{H}^2, \mathbf{H}^3, \mathbf{H}^4 \rbrace $. These features are then integrated into the encoder of the UNet as:
\begin{align}
\label{eq:1011}
\mathbf{H} &= \mathcal{A}(\mathbf{y}_{uf}^{nm}),\\
\widehat{\mathbf{F}}^i_{enc} &= \mathbf{F}^i_{enc} + \mathbf{H}^i, i\in \lbrace 1,2,3,4 \rbrace. 
\end{align}
Here, $\mathcal{A}$ denotes the Adapter module, $\mathbf{F}=\lbrace  \mathbf{F}^1_{enc}, \mathbf{F}^2_{enc}, \mathbf{F}^3_{enc}, \mathbf{F}^4_{enc} \rbrace$ represents the intermediate encoder features of the UNet. The dimensions of $\mathbf{H}^i$ are matched to those of $\mathbf{F}^i$ at each scale. The Adapter can incorporate either a single condition or multiple conditions. During training, we freeze the parameters of the score network and only optimize the Adapter module $\mathcal{A}$ using the objective in \eqref{eq:3}, where the original unconditional score model is extended to produce a condition-dependent score estimate $s_\theta(\mathbf{m}_t,t, \mathcal{A}(\mathbf{y}_{uf}^{nm}))$. Note that the Adapter architecture follows Mou et al. \cite{mou2024t2i}, and the score network adopts a UNet-based denoising architecture following Rombach et al. \cite{rombach2022high}.

In the proposed framework, the ILVR and Adapter-based conditioning can be used simultaneously when both model-domain and non-model-domain conditions are available. These two strategies are complementary because they handle different types of conditions. Nevertheless, within each strategy, we currently consider only a single condition: one model-domain reference condition for ILVR and one non-model-domain condition for Adapter-based conditioning. Incorporating multiple conditions within either strategy requires a carefully designed fusion mechanism to balance different constraints, which is beyond the scope of this work and will be explored in future studies.

\subsubsection{Computation of $\nabla_{\mathbf{m}_t} \log p_t(\mathbf{y}_f|\mathbf{m}_t)$}
We can apply the DPS method to approximate the likelihood score $\nabla_{\mathbf{m}_t} \log p_t(\mathbf{y}_f|\mathbf{m}_t)$ according to \eqref{eq:7}. However, DPS introduces the condition through a gradient-based guidance term, which provides a soft constraint and may not guarantee strict consistency between the generated samples and the given condition \cite{chen2025deep}. This limitation becomes more pronounced for complex multi-parameter synthesis. Inspired by the work of \cite{chung2022improving}, we therefore introduce an additional projection onto the condition-consistent subspace to further enforce the condition constraint, expressed as
\begin{equation}
\label{eq:12}
\mathbf{m}_{t-1}
=
\arg\min_{\mathbf{m}}
\varepsilon(\mathbf{y}_f,\mathbf{m})
+
\mu \mathcal{R}(\mathbf{m}),
\quad
\mathbf{m}^{(0)}=\mathbf{m}_{t-1}^{\prime},
\end{equation}
where $\mathbf{m}_{t-1}^\prime$ denotes the intermediate output of the reverse process at step $(t-1)$ according to \eqref{eq:5}, and is used as the initial model for the projection optimization. The term $\mathcal{R}$ denotes a 2D total-variation (TV) regularization used to suppress noise-induced artifacts, and $\mu$ is the regularization parameter determined according to the noise level. When the condition is not dominated by measurement noise, the explicit regularization term can be removed. 

Different from the work in \cite{chung2022improving}, the conditioning operator in \eqref{eq:12} is not restricted to a linear operator, and the influence of noise is explicitly considered. The inverse problem in \eqref{eq:12} can be solved using Adaptive Moment Estimation (Adam), which is robust and well-suited for a wide range
of non-convex optimization problems in machine learning \cite{adam2014method}. 
It is worth noting that, when the conditions are noise-free, the projection strategy in \eqref{eq:12} with multiple optimization iterations may produce results comparable to those obtained by DPS-projection. However, when the conditions are contaminated by noise, combining DPS with projection tends to provide better robustness to noise. This is because the gradient-based guidance in DPS imposes the condition in a soft manner, which can reduce the influence of noisy conditions, while a small number of projection iterations is sufficient to enforce the condition constraint without overfitting the noise.

Based on the above derivations, the discrete reverse diffusion process at step $(t-1)$ can be written as
\begin{align}
\label{eq:1314}
\mathbf{m}_{t-1}^\prime & = h(\mathbf{m}_t, s_\theta) -\lambda \nabla_{\mathbf{m}_t}   \varepsilon (\mathbf{y}_f, \widehat{\mathbf{m}}_0(\mathbf{m}_t)) + g(t)\mathbf{z}, \\
\mathbf{m}_{t-1} & = \mathcal{P}(\mathbf{m}_{t-1}^\prime),
\end{align}
where $\mathcal{P}$ represents the projection operator obtained by solving the inverse problem in \eqref{eq:12}, $h(\mathbf{m}_t, s_\theta)=\frac{1}{\sqrt{\alpha_{t}}}(\mathbf{m}_t + (1- \alpha_{t})s_\theta(\mathbf{m}_t, t))$, $g(t) = \sqrt{\tilde{\beta}_t}$, and $\mathbf{z}\backsim \mathcal{N}(\mathbf{0}, \mathbf{I})$. The term $\tilde{\beta}_t$ denotes the variance. A detailed derivation can be found in \cite{nichol2021improved}. When multiple conditions $\lbrace\mathbf{y}_{f_1}, \cdots, \mathbf{y}_{f_n}\rbrace$ are considered, the energy functions $\varepsilon(\mathbf{y}_f, \widehat{\mathbf{m}}_0(\mathbf{m}_t))$ in \eqref{eq:1314} and $\varepsilon(\mathbf{y}_f, \mathbf{m})$ in \eqref{eq:12} are extended as weighted combinations of the individual condition-consistency terms
\begin{align}
\label{eq:15}
\varepsilon (\lbrace\mathbf{y}_{f_1},  \cdots, \mathbf{y}_{f_n}\rbrace, \widehat{\mathbf{m}}_0) & \approx \zeta_1 \varepsilon (\mathbf{y}_{f_1}, \widehat{\mathbf{m}}_0) +\cdots +\zeta_n \varepsilon (\mathbf{y}_{f_n}, \widehat{\mathbf{m}}_0),\\
\varepsilon (\lbrace\mathbf{y}_{f_1},  \cdots, \mathbf{y}_{f_n}\rbrace, \mathbf{m}) & \approx \zeta_1 \varepsilon (\mathbf{y}_{f_1}, \mathbf{m}) +\cdots +\zeta_n \varepsilon (\mathbf{y}_{f_n}, \mathbf{m}),  
\end{align}
where $\widehat{\mathbf{m}}_0(\mathbf{m}_t)$ is denoted by $\widehat{\mathbf{m}}_0$ for simplicity, $\zeta_i$ denotes the weight assigned to the $i$-th condition and $\sum_{i=1}^n \zeta_i = 1$. These weights control the relative contribution of different conditions during conditional sampling. Note that different misfit terms associated with $\mathbf{y}_{f_i}$ involve different conditioning operators.

\begin{algorithm}[htbp]
\caption{Unified Multi-Condition Guided Diffusion Framework}
\label{alg:cond_sampling}
\textbf{Input:} 
Trained diffusion model $s_{\theta}$, 
conditions without explicit conditioning operator $\left\lbrace  \mathbf{y}_{uf}^{m}, \mathbf{y}_{uf}^{nm} \right\rbrace  $, 
conditions with explicit conditioning operator $\mathbf{y}_{f}$, 
conditioning operator $f(\cdot)$, 
adapter module $\mathcal{A}(\cdot)$, 
and the number of diffusion steps $T$.

\textbf{Output:} 
Generated elastic parameter model $\mathbf{m}_0$.
\begin{algorithmic}[1]
\State Initialize $\mathbf{m}_T \sim \mathcal{N}(\mathbf{0},\mathbf{I})$

\For{$t = T, T-1, \ldots, 1$}
    \If{$\mathbf{y}_{uf}^{nm}$ is not None}
        \State Compute the conditional prior score: 
        \State  $\displaystyle 
        \mathbf{s}_t =
        s_{\theta}\left(\mathbf{m}_t,t,\mathcal{A}(\mathbf{y}_{uf}^{nm})\right)$
    \Else
        \State Compute the unconditional prior score:
        \State $\mathbf{s}_t =
        s_{\theta}\left(\mathbf{m}_t,t\right)$
    \EndIf
    \State Perform one reverse diffusion step:
    \State $
    \mathbf{m}_{t-1}^{\prime}
    = h(\mathbf{m}_t,\mathbf{s}_t)
    + g(t)\mathbf{z},
    \quad \mathbf{z}\sim\mathcal{N}(\mathbf{0},\mathbf{I})$
    \If{$\mathbf{y}_{uf}^{m}$ is not None}
        \State Apply ILVR method:
        \State $
        \mathbf{m}_{t-1}^{\prime}
        = \mathbf{m}_{t-1}^{\prime}
        - \phi_N(\mathbf{m}_{t-1}^{\prime})
        + \phi_N\left([\mathbf{y}_{uf}^{m}]_{t-1}\right)$
    \EndIf
    \If{$\mathbf{y}_{f}$ is not None}
        \State Update the intermediate using DPS-projection:
        \State $ \mathbf{m}_{t-1}
        = \mathcal{P}(\mathbf{m}_{t-1}^{\prime} - \lambda         
        \nabla_{\mathbf{m}_t}
        \varepsilon
       (\mathbf{y}_{f},
        \widehat{\mathbf{m}}_0(\mathbf{m}_t)) )$
    \Else
        \State Set $\mathbf{m}_{t-1}=\mathbf{m}_{t-1}^{\prime}$
    \EndIf
\EndFor
\State \Return $\mathbf{m}_0$
\end{algorithmic}
\end{algorithm}

\subsection{Generation of Training Datasets}
\label{sec:3B}
Diffusion models learn prior distributions of elastic parameter models from training datasets, which provides the basis for elastic parameter synthesis. To obtain geologically plausible synthesis results, the training datasets should reflect key statistical relationships and structural characteristics relevant to the target data. In this study, we construct training datasets for elastic multi-parameter synthesis by integrating BGMM-based parameter sampling, vertical shearing fields, and displacement vector fields. The main steps are as follows:

\textbf{Step 1: Generating parameter sequences.}
Given the available well logs, we first employ a BGMM to learn the joint statistical distribution of elastic parameters in a parametric manner. In this study, P-wave velocity ($v_p$), S-wave velocity ($v_s$), and density ($\rho$) are used as the basic elastic parameters, from which other elastic attributes can be further derived if needed. Accordingly, the elastic-parameter vector at each depth or time sample is defined as $\mathbf{x}=(v_p, v_s, \rho)^T$, which is modeled as a three-variate random vector. For this multivariate variable $\mathbf{x}$, the Gaussian mixture distribution can be written as
\begin{equation}
\label{eq:16}
p(\mathbf{x}) = \sum_{k=1}^K \pi_k \mathcal{N}(\mathbf{x}| \boldsymbol{\mu}_k, \boldsymbol{\Sigma}_k),
\end{equation}
where \(K\) denotes the number of Gaussian components, \(\pi_k\) is the mixing weight of the \(k\)-th component with \(\sum_{k=1}^K \pi_k=1\), and each component is characterized by a Gaussian distribution with mean vector \(\boldsymbol{\mu}_k\) and covariance matrix \(\boldsymbol{\Sigma}_k\).  Variational inference is then used to approximate posterior distributions over the parameters of a Gaussian mixture distribution. In this study, the BGMM is implemented using the \texttt{BayesianGaussianMixture} module in scikit-learn to approximate the probability density function of the high-frequency components extracted from well logs. High-frequency elastic-parameter samples are then drawn from the learned distribution and used to construct synthetic parameter sequences.

After generating the high-frequency parameter sequences, a random low-frequency trend is added to each sampled parameter sequence. Taking the P-wave velocity as an example, its low-frequency component \(\mathbf{v}_p^{LF}\) is modeled as a random linear function of depth or time \(\mathbf{z}\), expressed as
\begin{equation}
\label{eq:17}
\mathbf{v}_p^{LF} = \mathbf{v}_p^{0}+ s \mathbf{z},
\end{equation}
where \(\mathbf{v}_p^{0}\) and \(s\) denote the intercept and slope, respectively. They are independently sampled from uniform distributions within physically reasonable bounds of the P-wave velocity. For the remaining parameters, their low-frequency components are derived from \(\mathbf{v}_p^{LF}\) using empirical relationships.

\textbf{Step 2: Generating layered models.}
Considering the computational cost, we construct 2D horizontally layered models by laterally repeating the generated 1D parameter sequences. To simulate more realistic geological structures, folding and faulting are sequentially incorporated into the layered models following the structural modeling procedure proposed by Wu et al. \cite{wu2020building}. Specifically, folding is first simulated by vertically shearing the horizontally layered models. The corresponding shift field is defined as \(S(X,Z)\) in the global coordinate system \((X,Z)\):
\begin{equation}
\label{eq:18}
S(X,Z) = S_1(X,Z) + S_2(X,Z),
\end{equation}
where $S_1(X,Z)=aX+c$ is a linear function used to introduce dipping structures, and $S_2(X,Z) = \frac{1.5}{Z_{\max}} Z \sum_{k=1}^{N} b_k e^{\frac{-(X - c_k)^2}{2\sigma_k^2}}$ is a combination of N Gaussian functions used to generate bending structures with different shapes. After the folding deformation, 2D displacement vector fields are further applied to simulate faulting. The displacement field $d(x, z = 0)$ in the local coordinate system is defined as  
\begin{equation}
\label{eq:19}
\begin{aligned}
d(x, z = 0) &= 2d_{\max} \bigl[1 - r(x)\bigr]\quad \times \sqrt{\frac{\bigl[1 + r(x)\bigr]^2}{4} - r^2(x)},
\end{aligned}
\end{equation}
where $r(x) = \frac{|x-x_0|}{l_x}$. This displacement is then extrapolated to the hanging-wall and foot-wall blocks to obtain the complete fault displacement field. During the simulation, the parameters in the folding and faulting equations are randomly sampled from predefined ranges to generate diverse folding and faulting structures. To further increase the lateral variability of elastic parameters, additional linear and Gaussian perturbations are introduced along the horizontal direction. Finally, horizon positions and fault masks are extracted from the underlying coordinate mapping and used as conditional guidance for the diffusion model.

\section{Examples}
\subsection{Synthetic data examples}
\label{sec:IVA}
To validate the effectiveness of the proposed method, we conduct experiments on the Marmousi II model, including evaluation of the synthetic dataset, single-condition guided elastic parameter synthesis, and multi-condition guided elastic parameter synthesis. For diffusion-based synthesis, the number of diffusion steps $T$ is set to 200 to provide sufficient sampling steps for applying different types of conditioning guidance. Although fewer steps may be sufficient for relatively simple conditions, such as low-frequency guidance, we use the same setting of $T=200$ in most experiments for consistency. In the multi-condition guided Marmousi II seismic inversion experiment, the DPS-projection guidance is applied only during the last 33\% of the reverse sampling process to reduce computational cost, as it yields results comparable to those obtained with guidance applied throughout the entire reverse process. The incorporated conditions include seismic data, low-frequency models, well logs, interpolated well-log models, and structural information such as horizons and faults. In this study, seismic data, low-frequency models, and well logs are included in $\mathbf{y}_f$ because they can be related to the elastic parameter model through explicit operators, including seismic forward modeling, smoothing or low-pass filtering, and spatial sampling, respectively. Interpolated well-log models, horizons, and faults are included in $\mathbf{y}_{uf}$ because no directly explicit operator is available for these conditions. Among them, interpolated well-log models serve as model-domain reference data, whereas horizons and faults provide non-model-domain structural information. 

\subsubsection{Evaluation of the Synthetic Dataset}
\begin{figure}[htb!]
\setlength{\abovecaptionskip}{0.2cm}
 \centering
   \subfigure[]{\includegraphics[width=0.45\columnwidth]{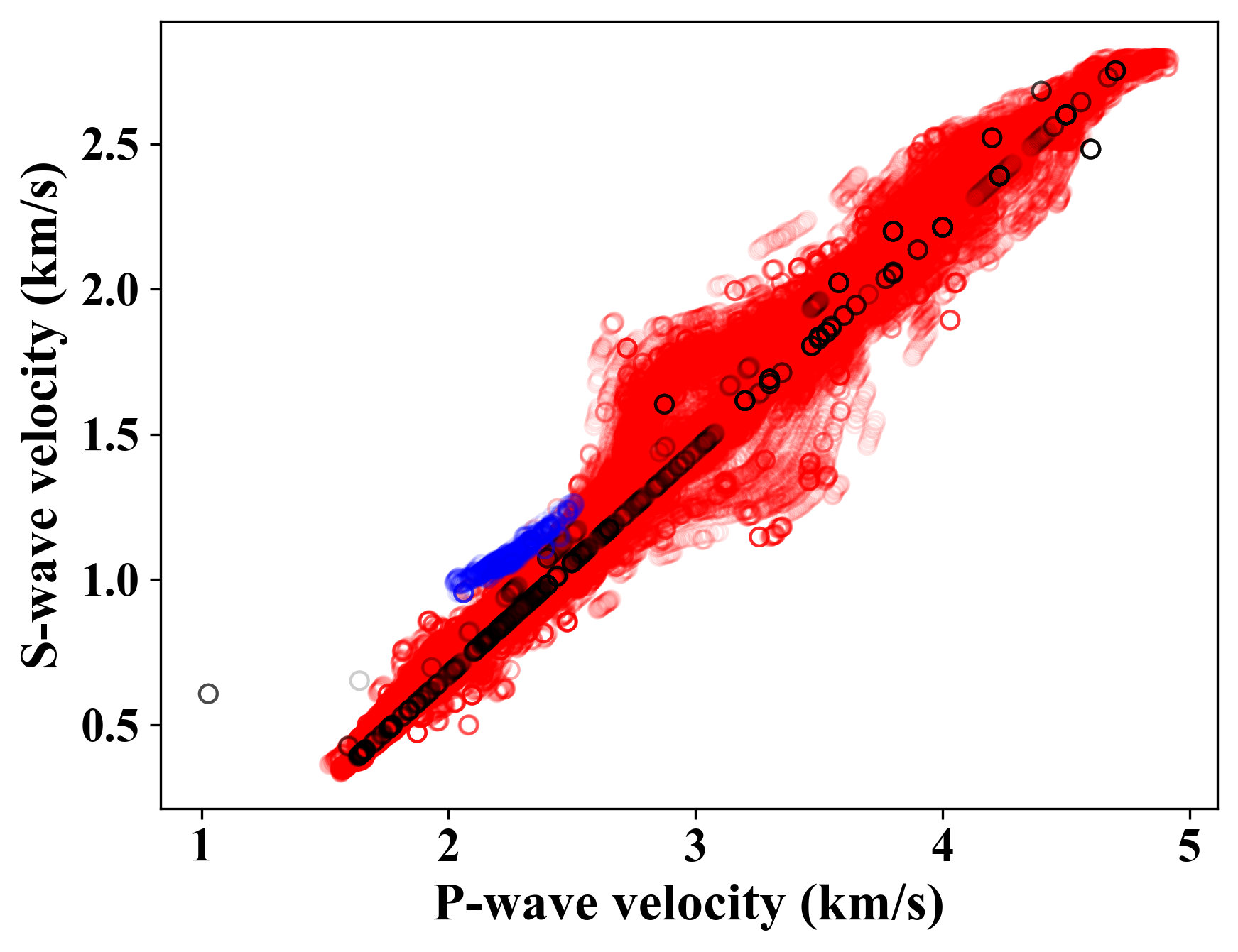}
 \label{fig:scatter_plotvp_vs}}
 \subfigure[]{\includegraphics[width=0.45\columnwidth]{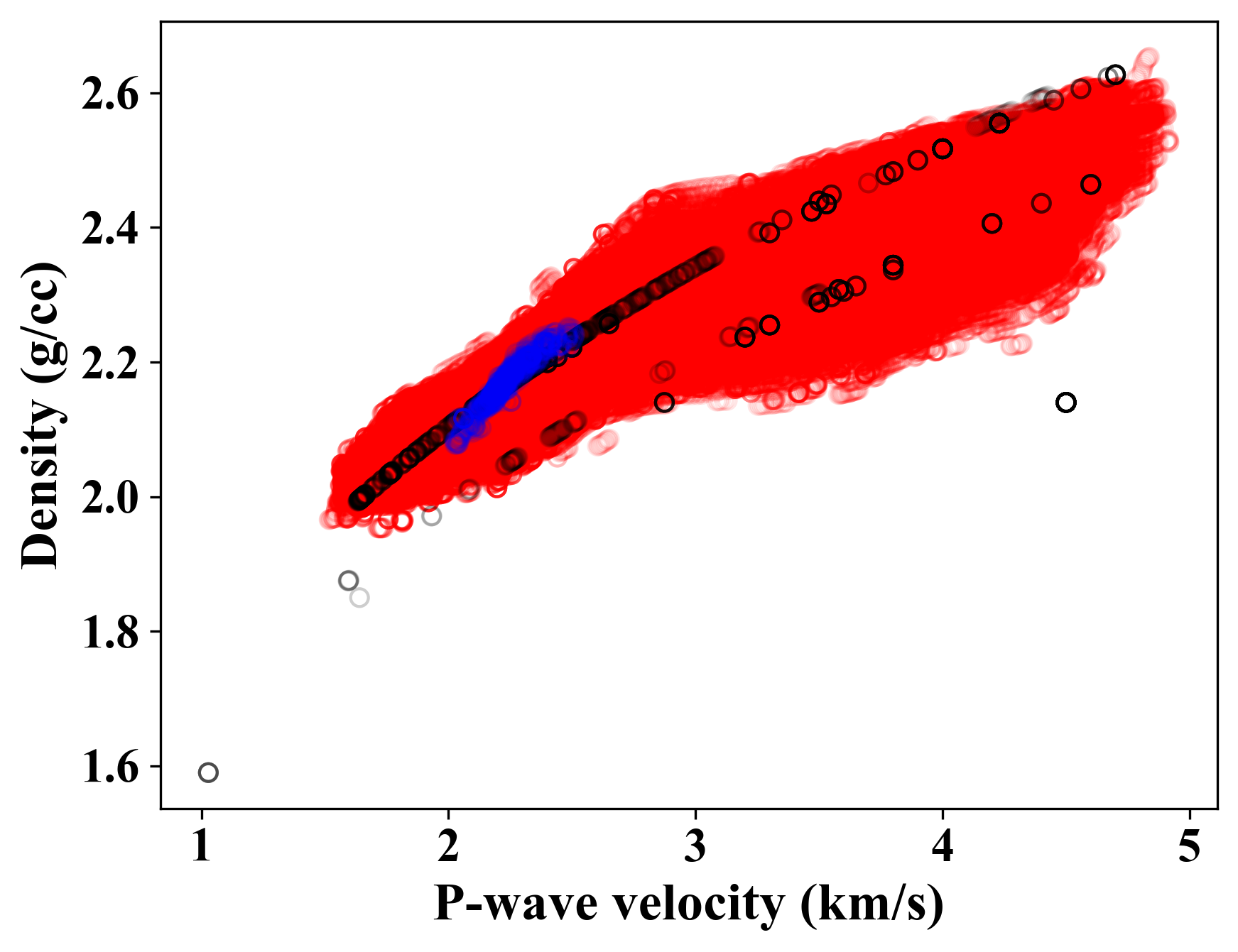}
 \label{fig:scatter_plotvp_rho}}  
 \caption{Scatter plots comparing the generated datasets with well log data: 
(a) P-wave velocity versus S-wave velocity and 
(b) P-wave velocity versus density. 
The red, black, and blue points denote the generated datasets, pseudo-well log data from the Marmousi II model, and measured well log data from the F3 project, respectively.}
\label{fig:scatter}
\end{figure}

Based on the log data extracted from three randomly placed pseudo-wells, we constructed training datasets to train the diffusion model. To assess the reliability of the generated datasets, Fig. \ref{fig:scatter} presents scatter plots of P-wave velocity versus S-wave velocity and P-wave velocity versus density. As shown in Fig. \ref{fig:scatter_plotvp_vs}, the generated datasets are broadly consistent with the pseudo-well logs and follow a similar diagonal trend, indicating that the linear relationship between P-wave and S-wave velocities is reasonably preserved. Fig. \ref{fig:scatter_plotvp_rho} further shows that the generated datasets generally reproduce the nonlinear increasing trend between P-wave velocity and density observed in the pseudo-well logs. These observations suggest that the generated datasets capture the main statistical relationships among different elastic parameters. 

\begin{figure*}[htb!]
\setlength{\abovecaptionskip}{0.2cm}
 \centering
   \subfigure[]{\includegraphics[width=0.65\columnwidth]{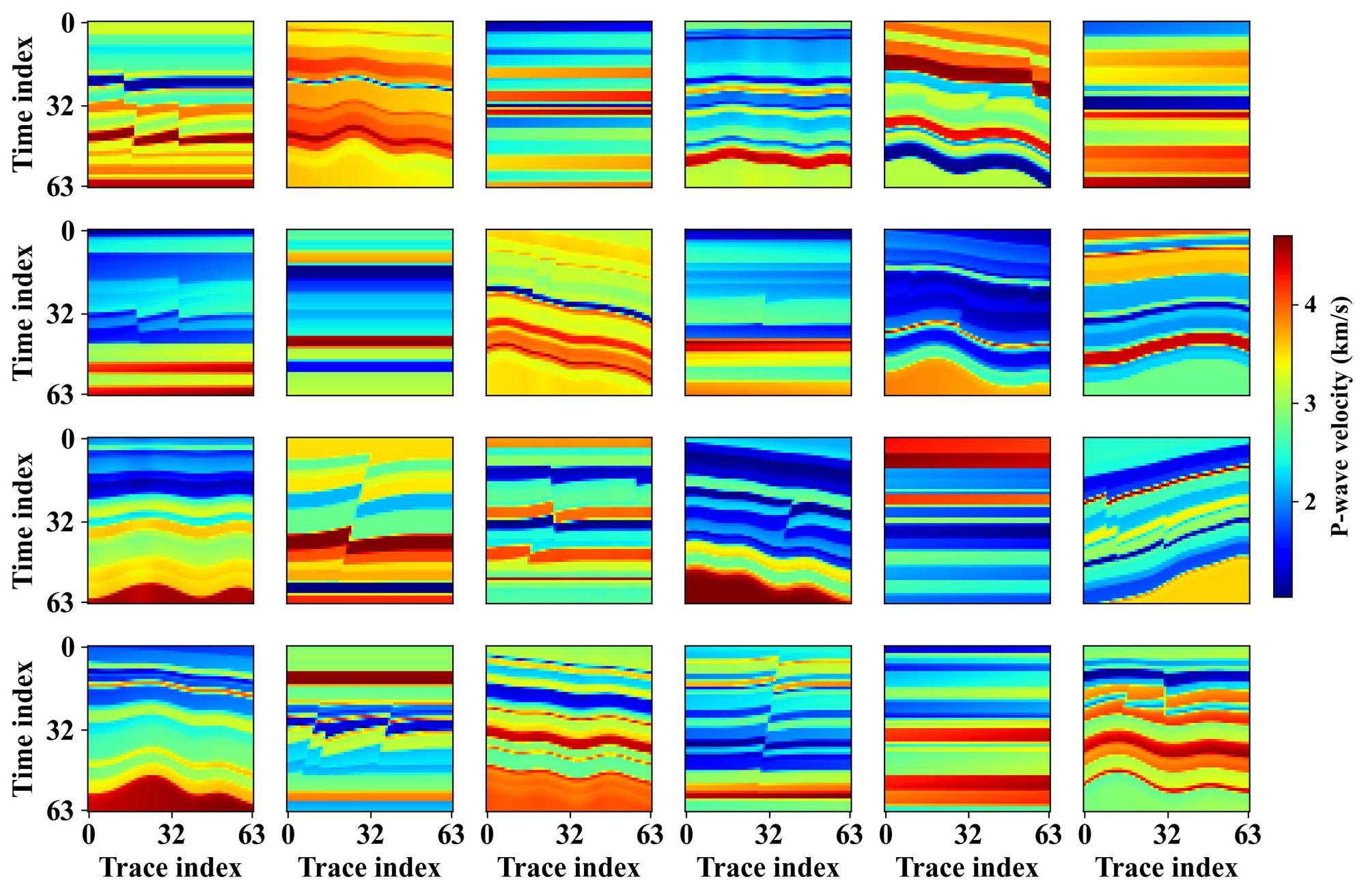}
 \label{fig:2Ddatasets-vp}}
 \subfigure[]{\includegraphics[width=0.65\columnwidth]{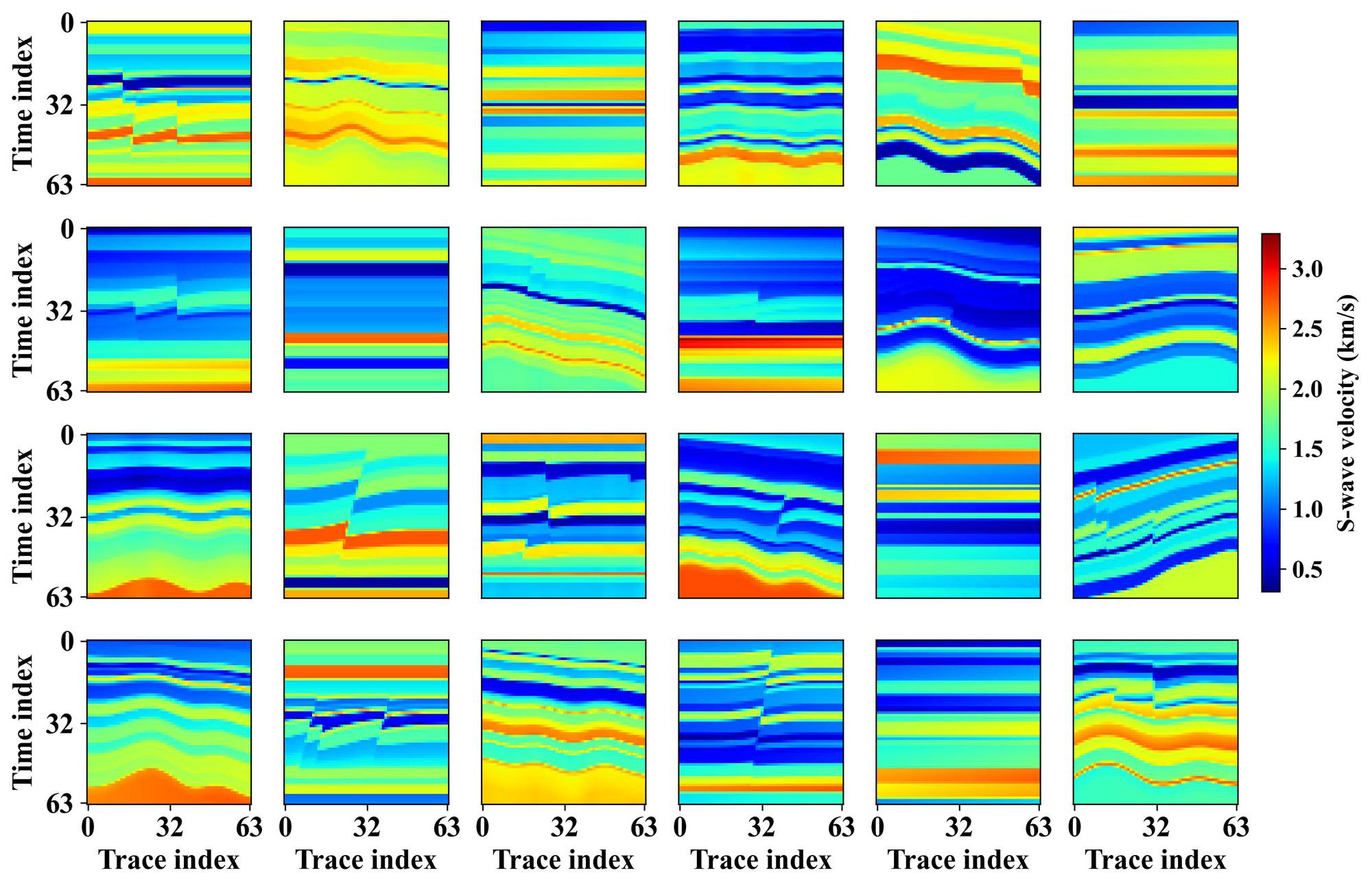}
 \label{fig:2Ddatasets-vs}} 
 \subfigure[]{\includegraphics[width=0.65\columnwidth]{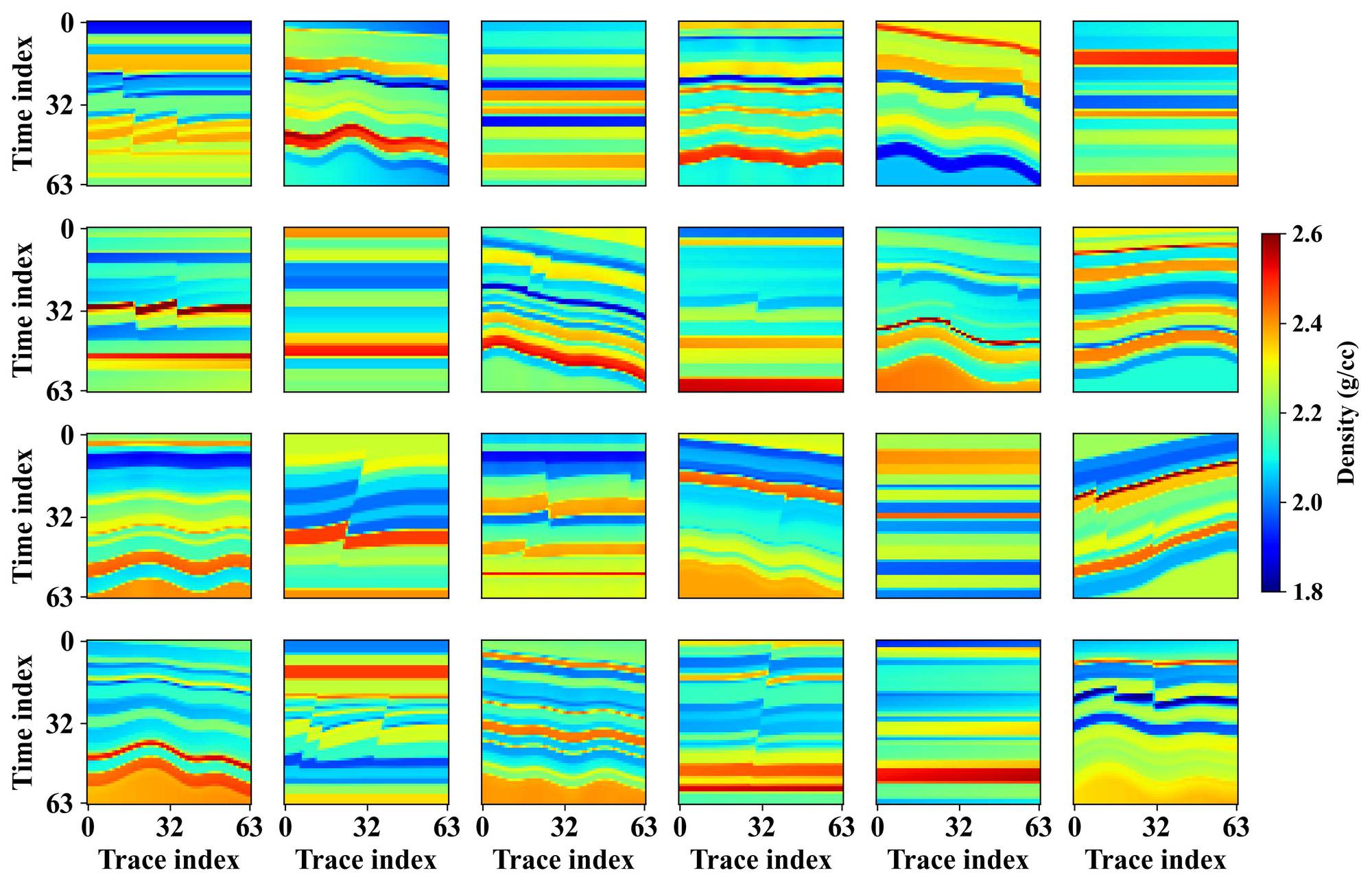}
 \label{fig:2Ddatasets-rho}} 
    \subfigure[]{\includegraphics[width=0.65\columnwidth]{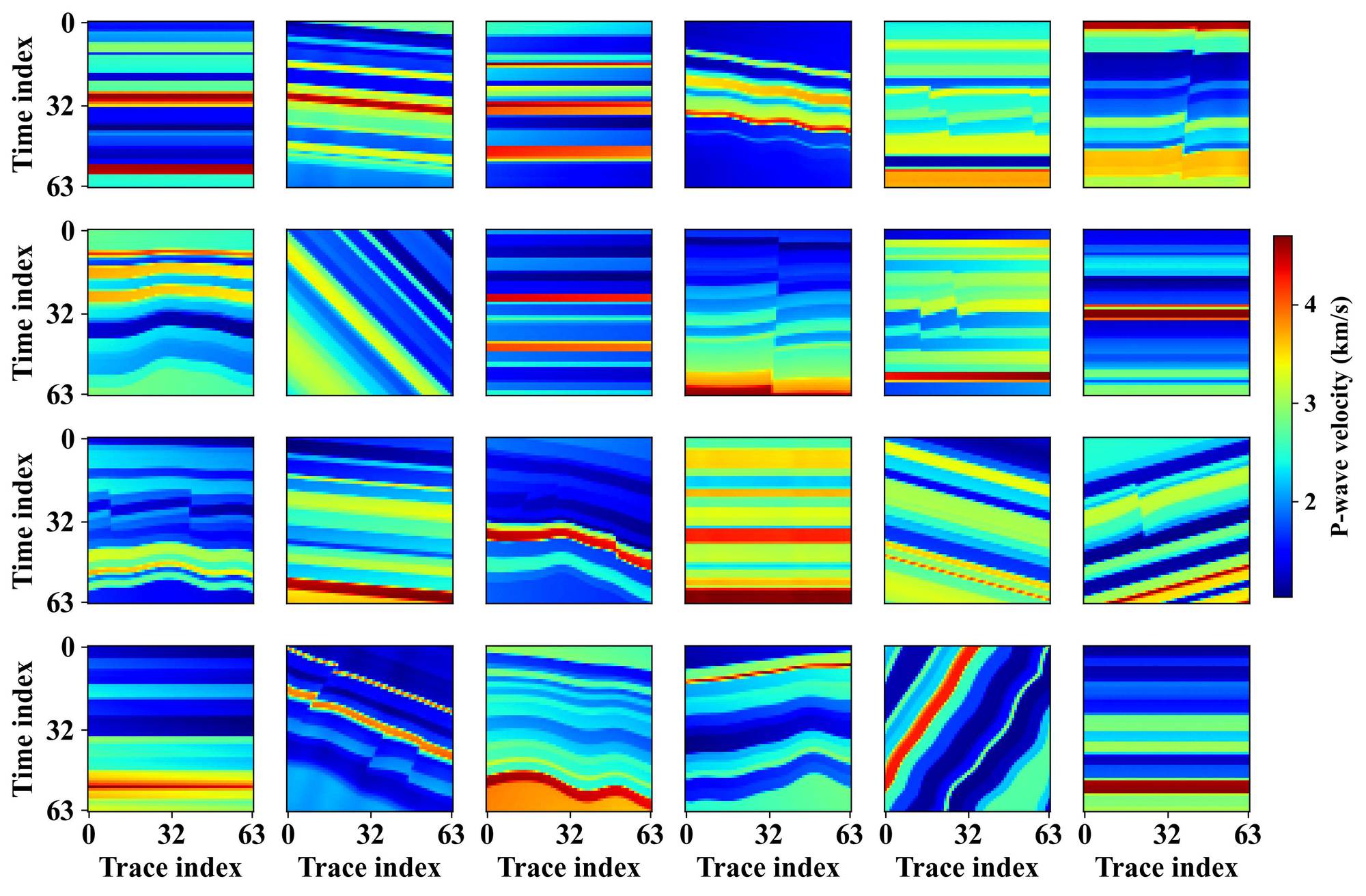}
 \label{fig:2Ddatasets-ddpm-vp}}
 \subfigure[]{\includegraphics[width=0.65\columnwidth]{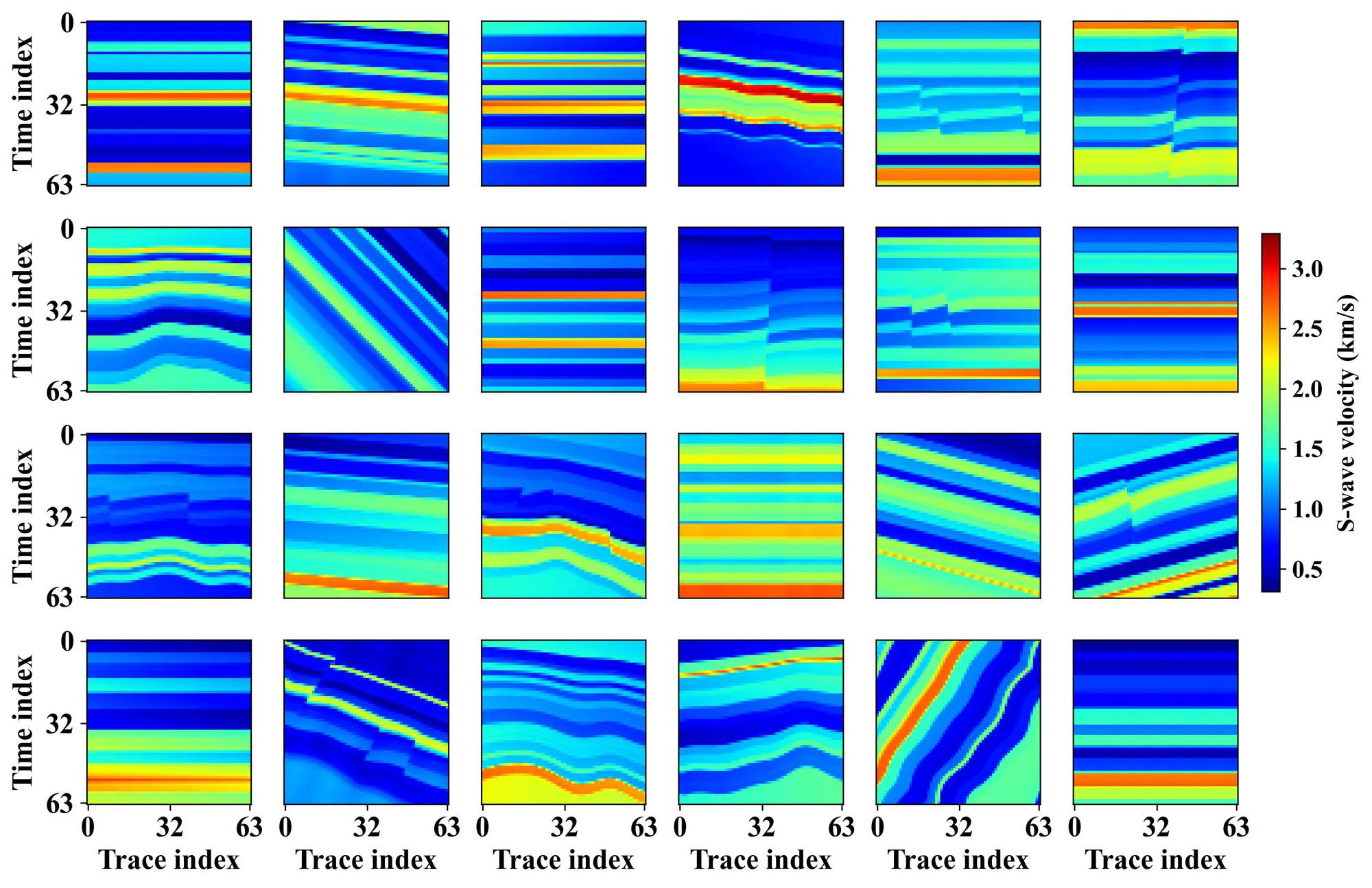}
 \label{fig:2Ddatasets-ddpm-vs}} 
 \subfigure[]{\includegraphics[width=0.65\columnwidth]{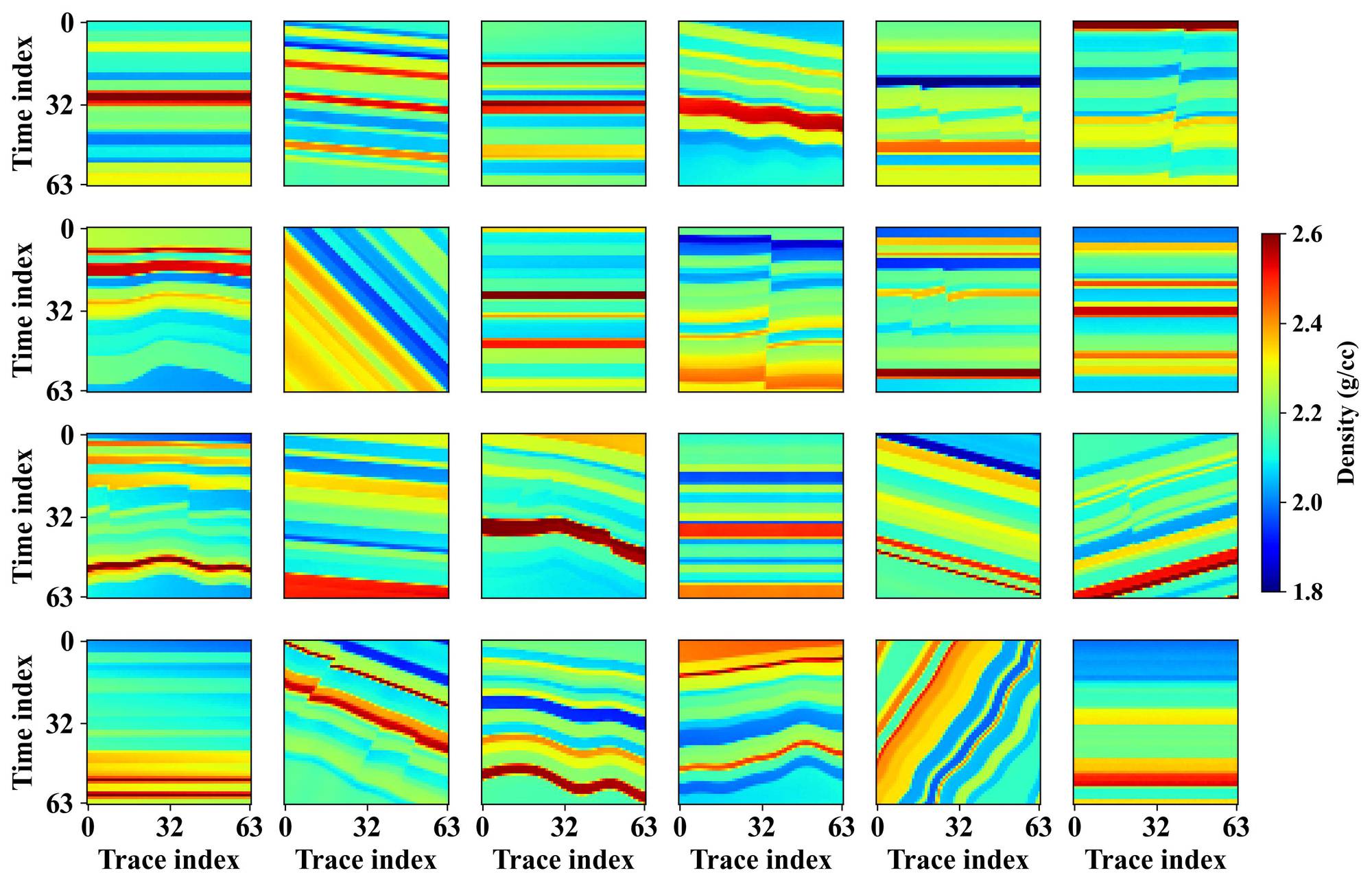}
 \label{fig:2Ddatasets-ddpm-rho}} 
 \caption{2D datasets of P-wave velocity, S-wave velocity, and density: (a)--(c) training samples constructed using the method described in Section \ref{sec:3B}, and (d)--(f) samples generated by the trained unconditional diffusion model.}
\label{fig:datasets}
\end{figure*}

Figs. \ref{fig:2Ddatasets-vp}--\ref{fig:2Ddatasets-rho} show the constructed 2D training datasets for P-wave velocity, S-wave velocity, and density, respectively. The datasets consist of four types of structural models: horizontal layered, faulted layered, folded layered, and folded-and-faulted layered models. Each structural category contains 2,500 samples, with each sample represented on a $64\times 64$ grid. Rotation-based data augmentation is further applied to increase structural variability. After augmentation, the training, validation, and testing datasets contain 12,000, 1,500, and 1,500 samples, respectively. The diffusion model is trained on these datasets using PyTorch Lightning, with a base learning rate of $2\times 10^{-6}$ and a maximum of 500 epochs. The training is performed on two NVIDIA GeForce RTX 3090 GPUs, and the learning rate is set according to the two-GPU training configuration. Figs. \ref{fig:2Ddatasets-ddpm-vp}--\ref{fig:2Ddatasets-ddpm-rho} show the elastic parameter samples synthesized by the trained unconditional diffusion model. The generated samples are generally consistent with the four structural types included in the training datasets. Moreover, owing to the rotation-based augmentation, the trained model is able to generate structurally diverse realizations with a wider range of dip angles, thereby further increasing the diversity of the generated elastic parameter models. Unless otherwise specified, the following diffusion-based synthesis experiments are performed on 12 samples randomly selected from the above testing dataset.

\subsubsection{Single-Condition Guided Elastic Parameter Synthesis} This section evaluates the effectiveness of the proposed method for elastic parameter synthesis under single-condition guidance. The considered conditions include seismic data, low-frequency models, well logs, interpolated well-log models, and structural information. The corresponding synthesis results are presented and analyzed in the following sections.

\begin{figure*}[htb!]
\setlength{\abovecaptionskip}{0.2cm}
 \centering
    \subfigure[]{\includegraphics[width=0.65\columnwidth]{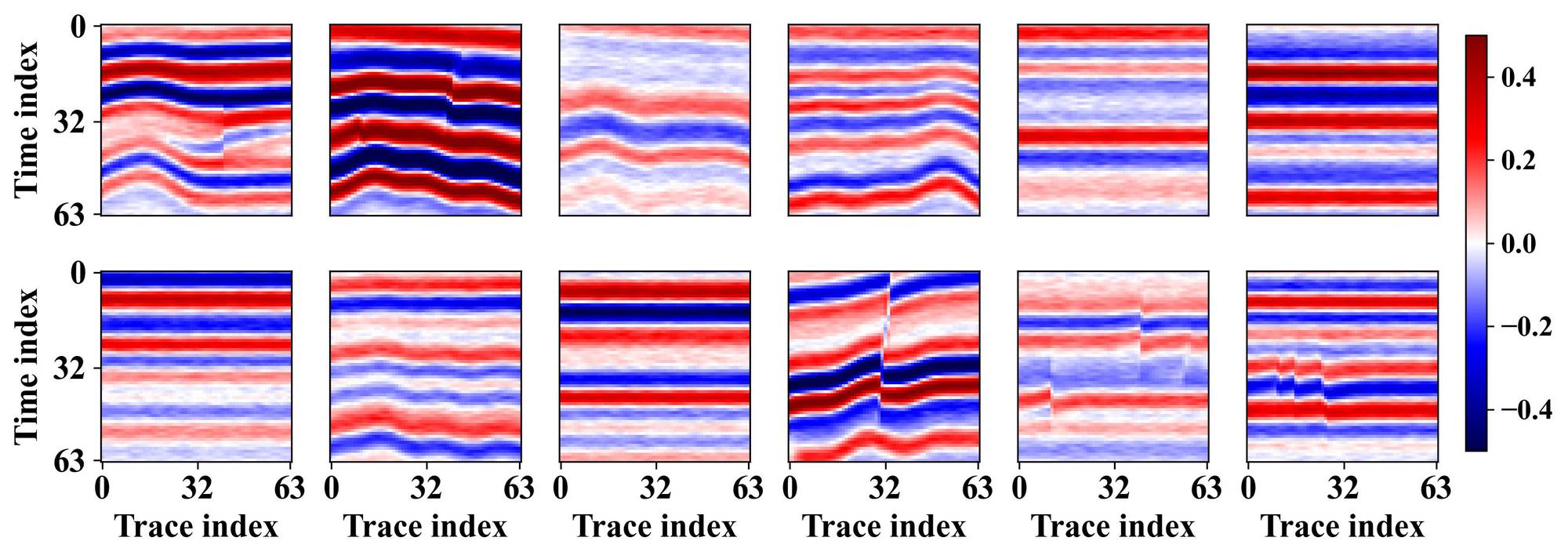}
 \label{fig:condseis-0}}
 \subfigure[]{\includegraphics[width=0.65\columnwidth]{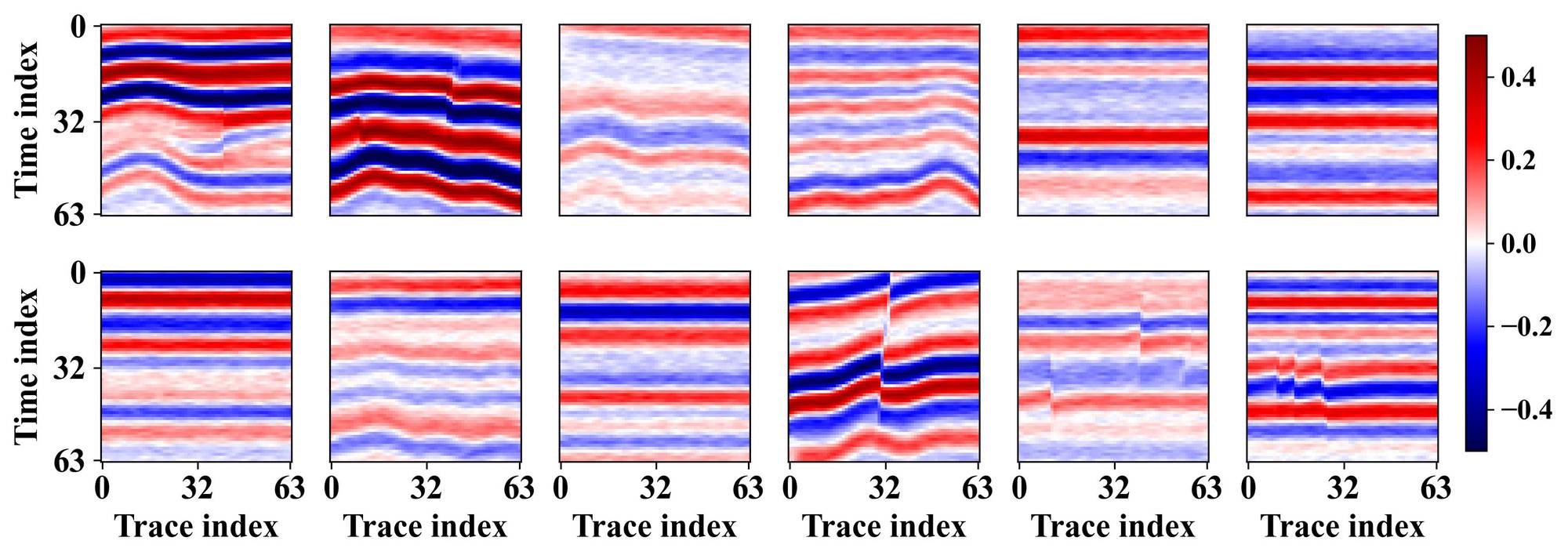}
 \label{fig:condseis-1}} 
  \subfigure[]{\includegraphics[width=0.65\columnwidth]{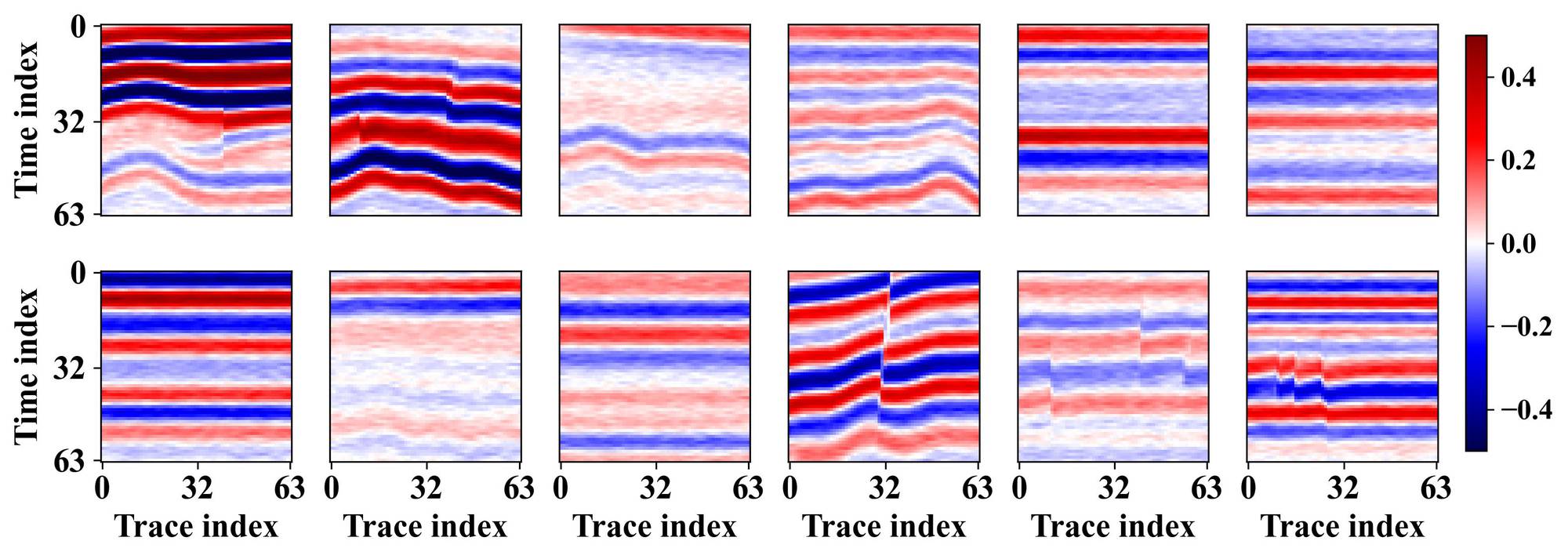}
 \label{fig:condseis-2}}
   \subfigure[]{\includegraphics[width=0.65\columnwidth]{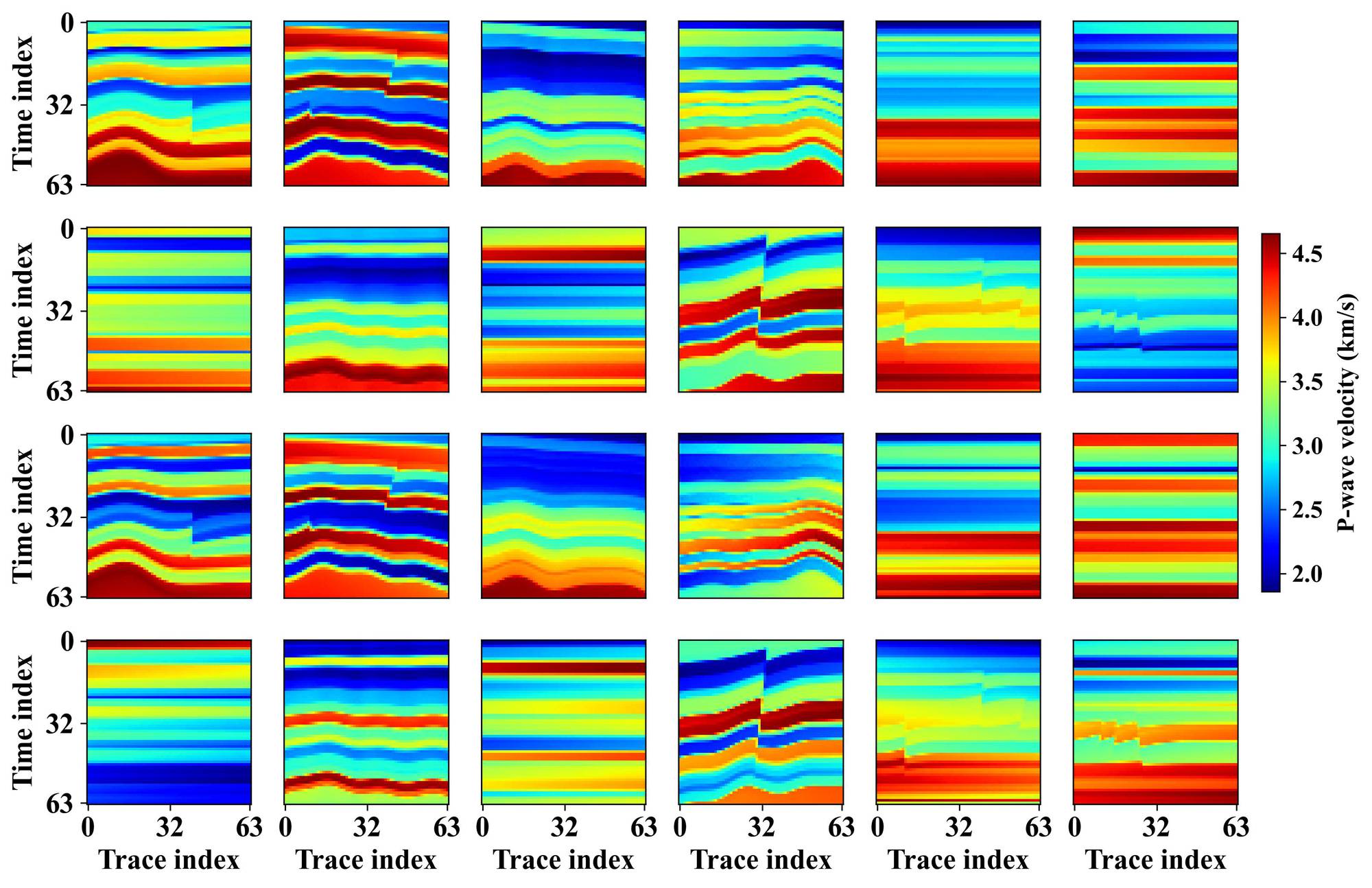}
 \label{fig:ddpm-condseis-vp}}
 \subfigure[]{\includegraphics[width=0.65\columnwidth]{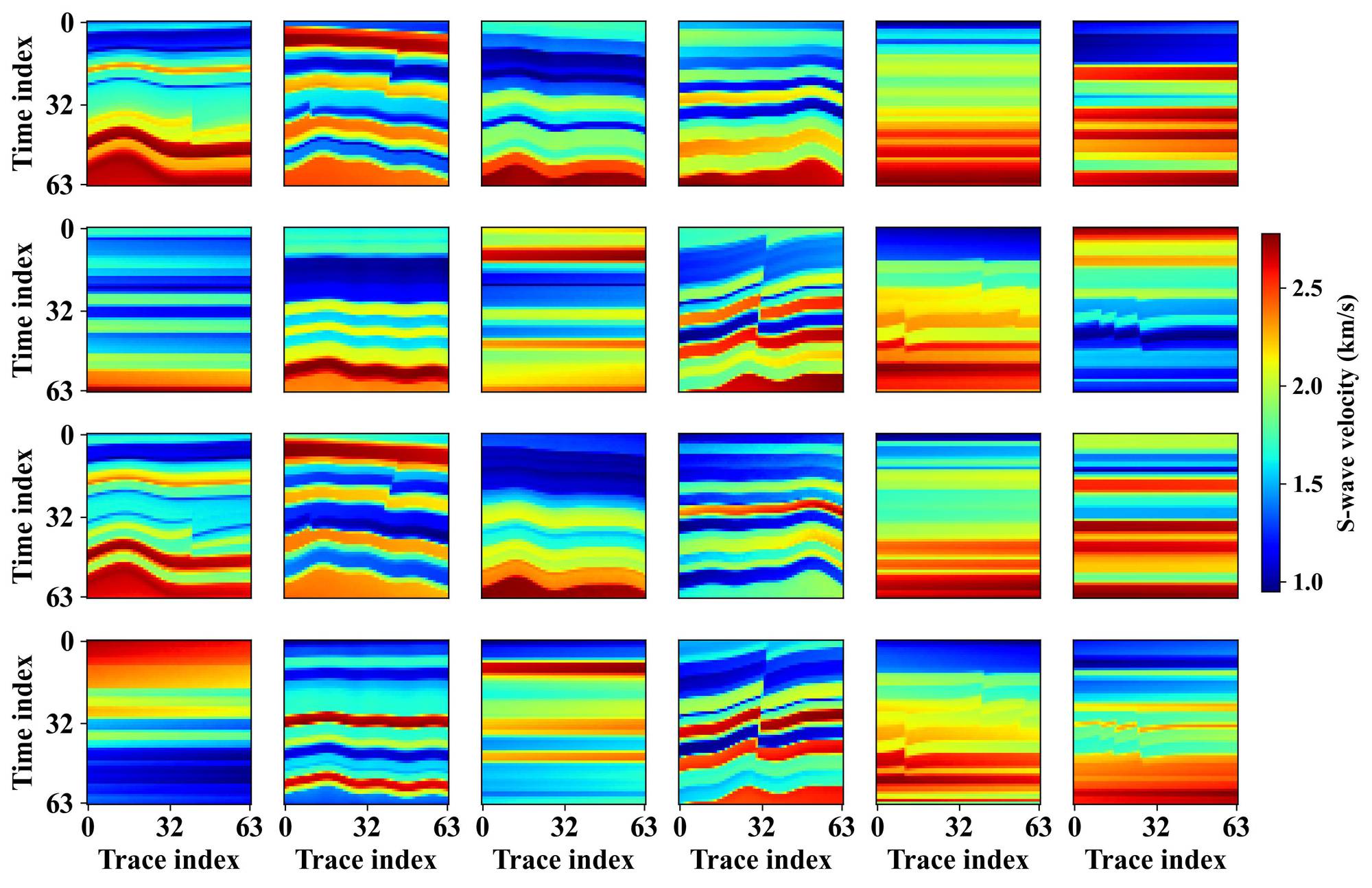}
 \label{fig:ddpm-condseis-vs}} 
  \subfigure[]{\includegraphics[width=0.65\columnwidth]{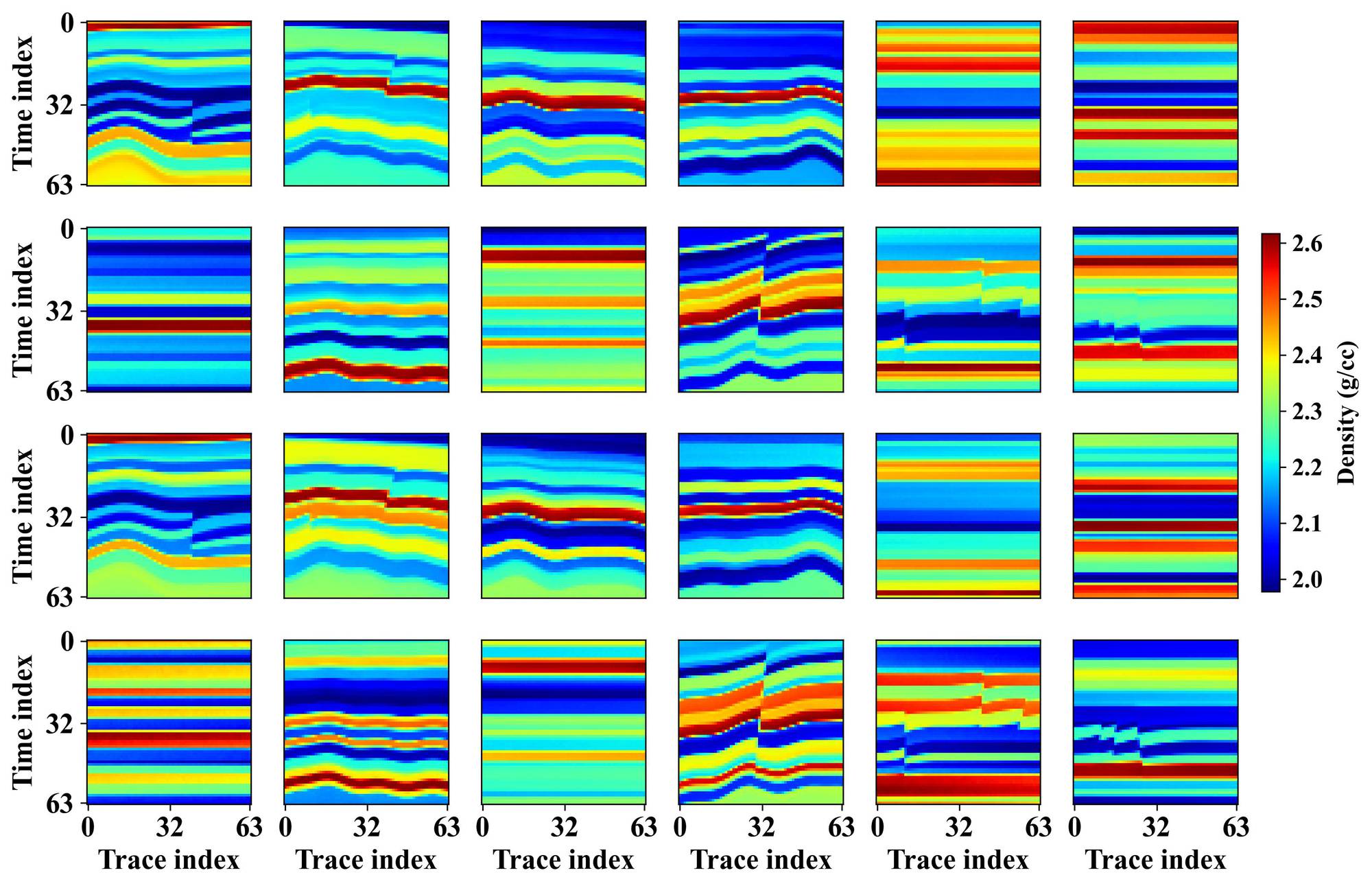}
 \label{fig:ddpm-condseis-rho}}  
   \subfigure[]{\includegraphics[width=0.65\columnwidth]{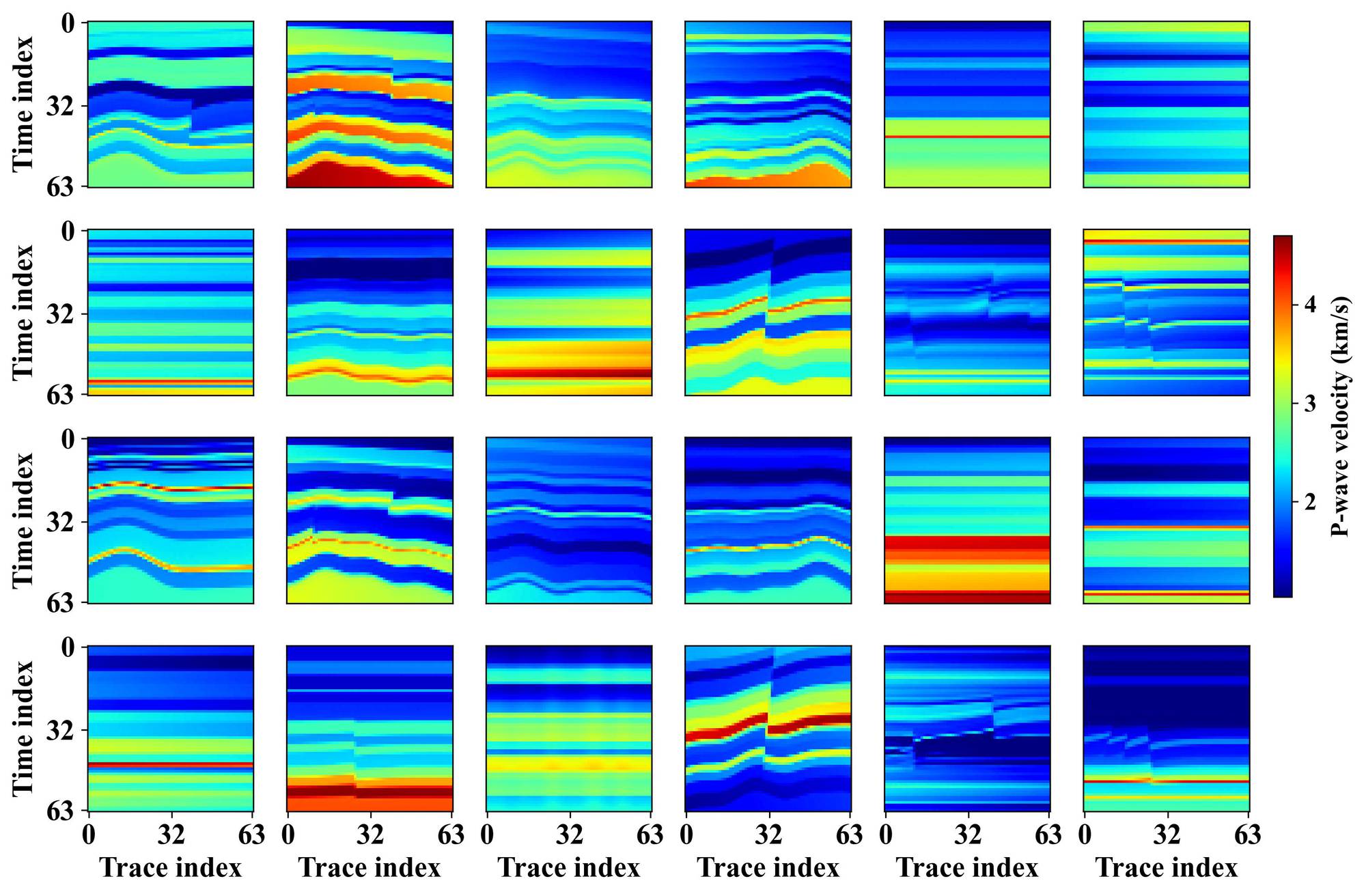}
 \label{fig:ddpm-condseisdps-vp}}
 \subfigure[]{\includegraphics[width=0.65\columnwidth]{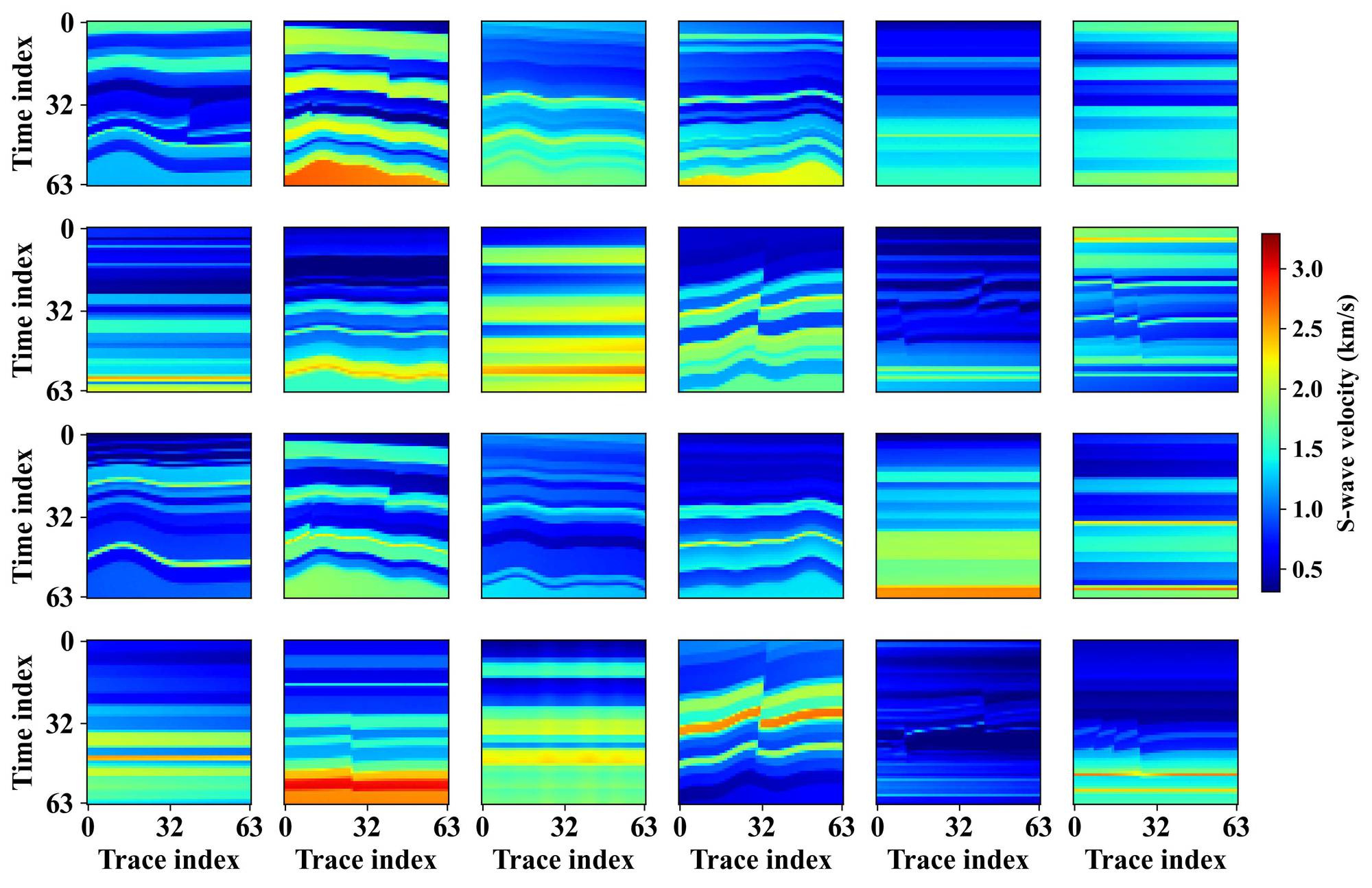}
 \label{fig:ddpm-condseisdps-vs}} 
  \subfigure[]{\includegraphics[width=0.65\columnwidth]{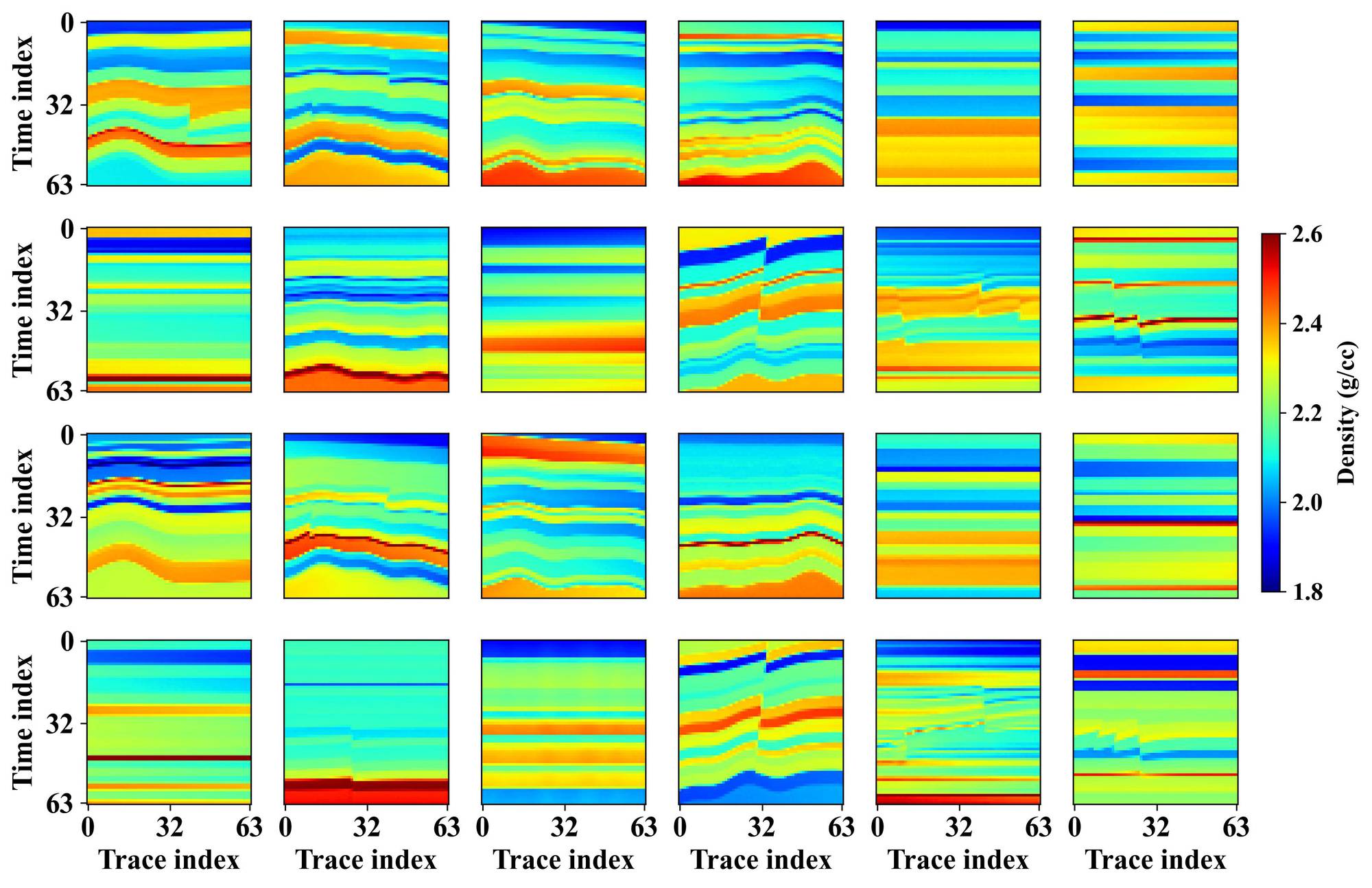}
 \label{fig:ddpm-condseisdps-rho}}  
 \caption{Elastic parameter synthesis conditioned on seismic data. 
(a)--(c) Noisy seismic angle gathers at incidence angles of $12^\circ$, $24^\circ$, and $36^\circ$, with an SNR of 22.36 dB. 
(d)--(f) Samples generated by the proposed DPS-projection. 
(g)--(i) Samples generated by DPS. 
In each panel, the first two rows show the results from the first sampling run, whereas the last two rows show the results from the second sampling run.}
\label{fig:seissamples}
\end{figure*}

\textbf{\textit{Elastic parameter synthesis conditioned on seismic data:}} To evaluate the ability of the proposed method to incorporate seismic-data constraints, seismic data are used as conditioning information for elastic parameter synthesis. This experiment can be interpreted as a seismic inverse problem regularized by the generative prior learned by the diffusion model. Figs. \ref{fig:condseis-0}--\ref{fig:condseis-2} show the synthetic seismic angle gathers at incidence angles of $12^\circ$, $24^\circ$, and $36^\circ$, with a signal-to-noise ratio (SNR) of 22.36 dB. These gathers are generated using the Shuey approximation and a 30 Hz Ricker wavelet, with a time sampling interval of 2 ms. Although the Shuey approximation is adopted, the forward mapping is treated as nonlinear with respect to the elastic parameters because the $v_s/v_p$ ratio is not fixed during synthesis. To mimic spatially correlated noise commonly observed in field data, random noise is first smoothed using a three-point moving-average operator and then added to the synthetic gathers.

\begin{figure*}[htb!]
\setlength{\abovecaptionskip}{0.2cm}
 \centering
    \subfigure[]{\includegraphics[width=0.65\columnwidth]{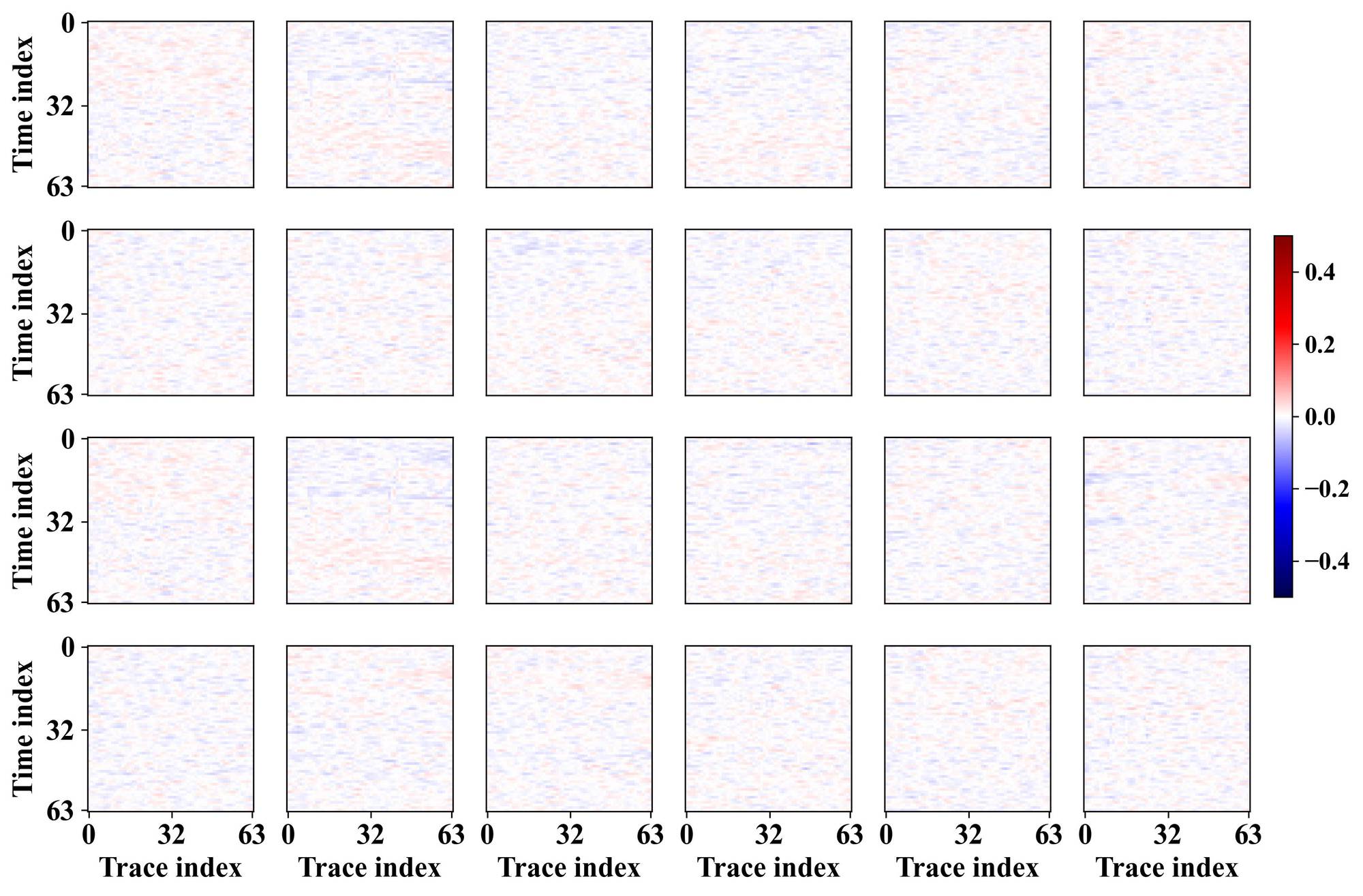}
 \label{fig:ddpm-condseis-obs0}}
 \subfigure[]{\includegraphics[width=0.65\columnwidth]{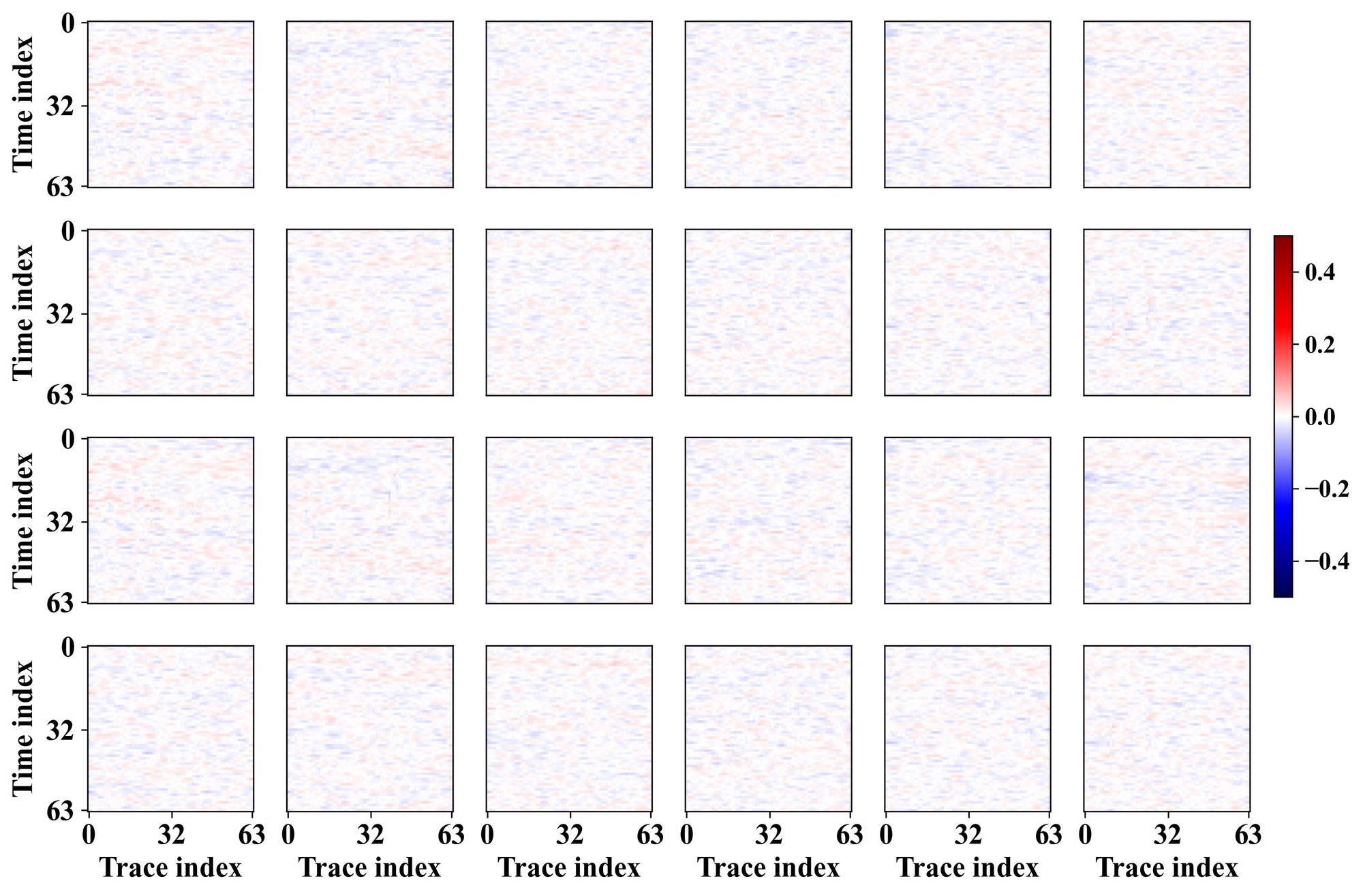}
 \label{fig:ddpm-condseis-obs1}} 
  \subfigure[]{\includegraphics[width=0.65\columnwidth]{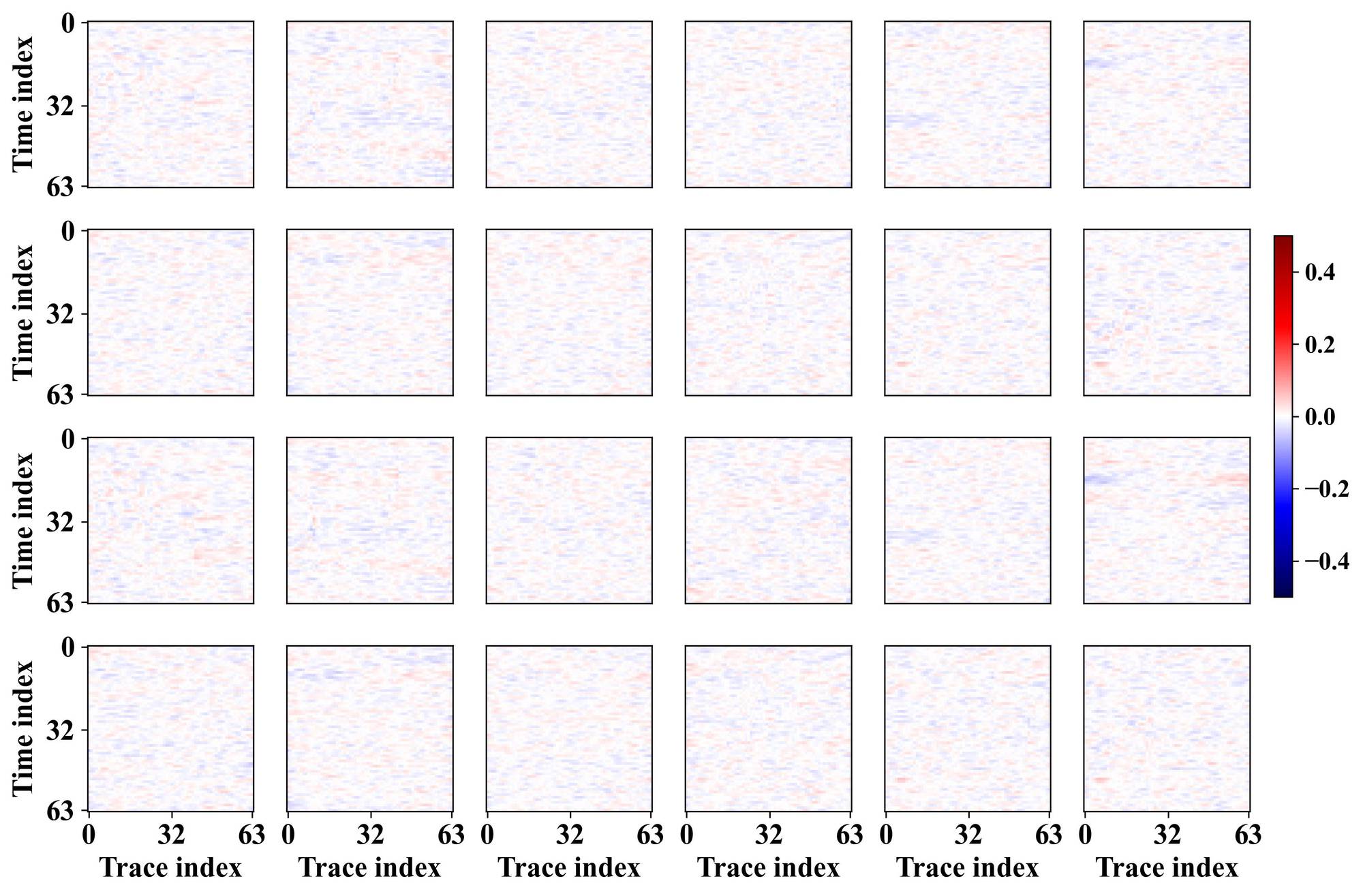}
 \label{fig:ddpm-condseis-obs2}} 
   \subfigure[]{\includegraphics[width=0.65\columnwidth]{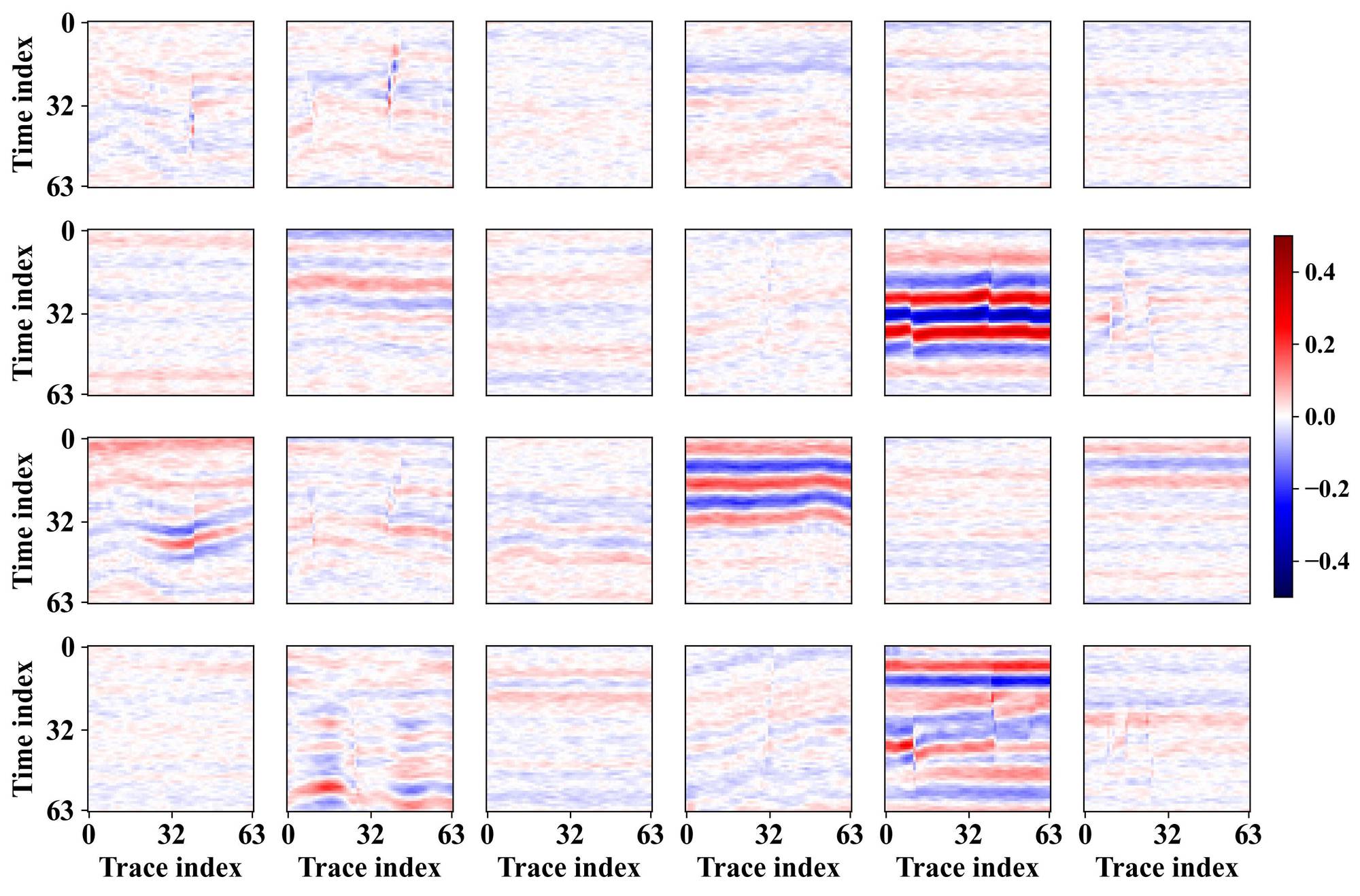}
 \label{fig:ddpm-condseisdps-obs0}}
 \subfigure[]{\includegraphics[width=0.65\columnwidth]{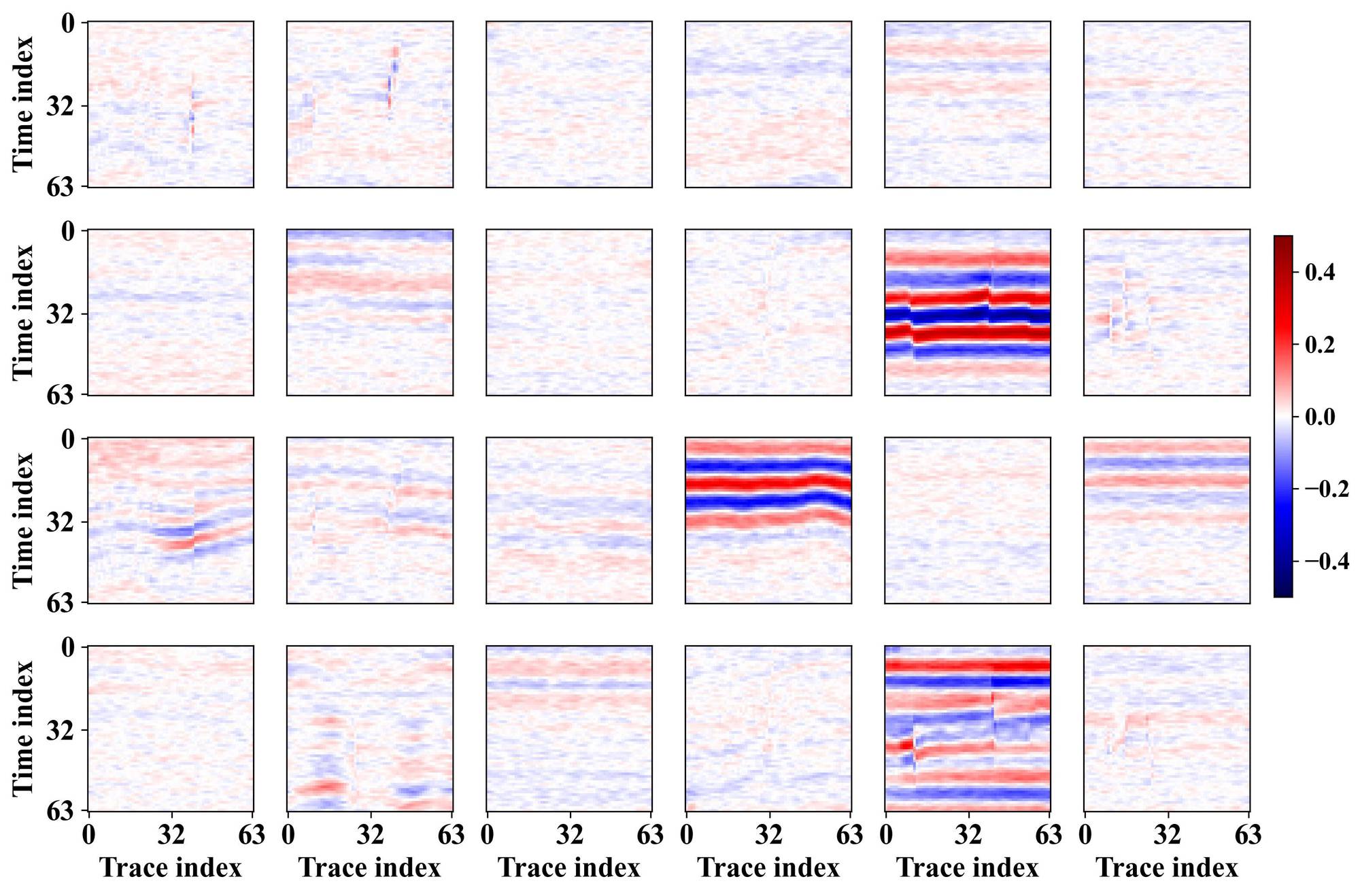}
 \label{fig:ddpm-condseisdps-obs1}} 
  \subfigure[]{\includegraphics[width=0.65\columnwidth]{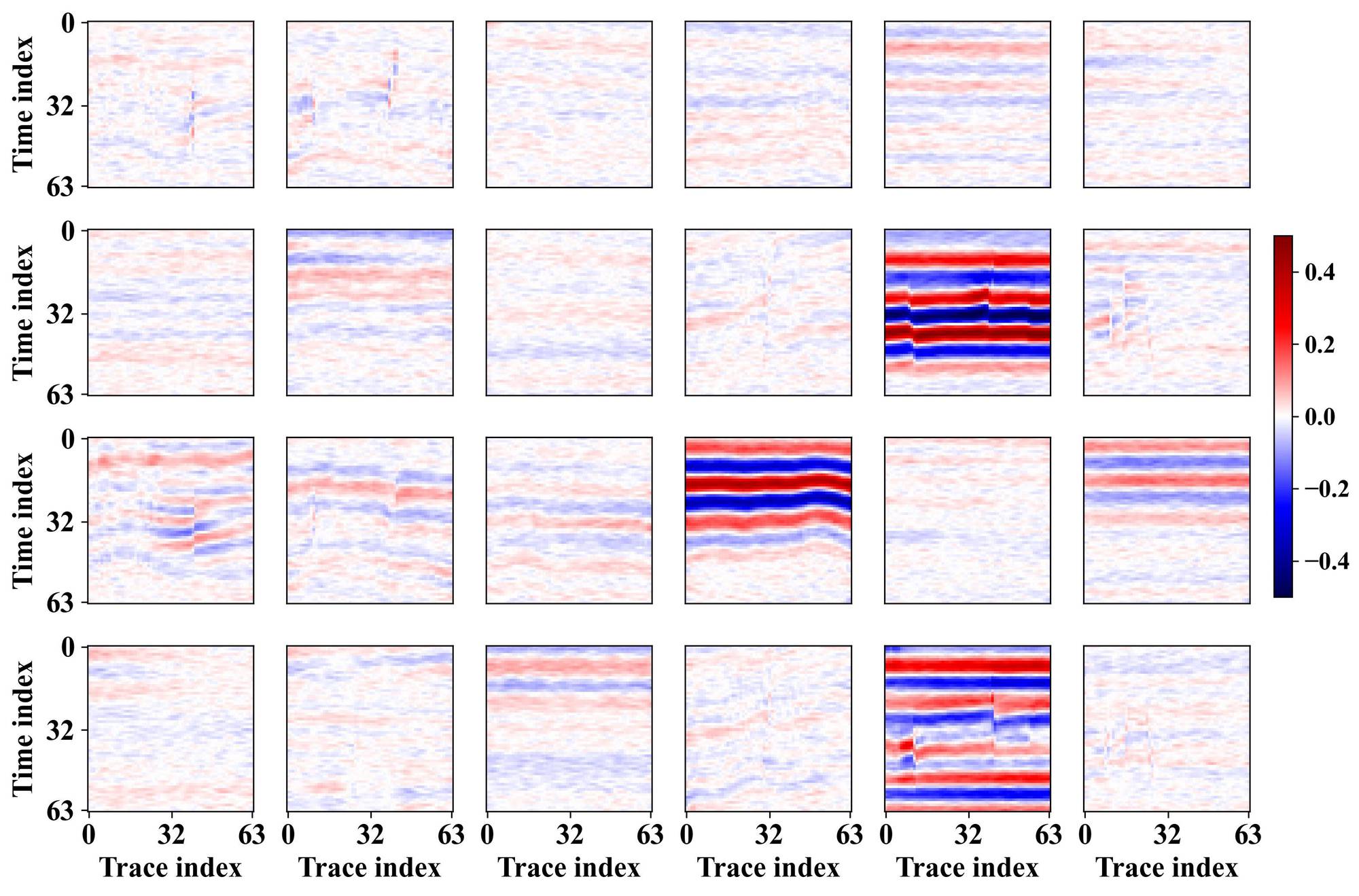}
 \label{fig:ddpm-condseisdps-obs2}}   
 \caption{Errors between the reconstructed seismic data and the synthetic observed seismic data shown in Figs. \ref{fig:condseis-0}--\ref{fig:condseis-2}. 
(a)--(c) Errors corresponding to the samples generated by the DPS-projection. 
(d)--(f) Errors corresponding to the samples generated by DPS.}
\label{fig:seissampleserro}
\end{figure*}

Figs. \ref{fig:ddpm-condseis-vp}--\ref{fig:ddpm-condseisdps-rho} show the samples synthesized by the proposed DPS-projection and DPS, respectively. The samples obtained by the proposed method exhibit structural patterns consistent with the input seismic data. In contrast, some samples obtained by DPS fail to preserve structures consistent with the seismic observations, indicating that the seismic data constraint is less effectively enforced in DPS. For a more intuitive comparison, Fig. \ref{fig:seissampleserro} shows the errors between the reconstructed seismic data, calculated from the synthesized samples, and the synthetic observed seismic data. The smaller reconstruction errors obtained by the proposed method demonstrate that the proposed DPS-projection incorporates seismic data more effectively and enforces data consistency more strongly than DPS. Meanwhile, the results from different sampling runs in Figs. \ref{fig:ddpm-condseis-vp}--\ref{fig:ddpm-condseisdps-rho} show noticeable differences in absolute parameter values, although they produce similar seismic responses. It indicates that seismic data alone cannot fully constrain the absolute values of elastic parameters. This limitation is mainly associated with the band-limited nature of seismic data, which contributes to the non-uniqueness of the inverse problem. Therefore, additional prior information is required to further reduce the solution ambiguity.

\begin{figure*}[htb!]
\setlength{\abovecaptionskip}{0.2cm}
 \centering
    \subfigure[]{\includegraphics[width=0.65\columnwidth]{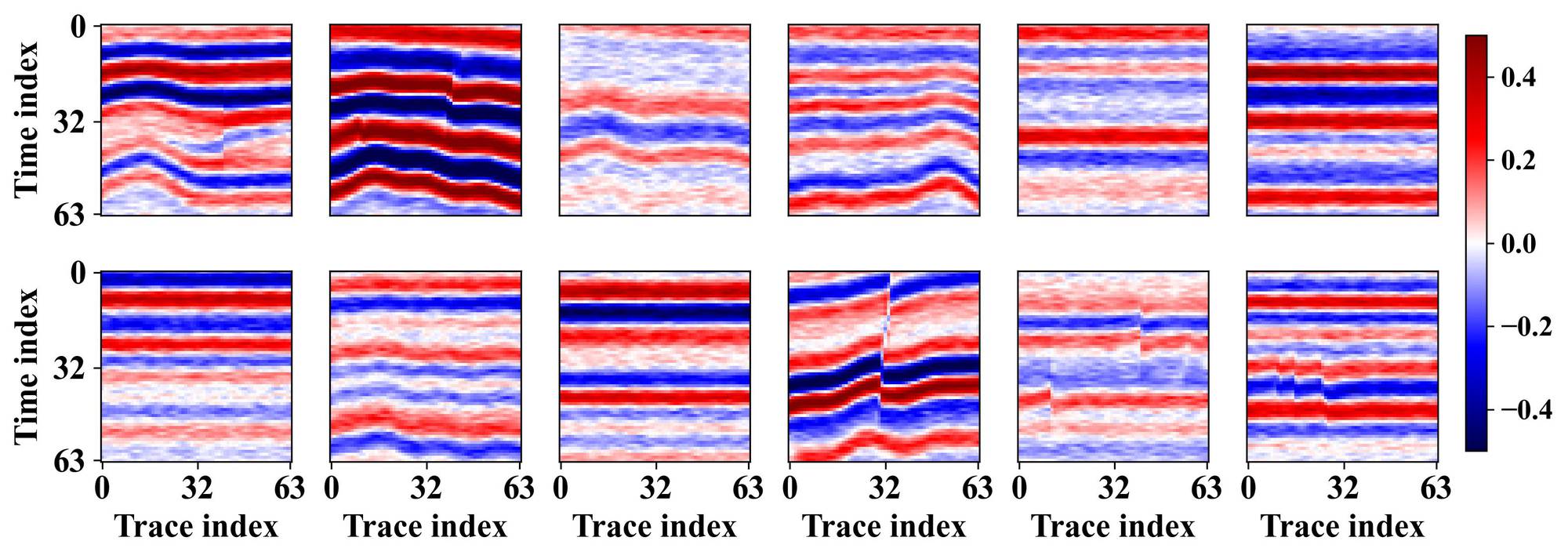}
 \label{fig:condseis16-0}}
 \subfigure[]{\includegraphics[width=0.65\columnwidth]{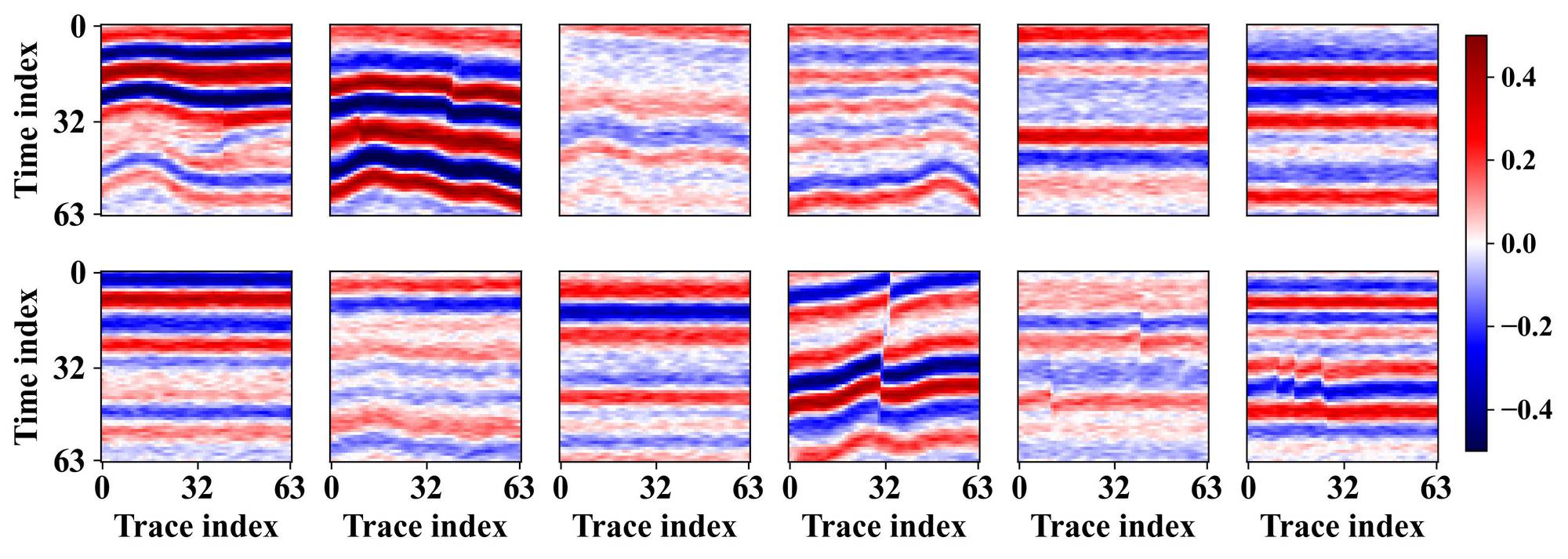}
 \label{fig:condseis16-1}} 
  \subfigure[]{\includegraphics[width=0.65\columnwidth]{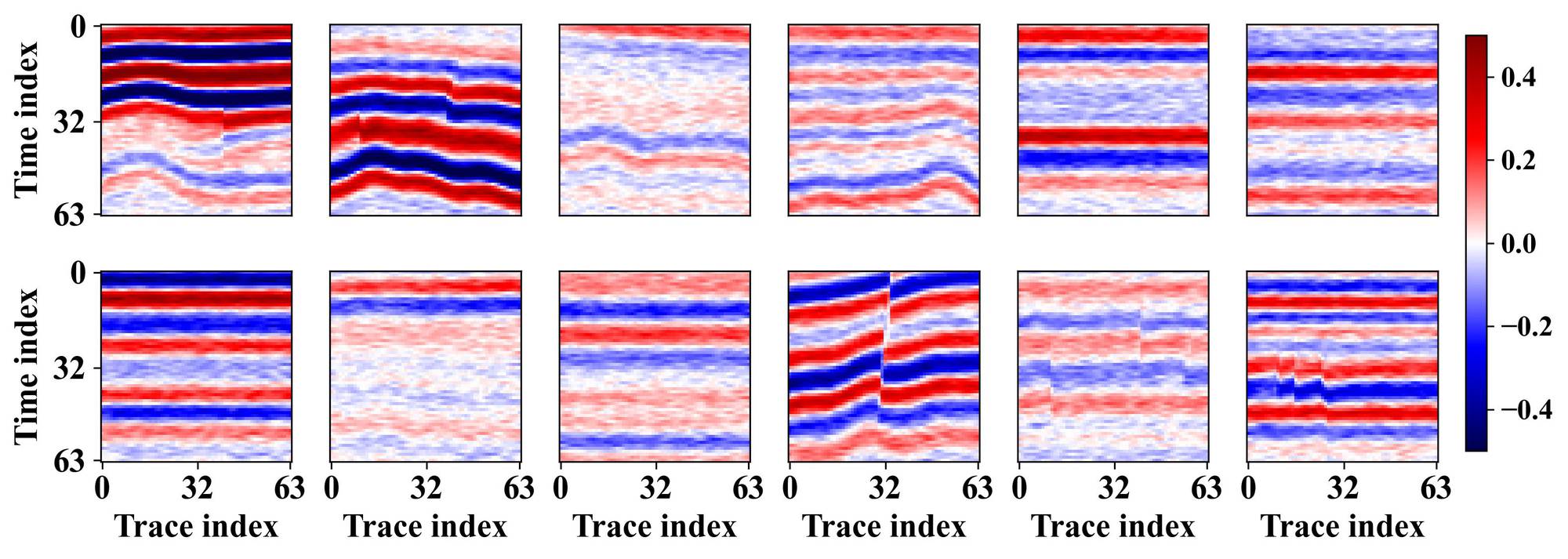}
 \label{fig:condseis16-2}}
   \subfigure[]{\includegraphics[width=0.65\columnwidth]{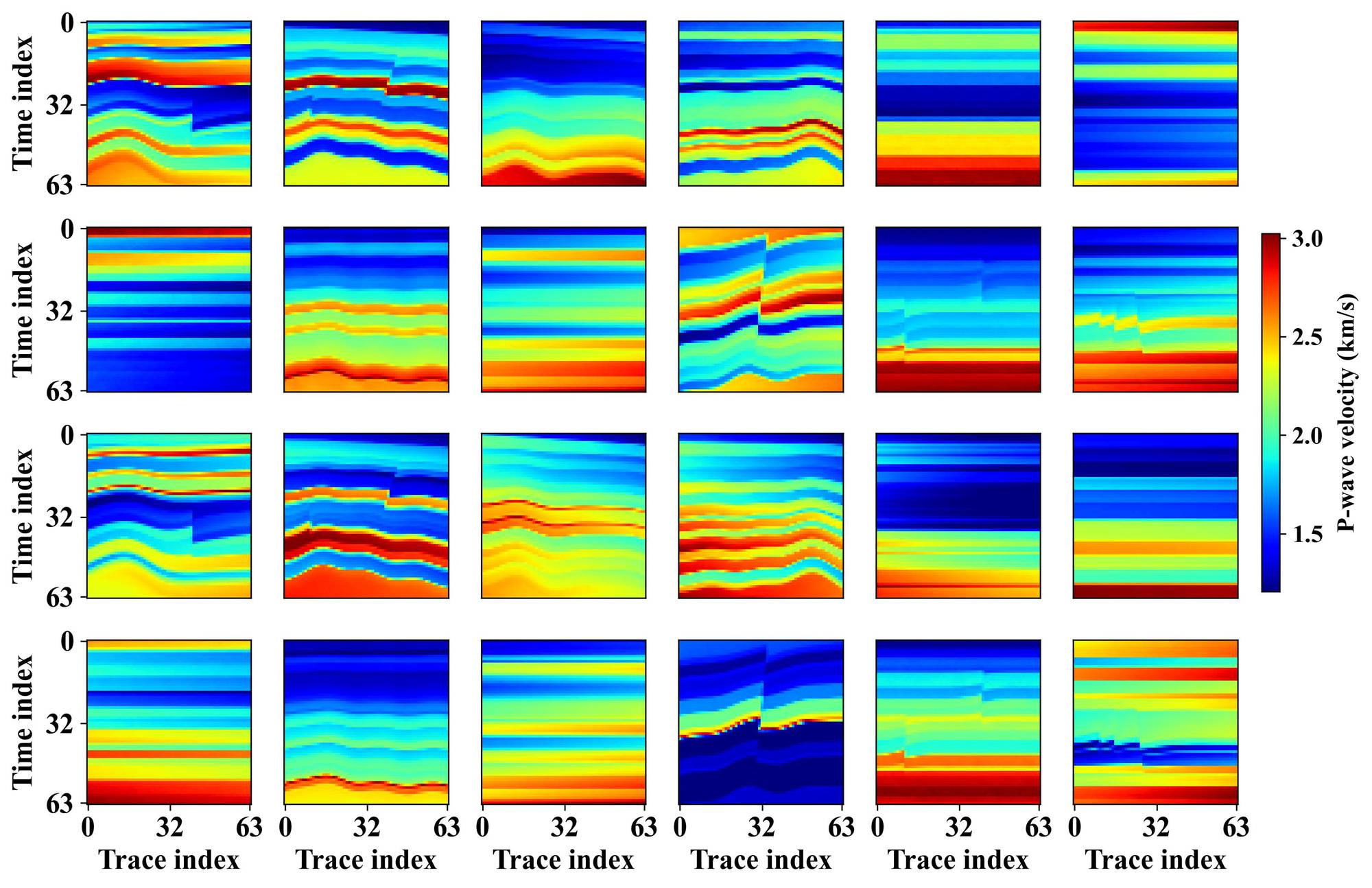}
 \label{fig:ddpm-condseis16-vp}}
 \subfigure[]{\includegraphics[width=0.65\columnwidth]{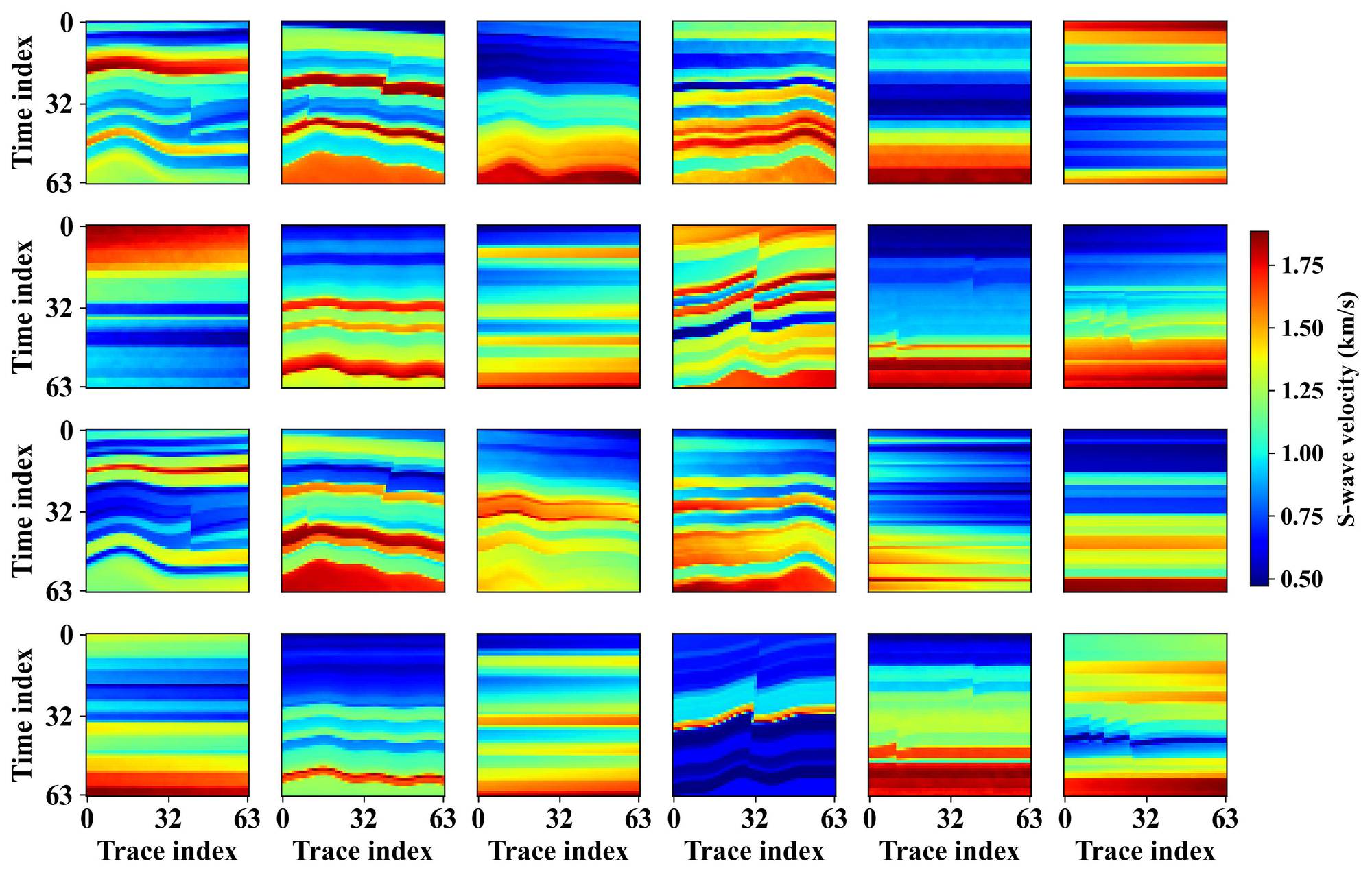}
 \label{fig:ddpm-condseis16-vs}} 
  \subfigure[]{\includegraphics[width=0.65\columnwidth]{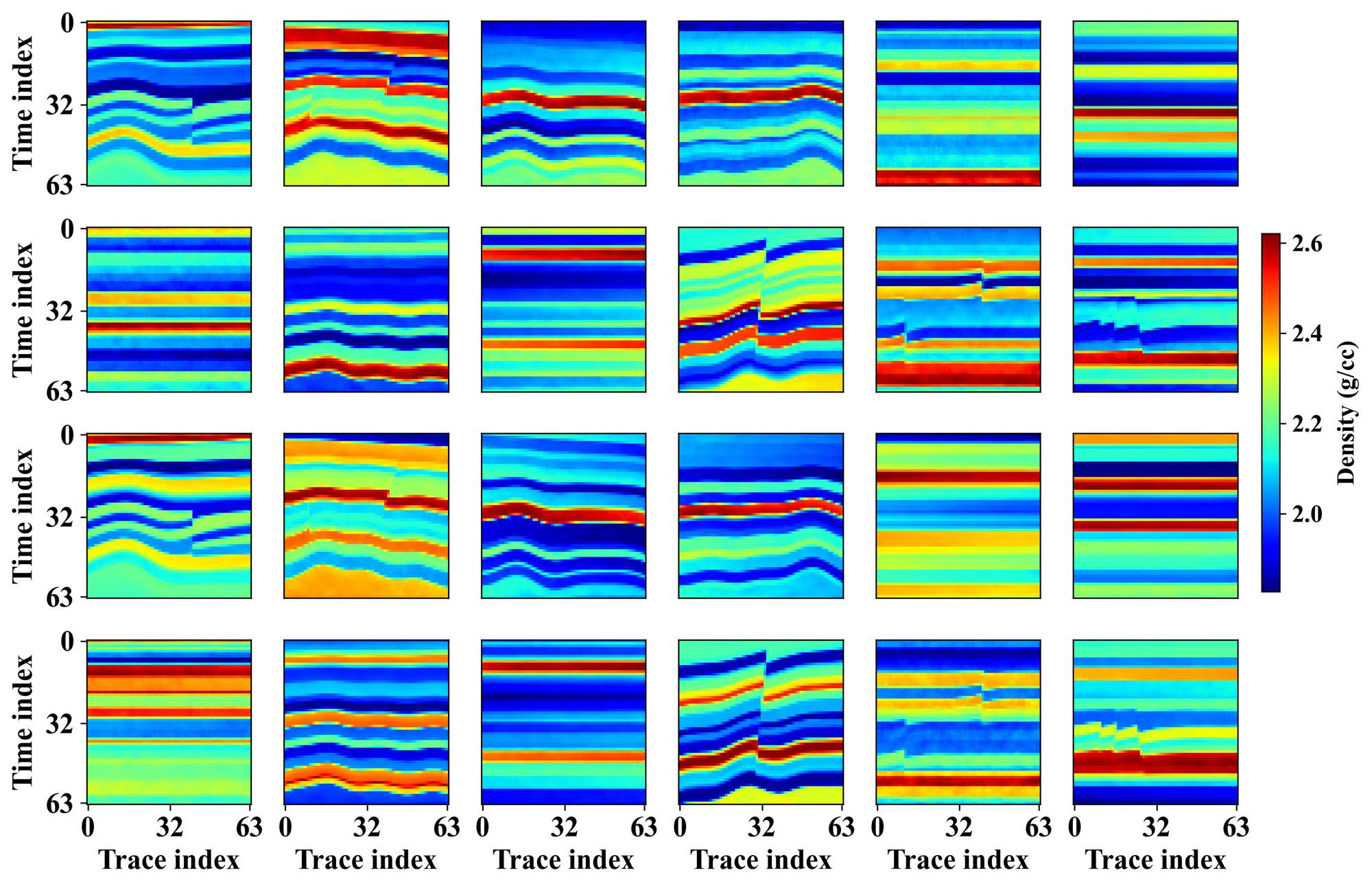}
 \label{fig:ddpm-condseis16-rho}} 
 \caption{ Elastic parameter synthesis conditioned on seismic data. 
(a)--(c) Noisy seismic angle gathers at incidence angles of $12^\circ$, $24^\circ$, and $36^\circ$, with an SNR of 16.34 dB. 
(d)--(f) Samples generated by the proposed DPS-projection. 
In each panel, the first two rows show the results from the first sampling run, whereas the last two rows show the results from the second sampling run.
}
\label{fig:seissamples16}
\end{figure*}

Finally, we increase the noise level in the synthetic seismic data to an SNR of 16.34 dB to evaluate the robustness of the proposed method to noise, as shown in Fig. \ref{fig:seissamples16}. To reduce the influence of noisy observations, the number of projection iterations is decreased from 30, as used in the 22.36 dB case, to 20, thereby relaxing the seismic data constraint. The results indicate that the proposed method maintains stable synthesis performance under increased noise levels. The main structural features are generally preserved, although some subtle fault details become less clearly delineated due to the weakened data constraint.

\begin{figure*}[htb!]
\setlength{\abovecaptionskip}{0.2cm}
 \centering
    \subfigure[]{\includegraphics[width=0.65\columnwidth]{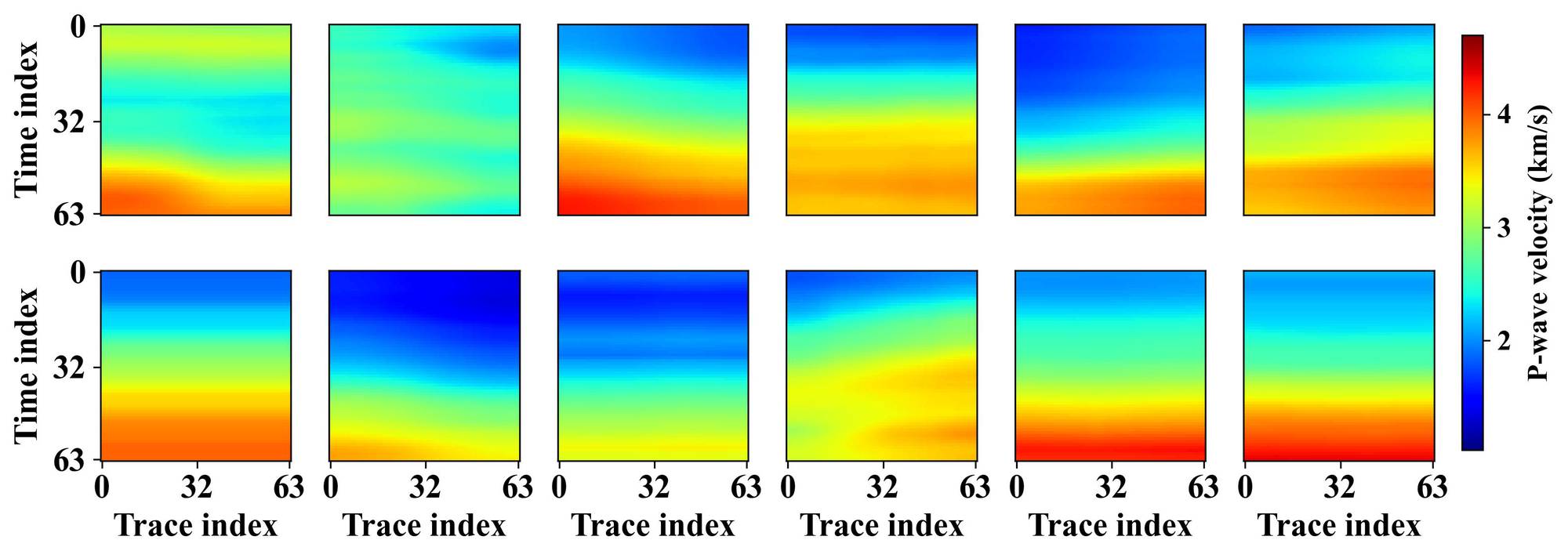}
 \label{fig:condlow-vp}}
 \subfigure[]{\includegraphics[width=0.65\columnwidth]{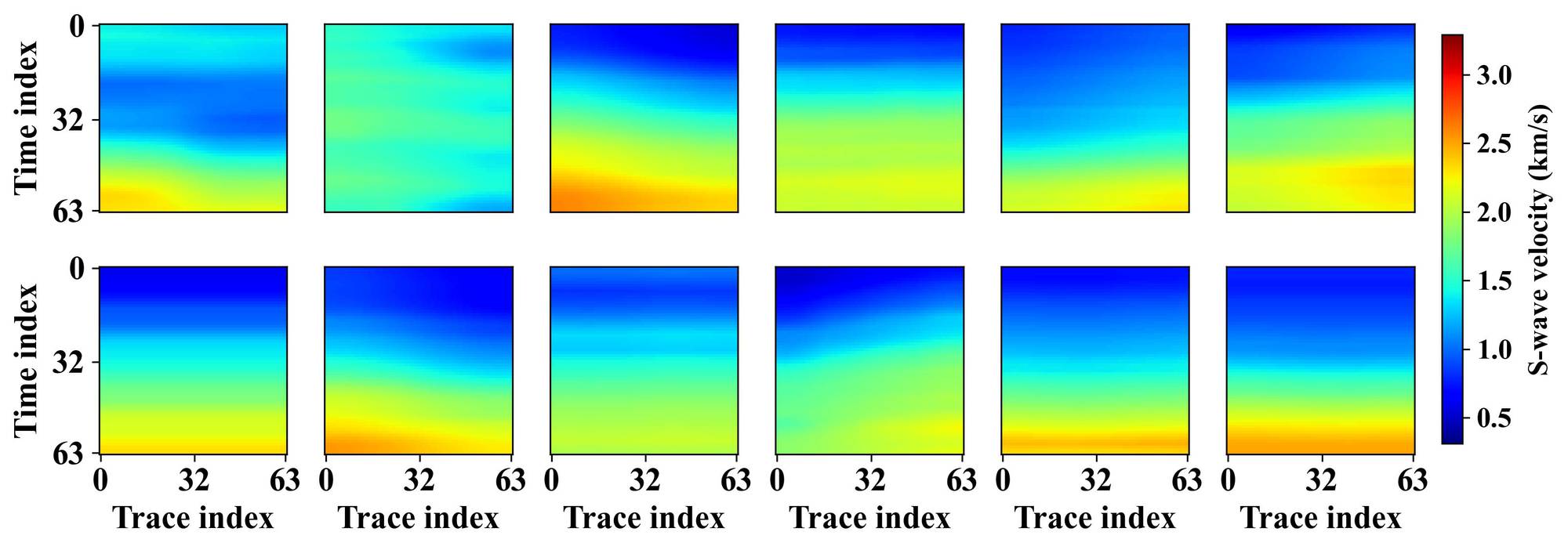}
 \label{fig:condlow-vs}} 
  \subfigure[]{\includegraphics[width=0.65\columnwidth]{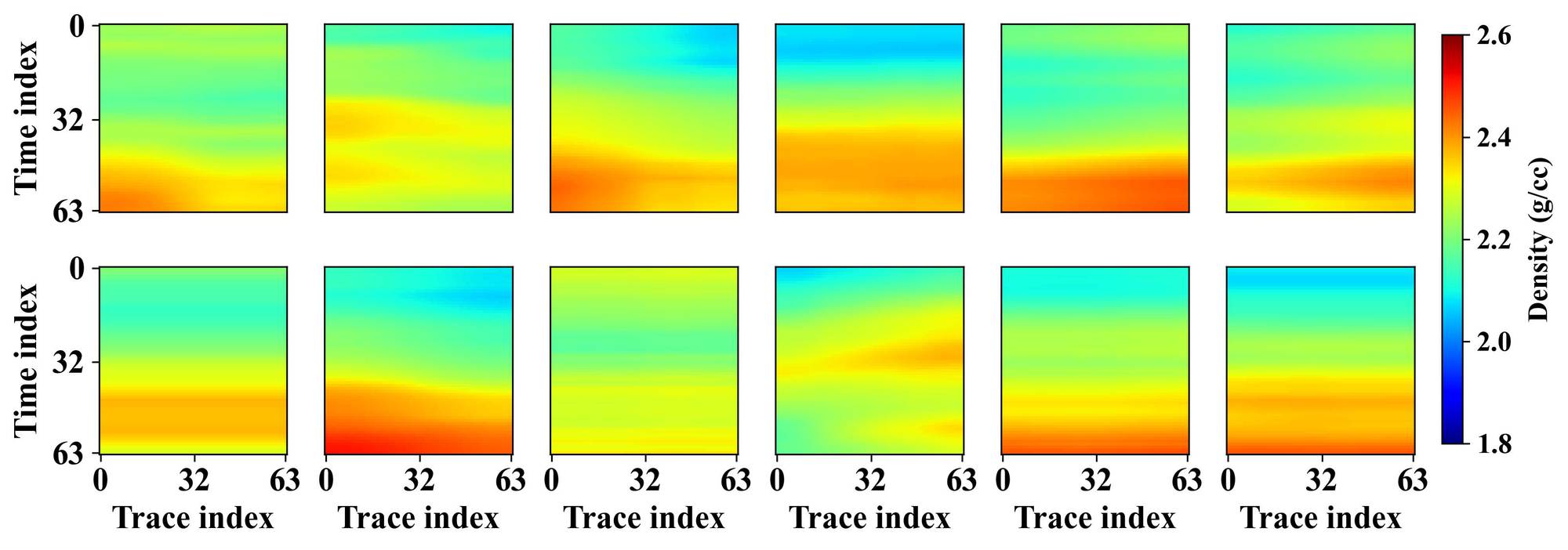}
 \label{fig:condlow-rho}}
   \subfigure[]{\includegraphics[width=0.65\columnwidth]{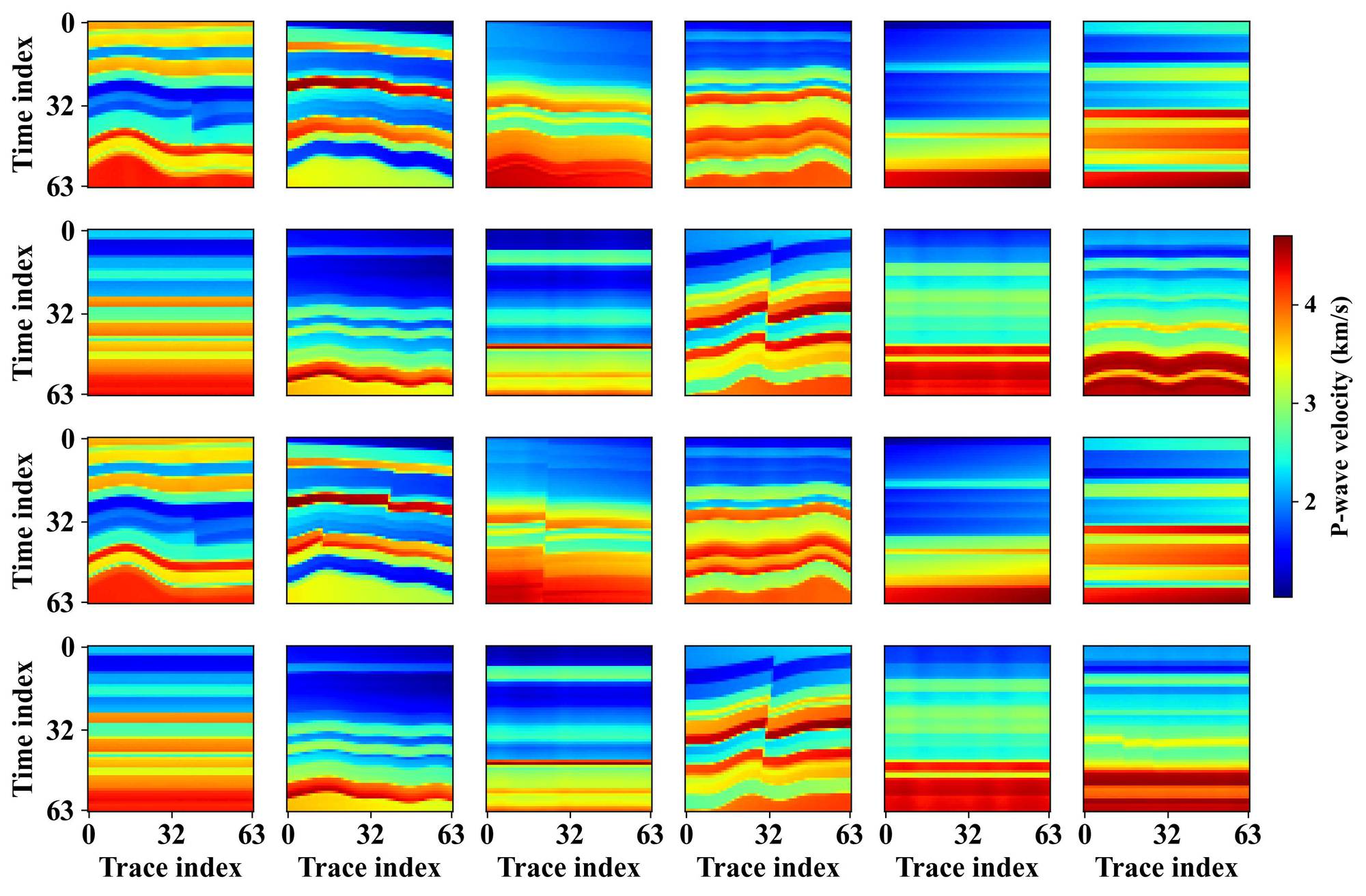}
 \label{fig:ddpm-condlow-vp}}
 \subfigure[]{\includegraphics[width=0.65\columnwidth]{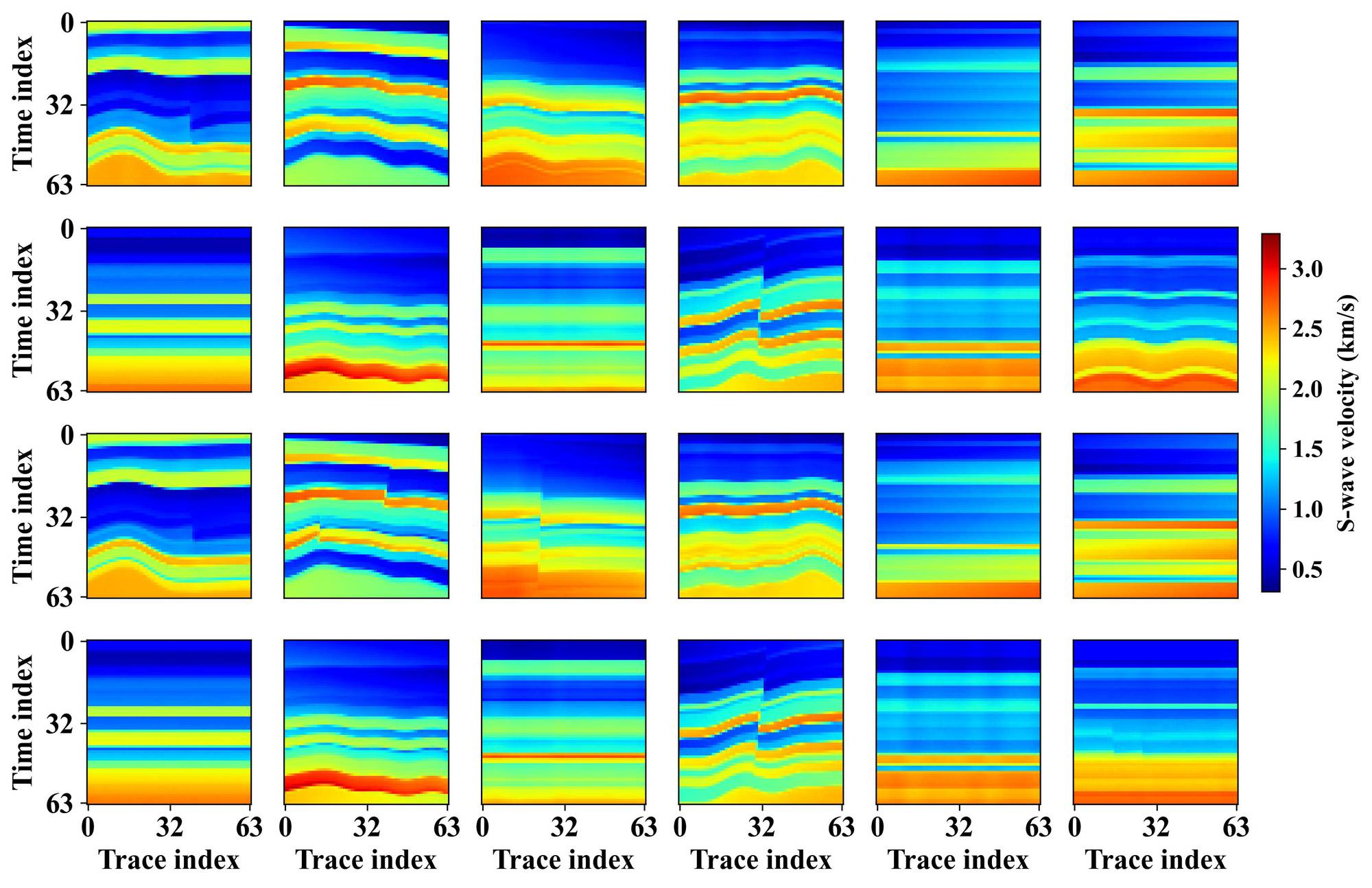}
 \label{fig:ddpm-condlow-vs}} 
  \subfigure[]{\includegraphics[width=0.65\columnwidth]{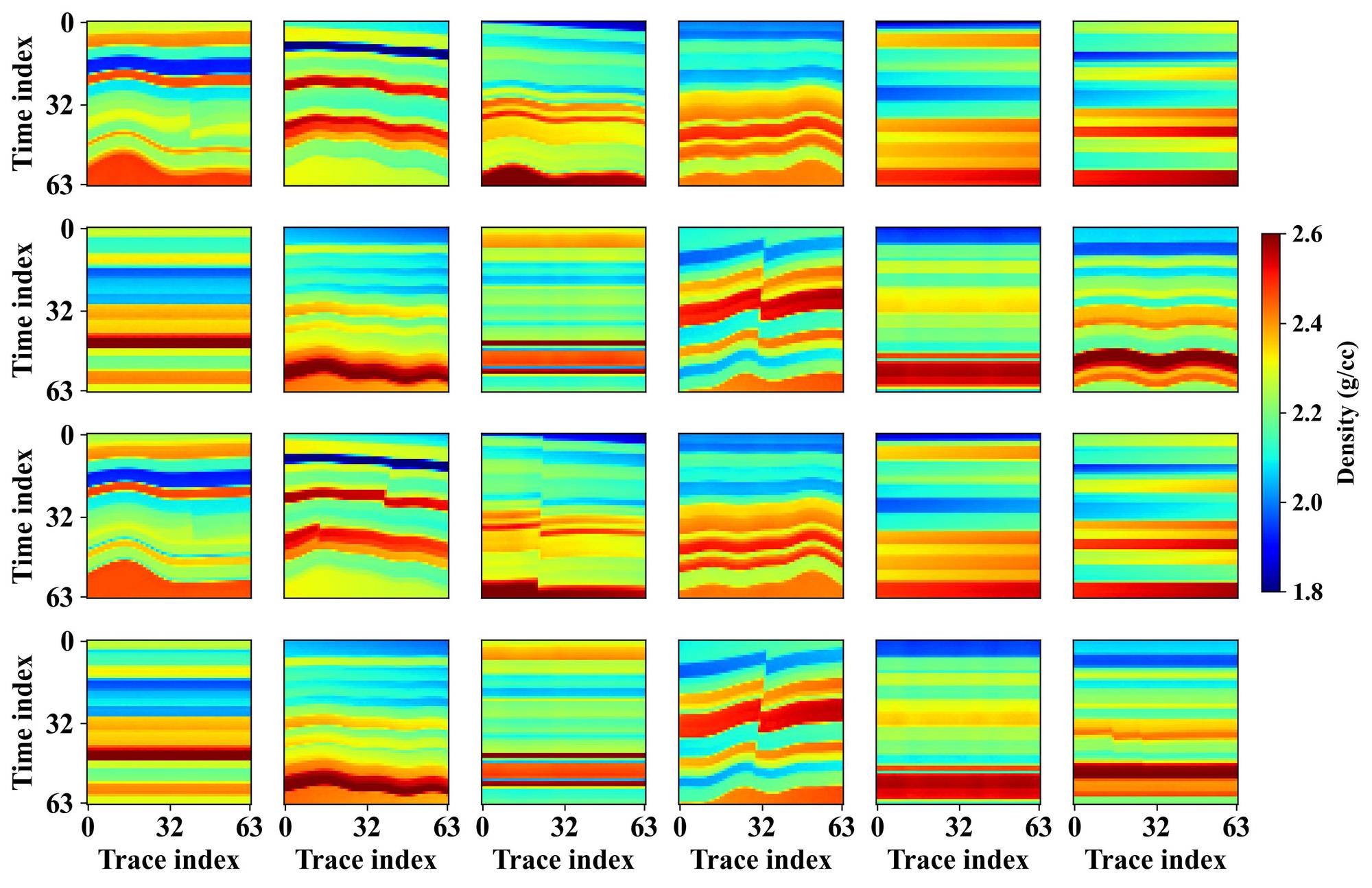}
 \label{fig:ddpm-condlow-rho}} 
   \subfigure[]{\includegraphics[width=0.65\columnwidth]{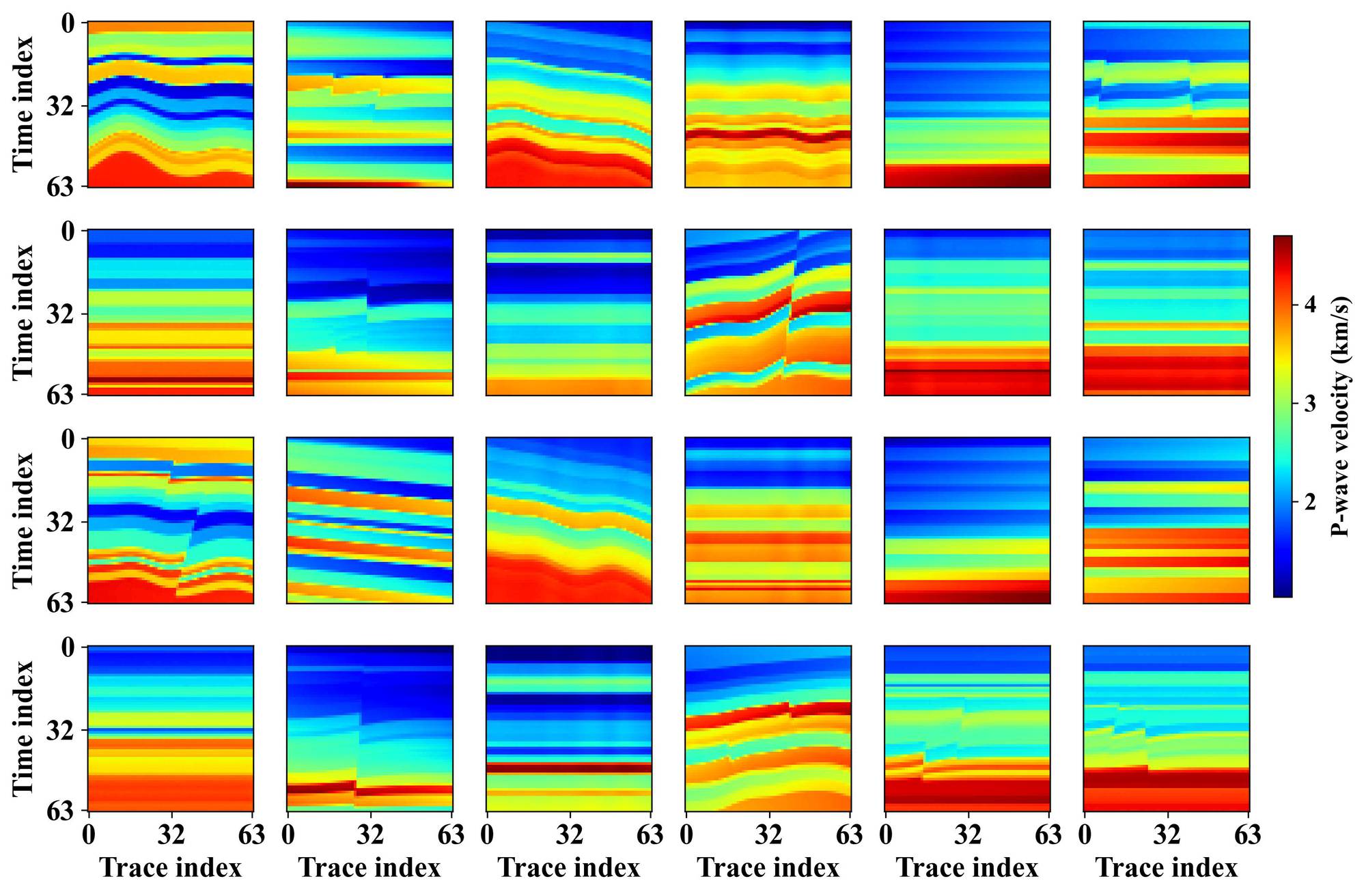}
 \label{fig:ddpm-condlowdps-vp}}
 \subfigure[]{\includegraphics[width=0.65\columnwidth]{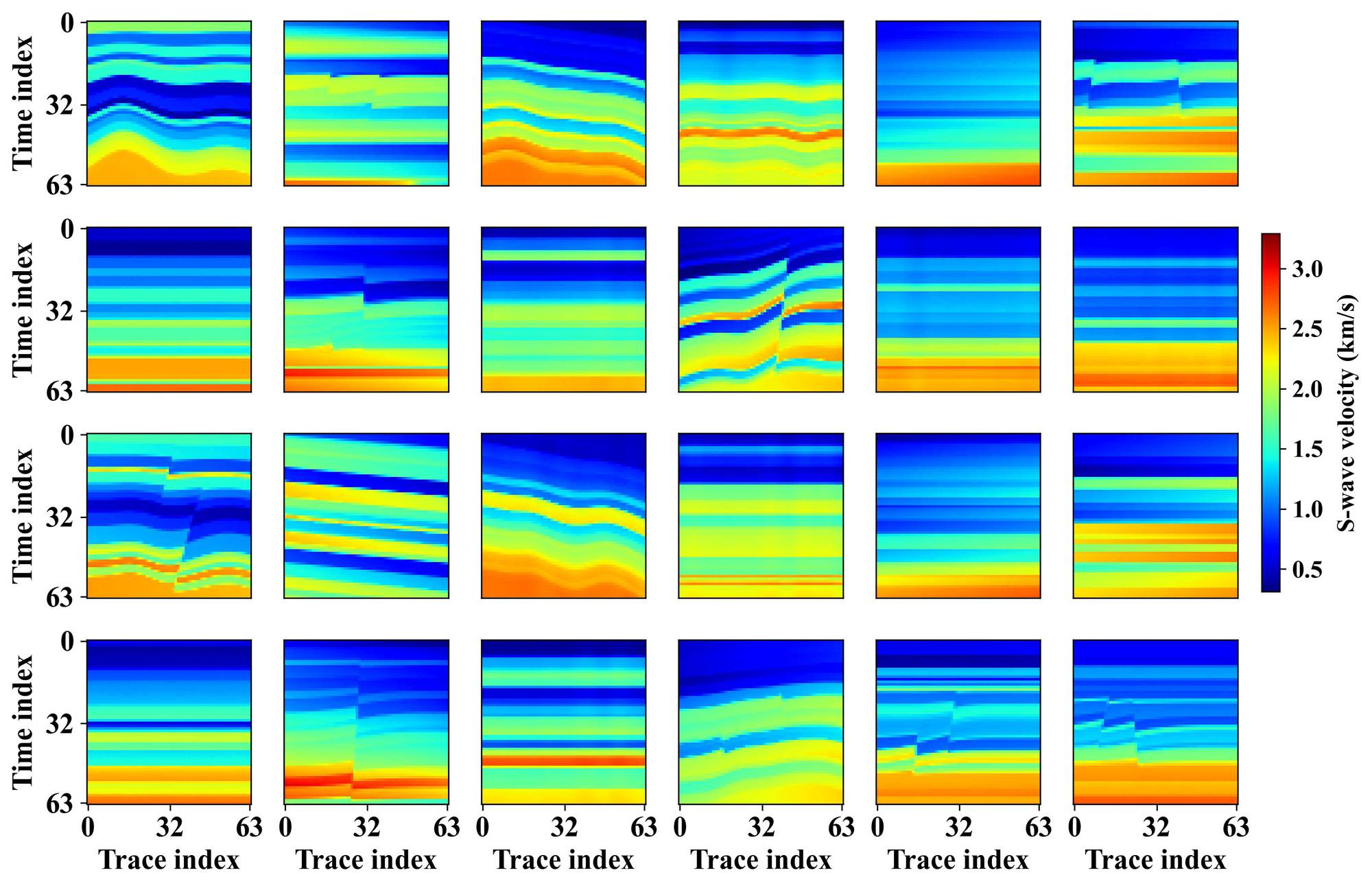}
 \label{fig:ddpm-condlowdps-vs}} 
  \subfigure[]{\includegraphics[width=0.65\columnwidth]{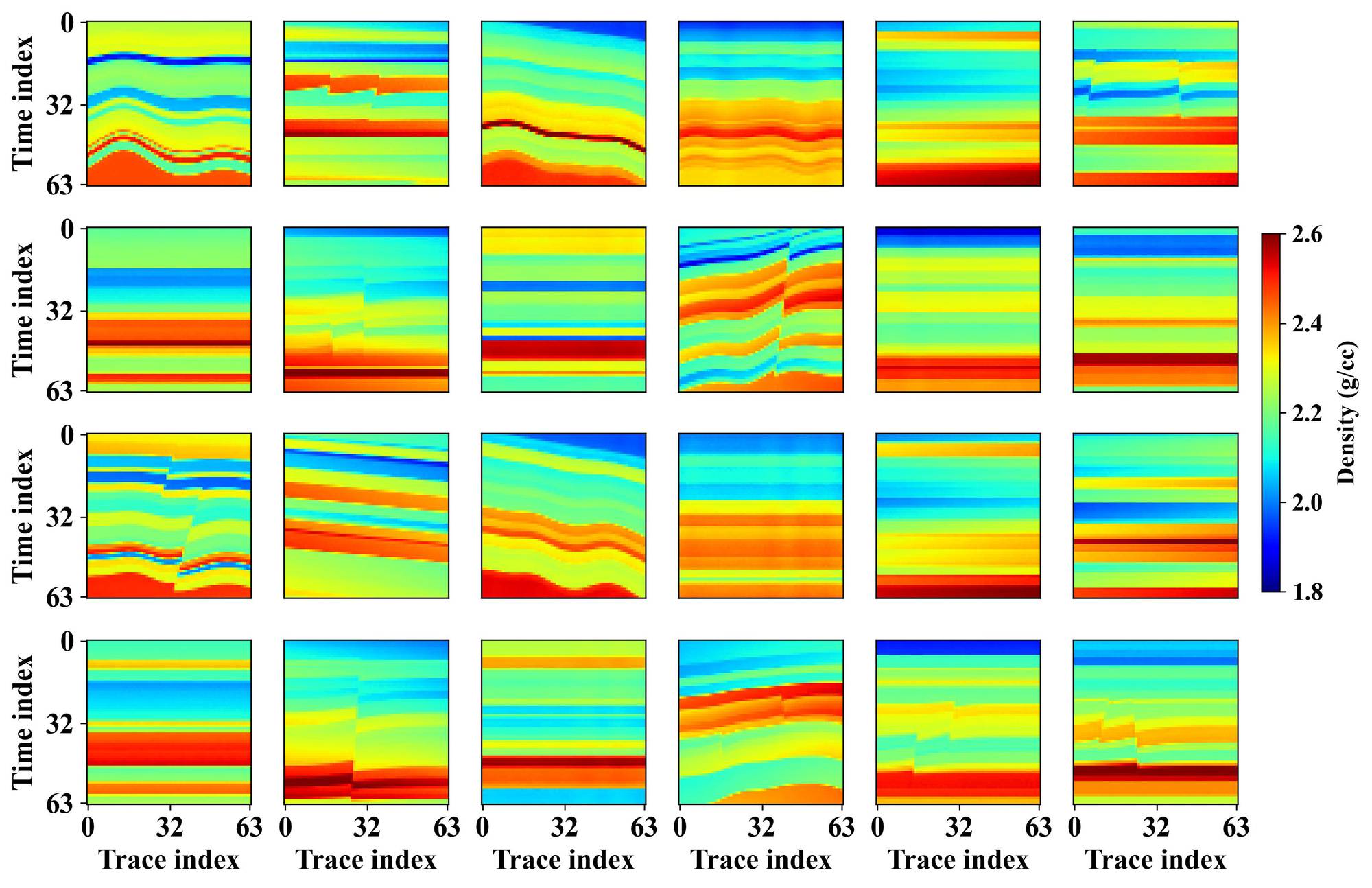}
 \label{fig:ddpm-condlowdps-rho}}  
 \caption{Elastic parameter synthesis conditioned on low-frequency models. 
(a)--(c) Low-frequency models derived using a $31 \times 31$ mean filter. 
(d)--(f) Samples generated by the proposed DPS-projection. 
(g)--(i) Samples generated by DPS. 
In each panel, the first two rows show the results from the first sampling run, whereas the last two rows show the results from the second sampling run.
 }
\label{fig:lowsamples31}
\end{figure*}

\begin{figure*}[htb!]
\setlength{\abovecaptionskip}{0.2cm}
 \centering
    \subfigure[]{\includegraphics[width=0.65\columnwidth]{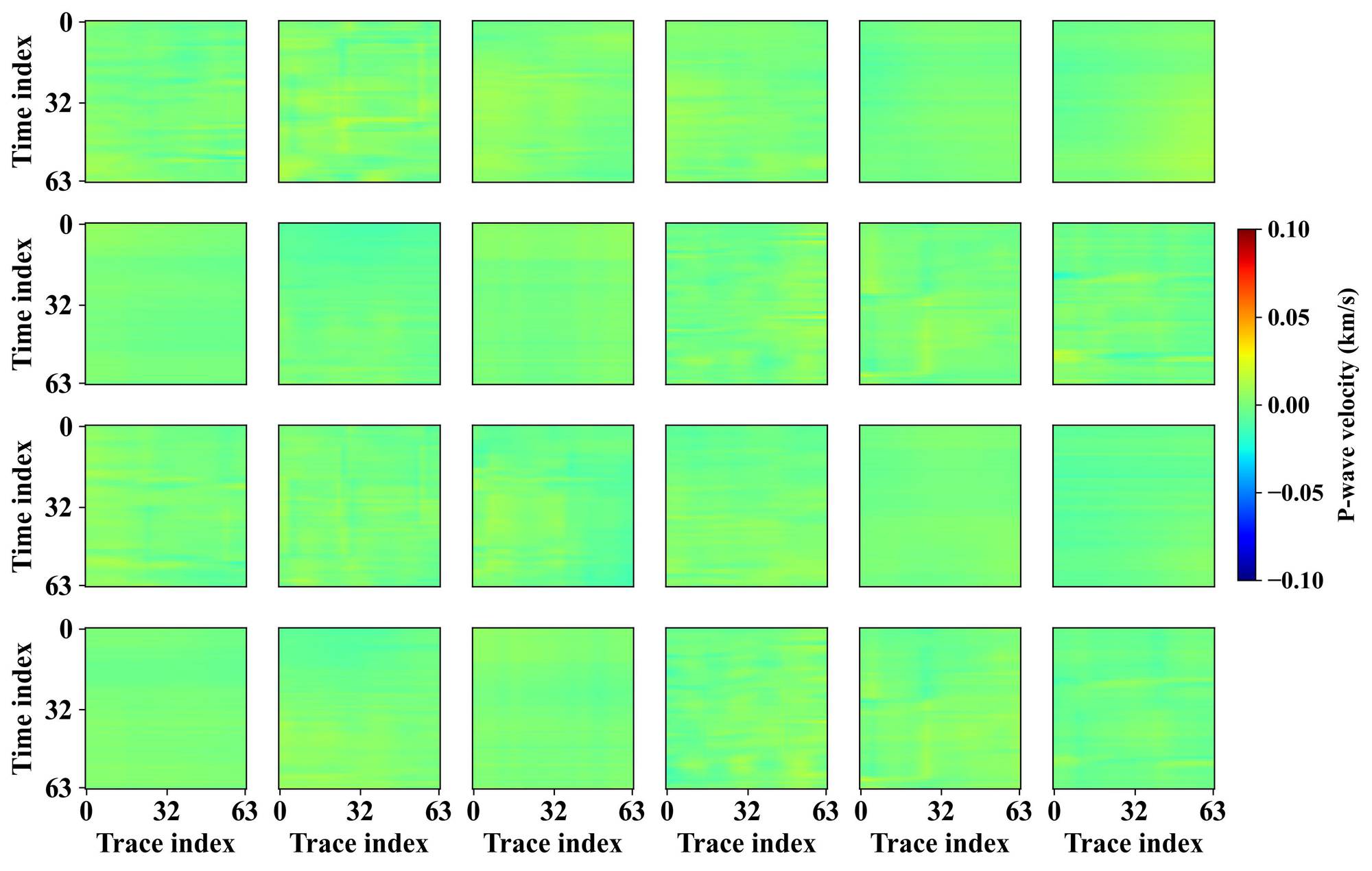}
 \label{fig:ddpm-condlow-vp-smooth}}
 \subfigure[]{\includegraphics[width=0.65\columnwidth]{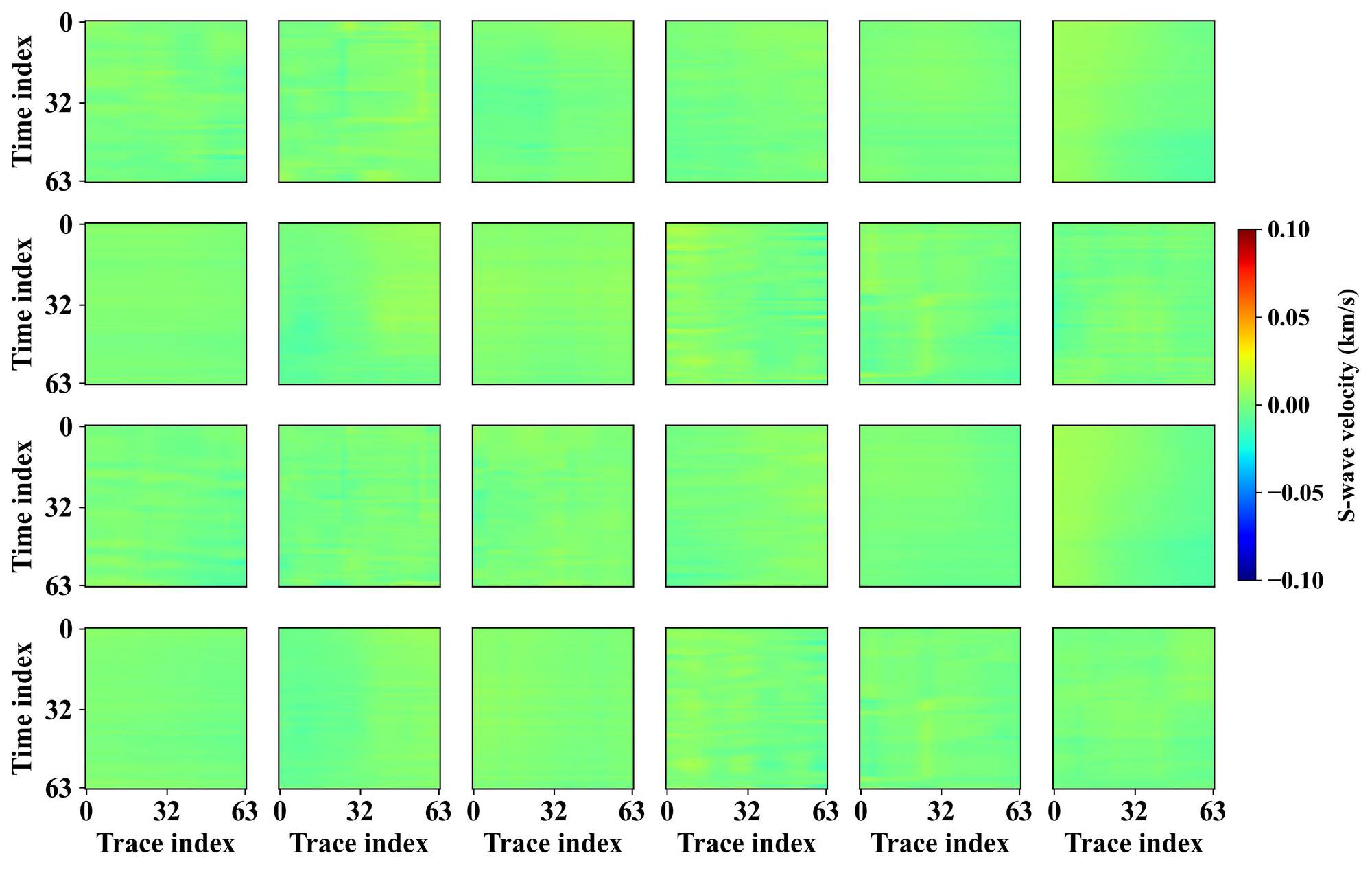}
 \label{fig:ddpm-condlow-vs-smooth}} 
  \subfigure[]{\includegraphics[width=0.65\columnwidth]{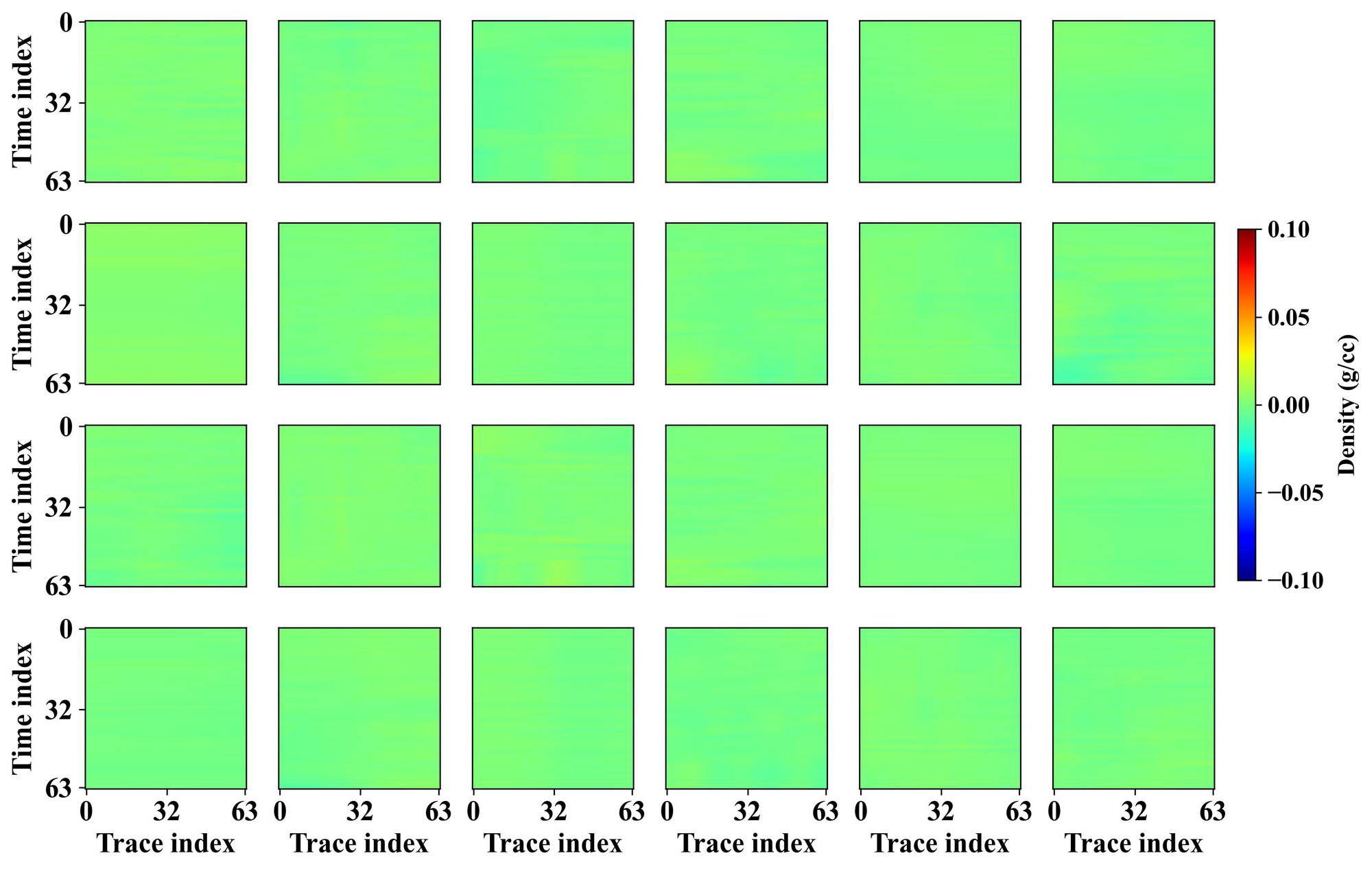}
 \label{fig:ddpm-condlow-rho-smooth}}
   \subfigure[]{\includegraphics[width=0.65\columnwidth]{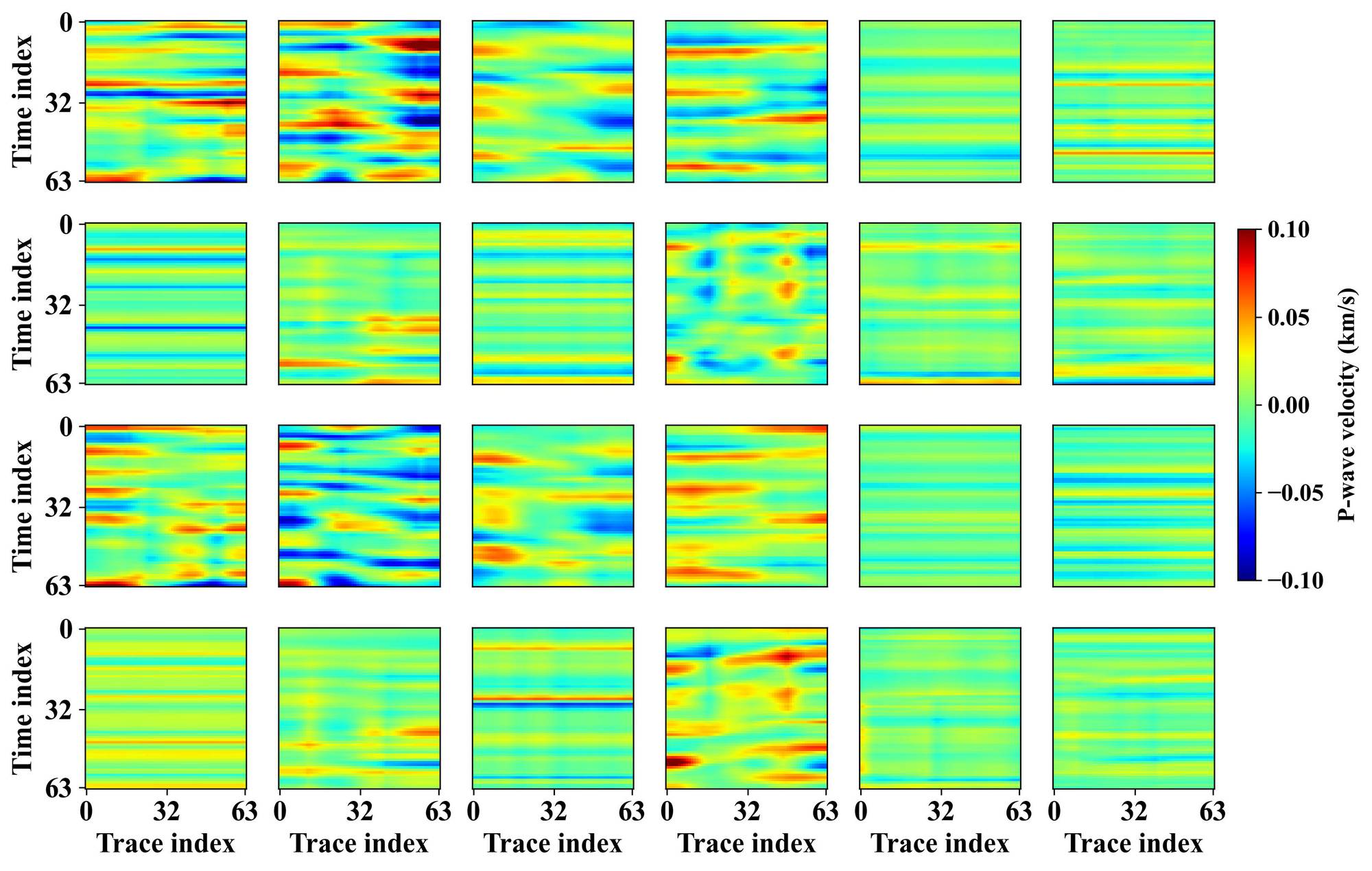}
 \label{fig:ddpm-condlowdps-vp-smooth}}
 \subfigure[]{\includegraphics[width=0.65\columnwidth]{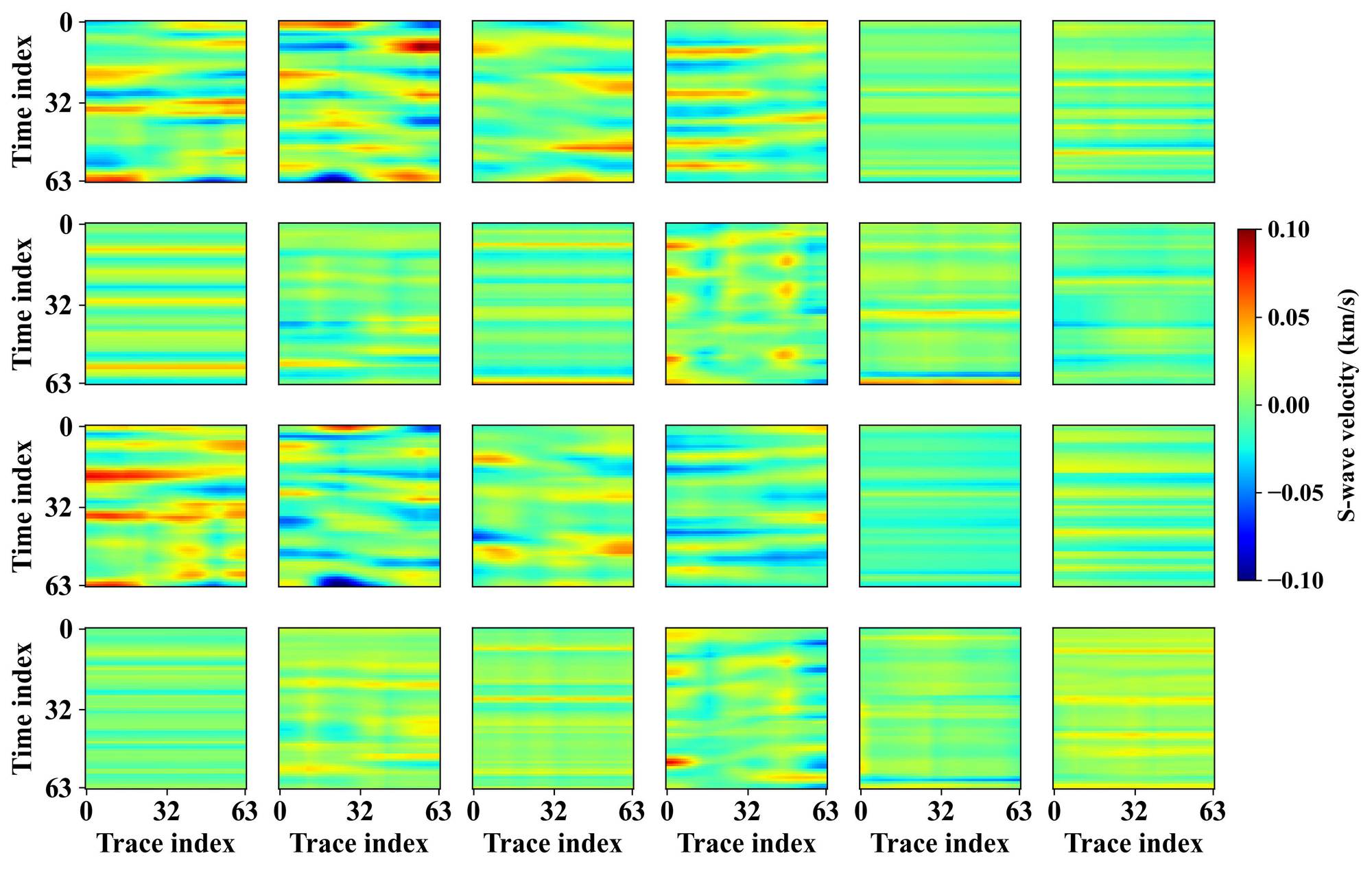}
 \label{fig:ddpm-condlowdps-vs-smooth}} 
  \subfigure[]{\includegraphics[width=0.65\columnwidth]{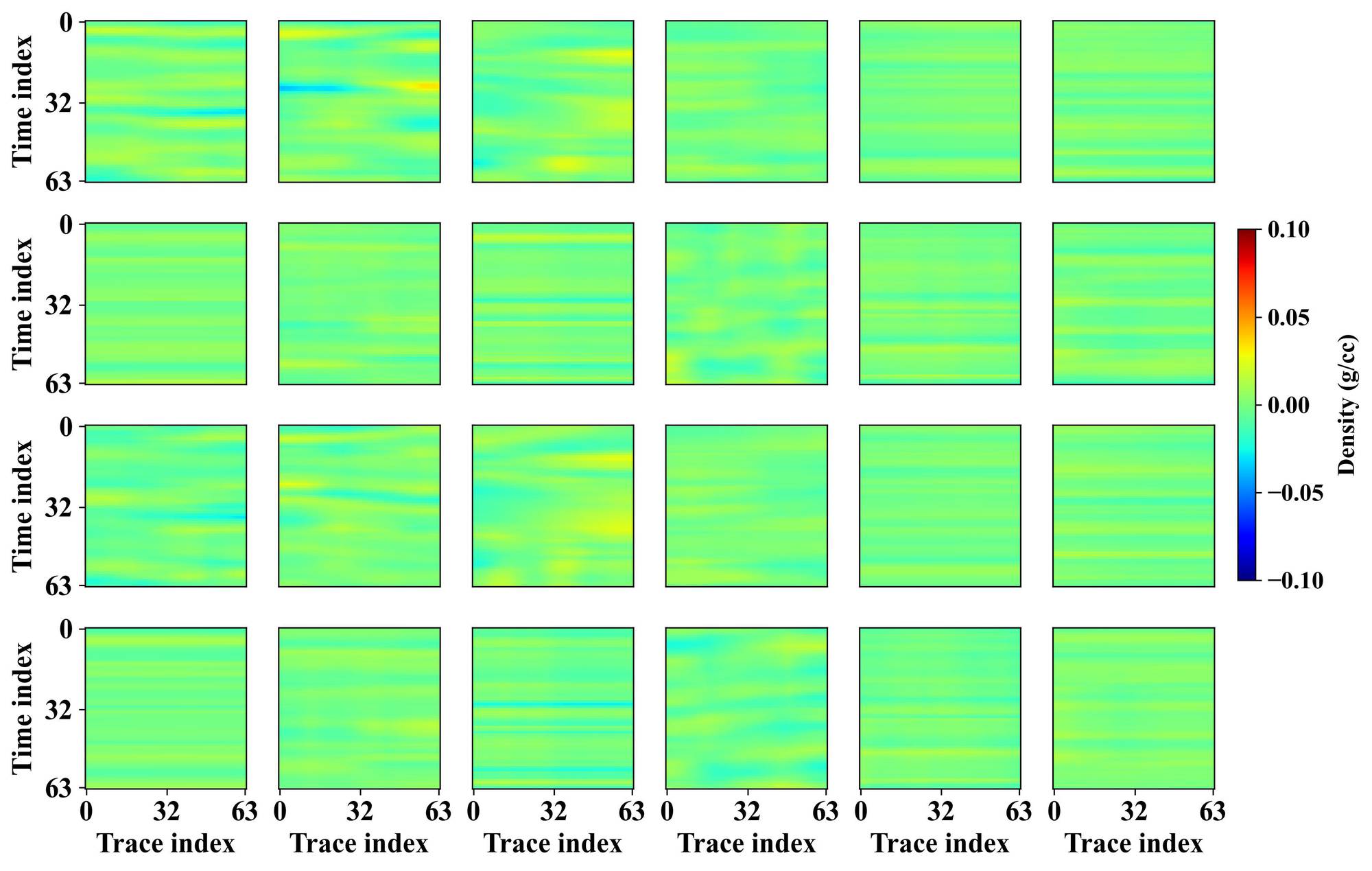}
 \label{fig:ddpm-condlowdps-rho-smooth}}   
 \caption{Errors between the low-frequency components of the generated samples and the provided low-frequency models shown in Figs. \ref{fig:condlow-vp}--\ref{fig:condlow-rho}. 
The low-frequency components are obtained by applying a $31 \times 31$ mean filter to the generated samples. 
(a)--(c) Errors corresponding to the samples generated by the proposed DPS-projection. 
(d)--(f) Errors corresponding to the samples generated by DPS.
}
\label{fig:lowsamples31erro}
\end{figure*}

\textit{\textbf{Elastic parameter synthesis conditioned on low-frequency models:}} This experiment evaluates the effectiveness of the proposed method for elastic parameter synthesis conditioned on low-frequency models. The low-frequency models are obtained by applying a $31 \times 31$ mean filter to the true elastic parameter models, as shown in Figs. \ref{fig:condlow-vp}--\ref{fig:condlow-rho}. The samples synthesized by the proposed DPS-projection are presented in Figs. \ref{fig:ddpm-condlow-vp}--\ref{fig:ddpm-condlow-rho}. These samples exhibit overall trends that are consistent with the prescribed low-frequency models. This consistency is further supported by the error analysis in Figs. \ref{fig:ddpm-condlow-vp-smooth}--\ref{fig:ddpm-condlow-rho-smooth}, where the errors are calculated between the low-frequency components of the synthesized samples, obtained using the same $31 \times 31$ mean filter, and the prescribed low-frequency models. Additionally, as observed from Figs. \ref{fig:ddpm-condlow-vp}--\ref{fig:ddpm-condlow-rho}, the proposed low-frequency conditioning strategy reduces the variability among different sampling runs, as reflected by the relatively small differences between repeated synthesis results. For comparison, samples synthesized by DPS are shown in Figs. \ref{fig:ddpm-condlowdps-vp}--\ref{fig:ddpm-condlowdps-rho}. The DPS results generally follow the large-scale trends of the prescribed low-frequency models, indicating that the low-frequency information is incorporated into the diffusion-based synthesis process to some extent. However, the two repeated DPS sampling results show more pronounced differences than those obtained by the proposed method, suggesting that the low-frequency constraint is not enforced as strongly in DPS. The larger errors in Figs. \ref{fig:ddpm-condlowdps-vp-smooth}--\ref{fig:ddpm-condlowdps-rho-smooth} further confirm this observation, showing that the mean-filtered DPS samples deviate more from the given low-frequency models.

\begin{figure*}[htb!]
\setlength{\abovecaptionskip}{0.2cm}
 \centering
    \subfigure[]{\includegraphics[width=0.65\columnwidth]{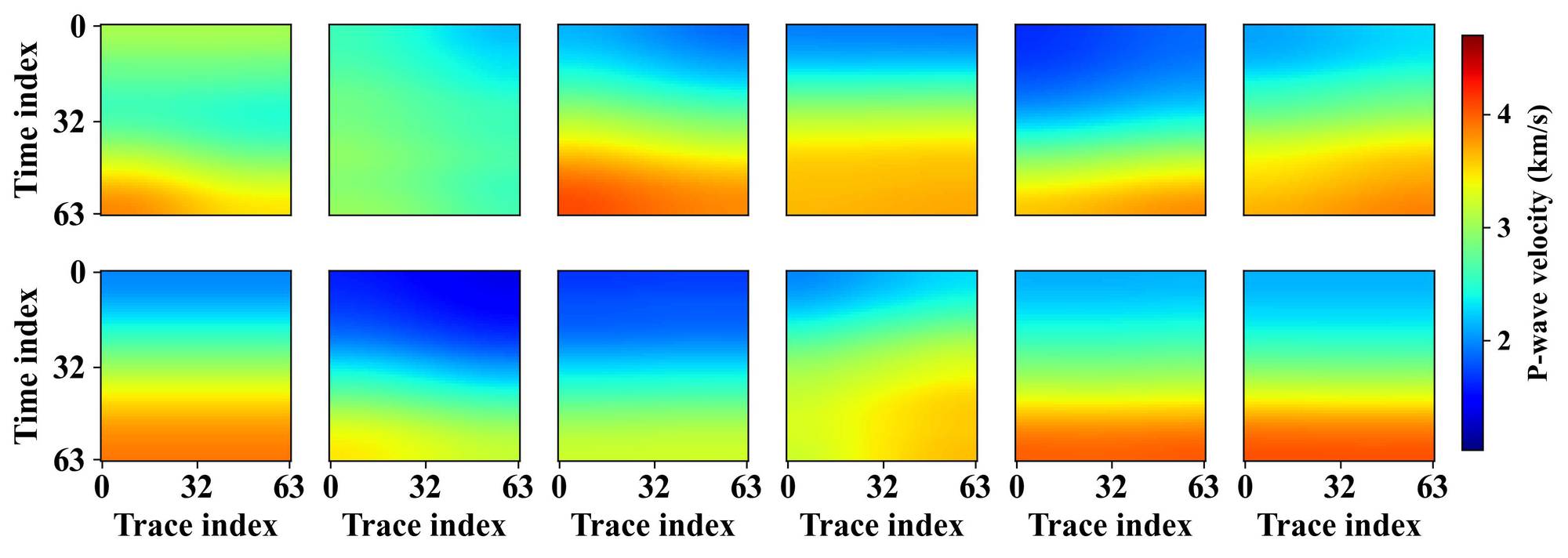}
 \label{fig:condlow2-vp}}
 \subfigure[]{\includegraphics[width=0.65\columnwidth]{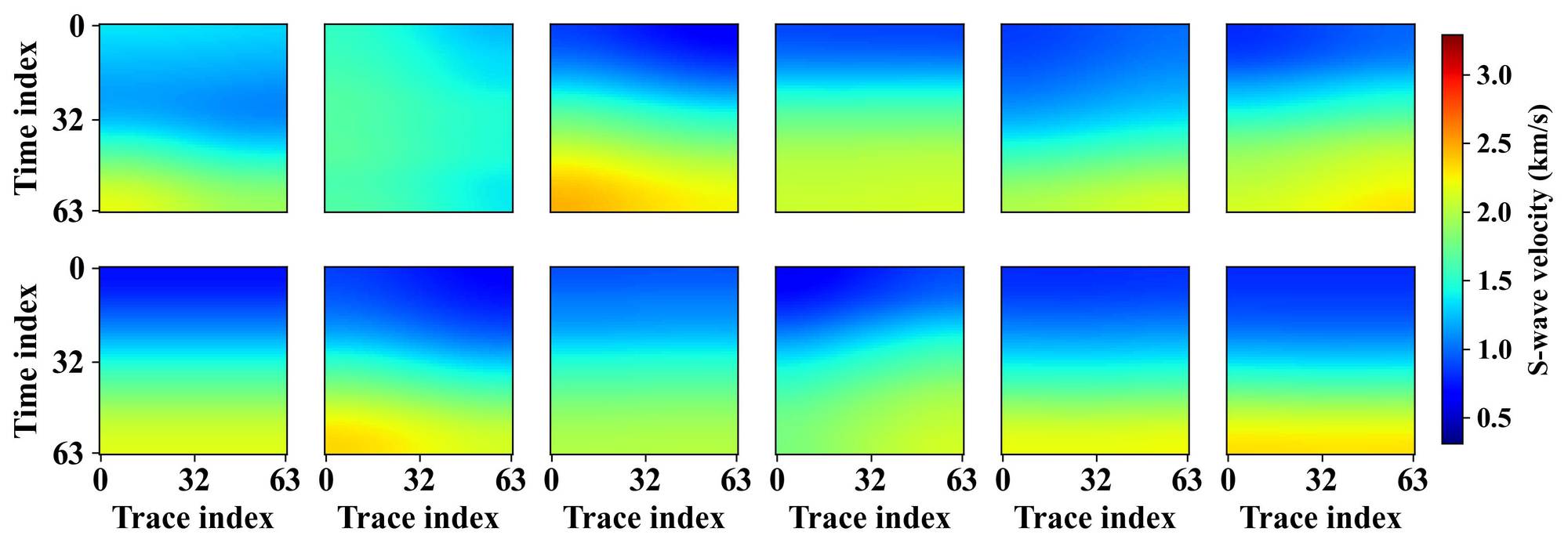}
 \label{fig:condlow2-vs}} 
  \subfigure[]{\includegraphics[width=0.65\columnwidth]{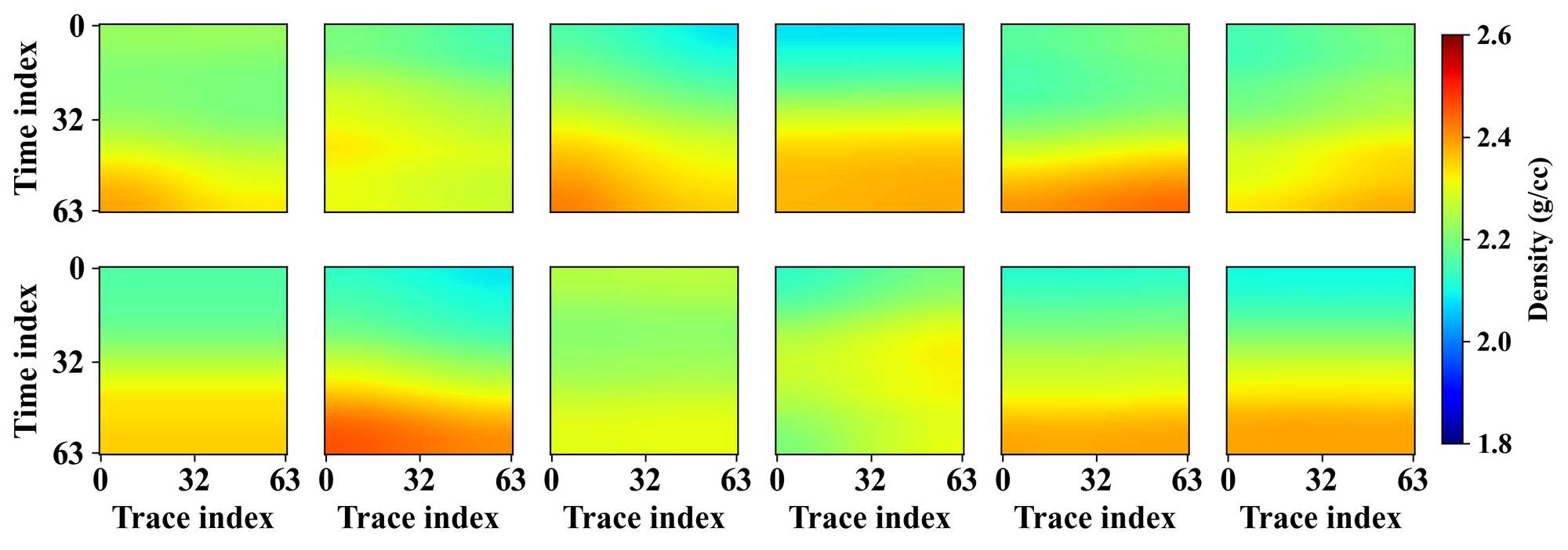}
 \label{fig:condlow2-rho}}
   \subfigure[]{\includegraphics[width=0.65\columnwidth]{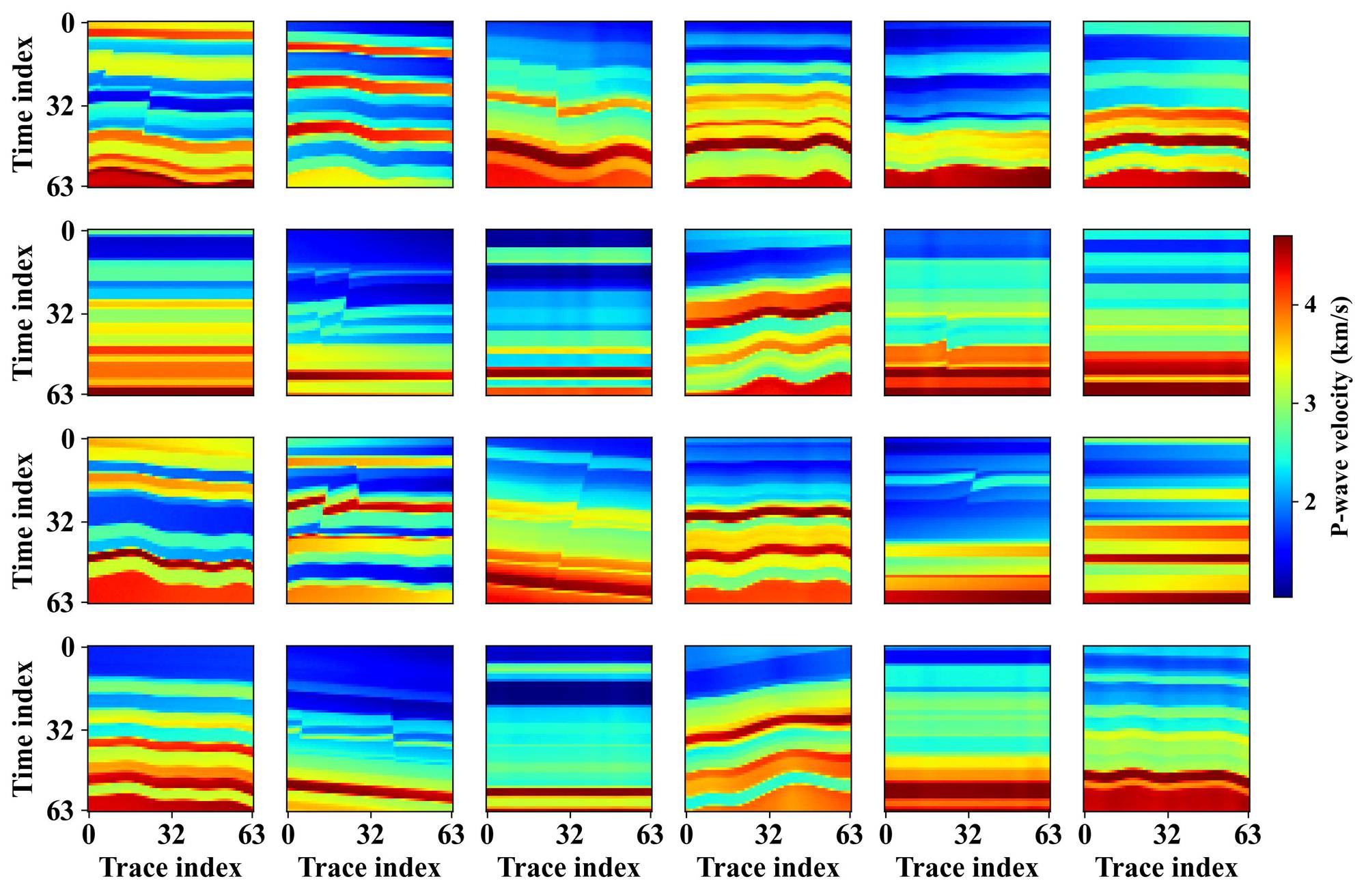}
 \label{fig:ddpm-condlow2-vp}}
 \subfigure[]{\includegraphics[width=0.65\columnwidth]{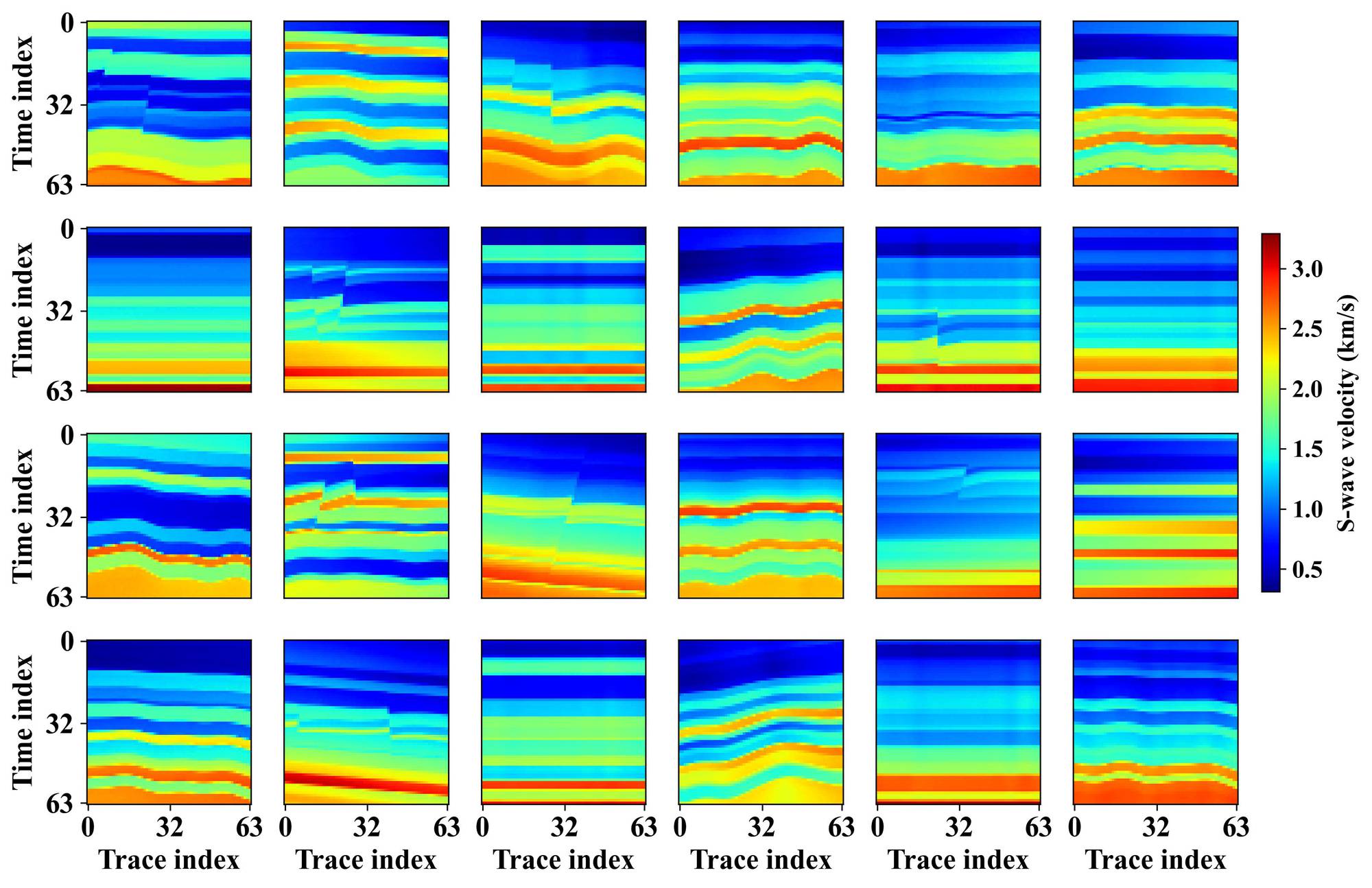}
 \label{fig:ddpm-condlow2-vs}} 
  \subfigure[]{\includegraphics[width=0.65\columnwidth]{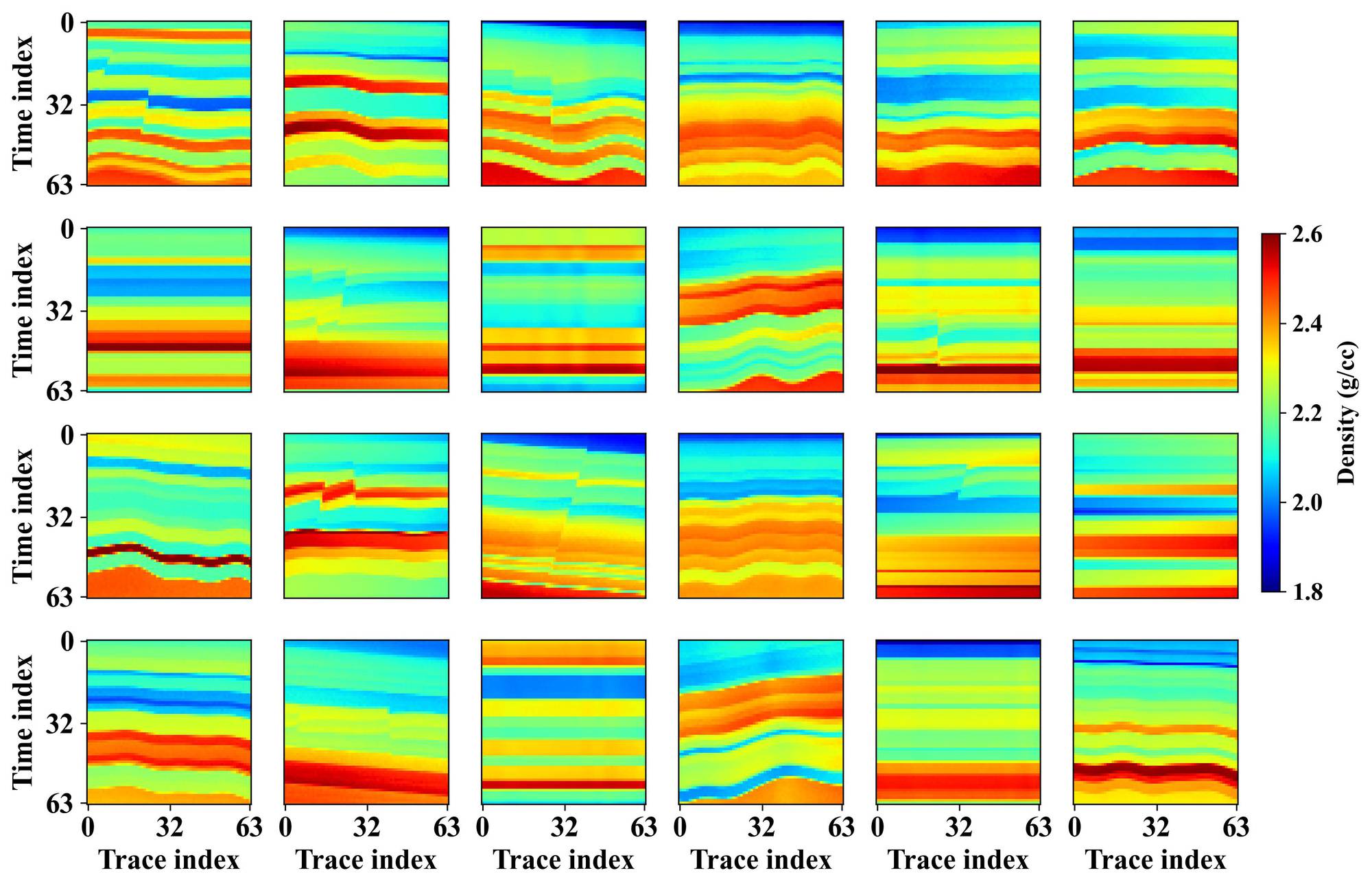}
 \label{fig:ddpm-condlow2-rho}} 
 \caption{Elastic parameter synthesis conditioned on low-frequency models. 
(a)--(c) Low-frequency models derived by applying a $31 \times 31$ mean filter twice. 
(d)--(f) Samples generated by the proposed DPS-projection. 
In each panel, the first two rows show the results from the first sampling run, whereas the last two rows show the results from the second sampling run.}
\label{fig:lowsamples312}
\end{figure*}

Finally, we further examine the influence of low-frequency model smoothness using smoother low-frequency models obtained by applying the $31 \times 31$ mean filter twice to the true elastic parameter models. As shown in Fig. \ref{fig:lowsamples312}, the samples synthesized by the proposed DPS-projection still follow the large-scale trends of the prescribed low-frequency models. However, compared with the results obtained using the low-frequency models filtered once, the repeated sampling results exhibit greater variability. This is because the smoother low-frequency models contain less local information and therefore impose weaker constraints on the synthesis process. These observations indicate that the informativeness of the low-frequency model directly affects the diversity of the synthesized samples, reflecting a trade-off between conditional control and sample variability.

\begin{figure*}[htb!]
\setlength{\abovecaptionskip}{0.2cm}
 \centering
    \subfigure[]{\includegraphics[width=0.65\columnwidth]{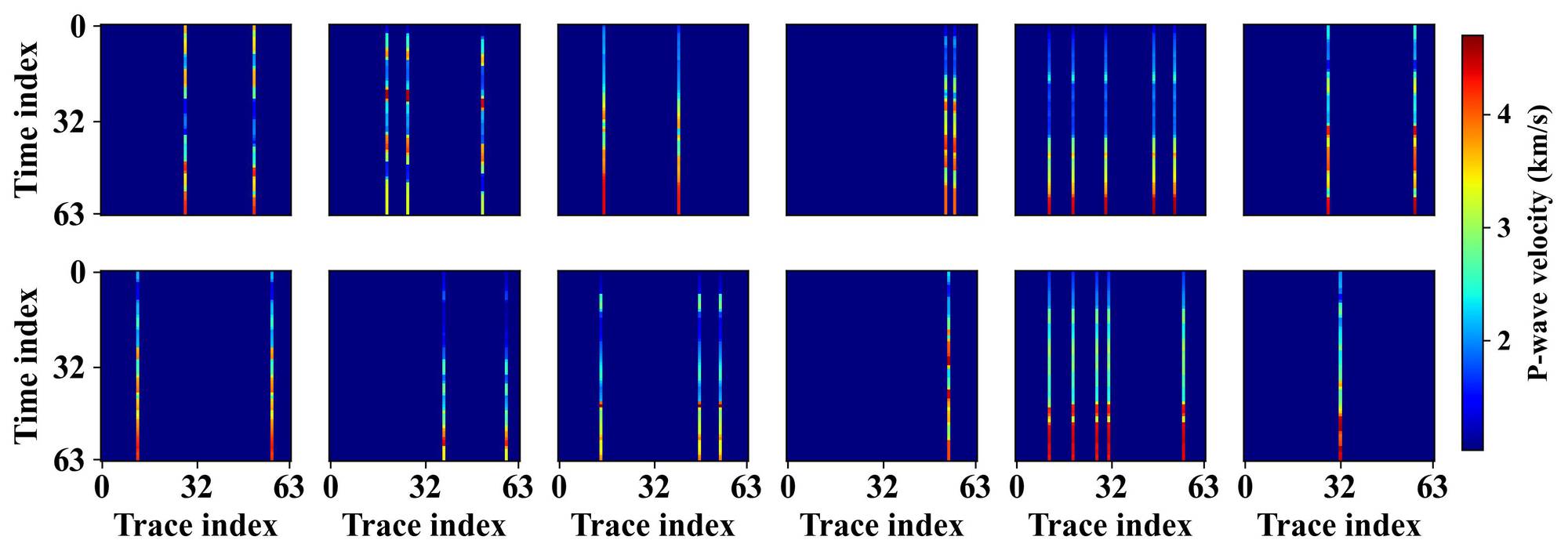}
 \label{fig:condlog-vp}}
 \subfigure[]{\includegraphics[width=0.65\columnwidth]{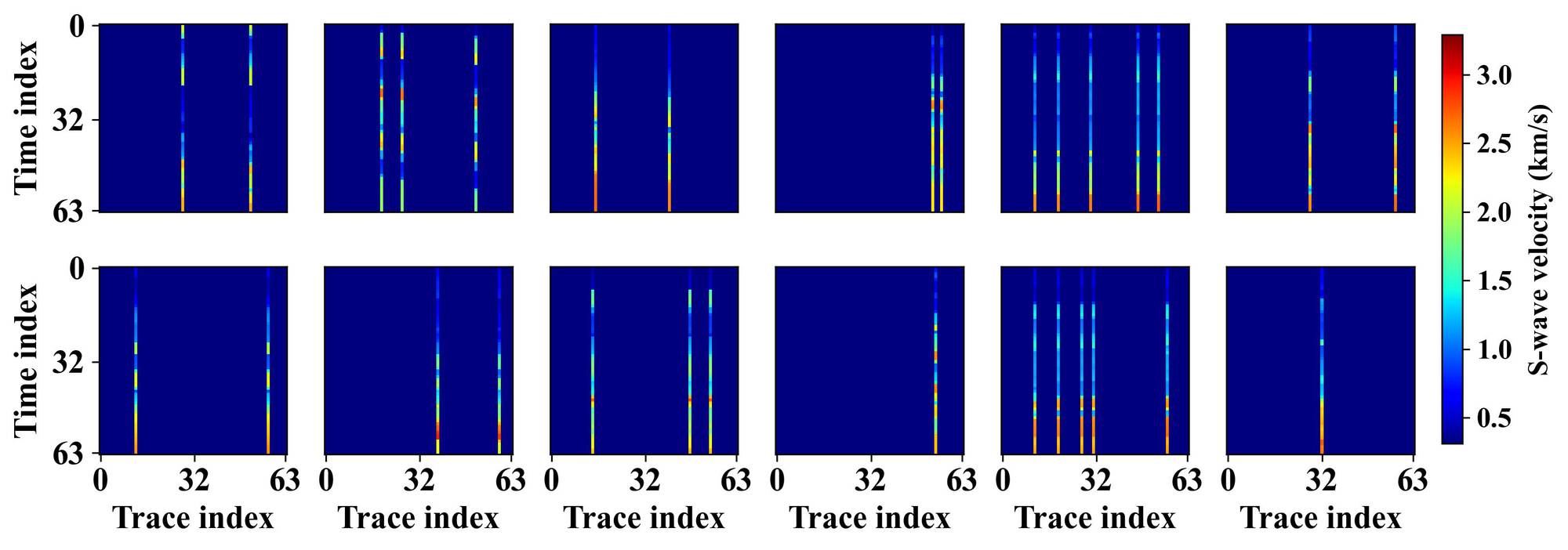}
 \label{fig:condlog-vs}} 
  \subfigure[]{\includegraphics[width=0.65\columnwidth]{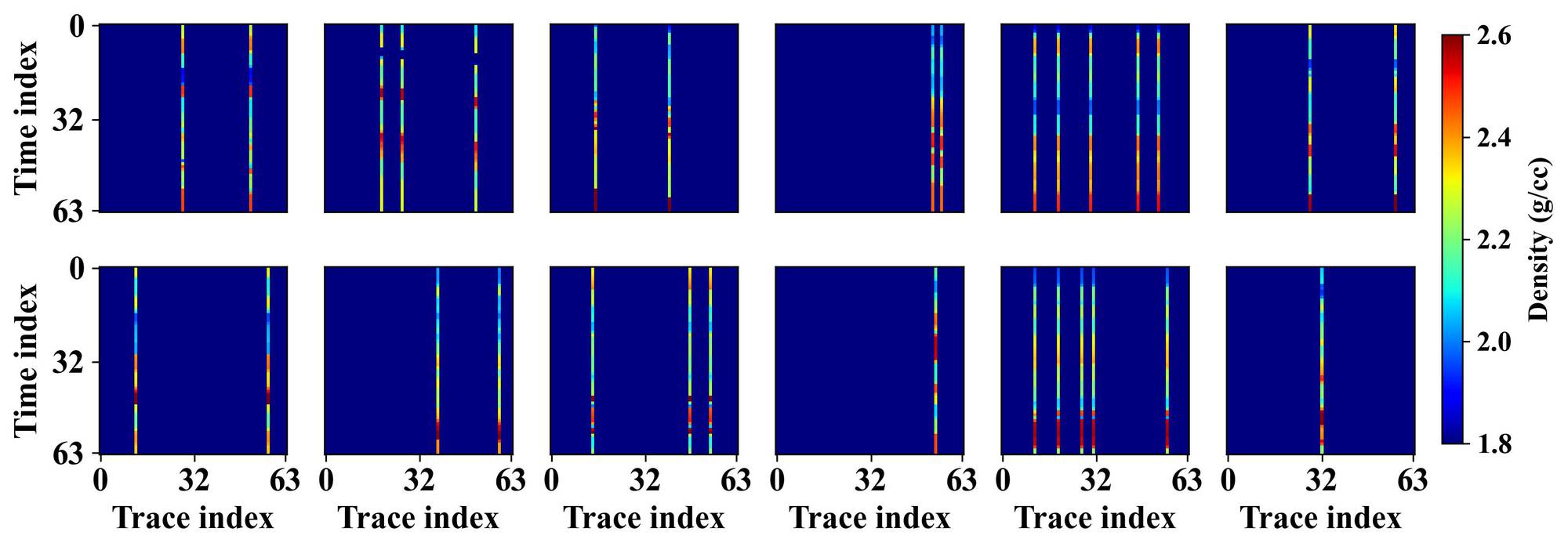}
 \label{fig:condlog-rho}}
   \subfigure[]{\includegraphics[width=0.65\columnwidth]{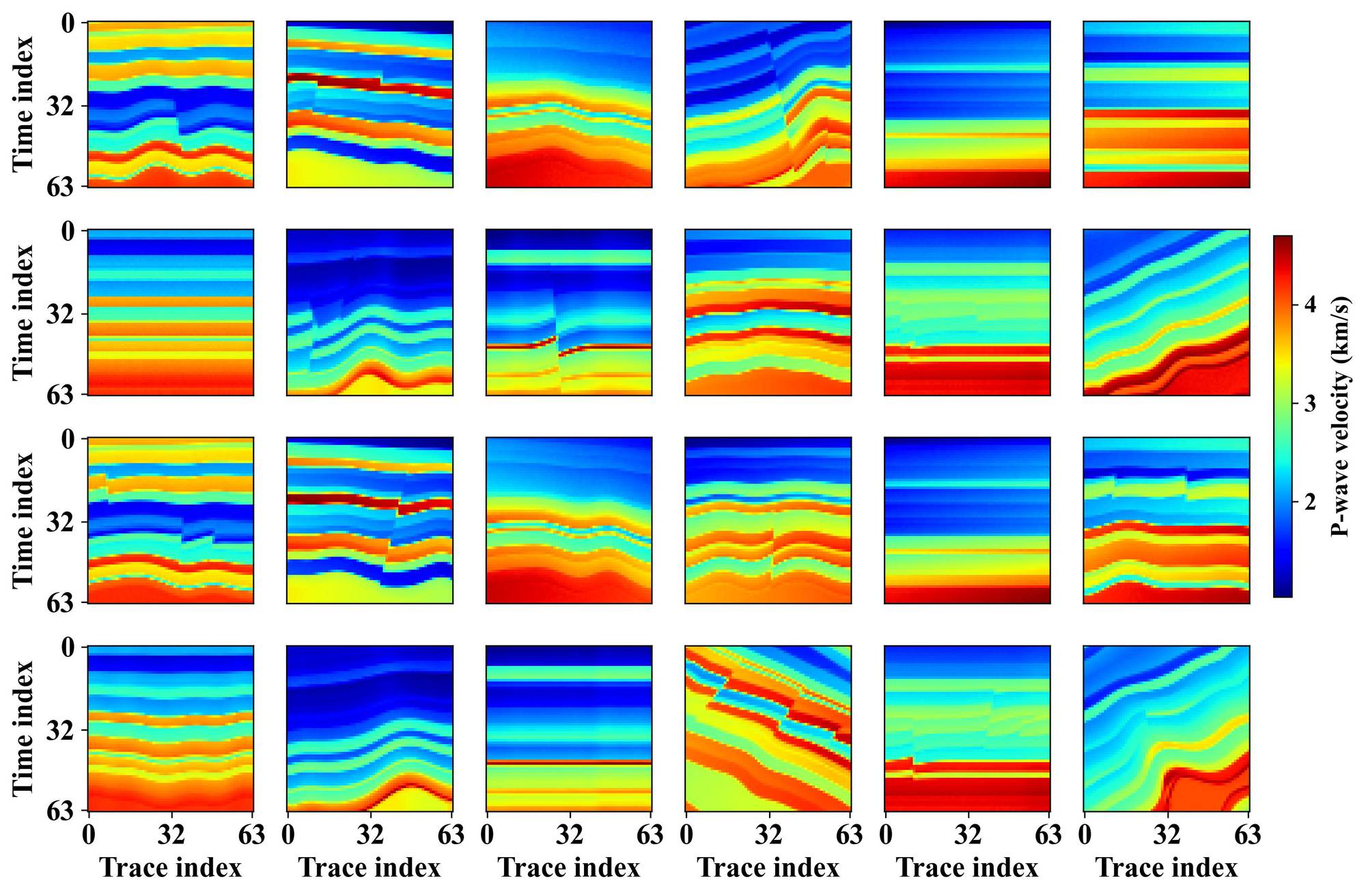}
 \label{fig:ddpm-condlog-vp}}
 \subfigure[]{\includegraphics[width=0.65\columnwidth]{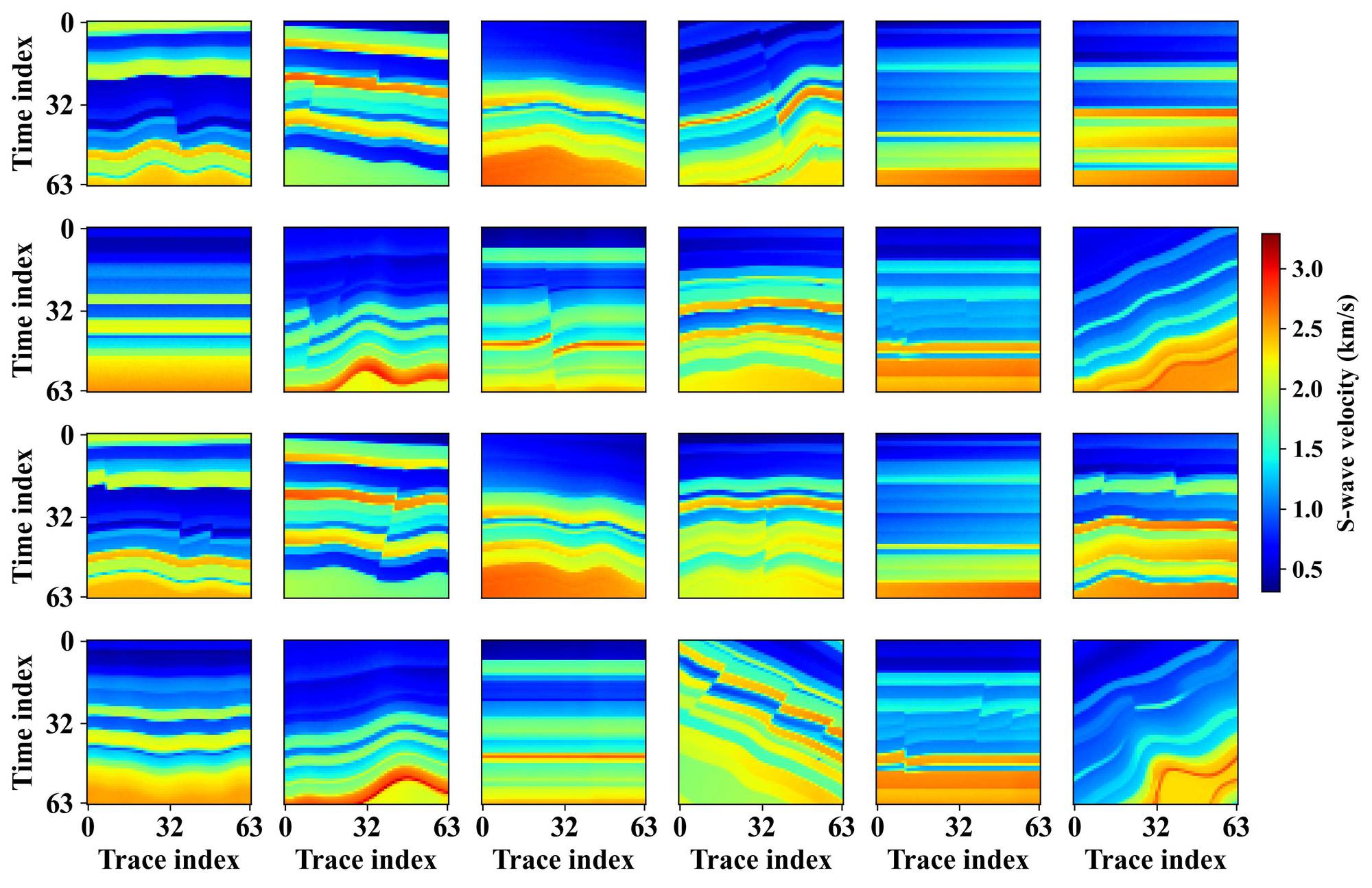}
 \label{fig:ddpm-condlog-vs}} 
  \subfigure[]{\includegraphics[width=0.65\columnwidth]{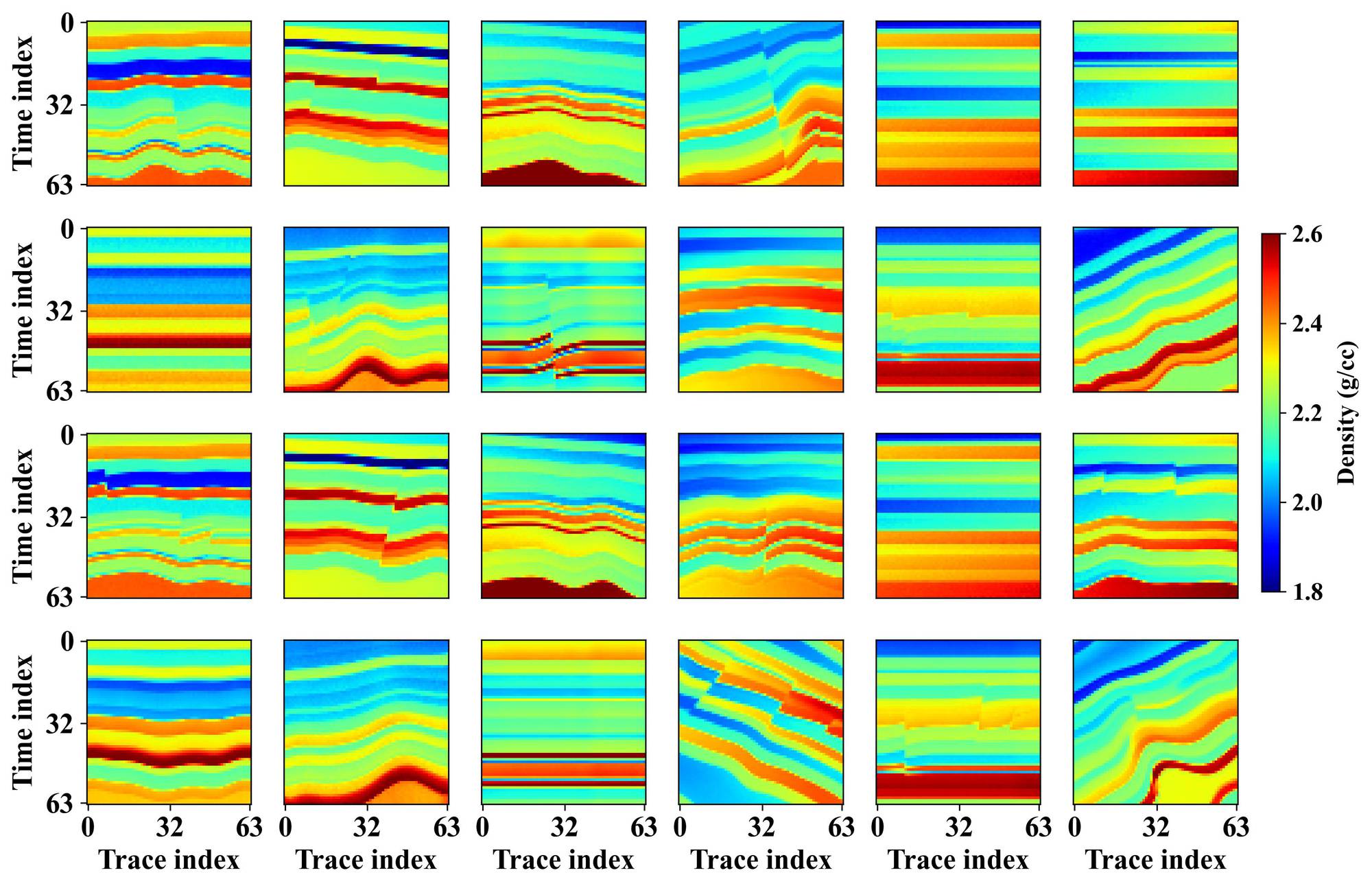}
 \label{fig:ddpm-condlog-rho}}  
   \subfigure[]{\includegraphics[width=0.65\columnwidth]{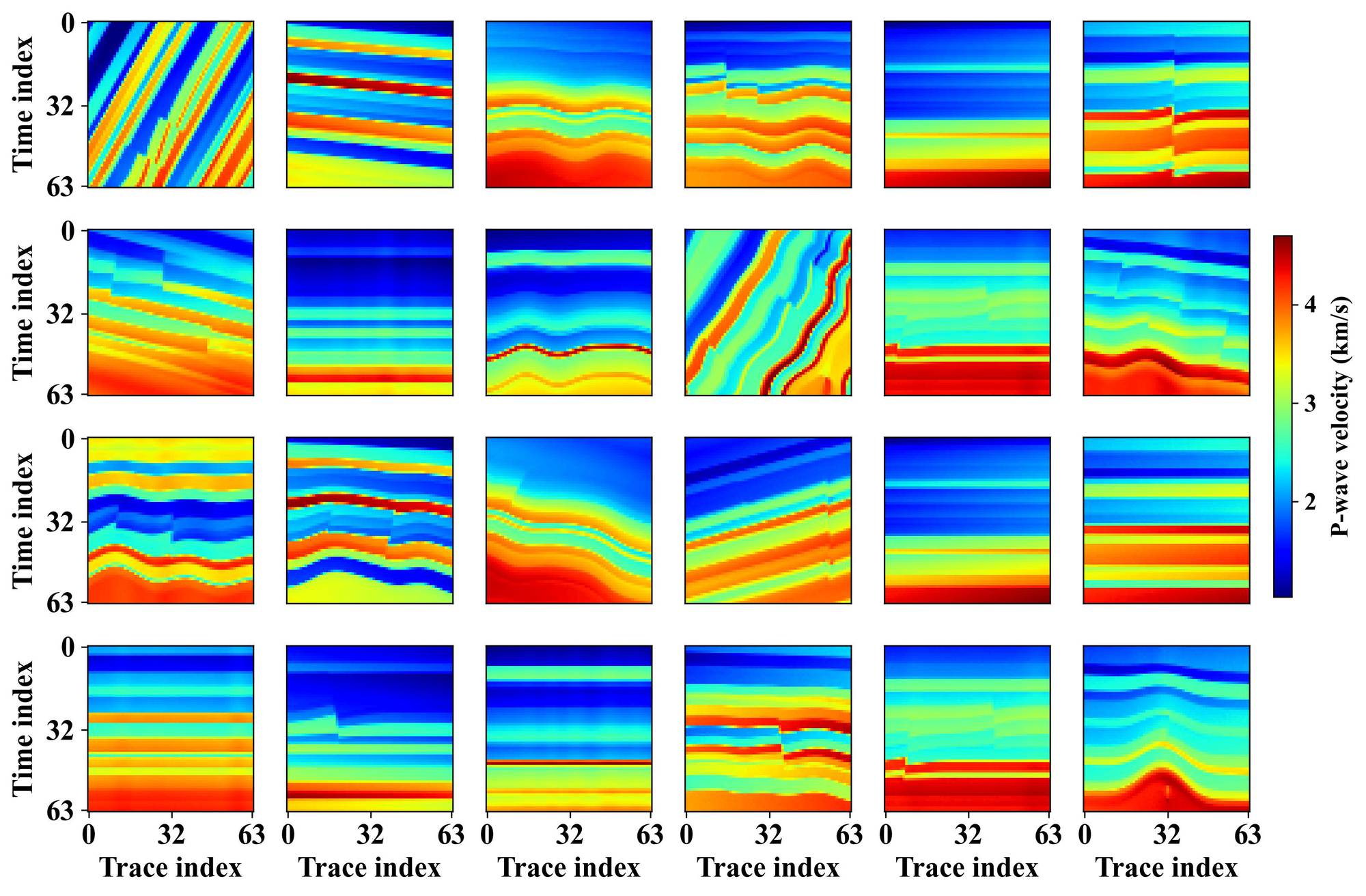}
 \label{fig:ddpm-condlogdps-vp}}
 \subfigure[]{\includegraphics[width=0.65\columnwidth]{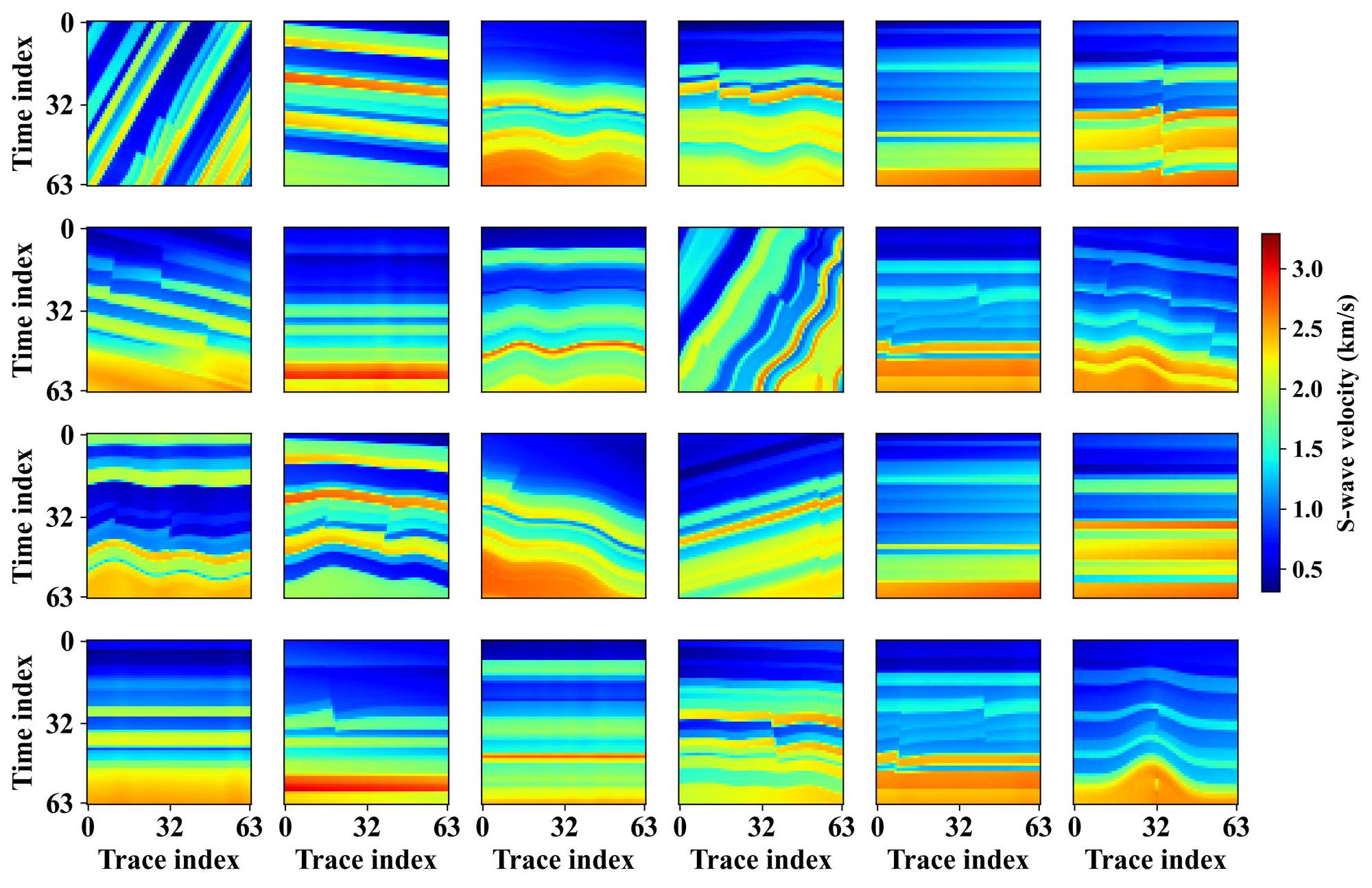}
 \label{fig:ddpm-condlogdps-vs}} 
  \subfigure[]{\includegraphics[width=0.65\columnwidth]{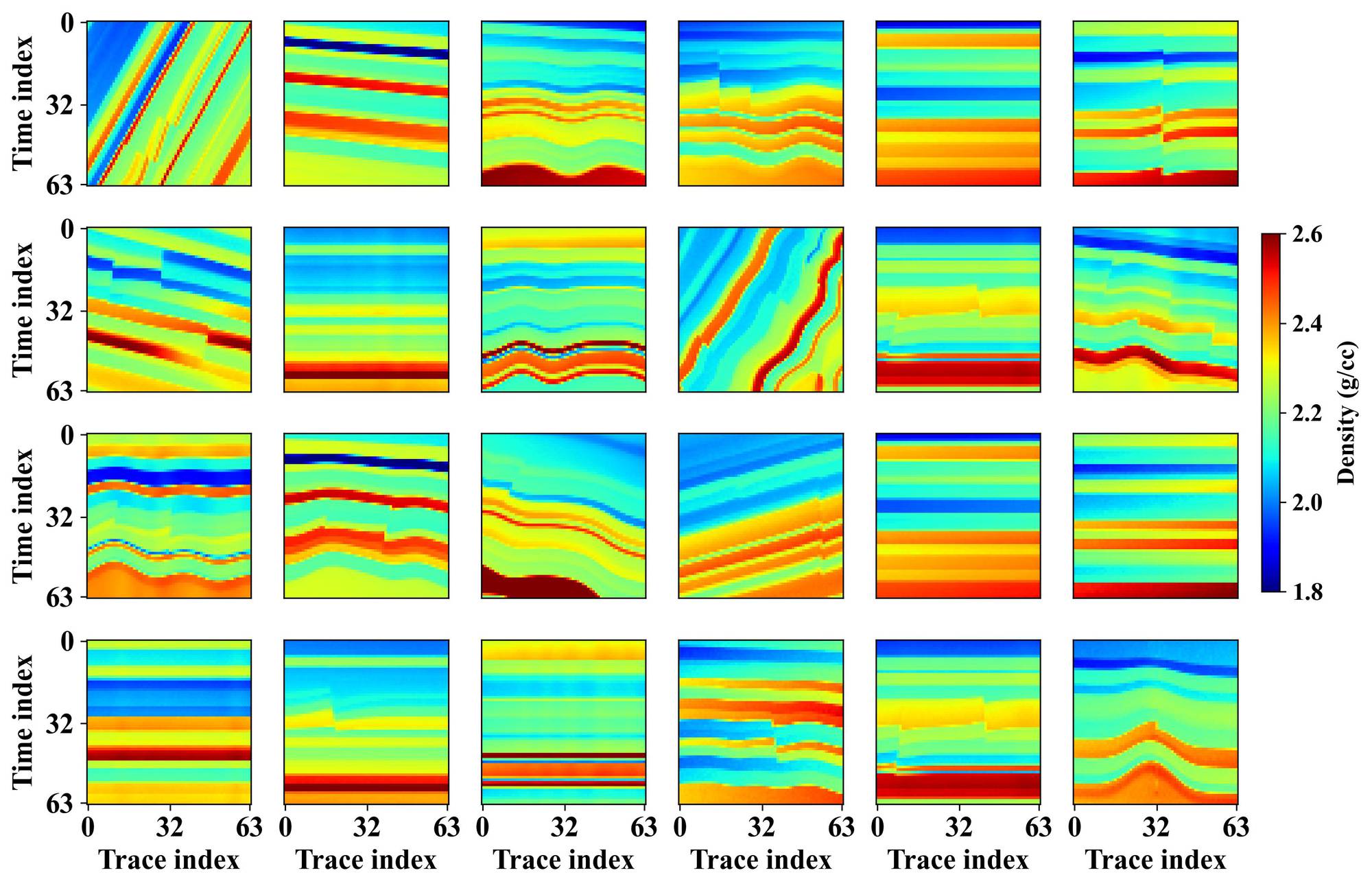}
 \label{fig:ddpm-condlogdps-rho}}  
 \caption{Elastic parameter synthesis conditioned on pseudo-well logs. 
(a)--(c) Pseudo-well logs of P-wave velocity, S-wave velocity, and density. 
(d)--(f) Samples generated by the proposed DPS-projection. 
(g)--(i) Samples generated by DPS. 
In each panel, the first two rows show the results from the first sampling run, whereas the last two rows show the results from the second sampling run.}
\label{fig:logsamples}
\end{figure*}

\begin{figure*}[htb!]
\setlength{\abovecaptionskip}{0.2cm}
 \centering
    \subfigure[]{\includegraphics[width=0.65\columnwidth]{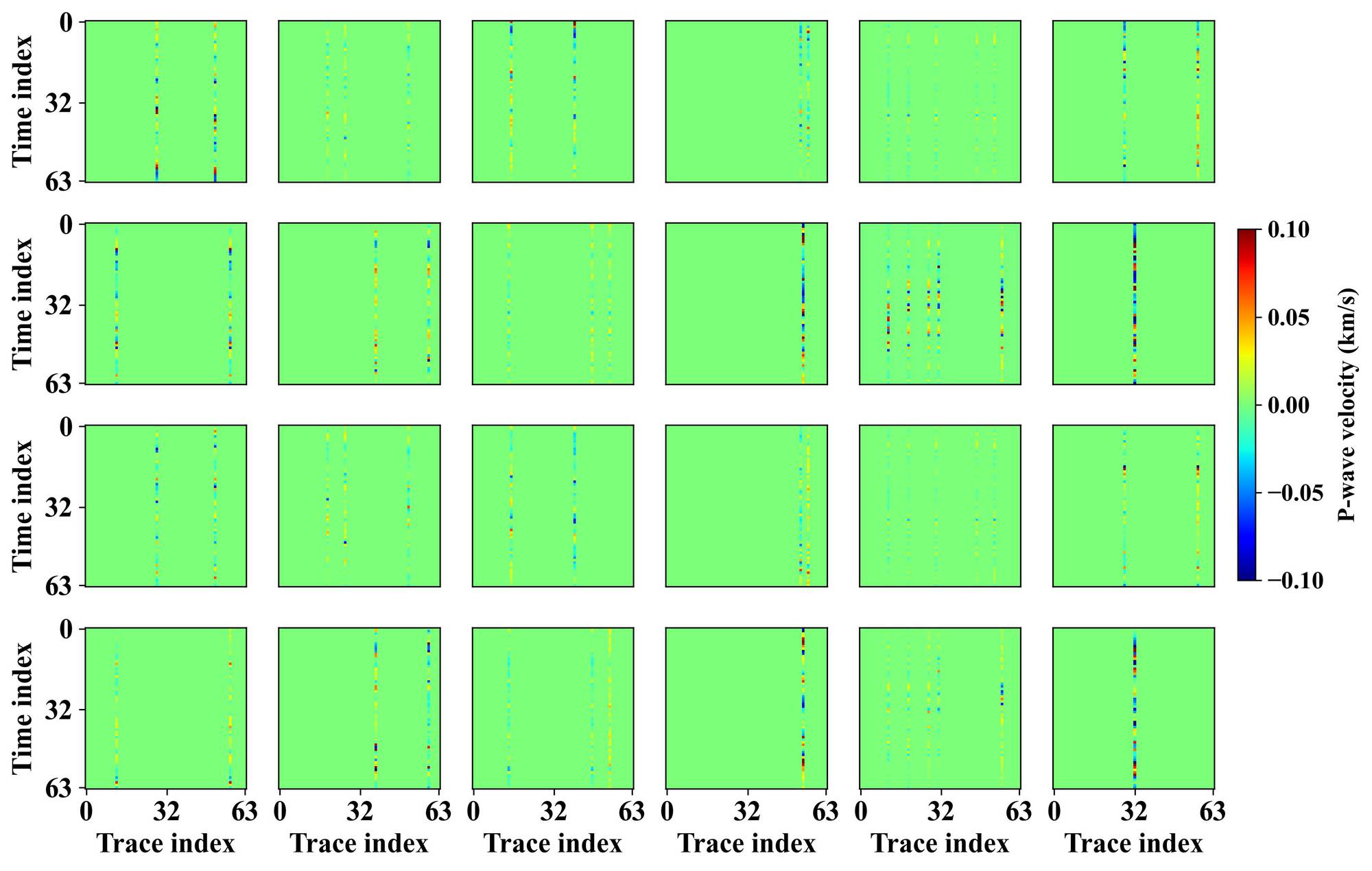}
 \label{fig:ddpm-condlog-vp-log}}
 \subfigure[]{\includegraphics[width=0.65\columnwidth]{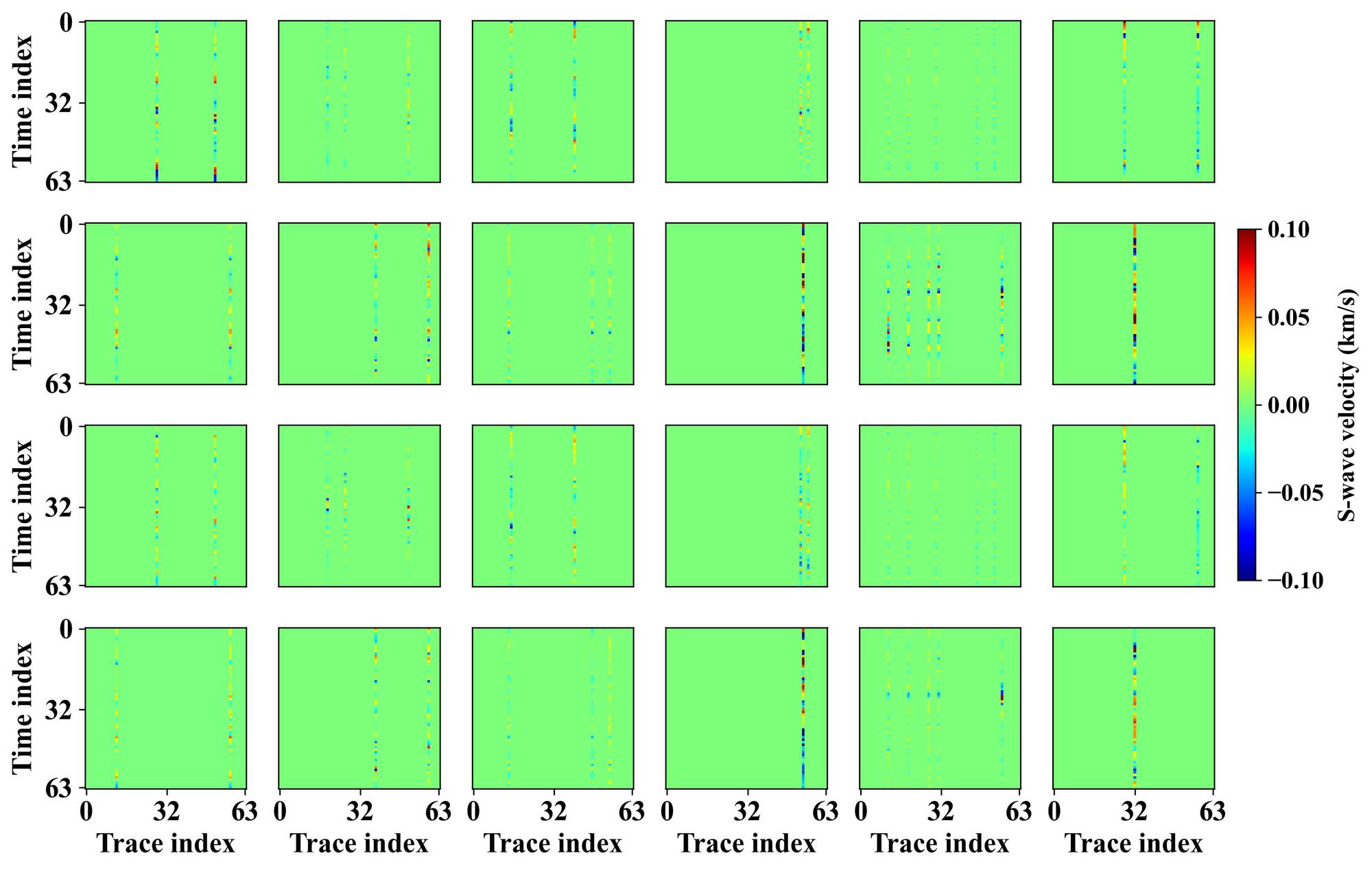}
 \label{fig:ddpm-condlog-vs-log}} 
  \subfigure[]{\includegraphics[width=0.65\columnwidth]{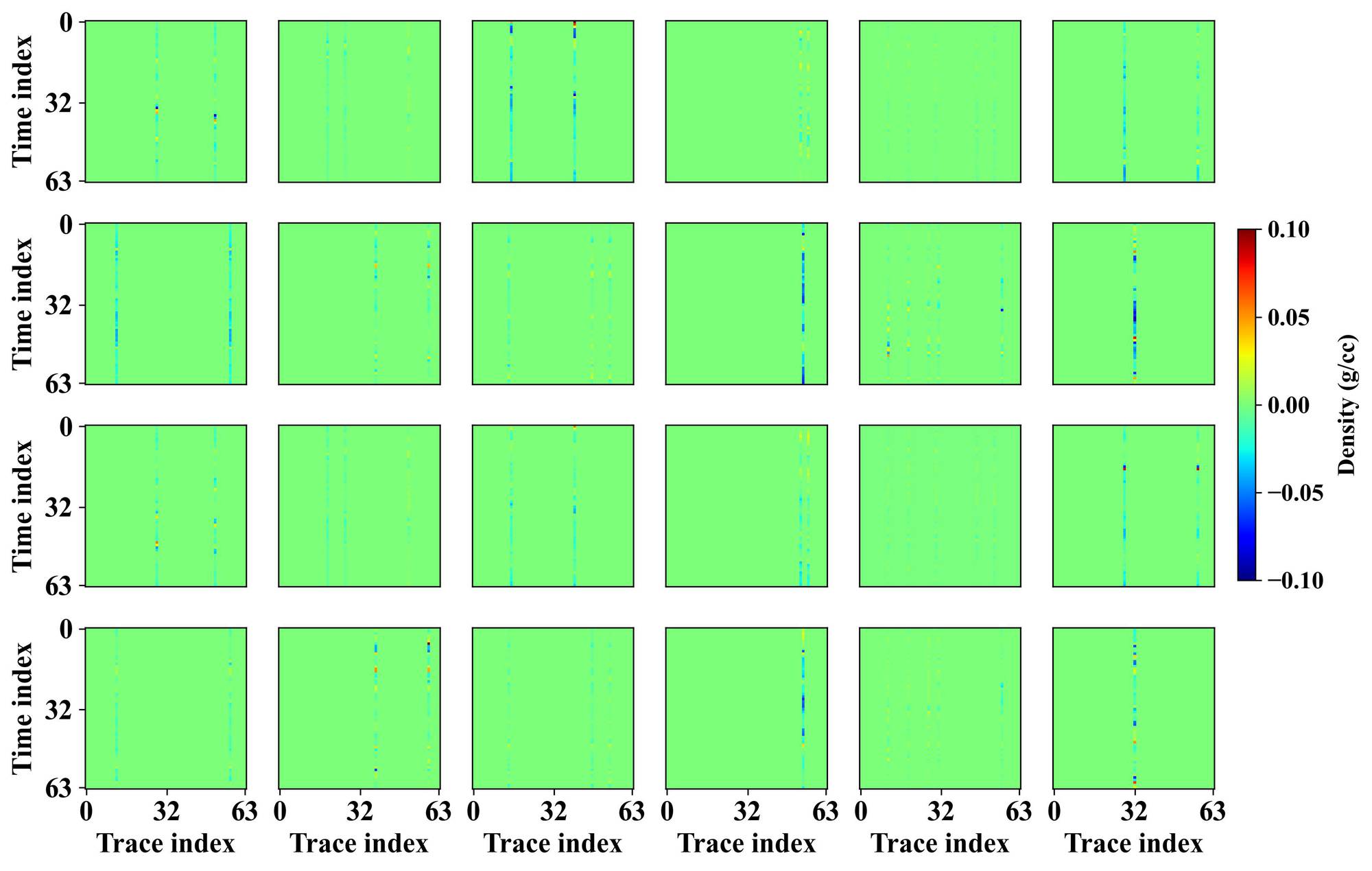}
 \label{fig:ddpm-condlog-rho-log}} 
   \subfigure[]{\includegraphics[width=0.65\columnwidth]{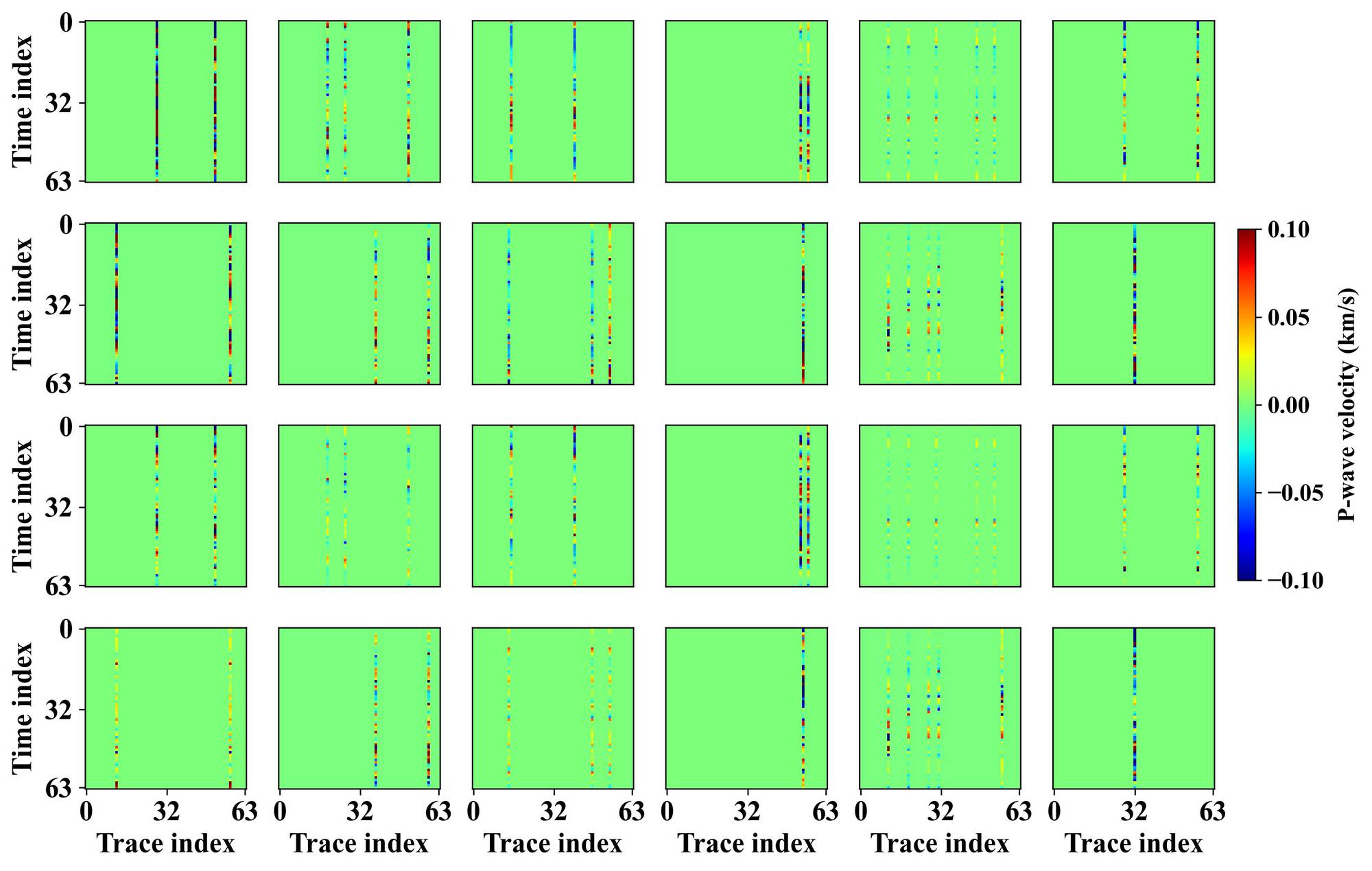}
 \label{fig:ddpm-condlogdps-vp-log}}
 \subfigure[]{\includegraphics[width=0.65\columnwidth]{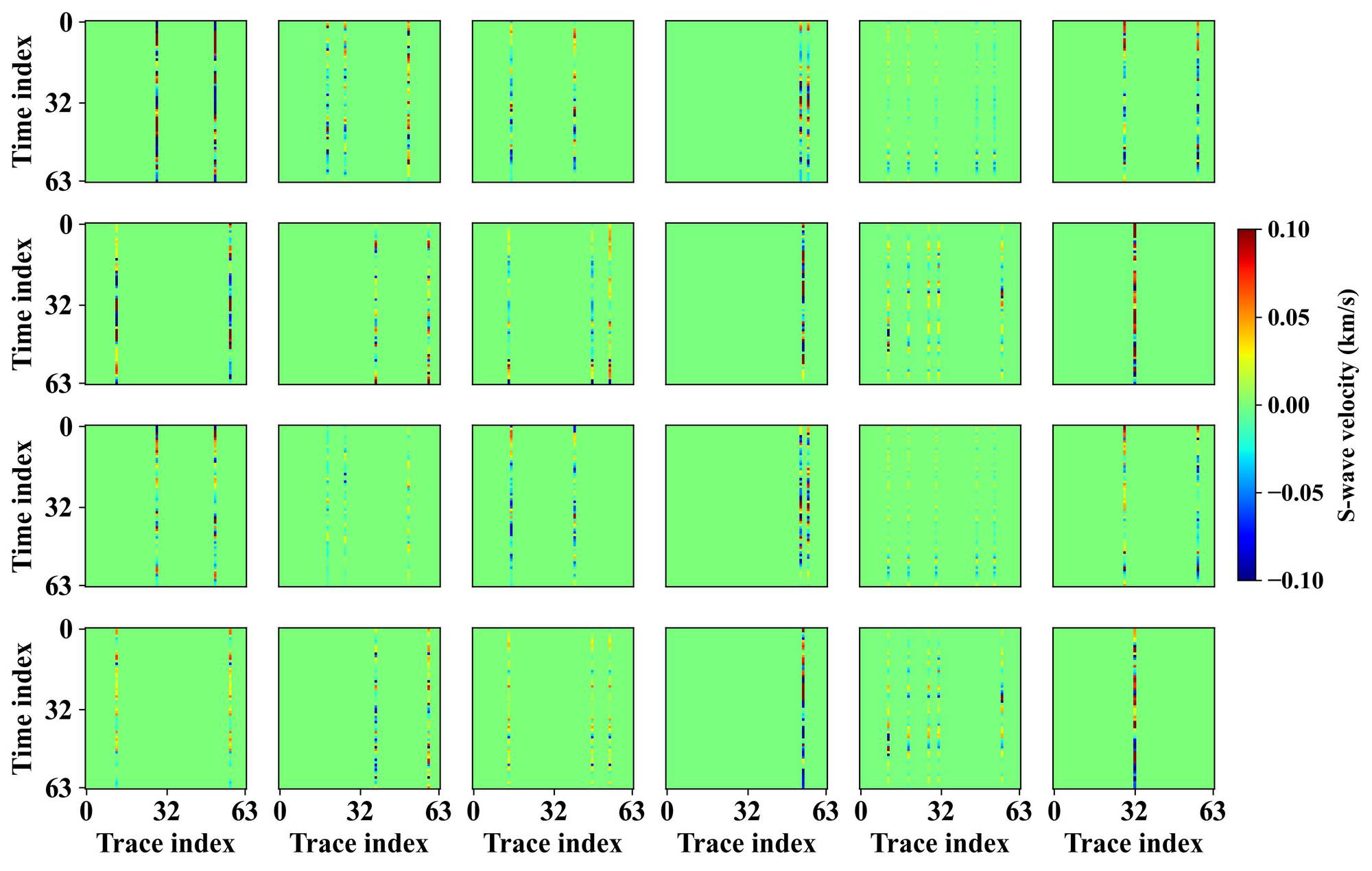}
 \label{fig:ddpm-condlogdps-vs-log}} 
  \subfigure[]{\includegraphics[width=0.65\columnwidth]{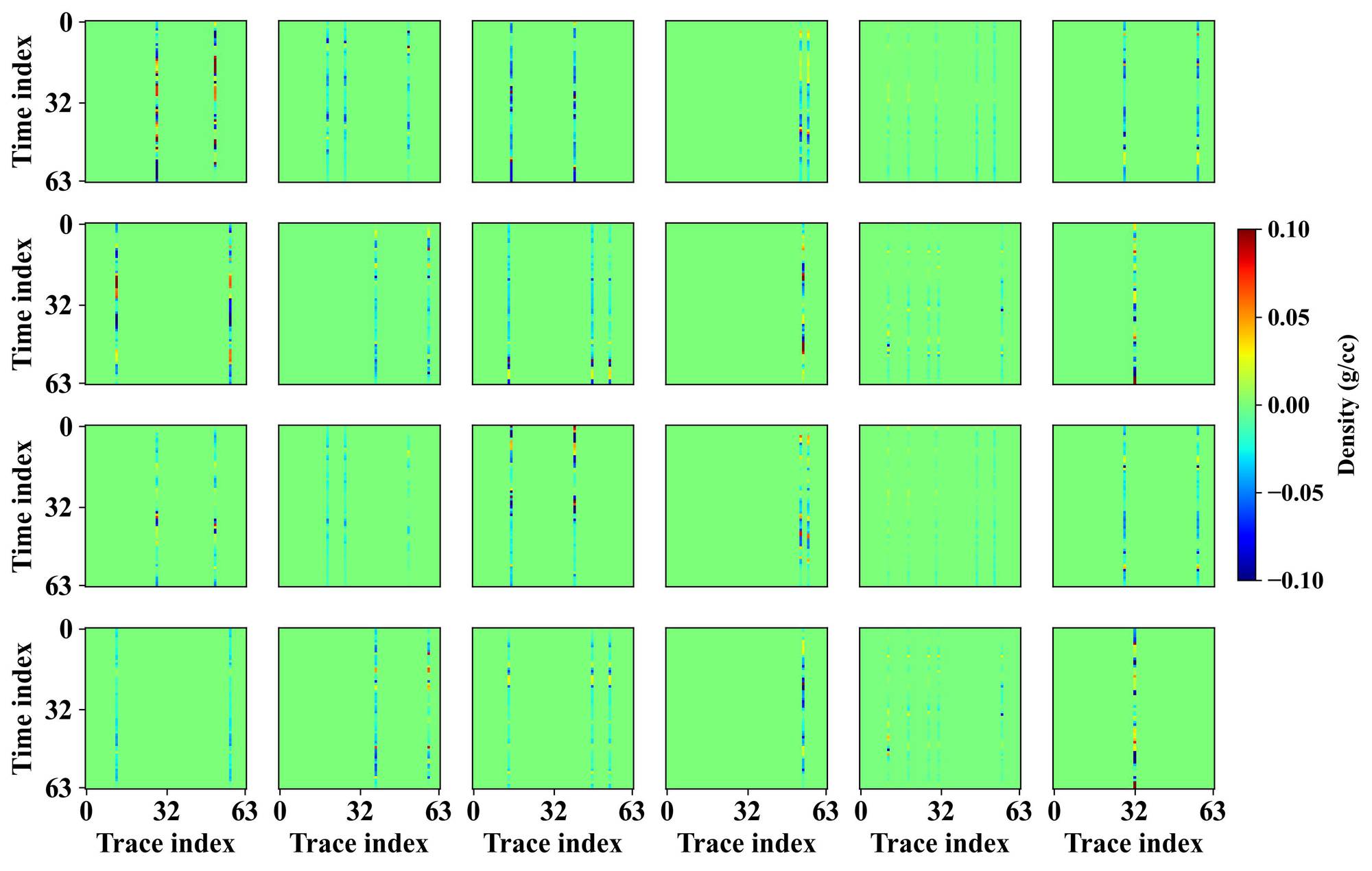}
 \label{fig:ddpm-condlogdps-rho-log}}   
 \caption{Errors between the logs extracted from the generated samples at the pseudo-well locations and the provided pseudo-well logs shown in Fig. \ref{fig:condlog-vp}--\ref{fig:condlog-rho}.
(a)--(c) Errors corresponding to the samples generated by the proposed DPS-projection.
(d)--(f) Errors corresponding to the samples generated by DPS.}
\label{fig:logsampleserro}
\end{figure*}

\textbf{\textit{Elastic parameter synthesis conditioned on well logs:}} This experiment is conducted to evaluate the capability of the proposed method to incorporate pseudo-well logs into the diffusion-based synthesis process. Figs. \ref{fig:condlog-vp}--\ref{fig:condlog-rho} show the pseudo-well logs of P-wave velocity, S-wave velocity, and density, where the nonzero values correspond to the logs at the pseudo-well locations. The samples generated by the proposed DPS-projection are shown in Figs. \ref{fig:ddpm-condlog-vp}--\ref{fig:ddpm-condlog-rho}, whereas those generated by DPS are shown in Figs. \ref{fig:ddpm-condlogdps-vp}--\ref{fig:ddpm-condlogdps-rho}. Both methods produce relatively diverse samples because the pseudo-well logs provide only sparse constraints. Specifically, the well log condition directly constrains the generated elastic parameters only at the pseudo-well locations, leaving considerable freedom in other regions. As a result, the two repeated sampling runs exhibit noticeable structural differences in regions away from the pseudo-well locations. Nevertheless, the samples generated by the proposed method maintain reasonable spatial continuity around the pseudo-well locations, indicating that the well log information is incorporated into the surrounding model space rather than being imposed only as isolated point constraints. Compared with the proposed method, the well log constraint appears to be less stably enforced in DPS. As a result, isolated amplitude anomalies can be observed near the pseudo-well locations in a few DPS-generated samples. Furthermore, Fig. \ref{fig:logsampleserro} shows the errors between the logs extracted from the generated samples at the pseudo-well locations and the given pseudo-well logs. The errors corresponding to DPS are larger than those obtained by the proposed method, further demonstrating that the proposed method can more effectively incorporate pseudo-well logs into diffusion-based elastic parameter synthesis.

\begin{figure*}[htb!]
\setlength{\abovecaptionskip}{0.2cm}
 \centering
    \subfigure[]{\includegraphics[width=0.65\columnwidth]{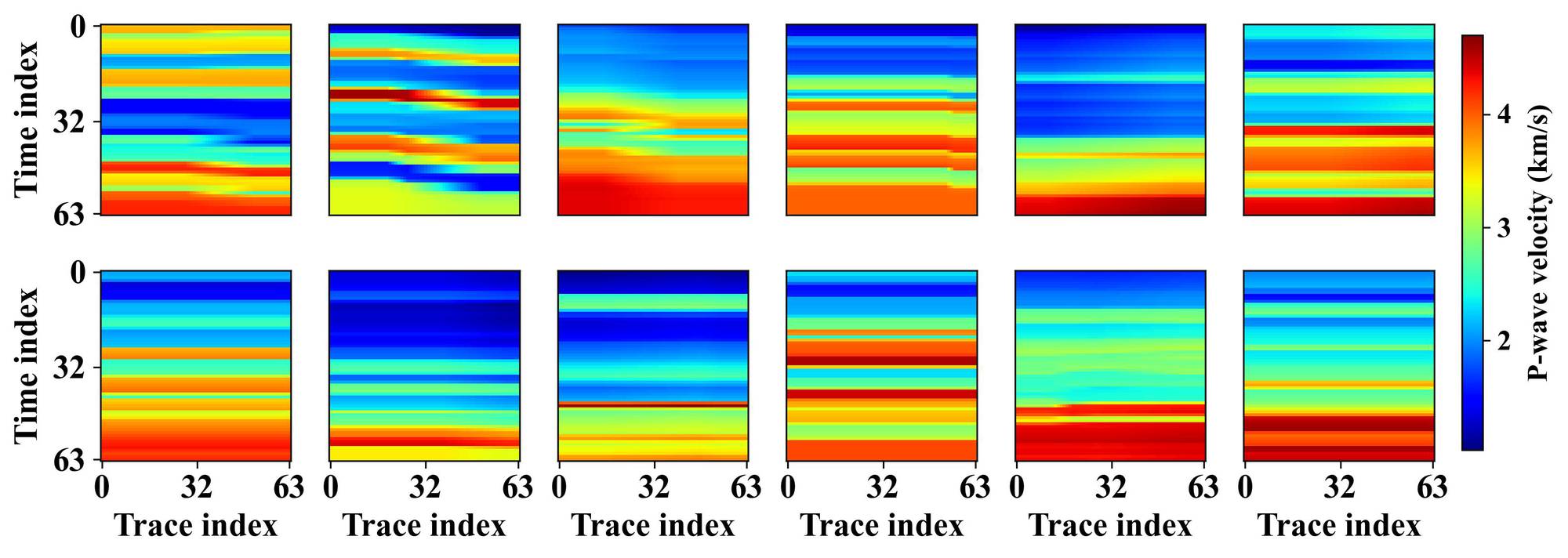}
 \label{fig:condsoft-vp}}
 \subfigure[]{\includegraphics[width=0.65\columnwidth]{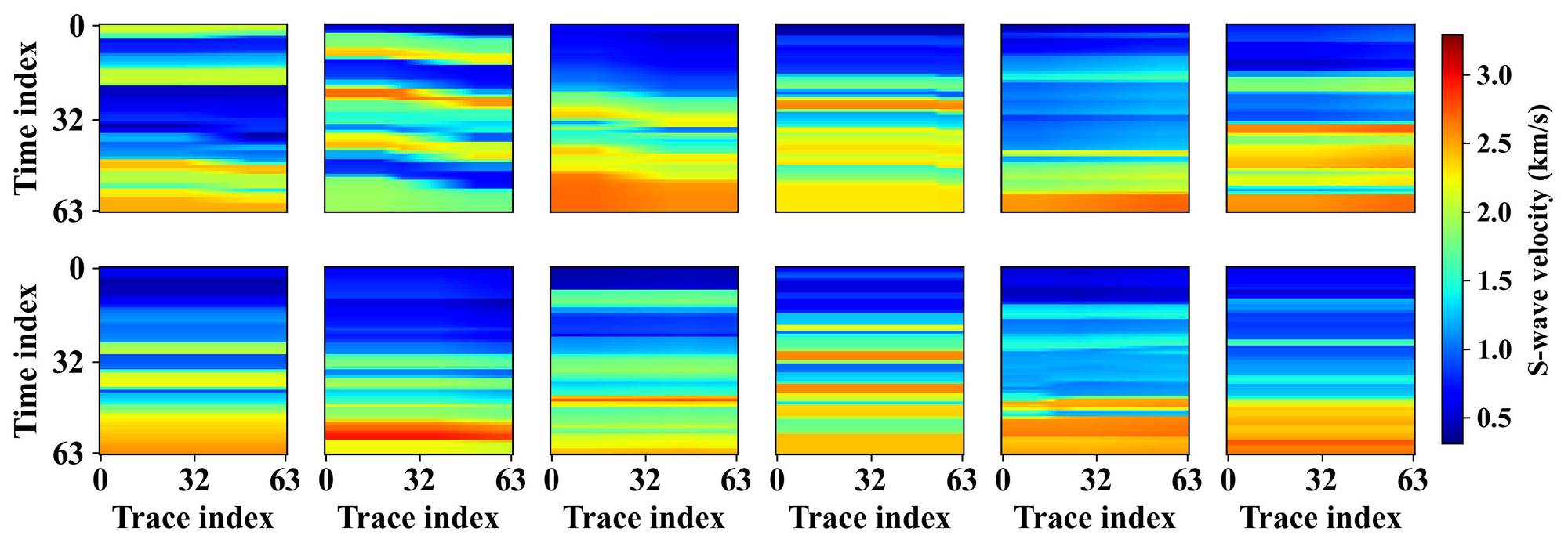}
 \label{fig:condsoft-vs}}
  \subfigure[]{\includegraphics[width=0.65\columnwidth]{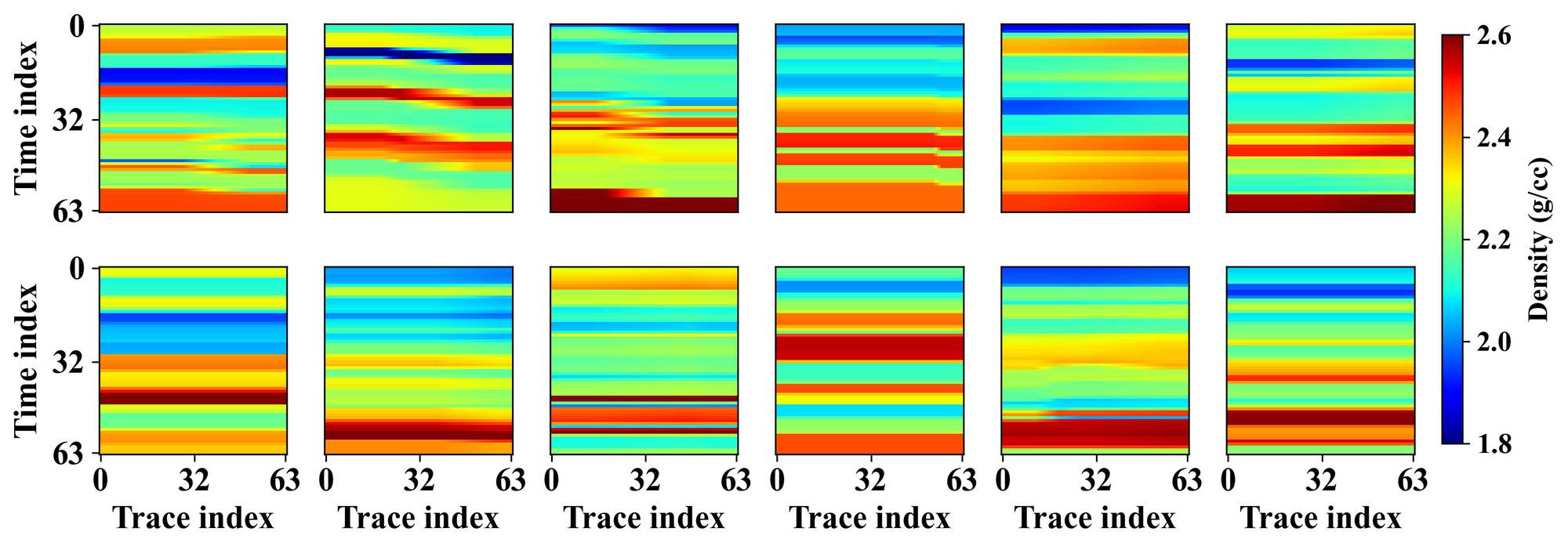}
 \label{fig:condsoft-rho}} 
  \subfigure[]{\includegraphics[width=0.65\columnwidth]{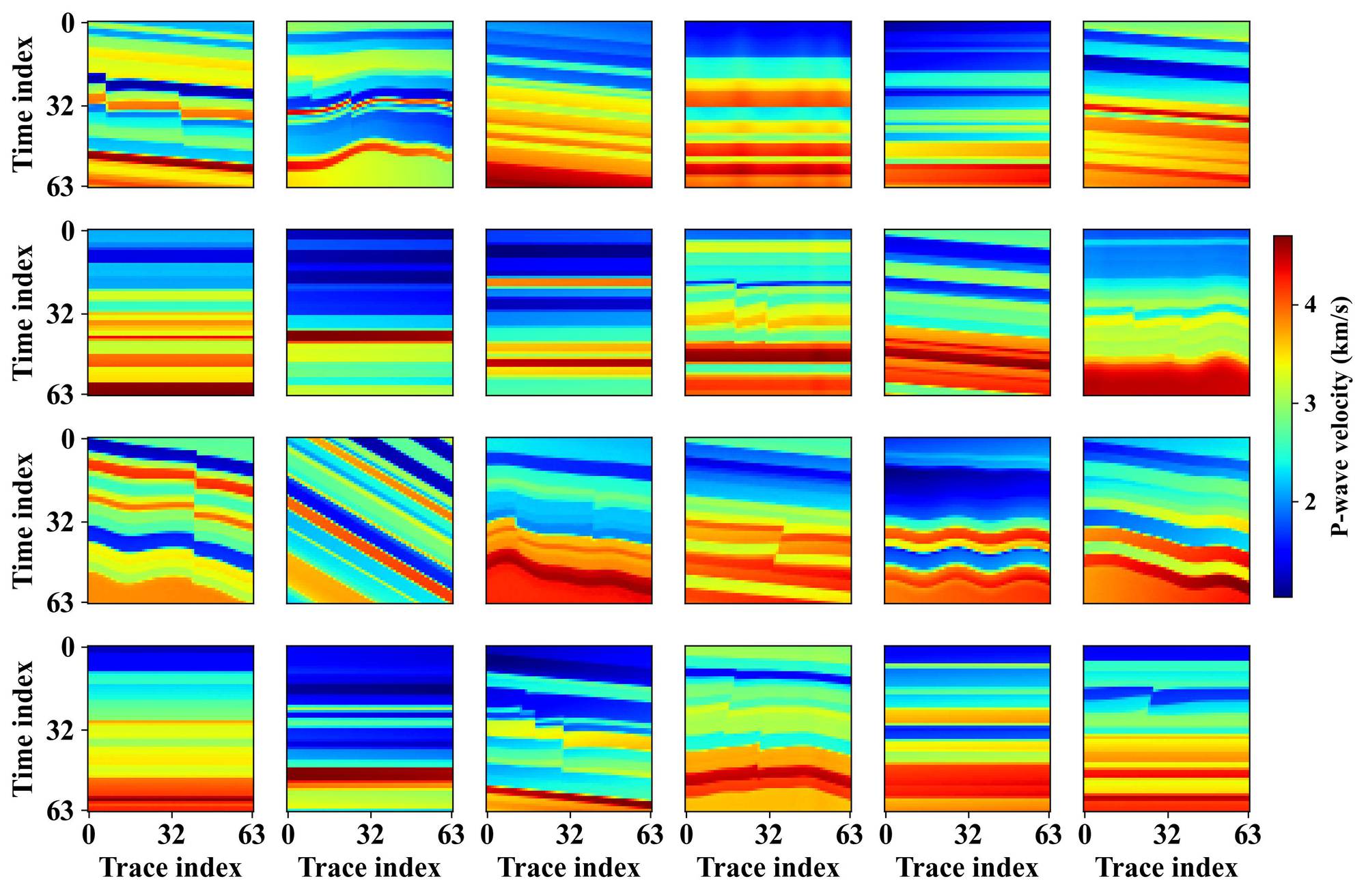}
 \label{fig:ddpm-condsoft-vp}}   
   \subfigure[]{\includegraphics[width=0.65\columnwidth]{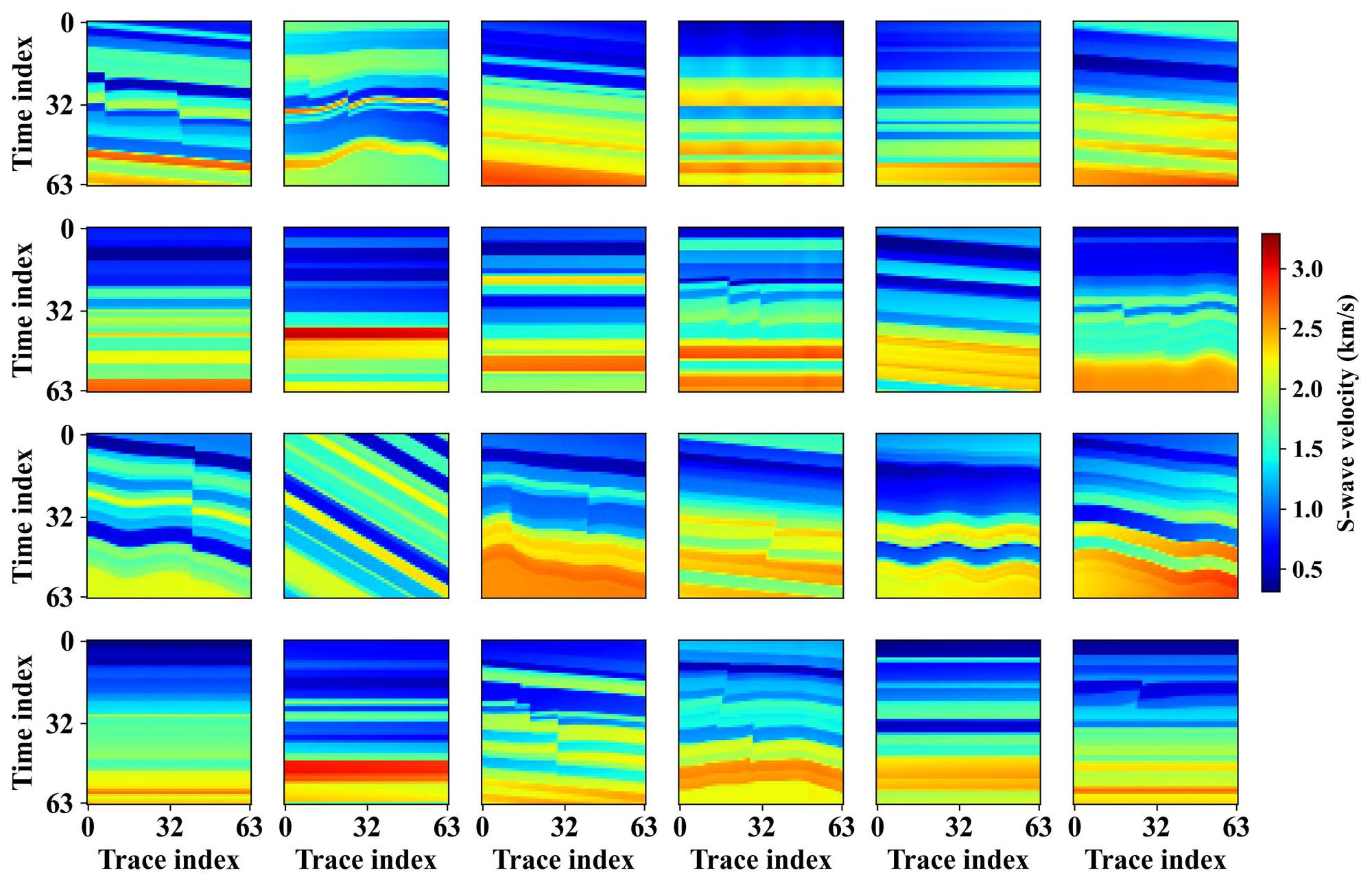}
 \label{fig:ddpm-condsoft-vs}} 
   \subfigure[]{\includegraphics[width=0.65\columnwidth]{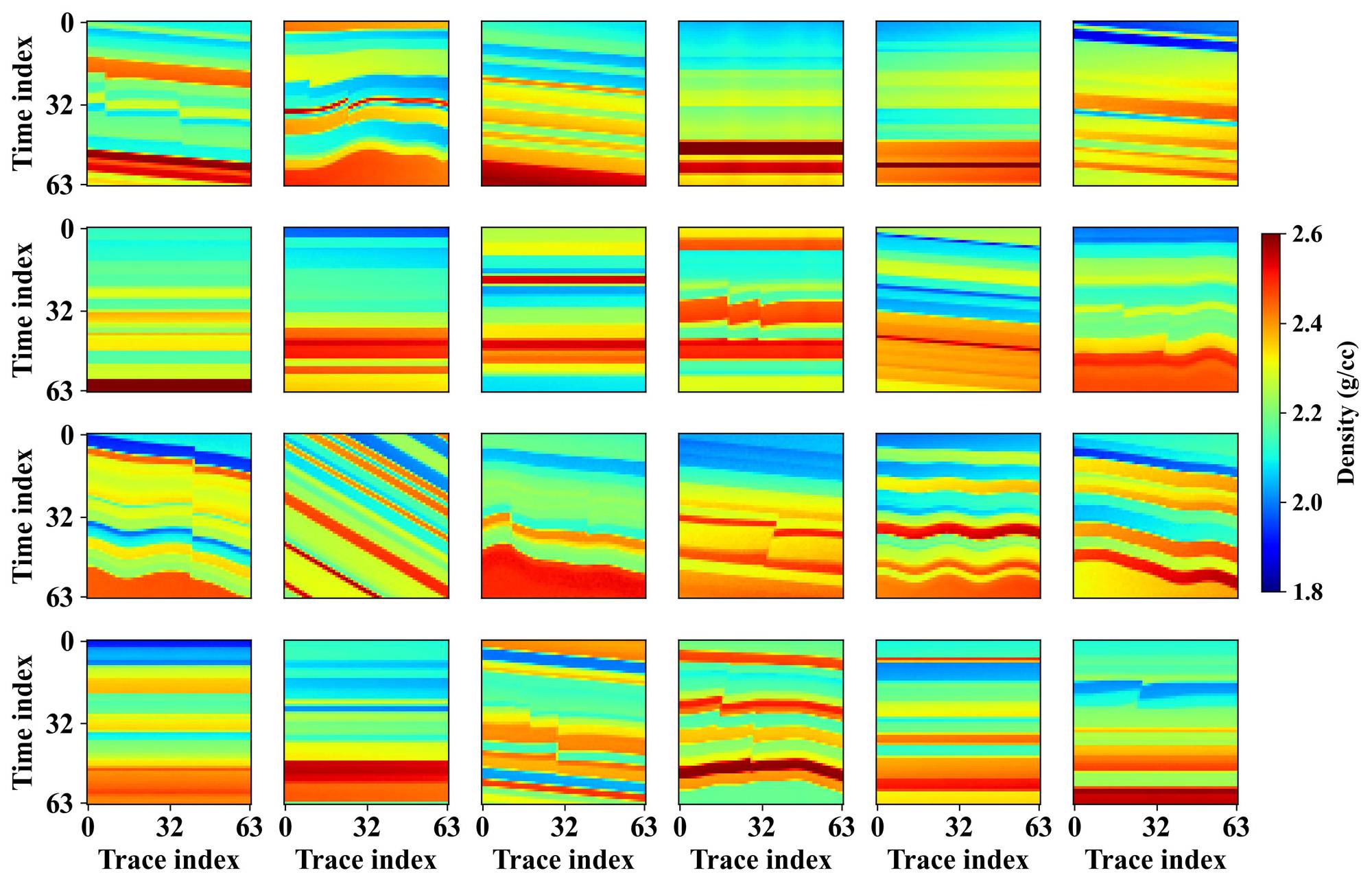}
 \label{fig:ddpm-condsoft-rho}} 
    \subfigure[]{\includegraphics[width=0.65\columnwidth]{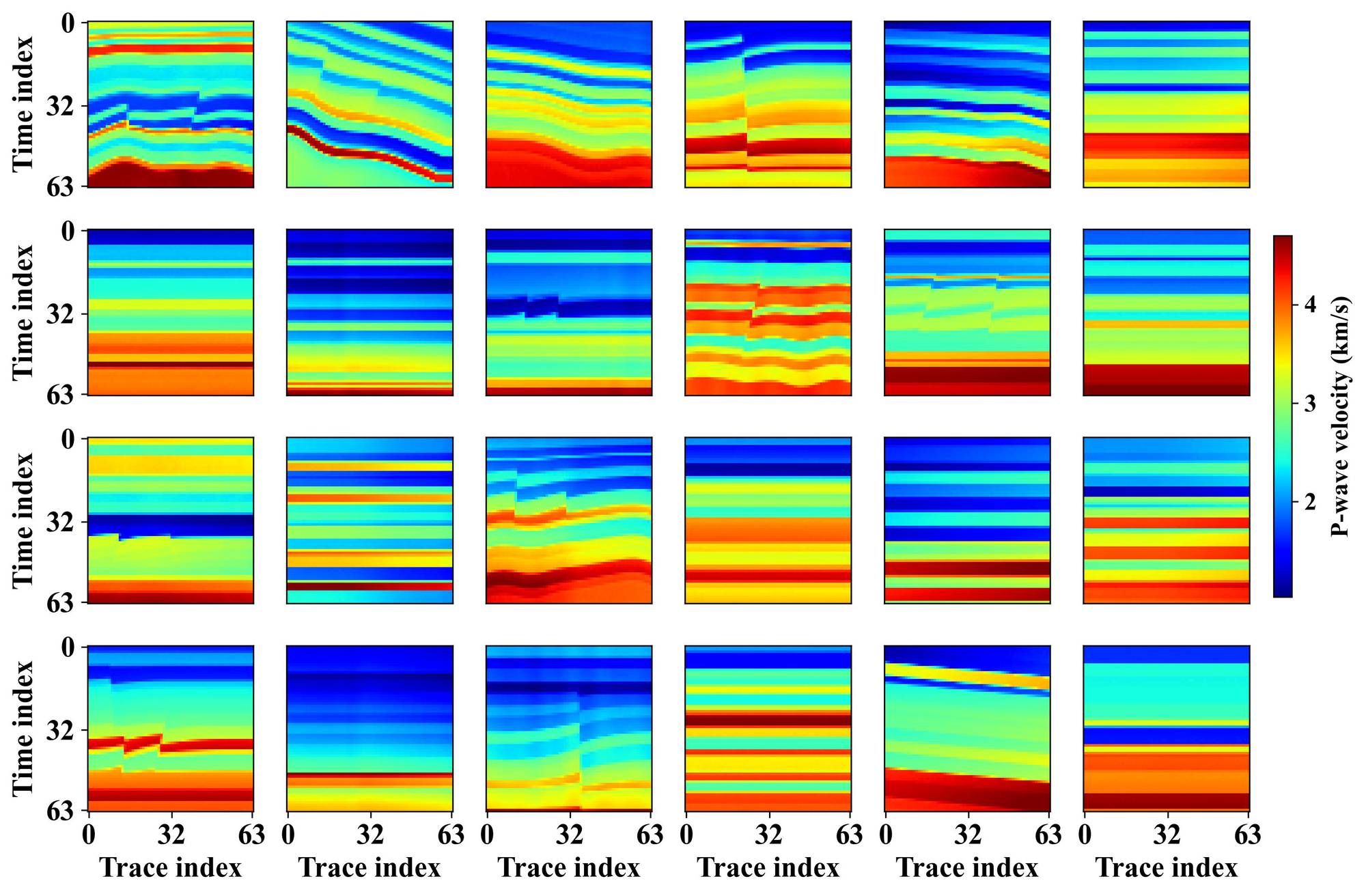}
 \label{fig:ddpm-condsoft16-vp}}   
   \subfigure[]{\includegraphics[width=0.65\columnwidth]{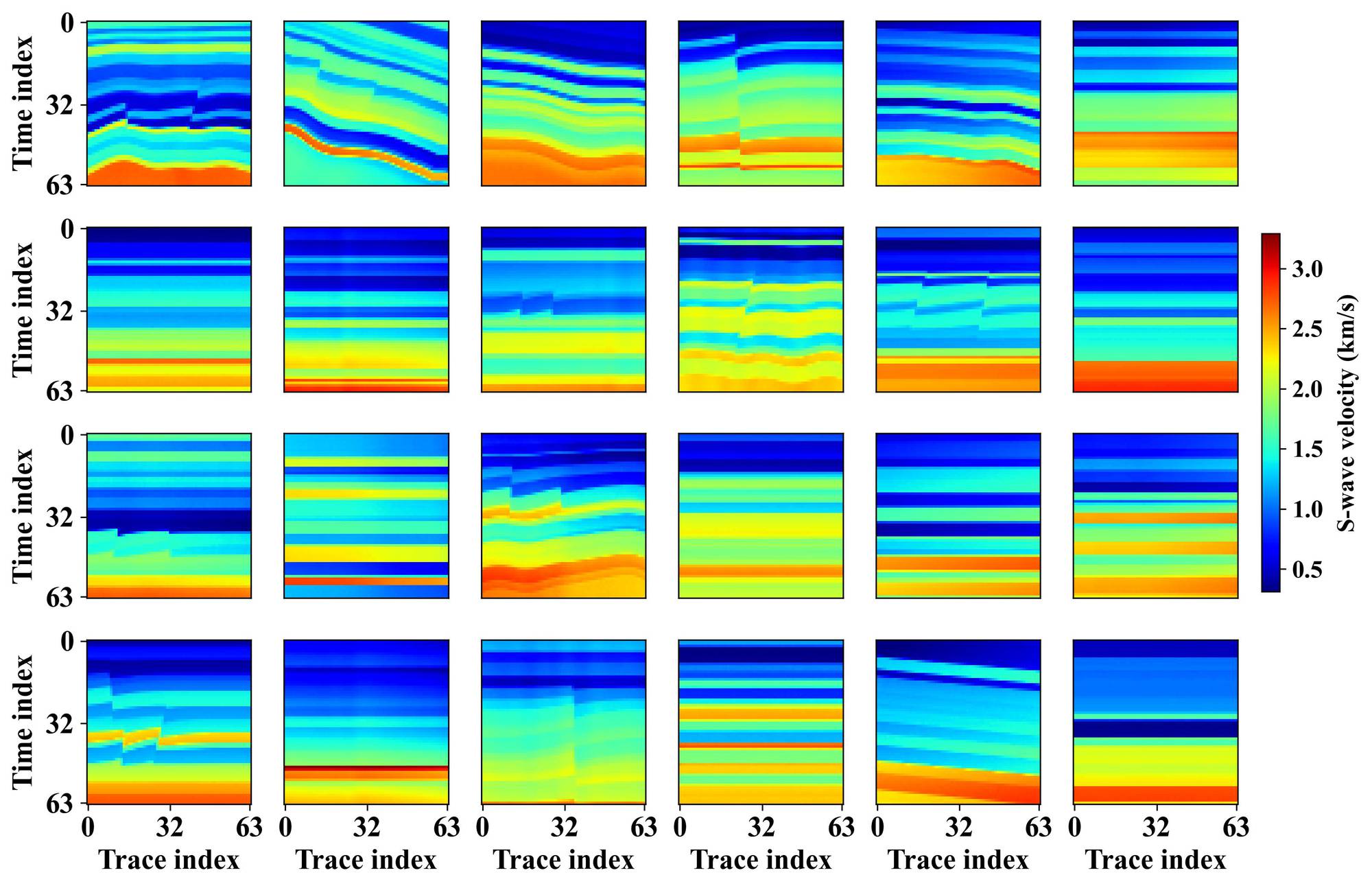}
 \label{fig:ddpm-condsoft16-vs}} 
   \subfigure[]{\includegraphics[width=0.65\columnwidth]{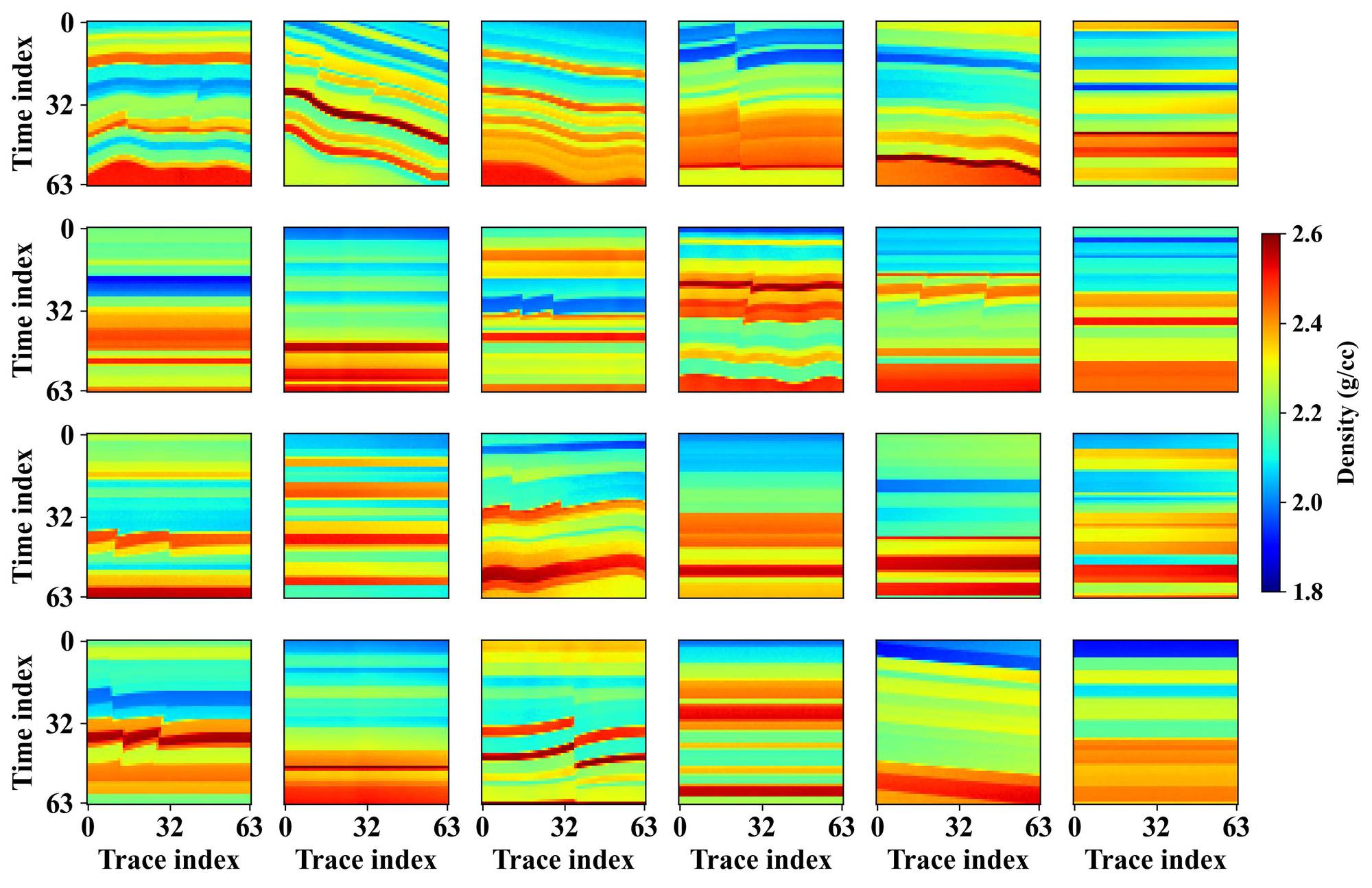}
 \label{fig:ddpm-condsoft16-rho}}
   \subfigure[]{\includegraphics[width=0.65\columnwidth]{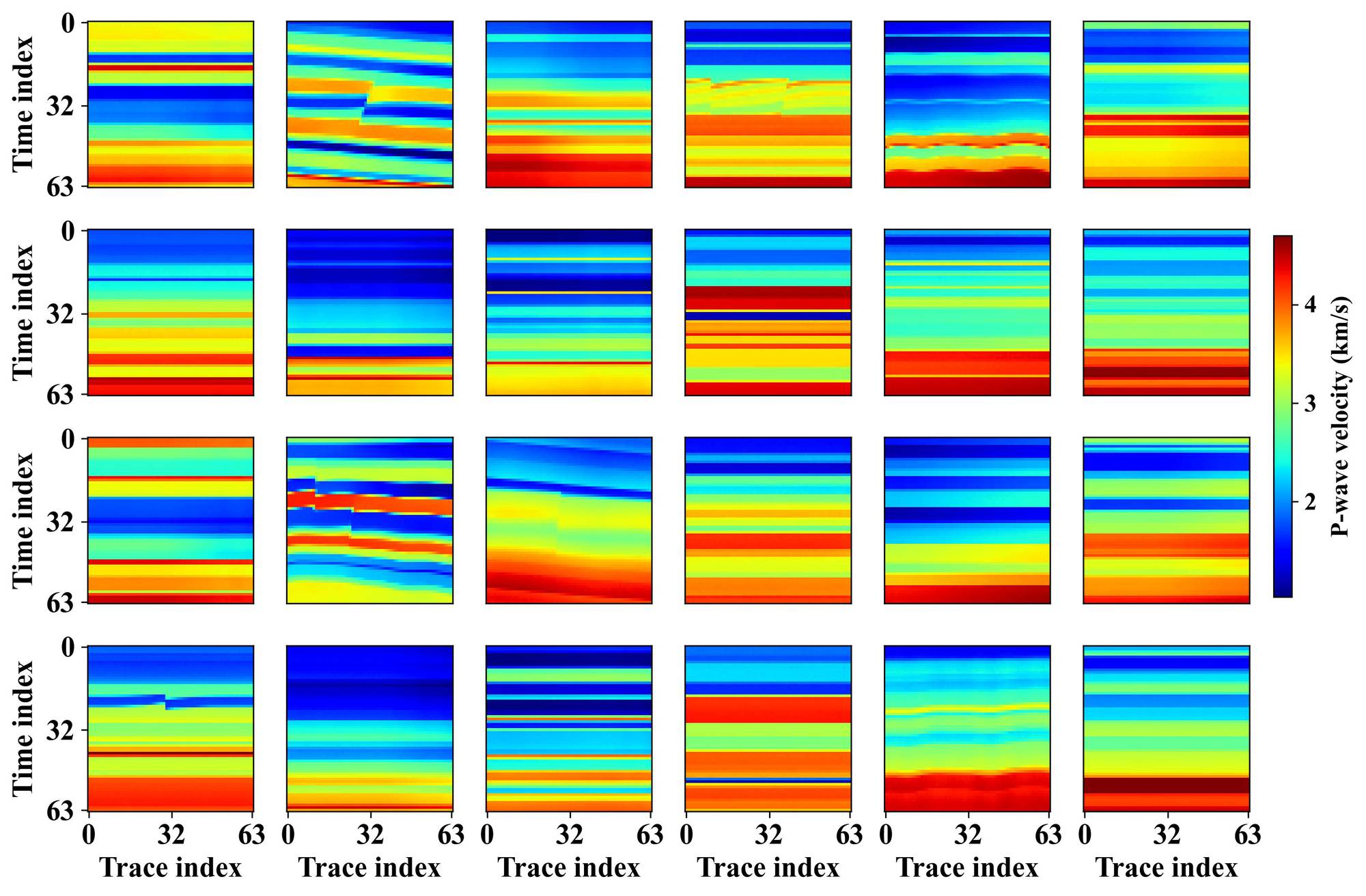}
 \label{fig:ddpm-condsoft8-vp}}   
   \subfigure[]{\includegraphics[width=0.65\columnwidth]{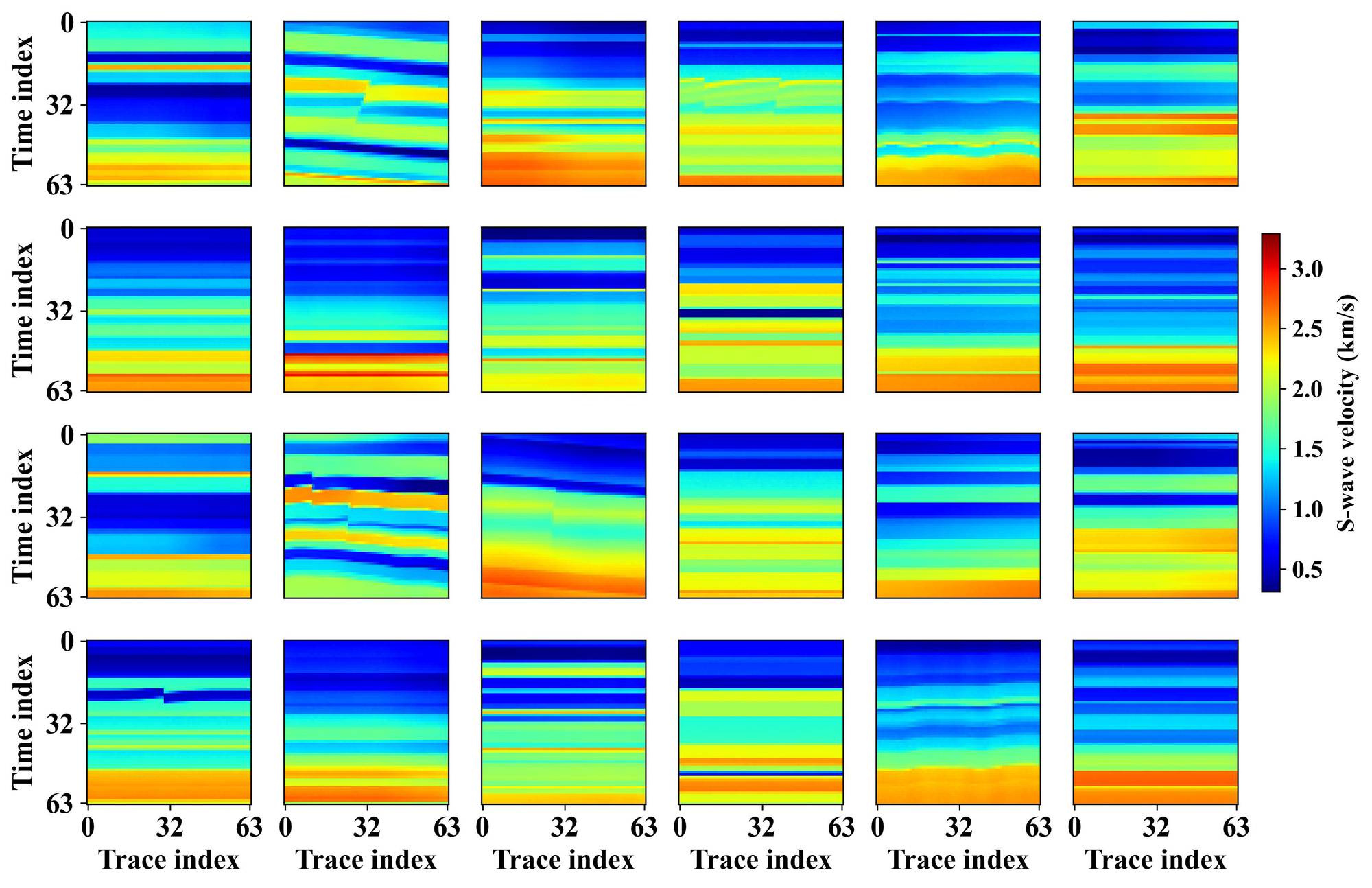}
 \label{fig:ddpm-condsoft8-vs}} 
   \subfigure[]{\includegraphics[width=0.65\columnwidth]{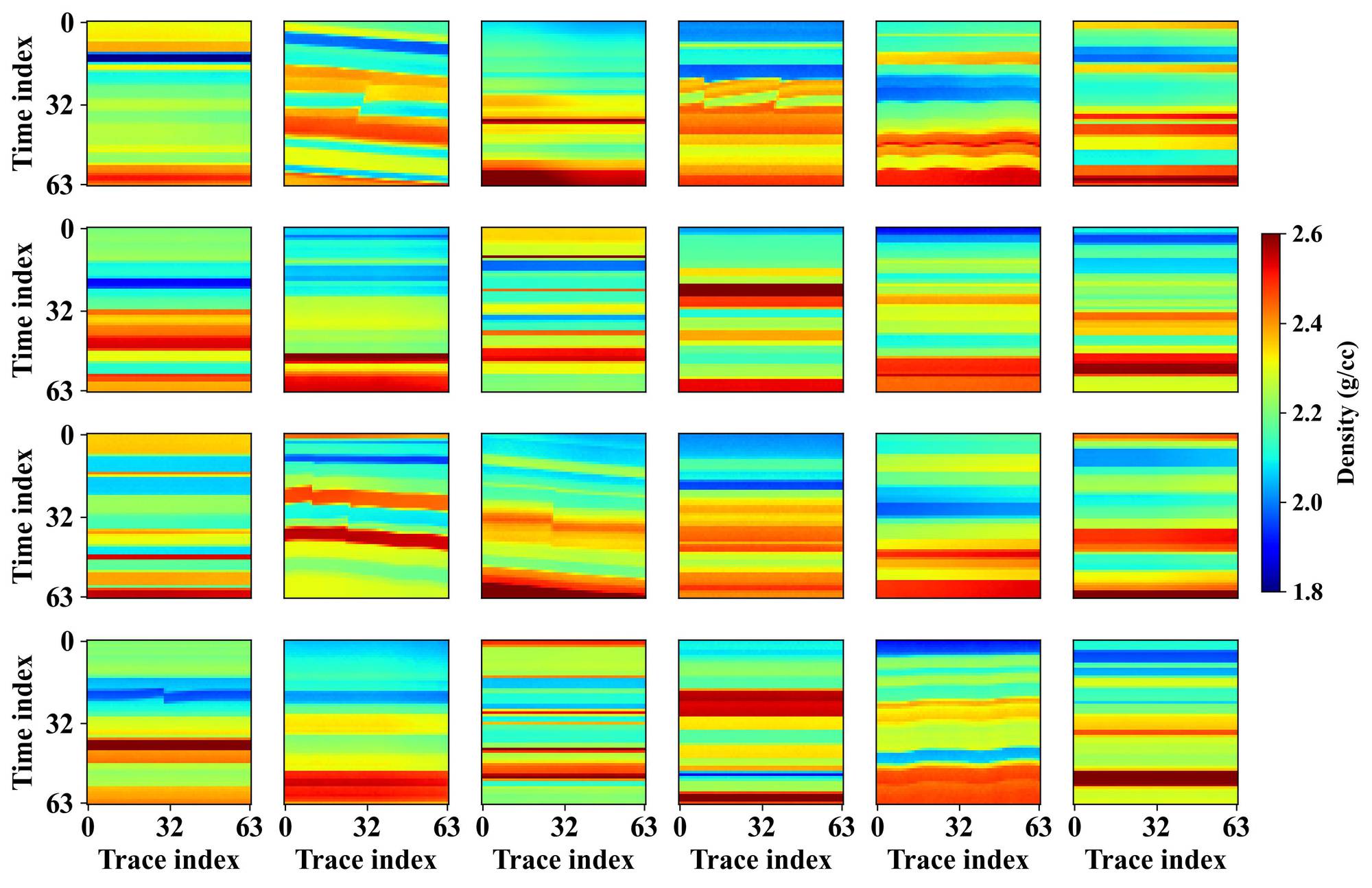}
 \label{fig:ddpm-condsoft8-rho}}
 \caption{Elastic parameter synthesis conditioned on interpolated well-log models. 
(a)--(c) Interpolated well-log models of P-wave velocity, S-wave velocity, and density. 
(d)--(f) Samples generated by ILVR with a factor of $N=32$. 
(g)--(i) Samples generated by ILVR with a factor of $N=16$. 
(j)--(l) Samples generated by ILVR with a factor of $N=8$. 
In each panel, the first two rows show the results from the first sampling run, whereas the last two rows show the results from the second sampling run.}
\label{fig:logsamplesILVR}
\end{figure*}

\textbf{\textit{Elastic parameter synthesis conditioned on interpolated well-log models:}} This experiment treats the interpolated well-log models as reference information and incorporates them into the diffusion-based synthesis process using the ILVR method. Figs. \ref{fig:condsoft-vp}--\ref{fig:condsoft-rho} show the interpolated well-log models obtained by applying 1D linear interpolation to the pseudo-well logs shown in Figs. \ref{fig:condlog-vp}--\ref{fig:condlog-rho}. The resulting interpolation profiles mainly exhibit nearly horizontal structures. Since there is no explicit relationship between the interpolation models and the elastic parameters, ILVR provides a flexible strategy for incorporating such reference information into the diffusion-based synthesis process without retraining the diffusion model. Figs. \ref{fig:ddpm-condsoft-vp}--\ref{fig:ddpm-condsoft8-rho} show the samples generated using ILVR with factors of $N=32$, $N=16$, and $N=8$. The factor $N$ controls the strength of the reference-data constraint. A larger $N$ introduces the reference information only at a coarser scale, resulting in a weaker constraint on the generated samples. Consequently, the samples generated with $N=32$, as shown in Figs. \ref{fig:ddpm-condsoft-vp}--\ref{fig:ddpm-condsoft-rho}, are weakly constrained by the interpolation models. They therefore exhibit pronounced structural variations within individual samples, larger deviations from the reference interpolation models, and greater differences between repeated sampling runs. As $N$ decreases, the reference information is imposed at a finer scale, leading to stronger consistency between the generated samples and the interpolation models. This effect is particularly evident for $N=8$, as shown in Figs. \ref{fig:ddpm-condsoft8-vp}--\ref{fig:ddpm-condsoft8-rho}.  However, the interpolated well-log models are only an approximate reference condition and may not accurately represent complex subsurface structures. Therefore, overly strong enforcement of this condition may reduce the geological plausibility and diversity of the synthesized samples. The above results indicate that the factor $N$ should be selected according to the reliability and informativeness of the reference data, so as to balance conditional control and sample variability.

\begin{figure*}[htb!]
\setlength{\abovecaptionskip}{0.2cm}
 \centering
\includegraphics[width=1.5\columnwidth]{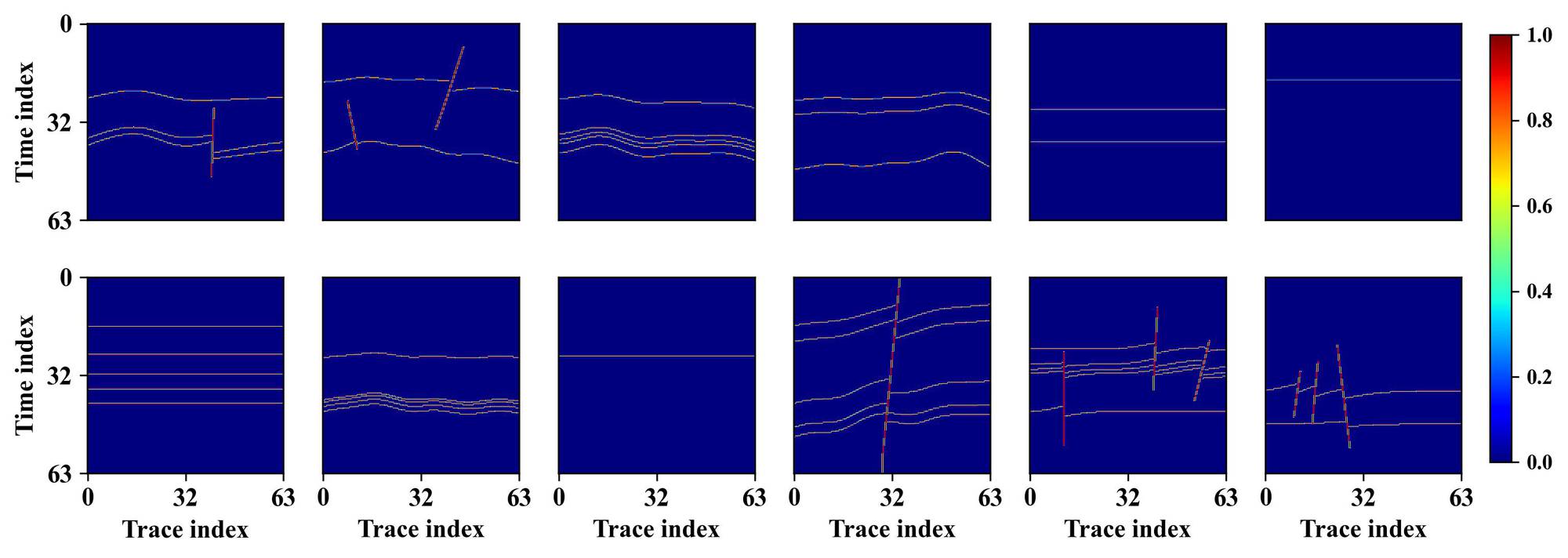}
 \caption{Binary structural masks used as conditioning information. Horizon and fault locations are assigned a value of 1, and the remaining regions are set to 0.}
\label{fig:condstruc}
\end{figure*}

\begin{figure*}[htb!]
\setlength{\abovecaptionskip}{0.2cm}
 \centering
   \subfigure[]{\includegraphics[width=0.65\columnwidth]{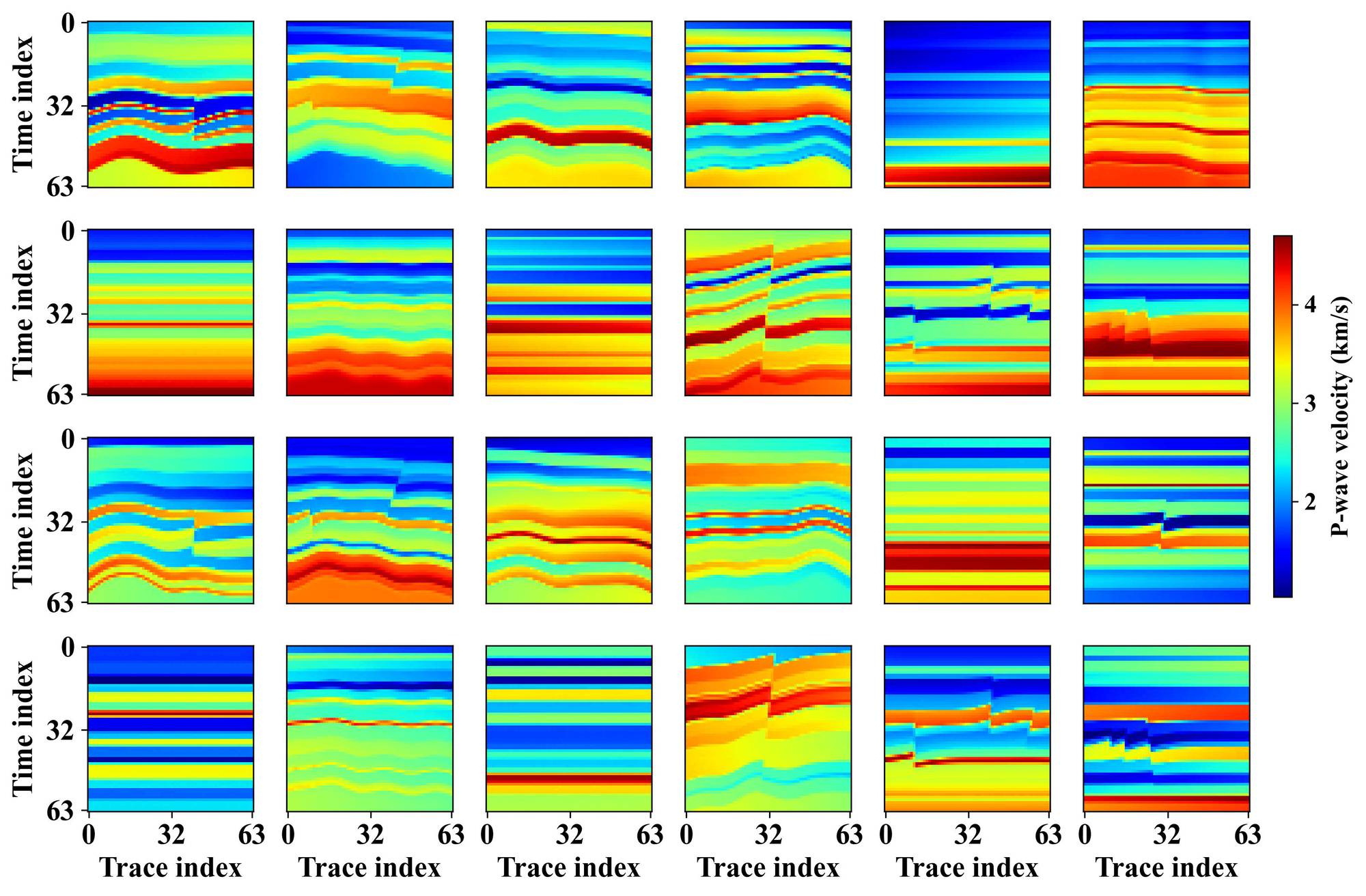}
 \label{fig:ddpm-condstruc-vp}}
 \subfigure[]{\includegraphics[width=0.65\columnwidth]{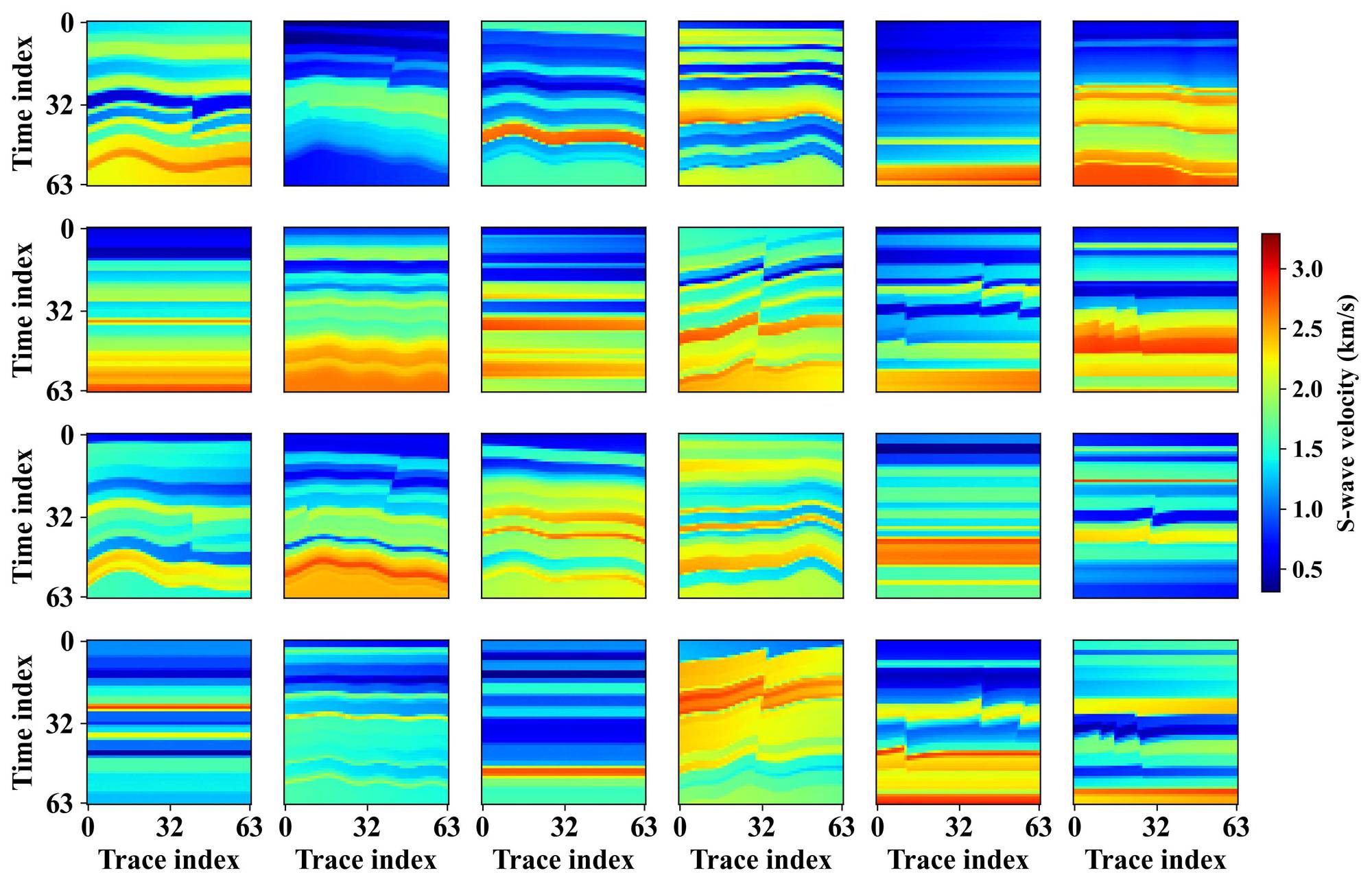}
 \label{fig:ddpm-condstruc-vs}} 
  \subfigure[]{\includegraphics[width=0.65\columnwidth]{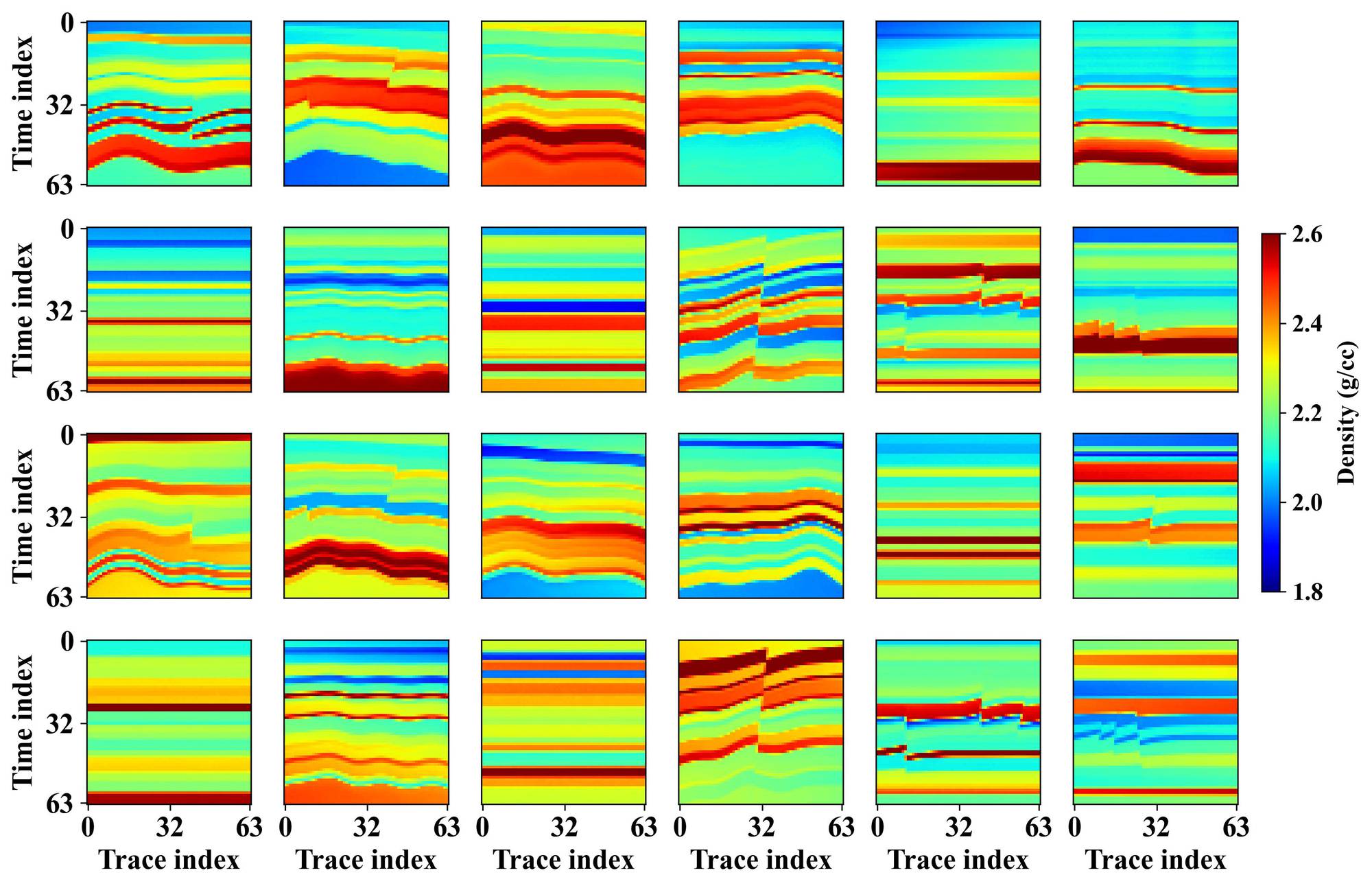}
 \label{fig:ddpm-condstruc-rho}}  
 \caption{Elastic parameter synthesis conditioned on structural information. 
(a)--(c) Samples generated using Adapter-based conditioning. 
In each panel, the first two rows show the results from the first sampling run, whereas the last two rows show the results from the second sampling run.}
\label{fig:strucsamples}
\end{figure*}

\textbf{\textit{Elastic parameter synthesis conditioned on structural information:}} This experiment further evaluates the effectiveness of Adapter-based conditioning in incorporating structural information, including horizons and faults, into the proposed diffusion-based synthesis process. Fig. \ref{fig:condstruc} shows the binary structural masks used as conditioning information. These masks are generated by randomly extracting horizon and fault locations during dataset construction, with structural locations assigned a value of 1 and the remaining regions set to 0. They cover different structural patterns, including flat layers, folded layers, faulted flat layers, and folded-and-faulted layers. Fig. \ref{fig:strucsamples} shows the samples synthesized under these structural conditions. These samples generally exhibit structural patterns consistent with the given horizons and faults, indicating that the Adapter-based conditioning can effectively guide the geometric structures of the synthesized elastic parameter models. Compared with seismic data, structural masks provide sparse but explicit geometric constraints. They mainly control the locations and shapes of horizons and faults, while still allowing variability in regions not directly constrained by the masks. Consequently, the generated samples can remain more structurally diverse than those conditioned on seismic data. In addition, structural masks are generally less sensitive to noise contamination than seismic observations and can therefore provide more explicit and direct constraints on geological geometry. However, they do not fully constrain the absolute values of elastic parameters. These results suggest that structural information provides a useful complementary constraint for improving the geological consistency of synthesized elastic parameter models.

%施加层位约束

\subsubsection{Multi-Condition Guided Elastic Parameter Synthesis} This section evaluates the effectiveness of the proposed method for elastic parameter synthesis under multi-condition guidance. The experiments are first conducted on $64 \times 64$ model patches, which have the same size as the training samples, and the adopted conditioning information is consistent with that used in the preceding single-condition experiments. To further examine whether the proposed method can be applied to larger model areas, we perform patch-wise synthesis on the Marmousi II model. Since the training datasets are constructed from three pseudo-well logs extracted from the Marmousi II model, this experiment also investigates whether the learned prior can be effectively used for patch-wise synthesis over a larger spatial extent.

\begin{figure*}[htb!]
\setlength{\abovecaptionskip}{0.2cm}
 \centering
    \subfigure[]{\includegraphics[width=0.65\columnwidth]{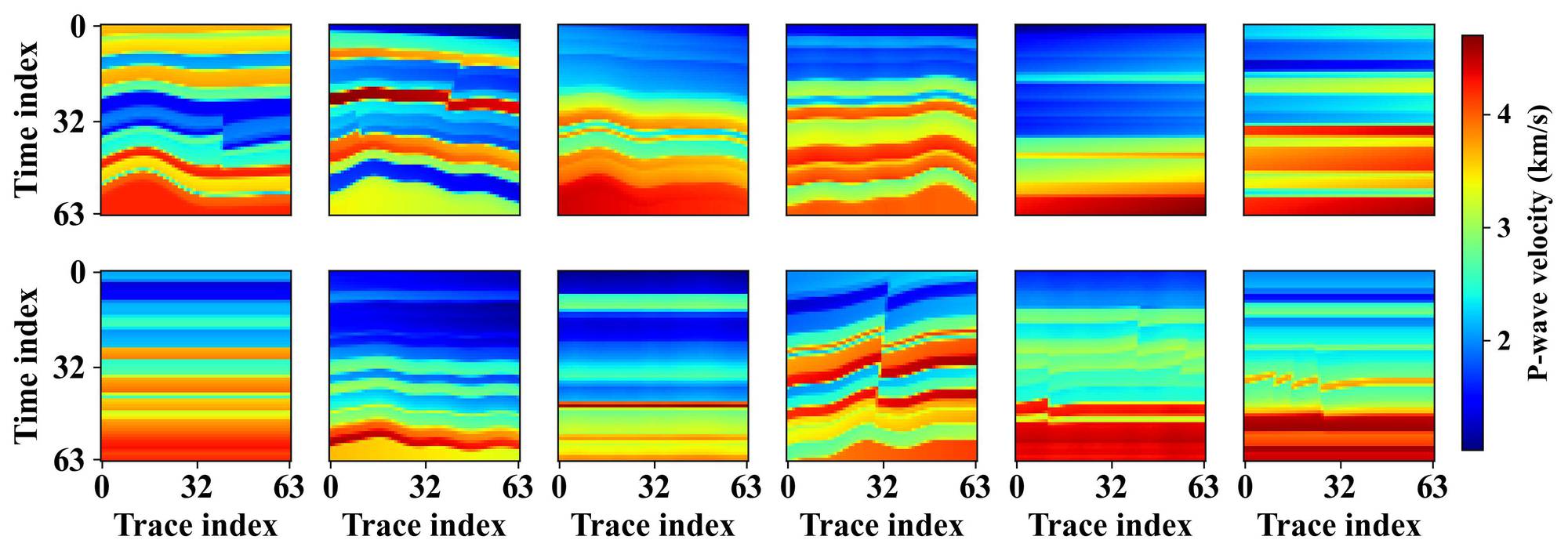}
 \label{fig:true-vp}}
 \subfigure[]{\includegraphics[width=0.65\columnwidth]{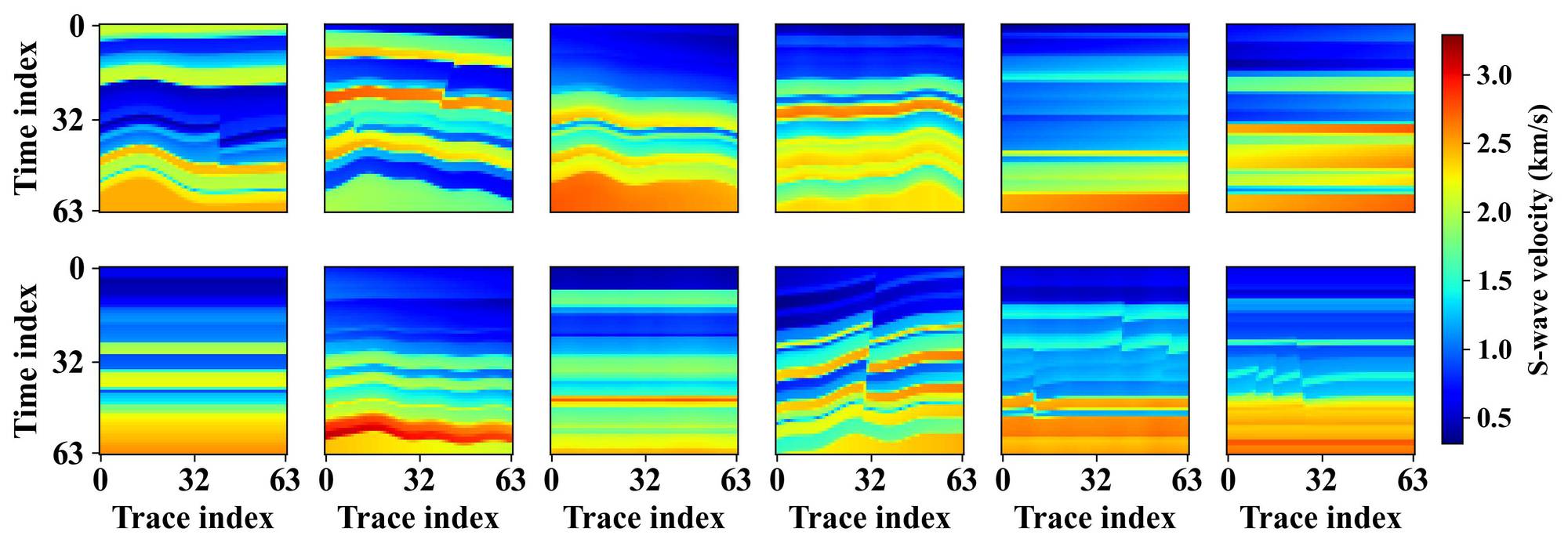}
 \label{fig:true-vs}} 
  \subfigure[]{\includegraphics[width=0.65\columnwidth]{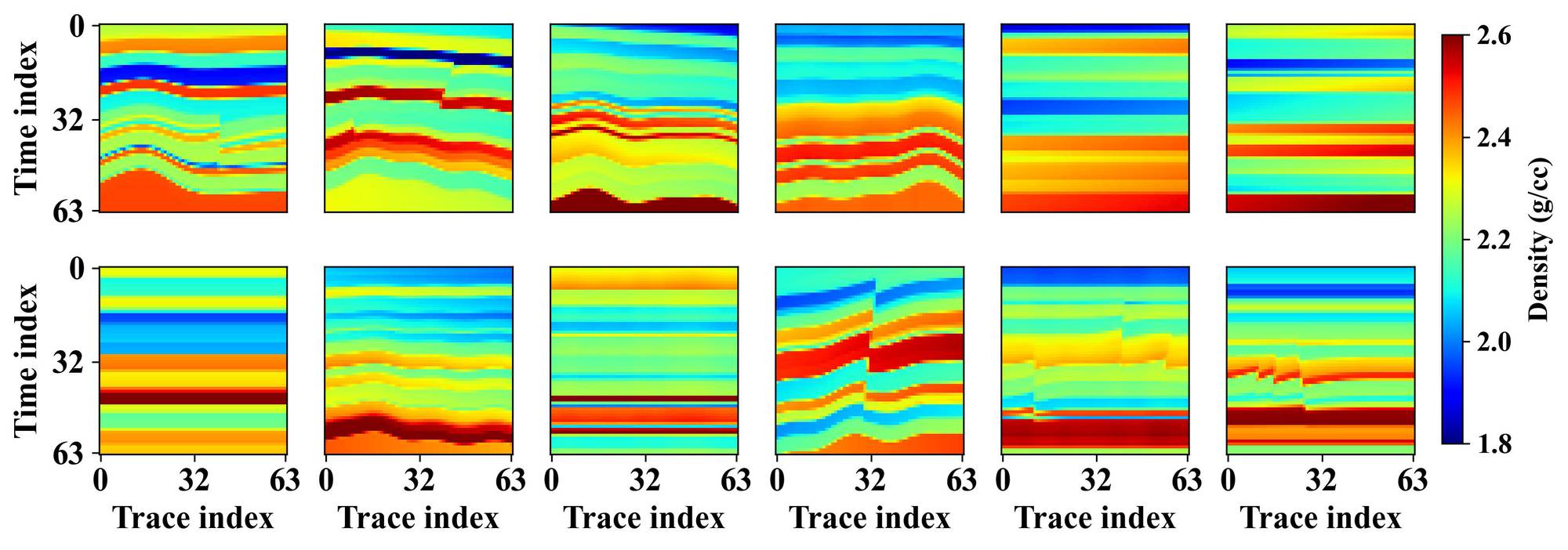}
 \label{fig:true-rho}}
  \subfigure[]{\includegraphics[width=0.65\columnwidth]{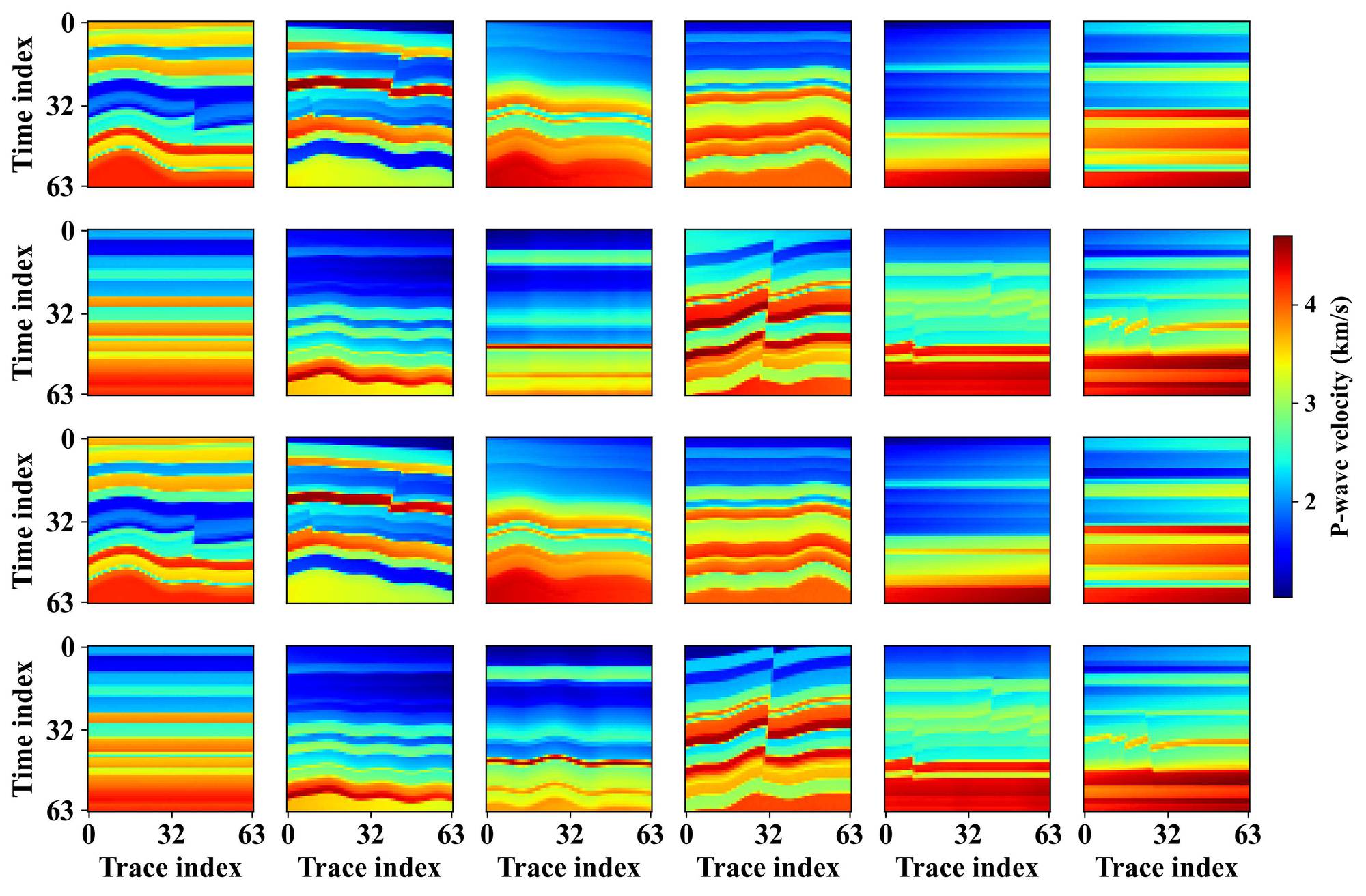}
 \label{fig:ddpm-condstruclog-vp}}
 \subfigure[]{\includegraphics[width=0.65\columnwidth]{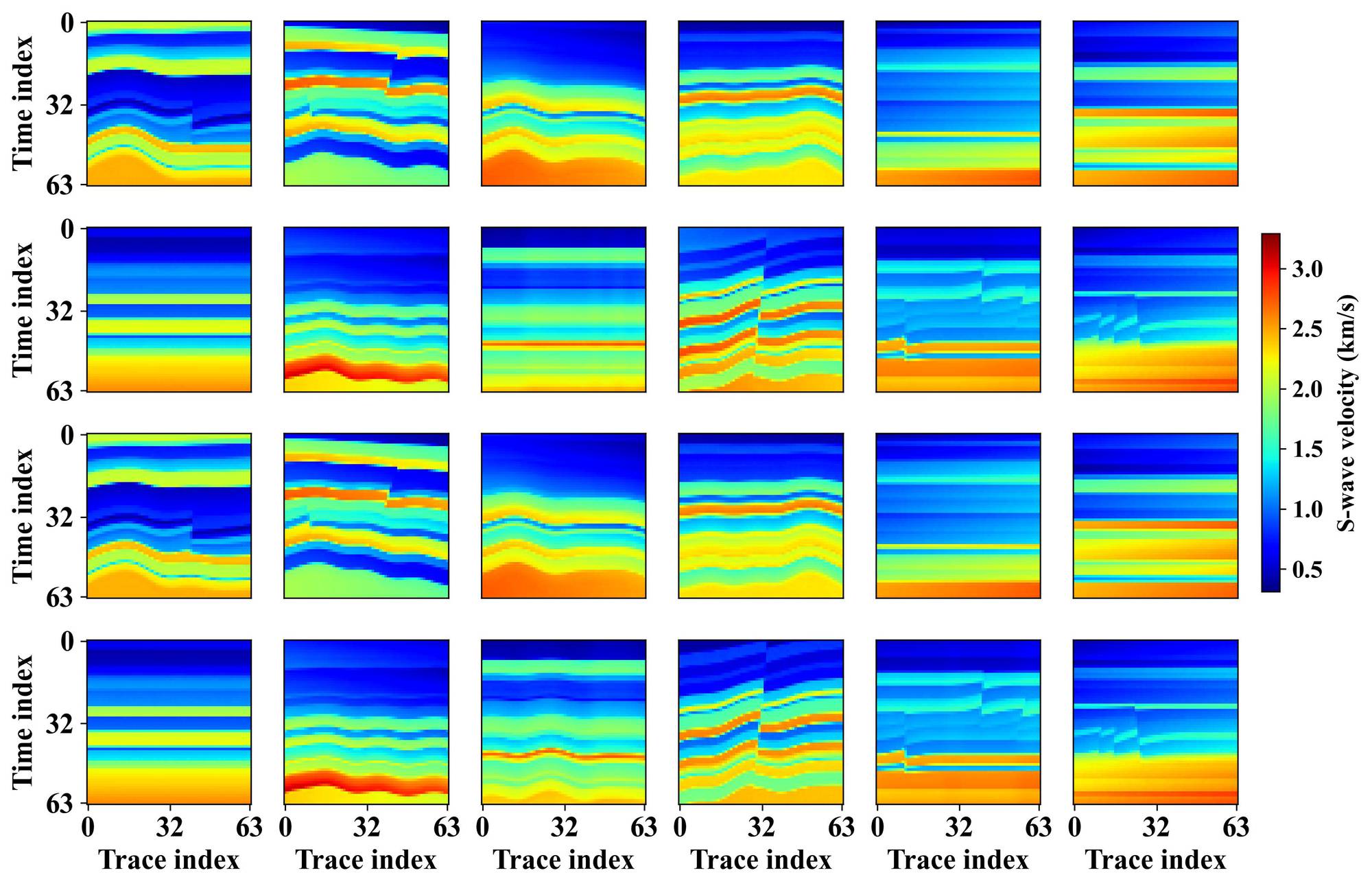}
 \label{fig:ddpm-condstruclog-vs}} 
  \subfigure[]{\includegraphics[width=0.65\columnwidth]{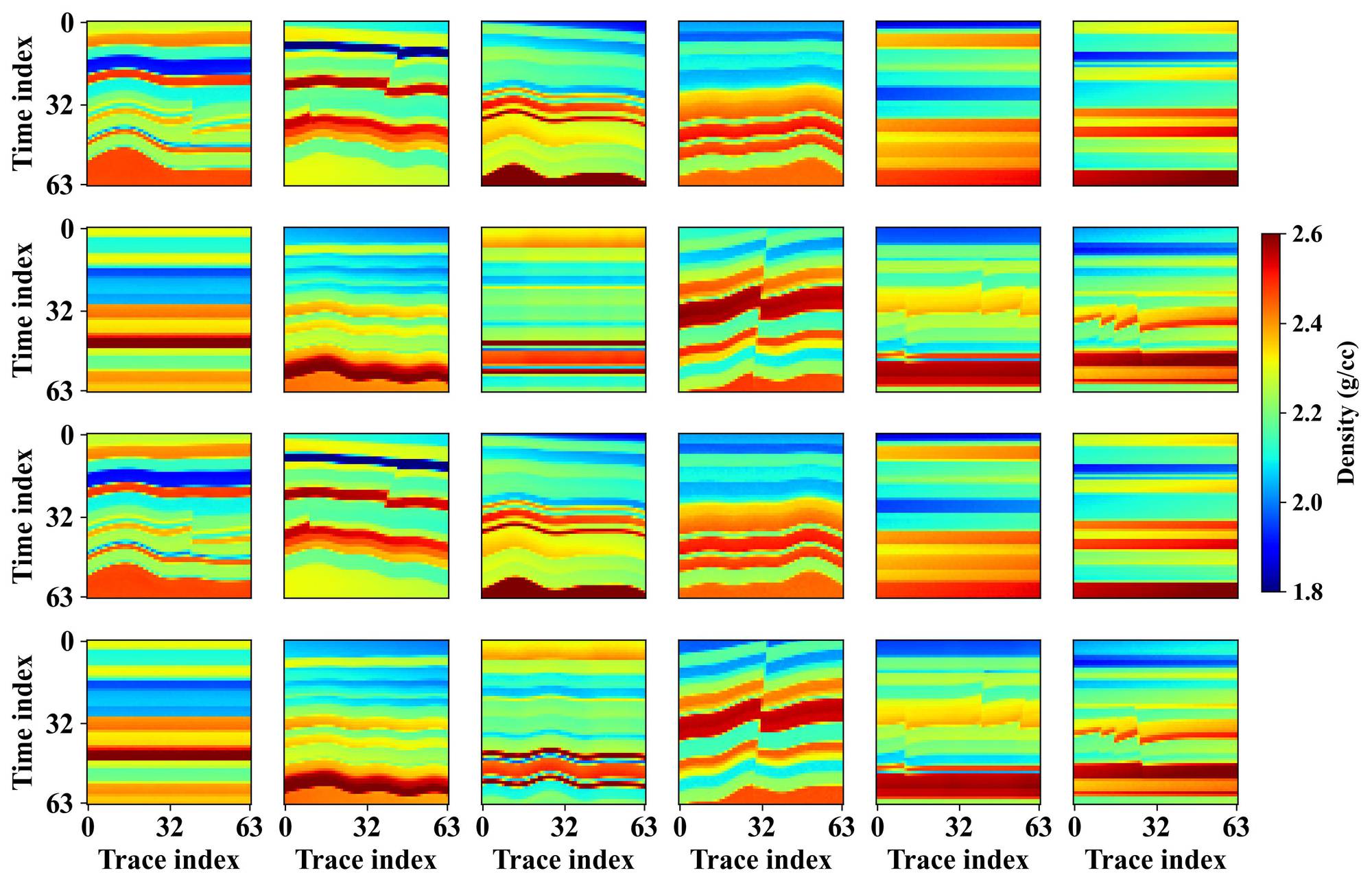}
 \label{fig:ddpm-condstruclog-rho}}
    \subfigure[]{\includegraphics[width=0.65\columnwidth]{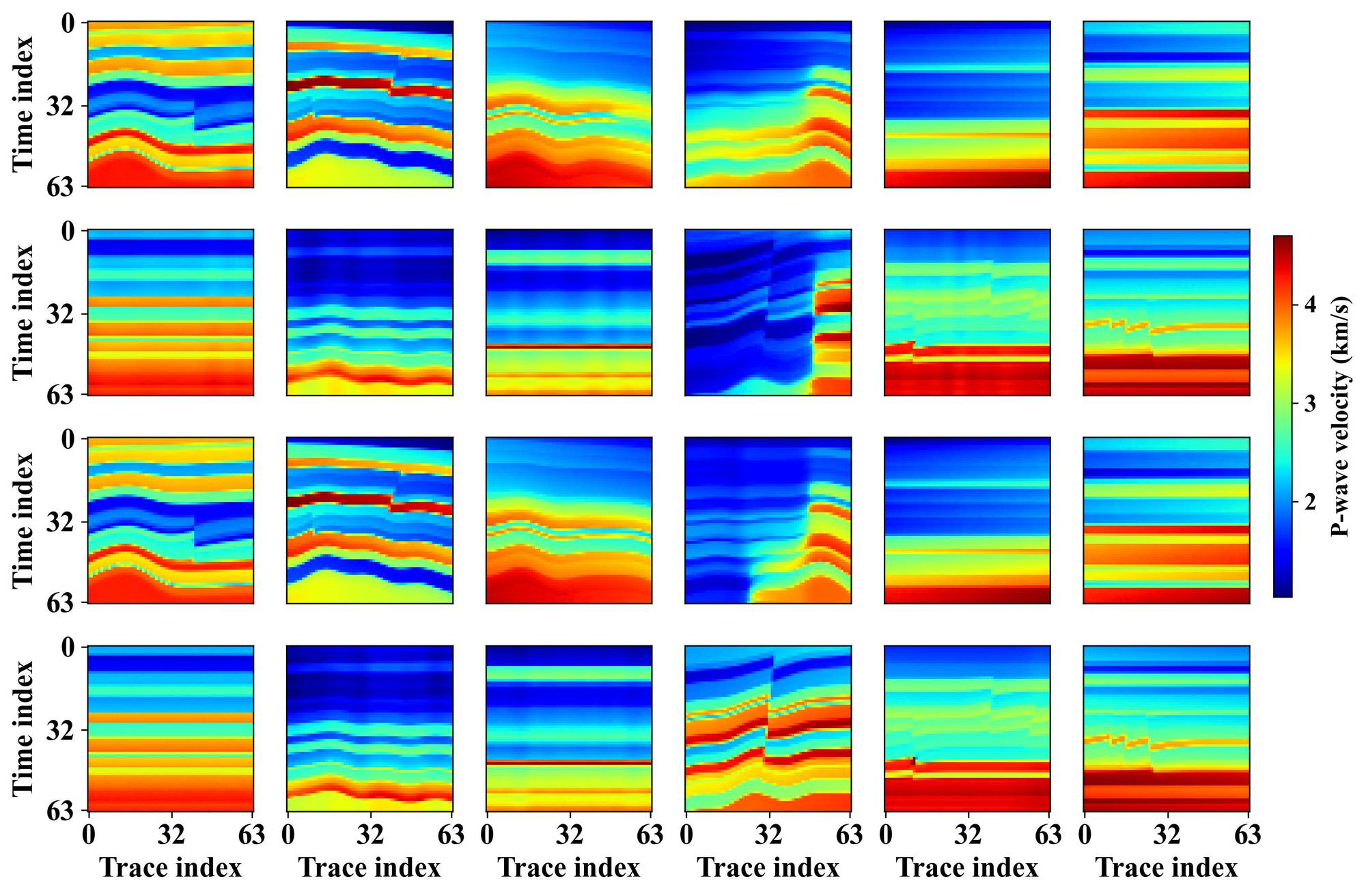}
 \label{fig:ddpm-condlogseis-vp}}
 \subfigure[]{\includegraphics[width=0.65\columnwidth]{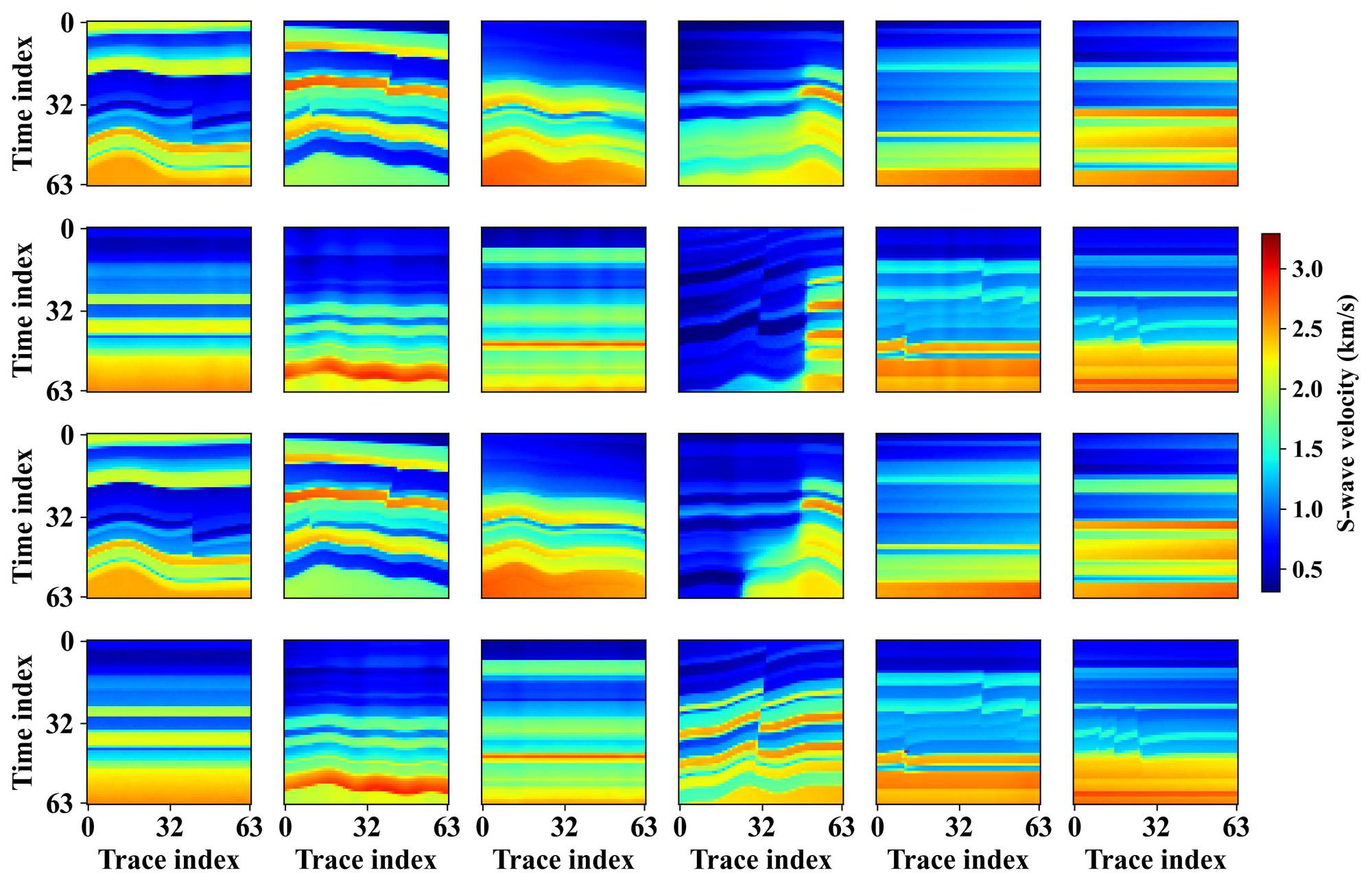}
 \label{fig:ddpm-condlogseis-vs}} 
  \subfigure[]{\includegraphics[width=0.65\columnwidth]{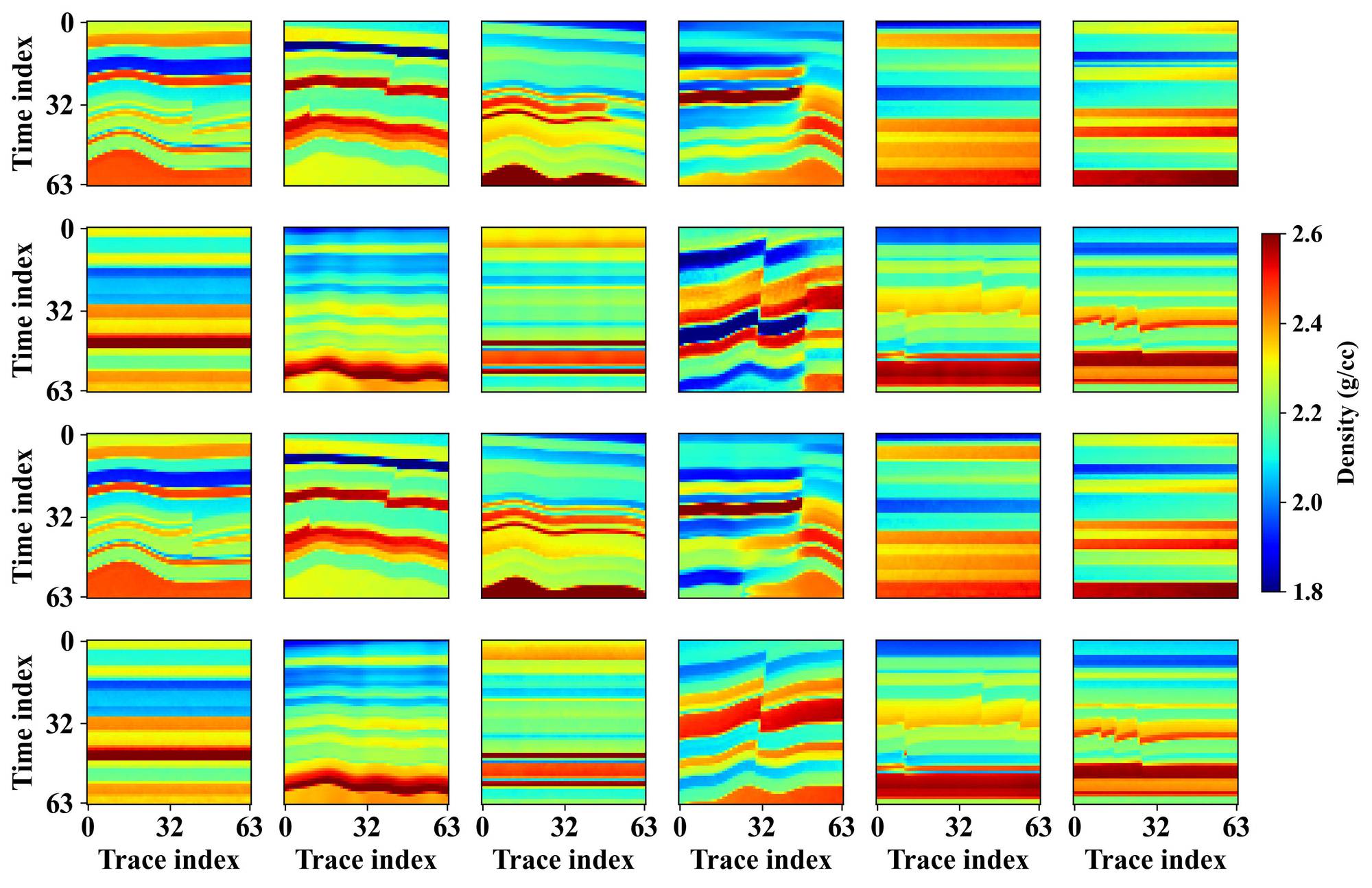}
 \label{fig:ddpm-condlogseis-rho}}     
  \subfigure[]{\includegraphics[width=0.65\columnwidth]{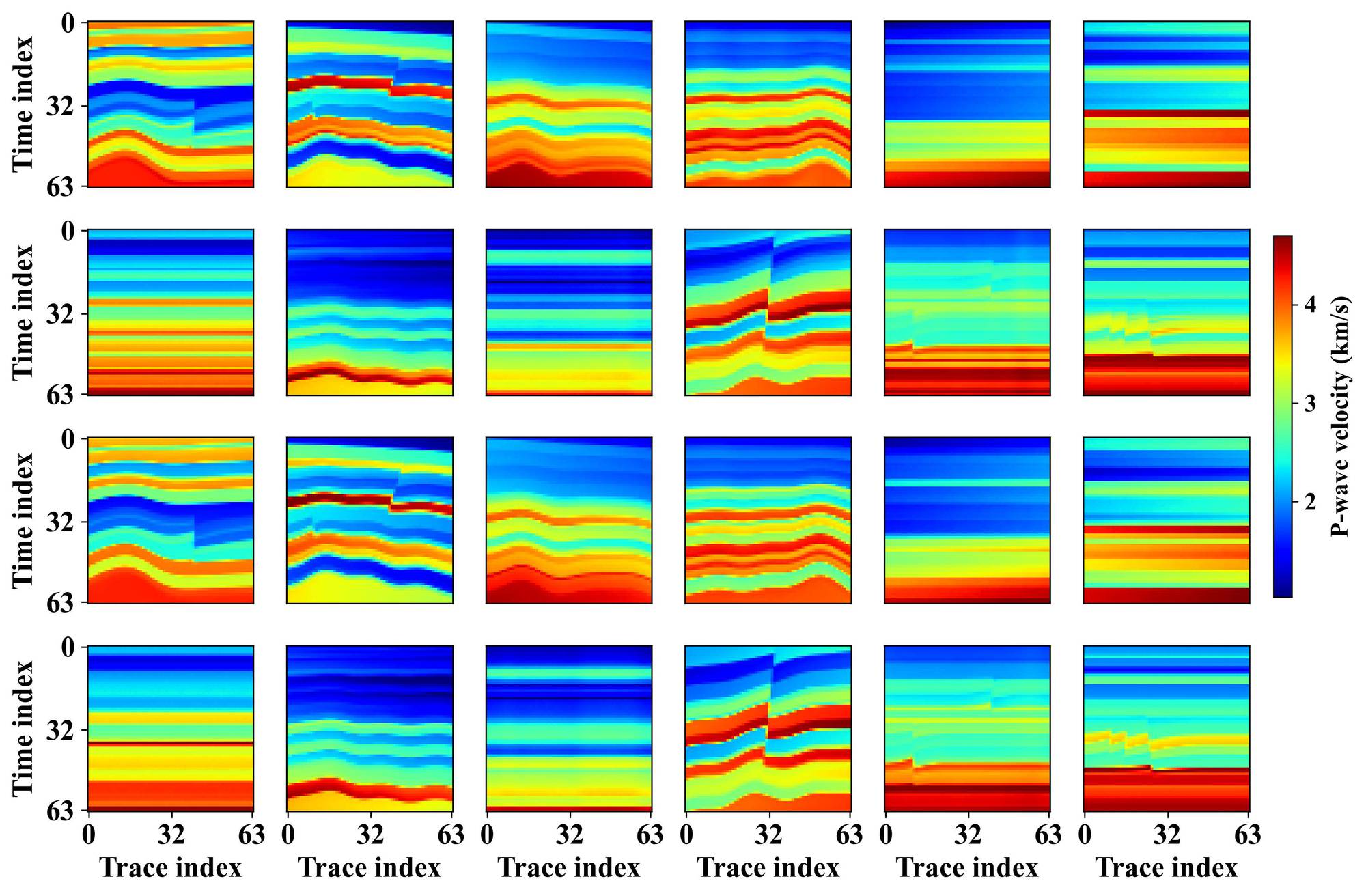}
 \label{fig:ddpm-condlow2seis-vp}}
 \subfigure[]{\includegraphics[width=0.65\columnwidth]{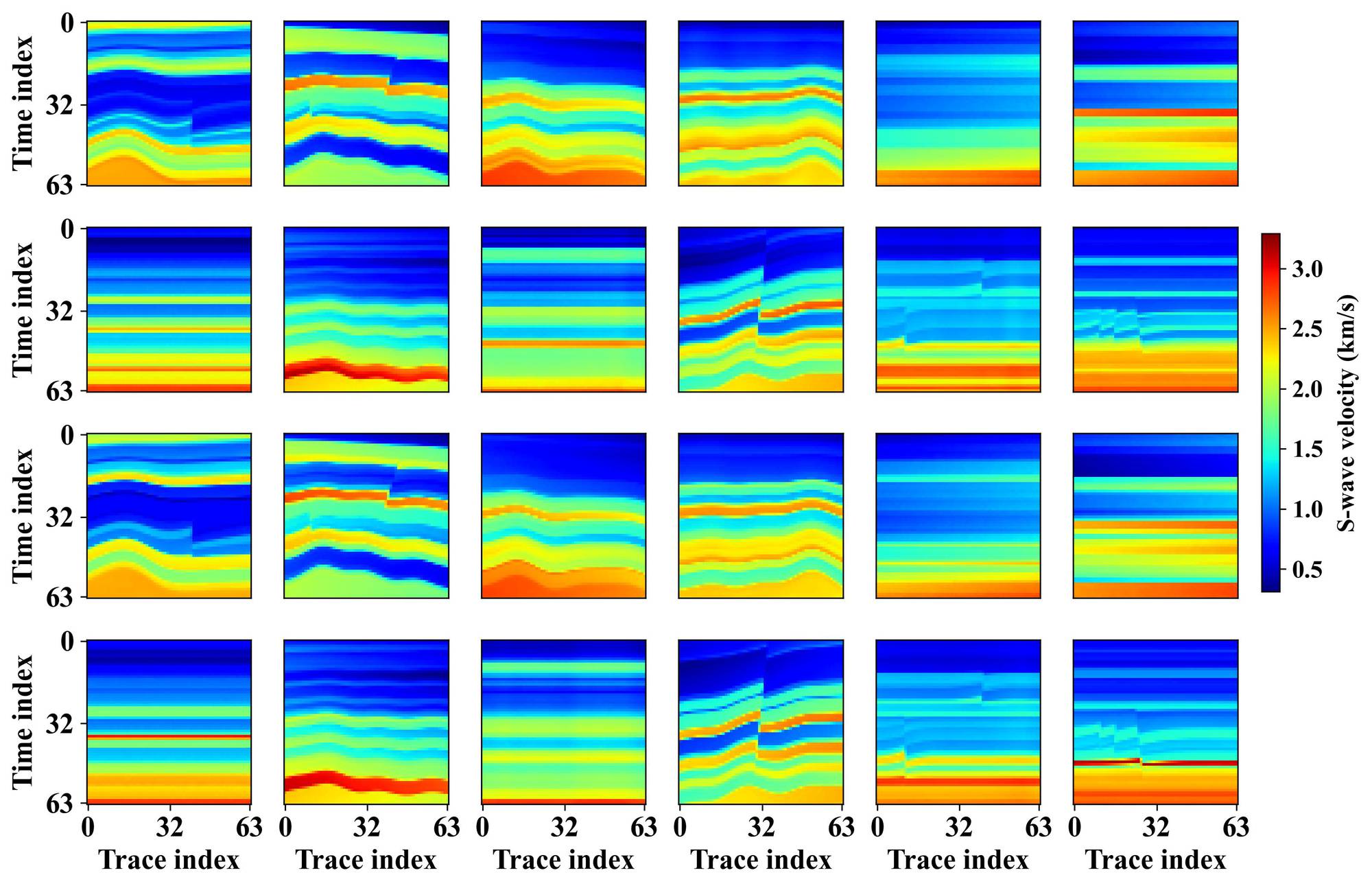}
 \label{fig:ddpm-condlow2seis-vs}} 
  \subfigure[]{\includegraphics[width=0.65\columnwidth]{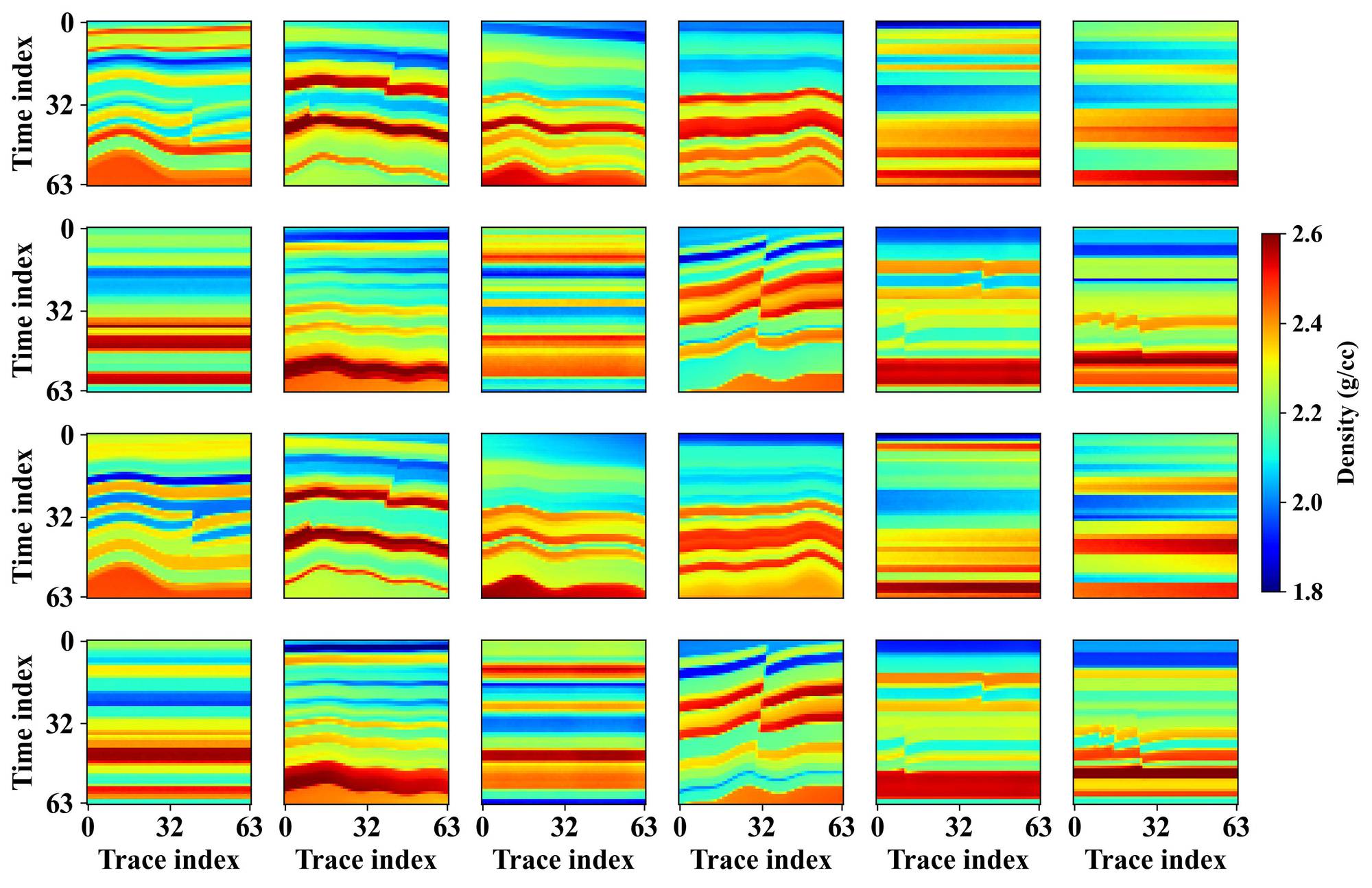}
 \label{fig:ddpm-condlow2seis-rho}}  
 \caption{Elastic parameter synthesis by using the proposed method under dual-condition guidance. 
(a)--(c) True P-wave velocity, S-wave velocity, and density models. 
(d)--(f) Samples conditioned on well logs and structural information. 
(g)--(i) Samples conditioned on seismic data and well logs. 
(j)--(l) Samples conditioned on seismic data and low-frequency models. 
In each panel, the first two rows show the results from the first sampling run, whereas the last two rows show the results from the second sampling run.
 }
 \label{fig:multi2samples}
\end{figure*}

\textbf{\textit{Patch-scale synthesis on $64 \times 64$ models:}} To analyze the differences between the synthesized samples and the true models, Figs. \ref{fig:true-vp}--\ref{fig:true-rho} show the true P-wave velocity, S-wave velocity, and density models. We first perform elastic parameter synthesis under several dual-condition settings, including pseudo-well logs combined with structural information, pseudo-well logs combined with seismic data (22.36 dB), and seismic data (22.36 dB) combined with low-frequency models. The corresponding synthesized samples are shown in Fig. \ref{fig:multi2samples}. When pseudo-well logs and structural information are jointly used as conditions, the structural masks guide the spatial propagation of the sparse pseudo-well log information along the prescribed horizons and faults. As a result, some generated samples exhibit structures and parameter distributions similar to those of the true models. However, this consistency depends strongly on the number of available pseudo-well logs and the richness of the structural information. This explains why the samples in the third column show structural differences from the true models and why some parameter values deviate from the true values when only a limited number of pseudo-well logs are available.

Compared with the combination of pseudo-well logs and structural information, the joint guidance of pseudo-well logs and seismic data provides stronger constraints on structures consistent with the seismic data, enabling the synthesized samples to better preserve structures consistent with the true models. However, some generated samples exhibit local block-like amplitude anomalies, especially when only a few pseudo-well logs are available. In these regions, the parameter values show a noticeable amplitude-level shift relative to the surrounding background. This is mainly because well logs provide sparse pointwise constraints, whereas seismic data provide band-limited spatial observations. Therefore, their combination may still be insufficient to fully recover the absolute parameter values in regions far from the wells. We further show the synthesized results conditioned on seismic data and low-frequency models, where the low-frequency models are obtained by applying the smoothing operator twice. The generated samples exhibit structures consistent with the true models. However, the prediction accuracy remains limited because the overly smoothed low-frequency models provide insufficient information about the true background variations. 

\begin{figure*}[htb!]
\setlength{\abovecaptionskip}{0.2cm}
 \centering
    \subfigure[]{\includegraphics[width=0.65\columnwidth]{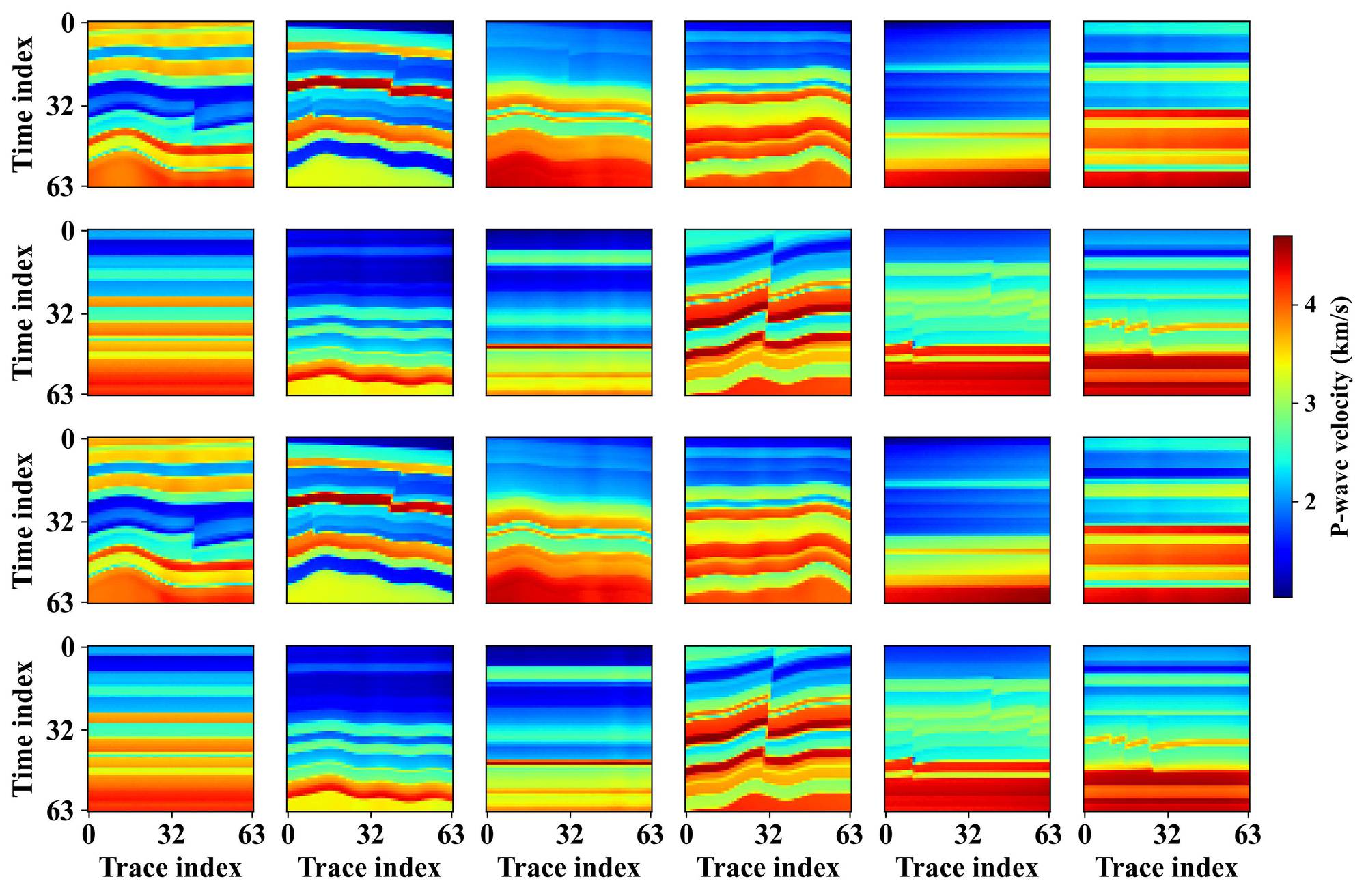}
 \label{fig:ddpm-condlogseis_impr-vp}}
 \subfigure[]{\includegraphics[width=0.65\columnwidth]{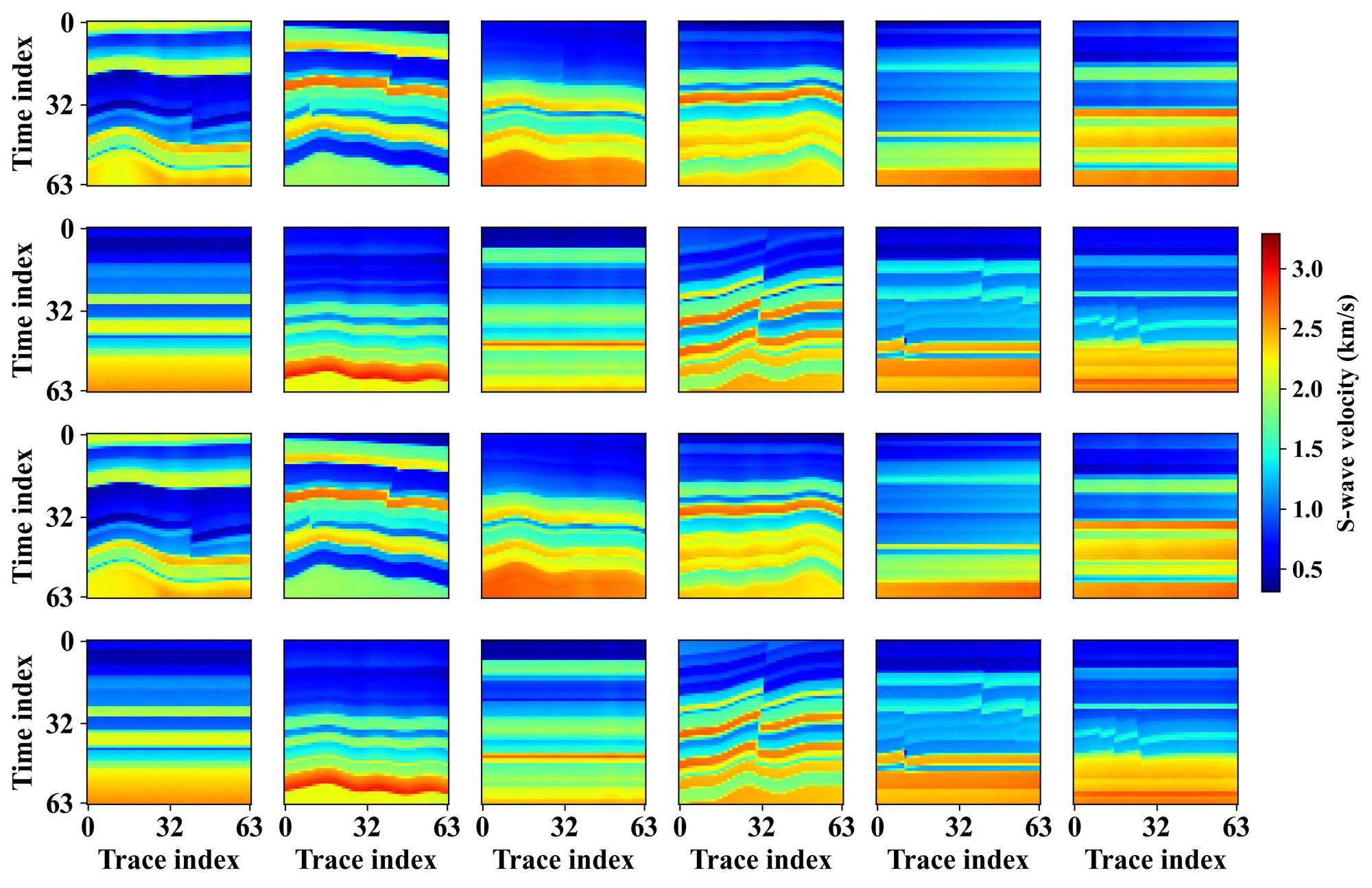}
 \label{fig:ddpm-condlogseis_impr-vs}} 
  \subfigure[]{\includegraphics[width=0.65\columnwidth]{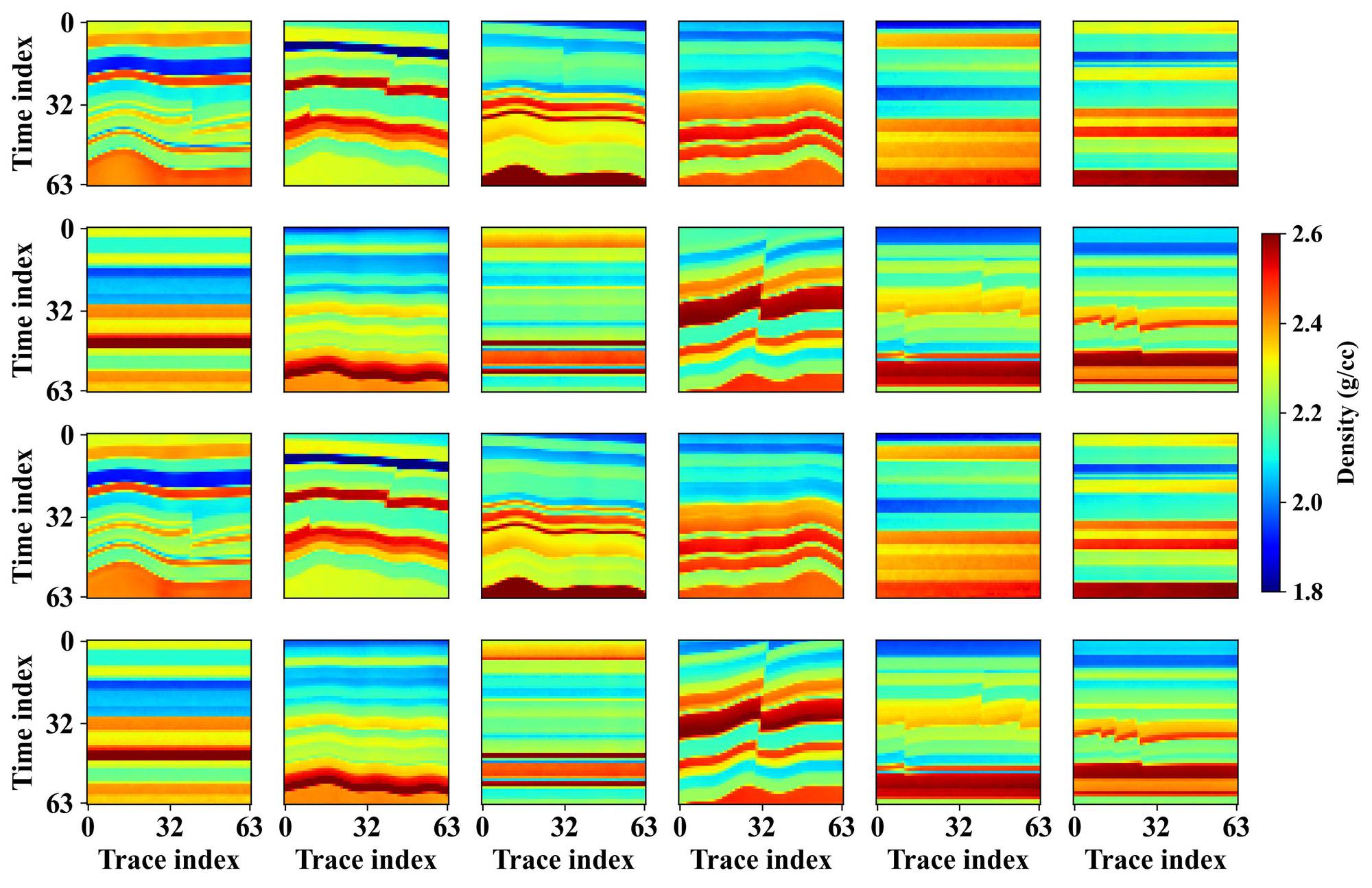}
 \label{fig:ddpm-condlogseis_impr-rho}} 
   \subfigure[]{\includegraphics[width=0.65\columnwidth]{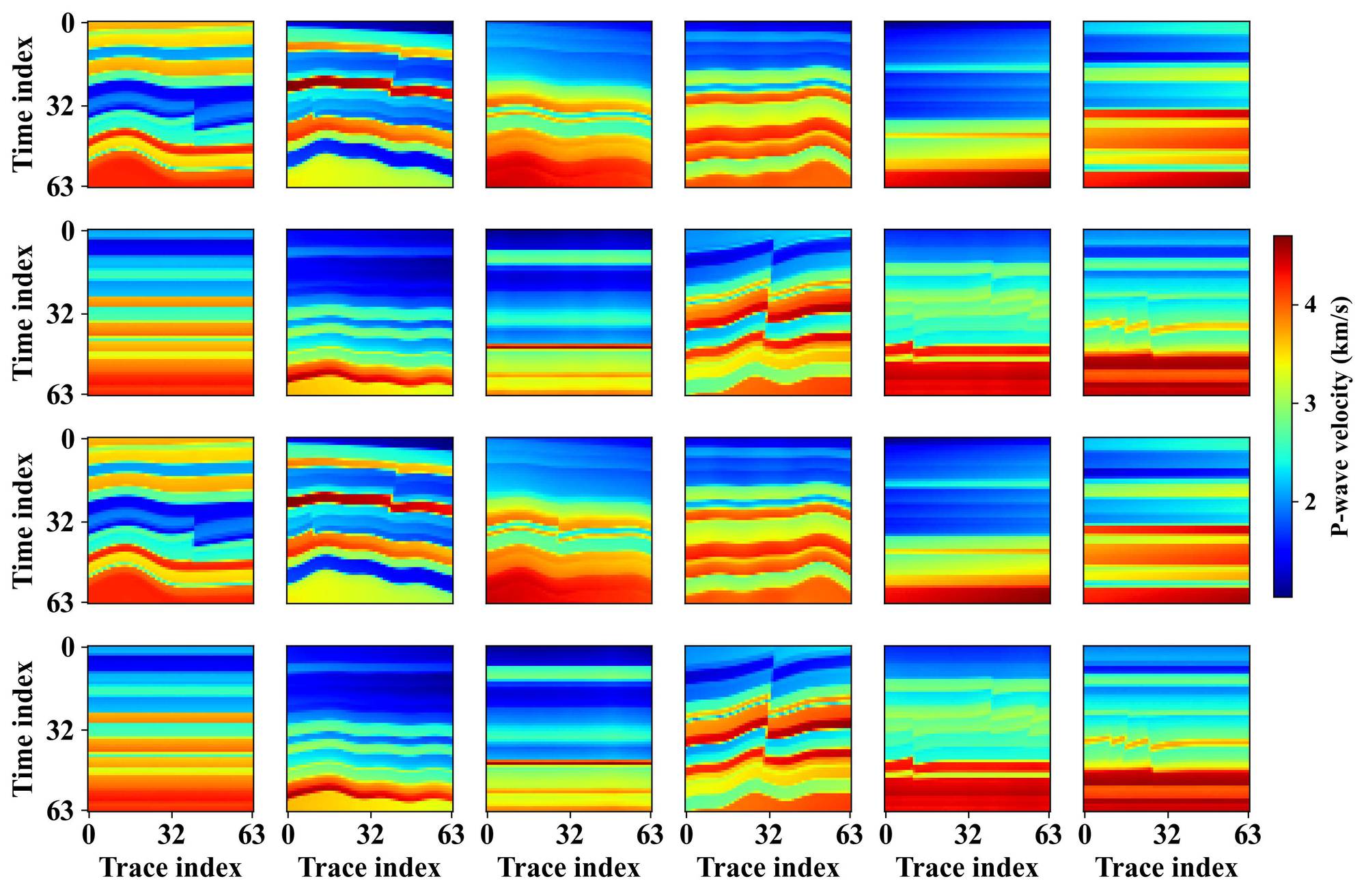}
 \label{fig:ddpm-condlow2logseis-vp}}
 \subfigure[]{\includegraphics[width=0.65\columnwidth]{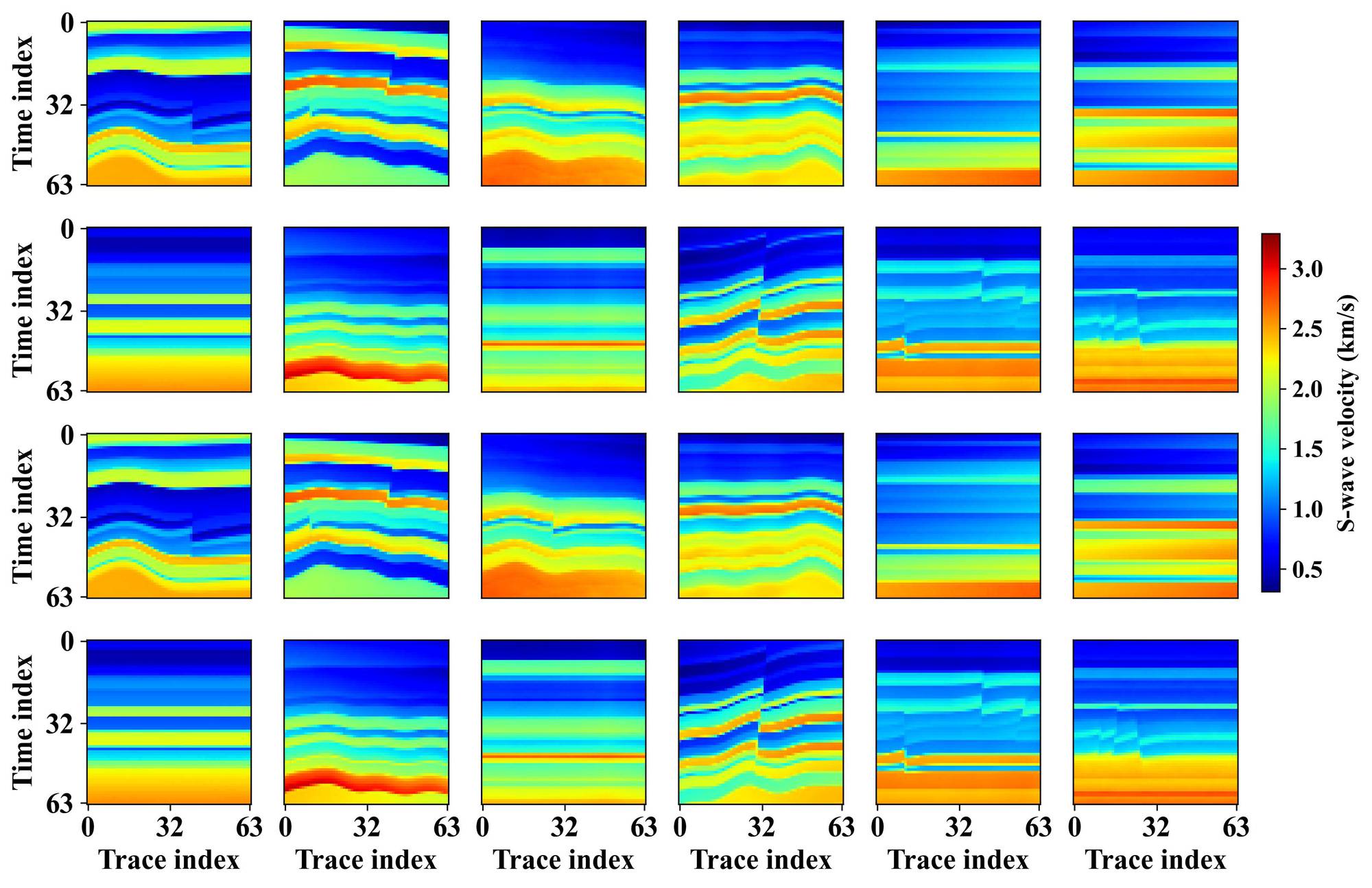}
 \label{fig:ddpm-condlow2logseis-vs}} 
  \subfigure[]{\includegraphics[width=0.65\columnwidth]{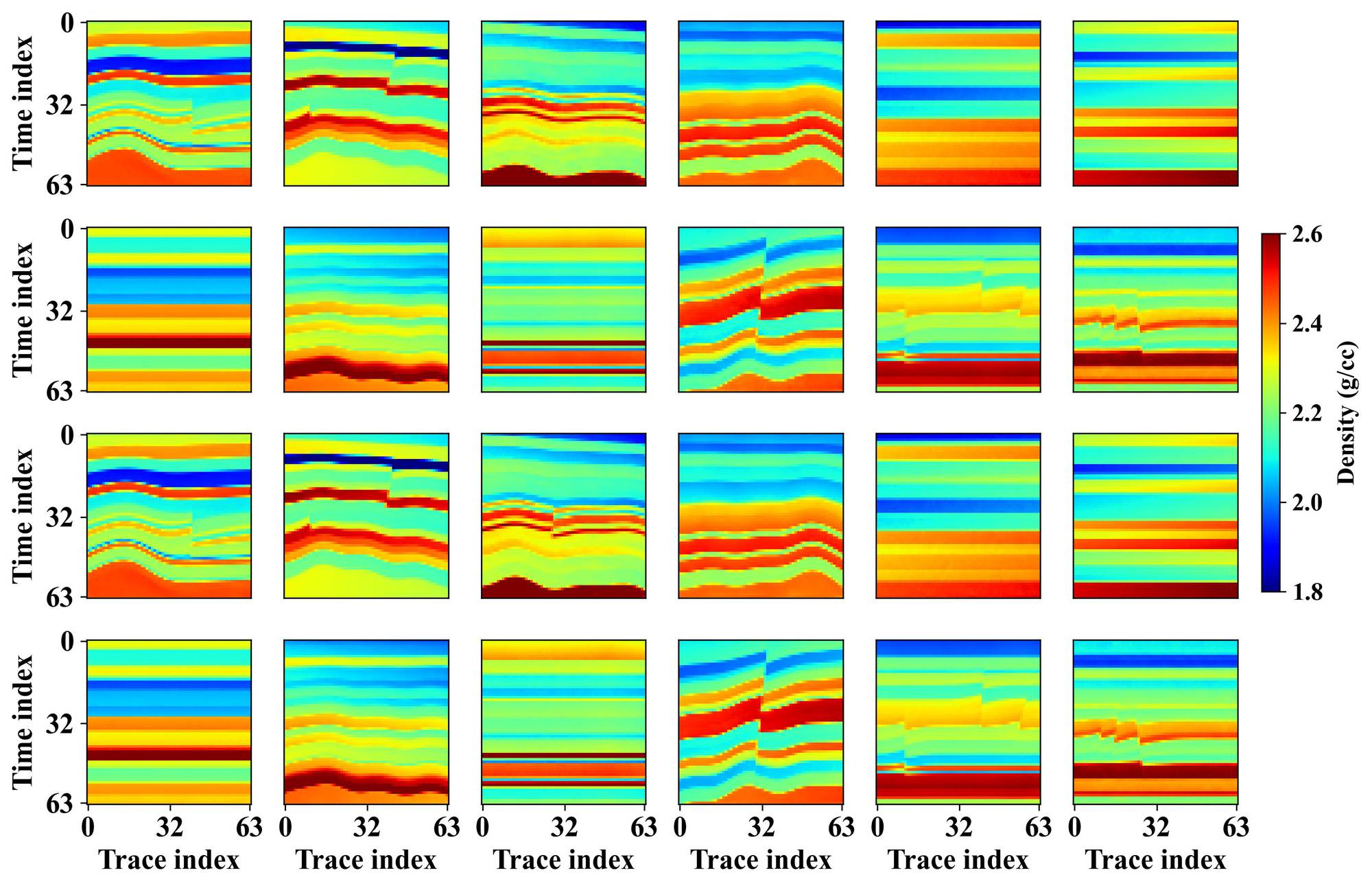}
 \label{fig:ddpm-condlow2logseis-rho}}  
 \caption{Elastic parameter synthesis under triple-condition guidance. 
(a)--(c) Samples conditioned on seismic data, well logs, and interpolated well-log models. 
(d)--(f) Samples conditioned on seismic data, low-frequency models, and well logs. 
In each panel, the first two rows show the results from the first sampling run, whereas the last two rows show the results from the second sampling run.}
\label{fig:multi3samples}
\end{figure*}

\begin{figure*}[htb!]
\setlength{\abovecaptionskip}{0.2cm}
 \centering
     \subfigure[]{\includegraphics[width=0.65\columnwidth]{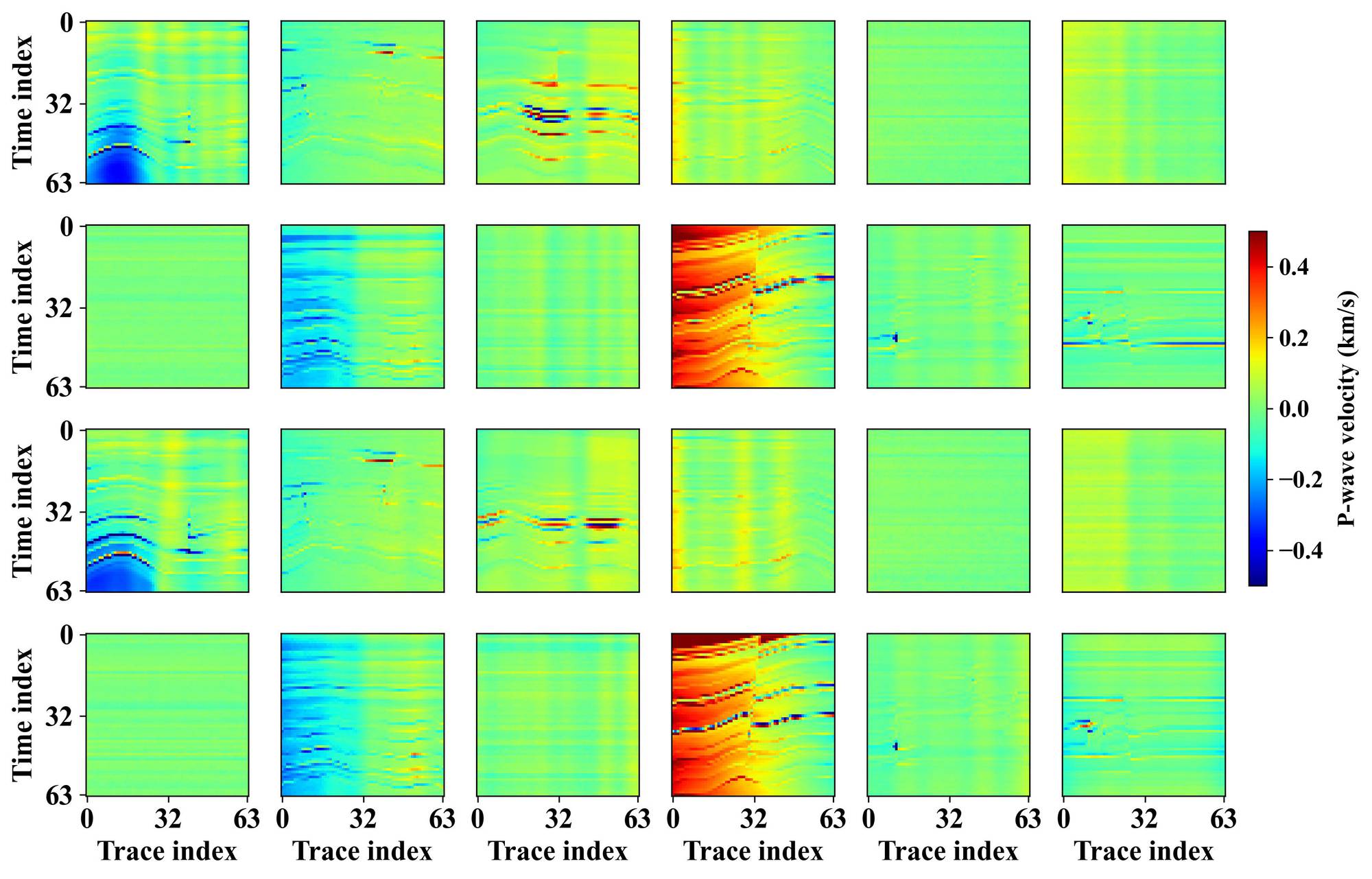}
 \label{fig:ddpm-condlogseis_impr_erro-vp}}
 \subfigure[]{\includegraphics[width=0.65\columnwidth]{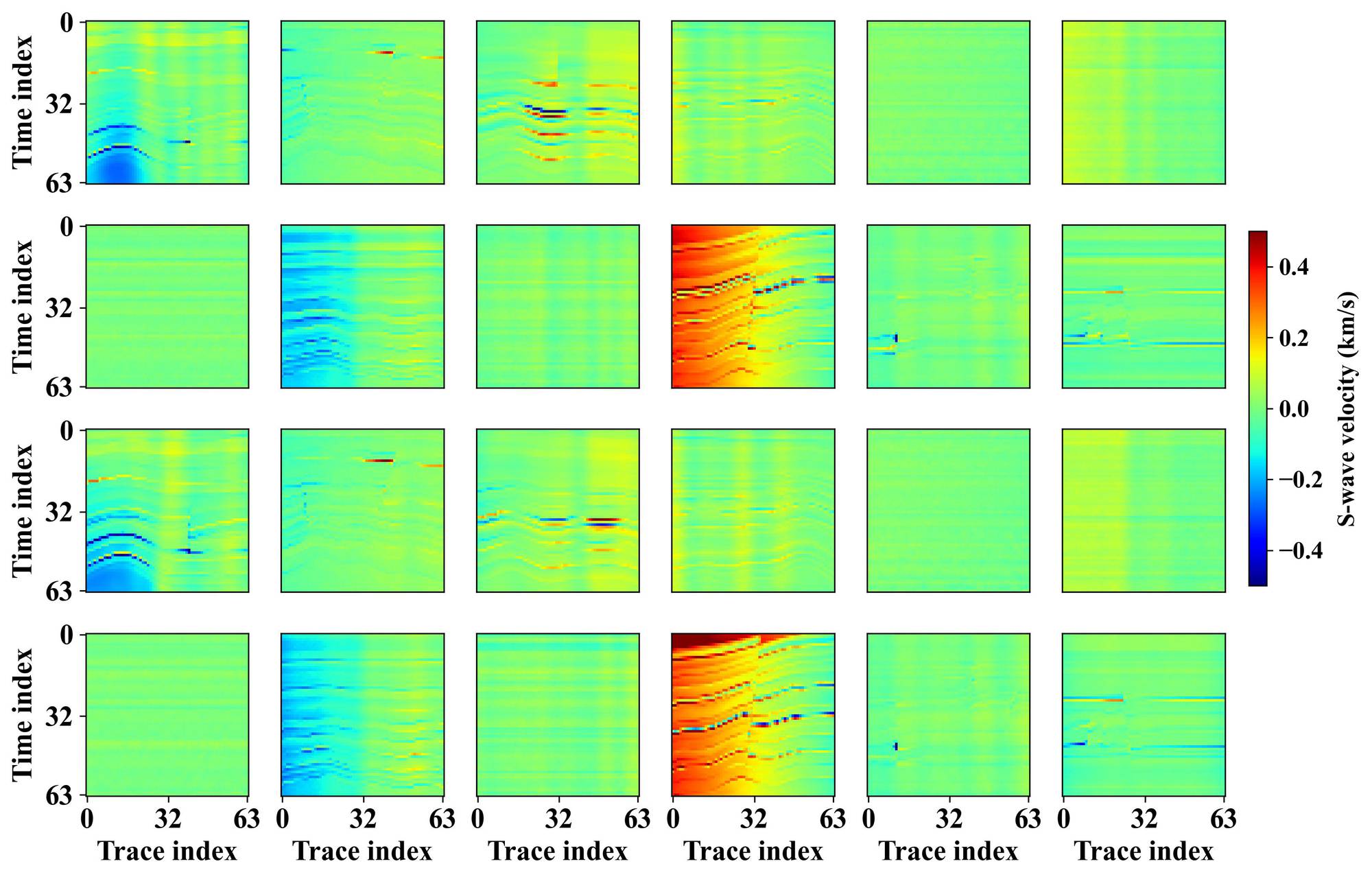}
 \label{fig:ddpm-condlogseis_impr_erro-vs}} 
  \subfigure[]{\includegraphics[width=0.65\columnwidth]{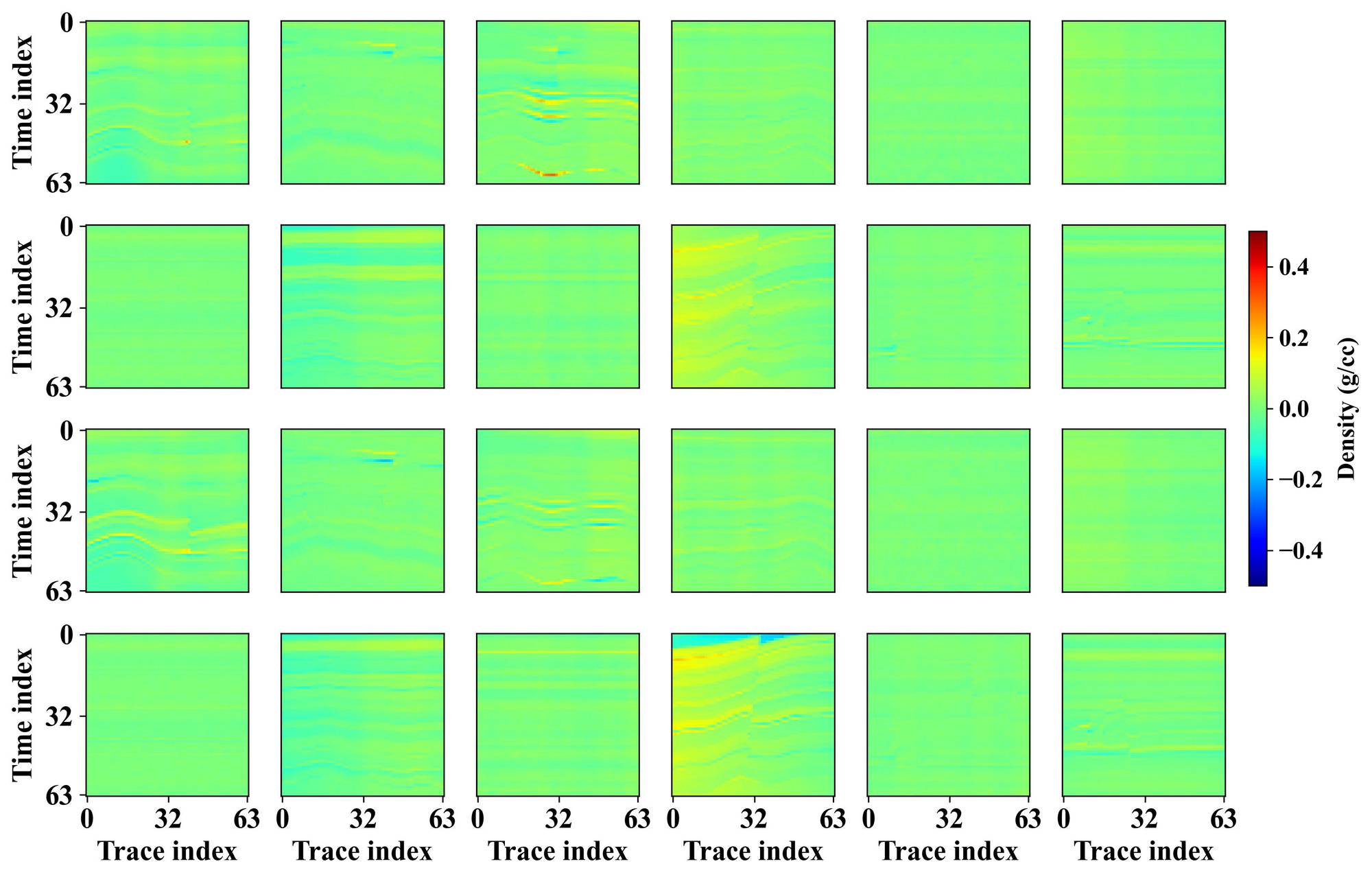}
 \label{fig:ddpm-condlogseis_impr_erro-rho}} 
   \subfigure[]{\includegraphics[width=0.65\columnwidth]{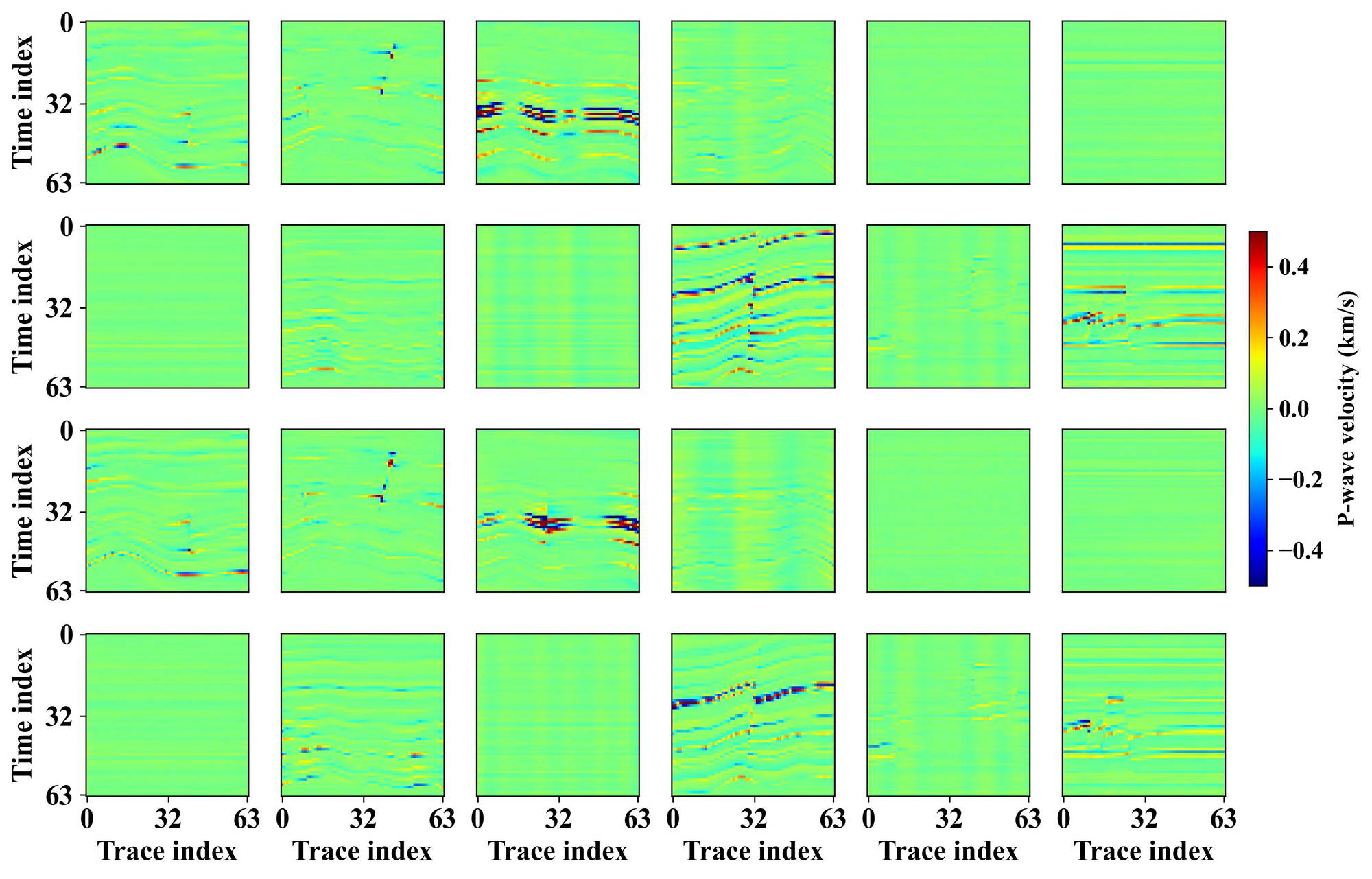}
 \label{fig:ddpm-condlow2logseis_erro-vp}}
 \subfigure[]{\includegraphics[width=0.65\columnwidth]{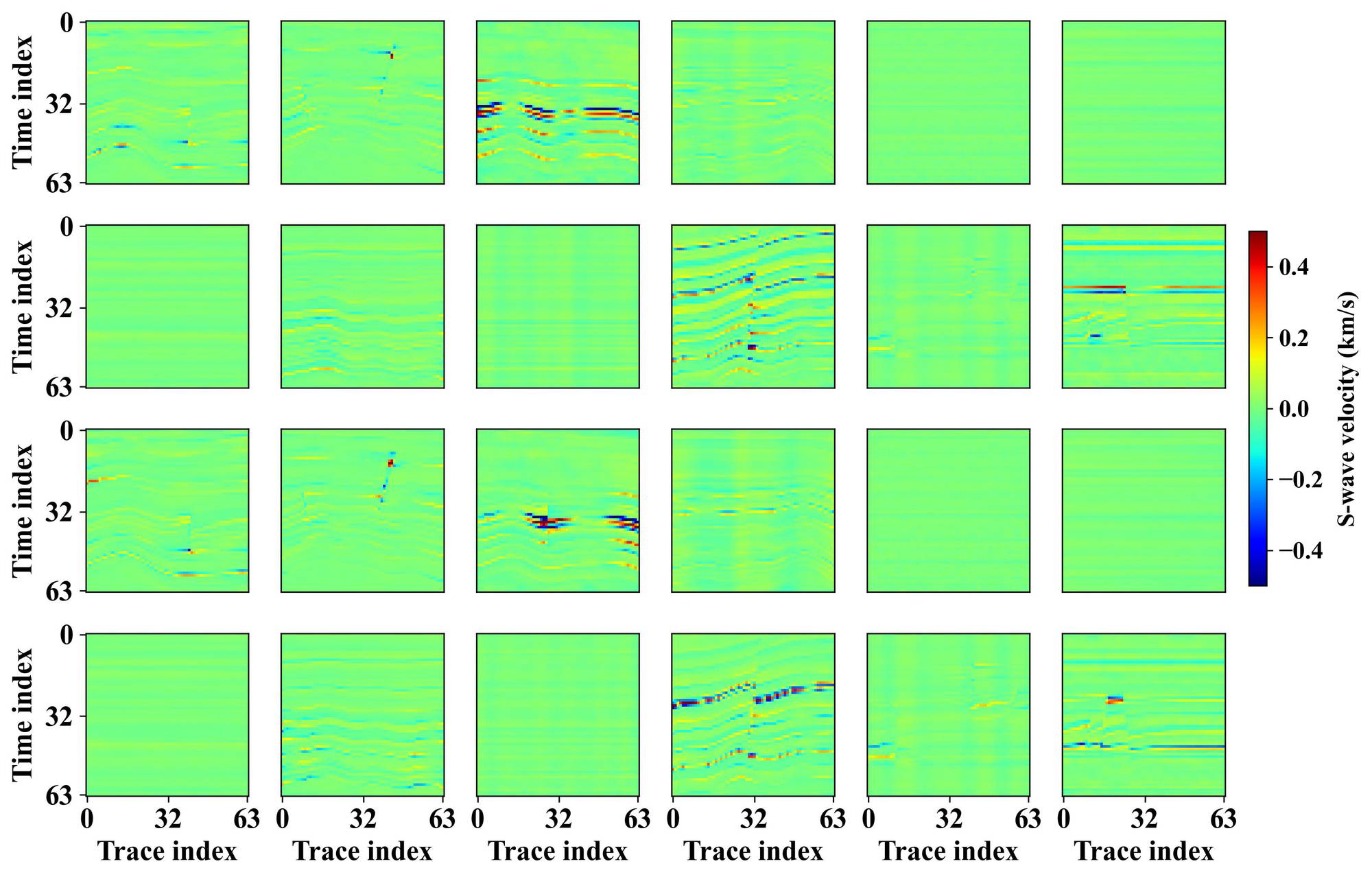}
 \label{fig:ddpm-condlow2logseis_erro-vs}} 
  \subfigure[]{\includegraphics[width=0.65\columnwidth]{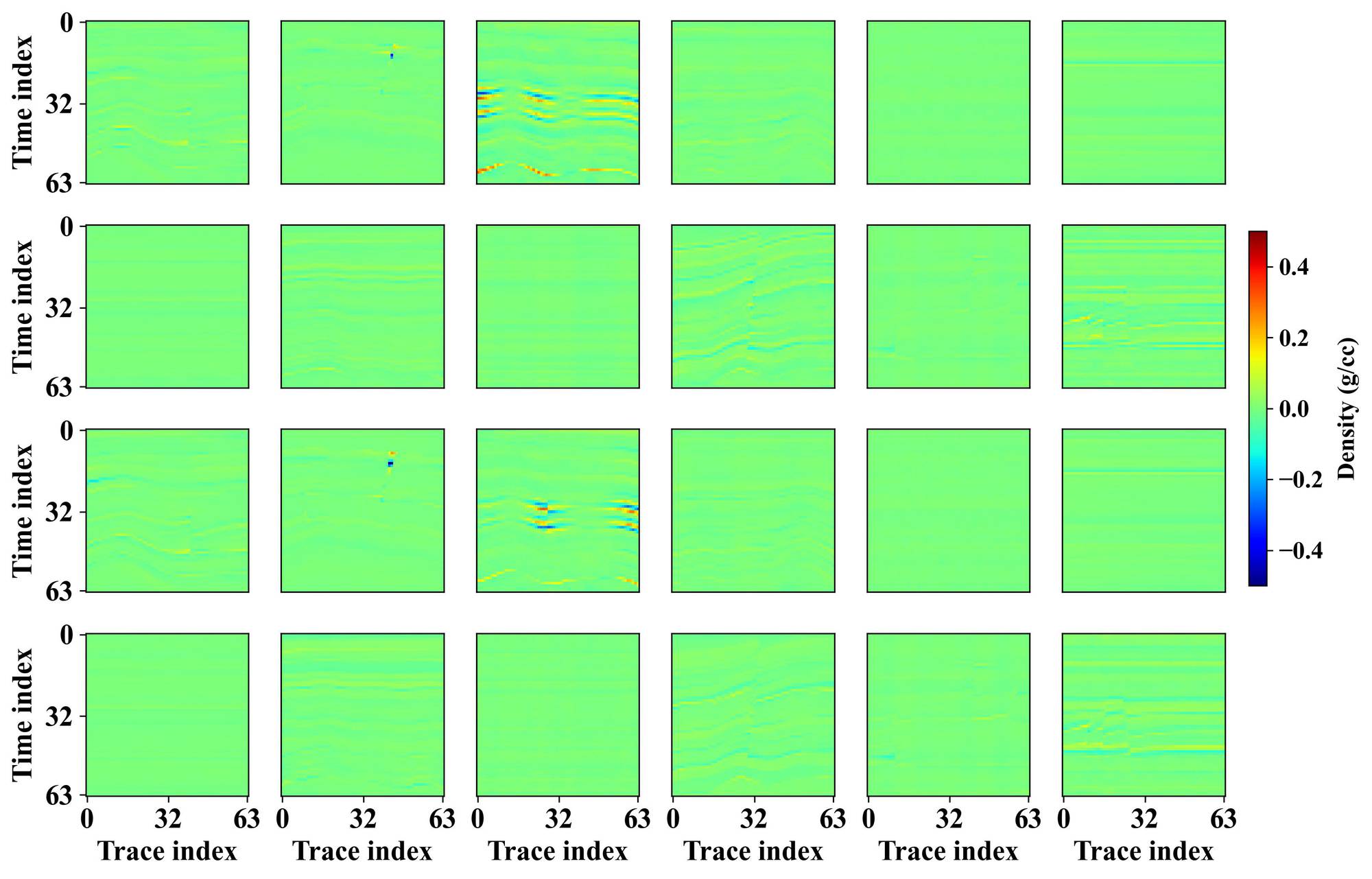}
 \label{fig:ddpm-condlow2logseis_erro-rho}}  
 \caption{Errors between the generated samples and the true models shown in Fig. \ref{fig:multi2samples}. 
%(a)--(c) Errors corresponding to the samples conditioned on seismic data and low-frequency models. 
(a)--(c) Errors corresponding to the samples conditioned on seismic data, well logs, and interpolated well-log models. 
(d)--(f) Errors corresponding to the samples conditioned on seismic data, well logs, and low-frequency models.}
\label{fig:multsampleserro}
\end{figure*}

Finally, we investigate the elastic parameter synthesis under triple-condition guidance, as shown in Fig. \ref{fig:multi3samples}. Figs. \ref{fig:ddpm-condlogseis_impr-vp}--\ref{fig:ddpm-condlogseis_impr-rho} show the samples conditioned on seismic data, pseudo-well logs, and interpolated well-log models, where the factor in ILVR is set to $N=32$. Figs. \ref{fig:ddpm-condlow2logseis-vp}--\ref{fig:ddpm-condlow2logseis-rho} show the samples conditioned on seismic data, pseudo-well logs, and low-frequency models. Compared with the samples conditioned only on seismic data and pseudo-well logs, the introduction of interpolated well-log models helps mitigate block-like amplitude anomalies. This is because the interpolation models provide additional spatial constraints between well locations and thus compensate, to some extent, for the sparsity of well log constraints.  However, small amplitude deviations can still be observed because ILVR with a relatively large factor $N$ imposes only a soft coarse-scale constraint from the interpolation models, which does not fully control local amplitude variations. Furthermore, the samples conditioned on seismic data, pseudo-well logs, and low-frequency models show better agreement with the true models. Compared with the results conditioned on seismic data and low-frequency models, the additional pseudo-well log constraints improve the local calibration of elastic parameter values, thereby enhancing the prediction accuracy. For a more intuitive comparison, Fig. \ref{fig:multsampleserro} shows the errors between the generated samples and the true models. The error maps support the above observations, showing that the joint use of seismic data, pseudo-well logs, and low-frequency models provides more accurate and stable synthesized results.

The above experiments demonstrate that different types of conditioning information play different roles in elastic parameter synthesis. Therefore, appropriate condition combinations should be selected according to the target application and the reliability of the available prior information, especially when the goal is to improve the prediction accuracy of elastic parameters.

\begin{figure*}[htb!]
\setlength{\abovecaptionskip}{0.2cm}
 \centering
   \subfigure[]{\includegraphics[width=0.65\columnwidth]{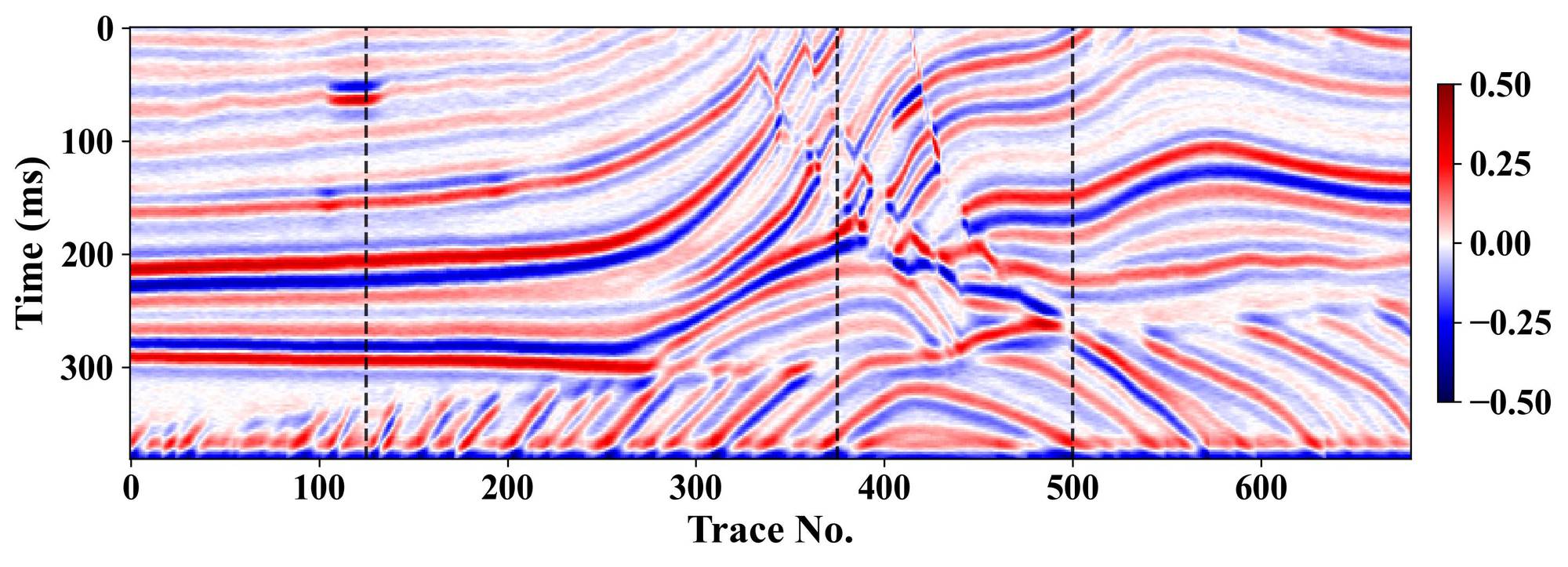}
 \label{fig:mar-seis0}}
    \subfigure[]{\includegraphics[width=0.65\columnwidth]{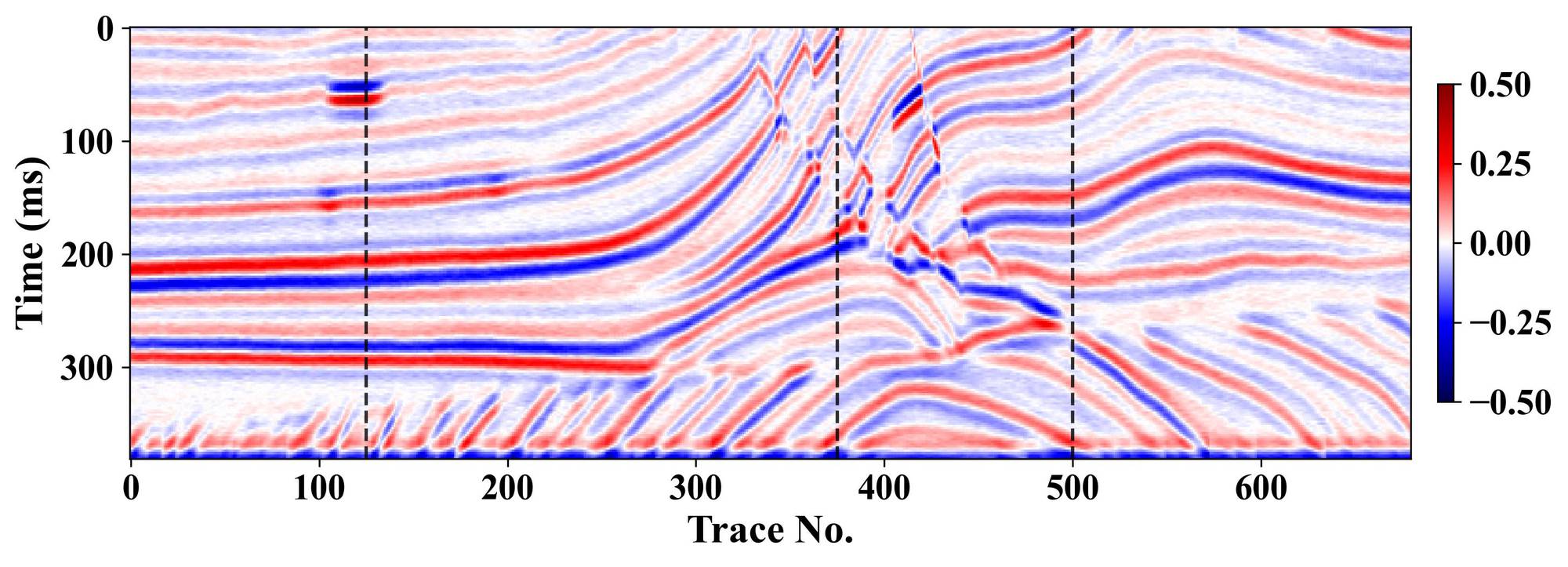}
 \label{fig:mar-seis1}}
    \subfigure[]{\includegraphics[width=0.65\columnwidth]{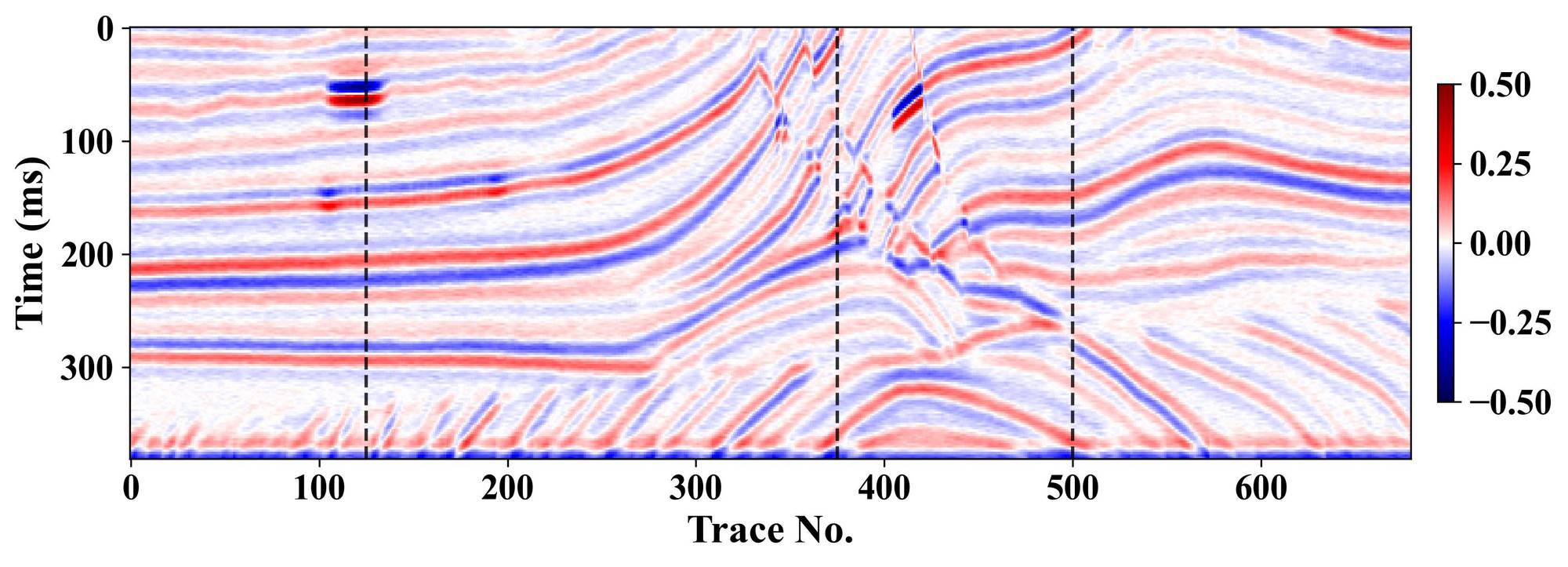}
 \label{fig:mar-seis2}}
    \subfigure[]{\includegraphics[width=0.65\columnwidth]{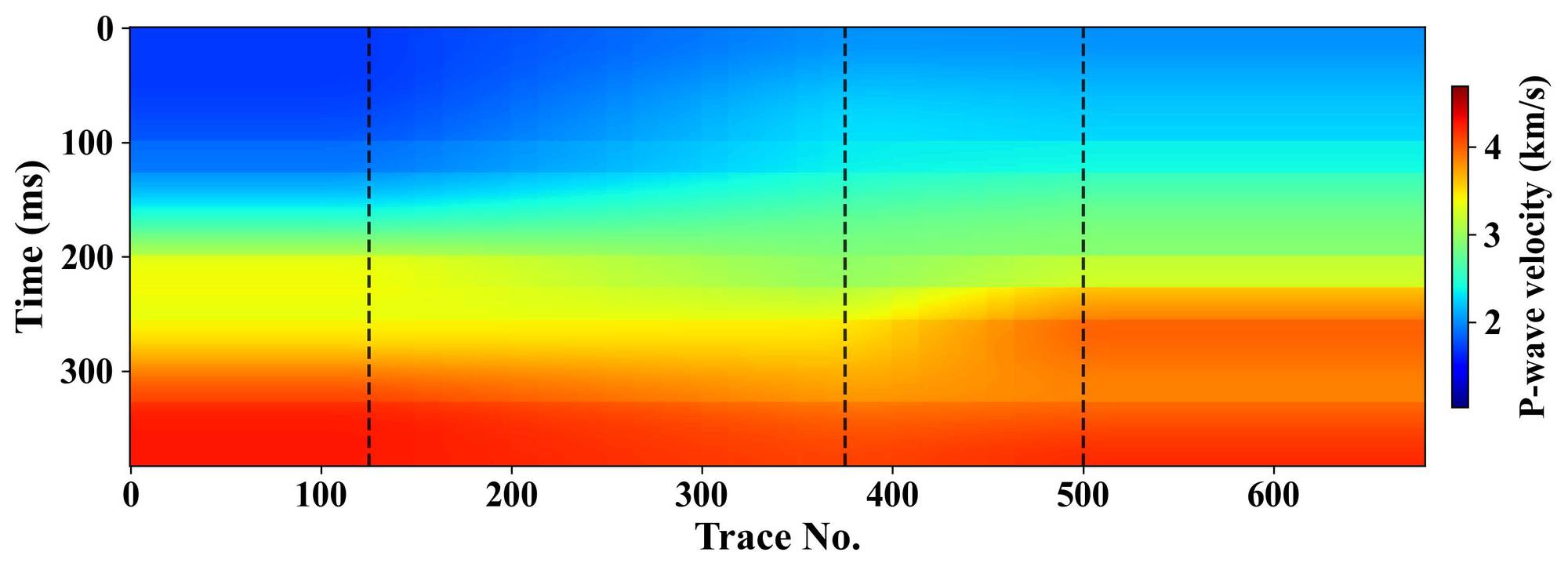}
 \label{fig:mar-lowvp}}  
     \subfigure[]{\includegraphics[width=0.65\columnwidth]{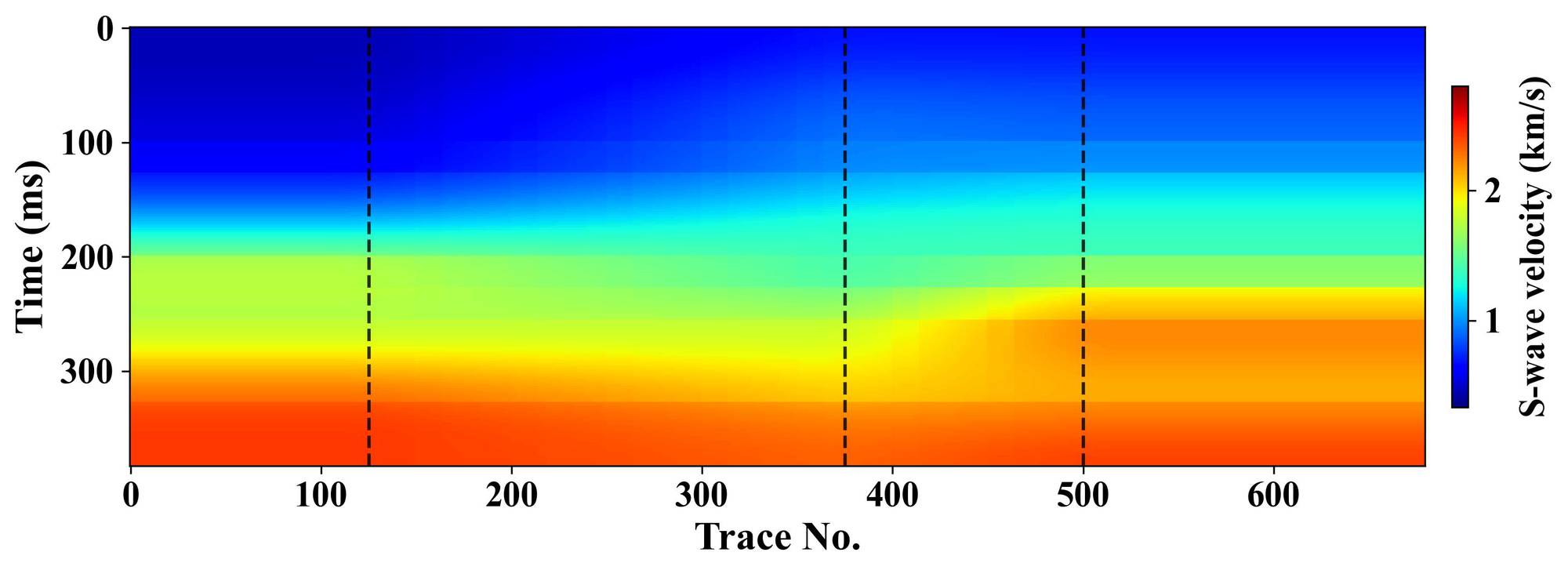}
 \label{fig:mar-lowvs}}
     \subfigure[]{\includegraphics[width=0.65\columnwidth]{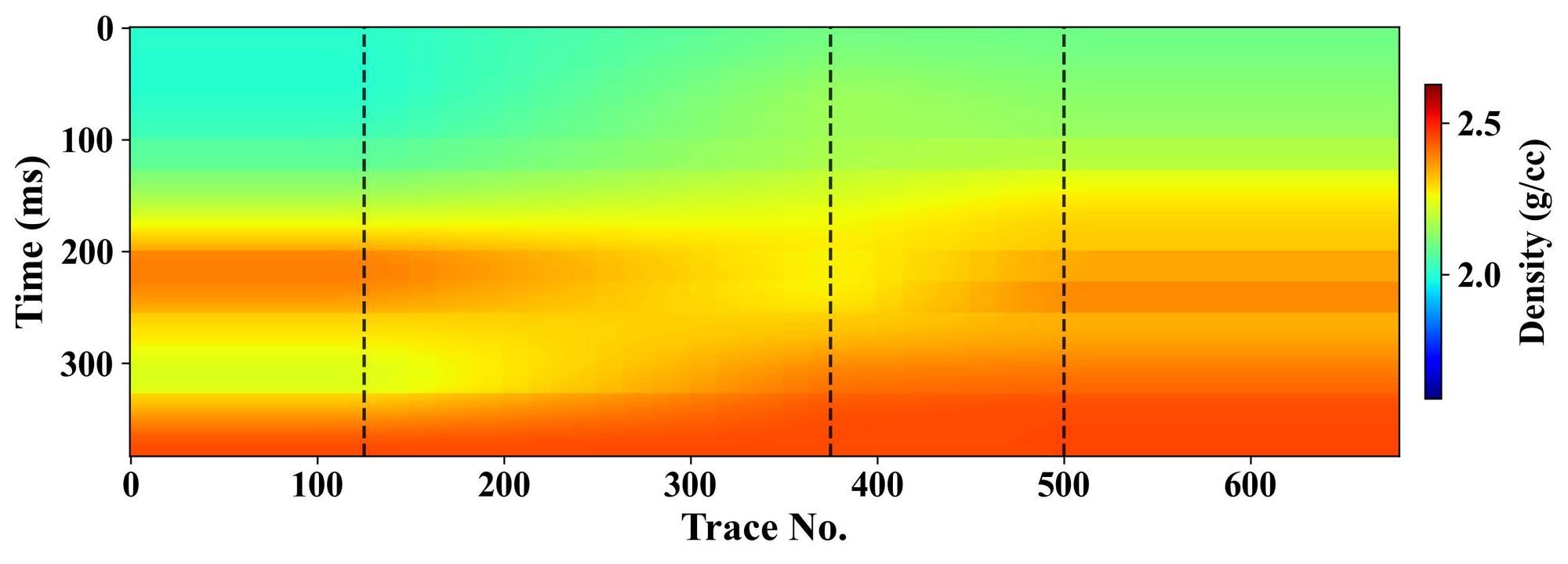}
 \label{fig:mar-lowrho}}
 \caption{Conditioning information used for elastic parameter synthesis on the Marmousi II model. 
(a)--(c) Synthetic seismic angle gathers at incidence angles of $12^\circ$, $24^\circ$, and $36^\circ$, with an SNR of 19.91 dB. 
(d)--(f) Low-frequency models of P-wave velocity, S-wave velocity, and density, obtained by applying 1D spatial interpolation to three pseudo-well logs and then applying a $31 \times 31$ mean filter three times.}
\label{fig:marseis}
\end{figure*}

\textbf{\textit{Large-scale synthesis on the Marmousi II model:}} We further apply the proposed method to a downsampled version of the Marmousi II model to evaluate its performance for elastic parameter synthesis over a larger spatial extent. When seismic data are included among the conditioning information, this synthesis task can also be viewed as a seismic inverse problem guided by the learned diffusion prior during the sampling process. Figs. \ref{fig:mar-seis0}--\ref{fig:mar-seis2} show the synthetic seismic angle gathers at incidence angles of $12^\circ$, $24^\circ$, and $36^\circ$, generated using the Shuey approximation and a 30 Hz Ricker wavelet. To mimic spatially correlated noise, random noise is first smoothed using a three-point moving-average operator and then added to the synthetic seismic data, resulting in an SNR of 19.91 dB. In addition to seismic data, the same three randomly selected pseudo-well logs used for dataset construction are extracted from the Marmousi II models and interpolated to obtain interpolated well-log models. The interpolated data are further smoothed by applying a $31 \times 31$ mean filter three times, yielding the low-frequency models shown in Figs. \ref{fig:mar-lowvp}--\ref{fig:mar-lowrho}.

\begin{figure*}[htb!]
\setlength{\abovecaptionskip}{0.2cm}
 \centering
   \subfigure[]{\includegraphics[width=0.65\columnwidth]{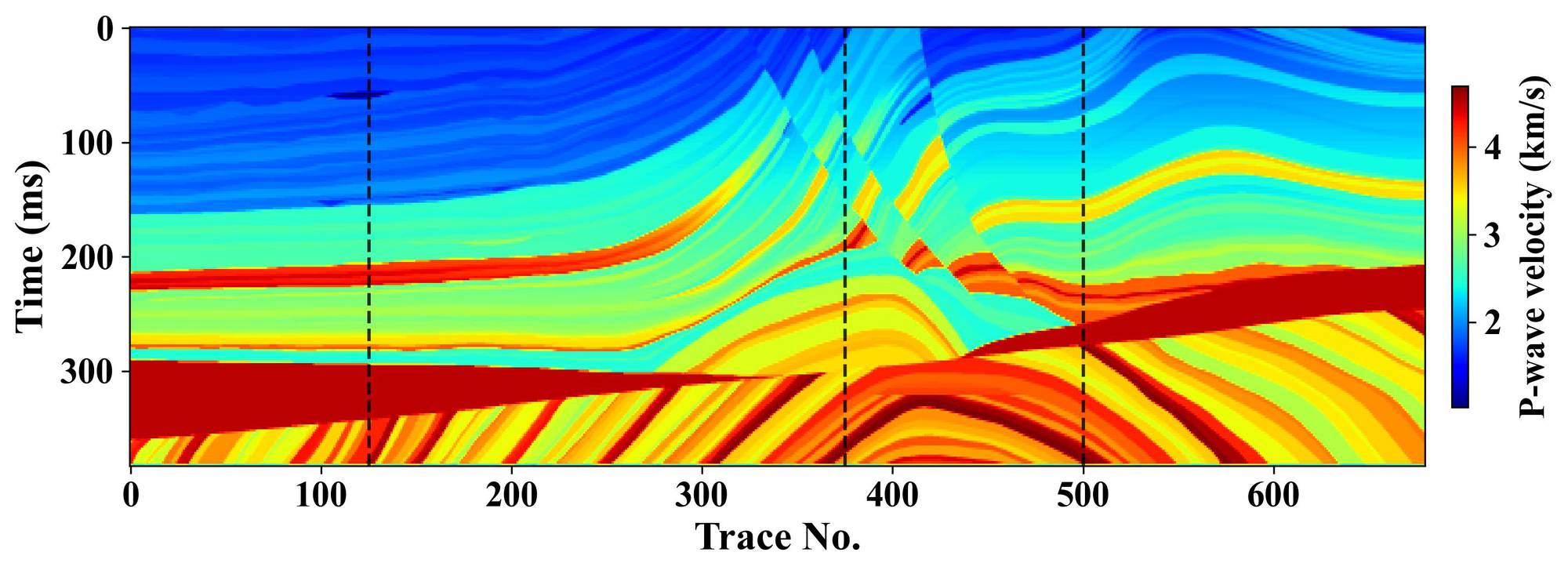}
 \label{fig:mar-vp}} 
     \subfigure[]{\includegraphics[width=0.65\columnwidth]{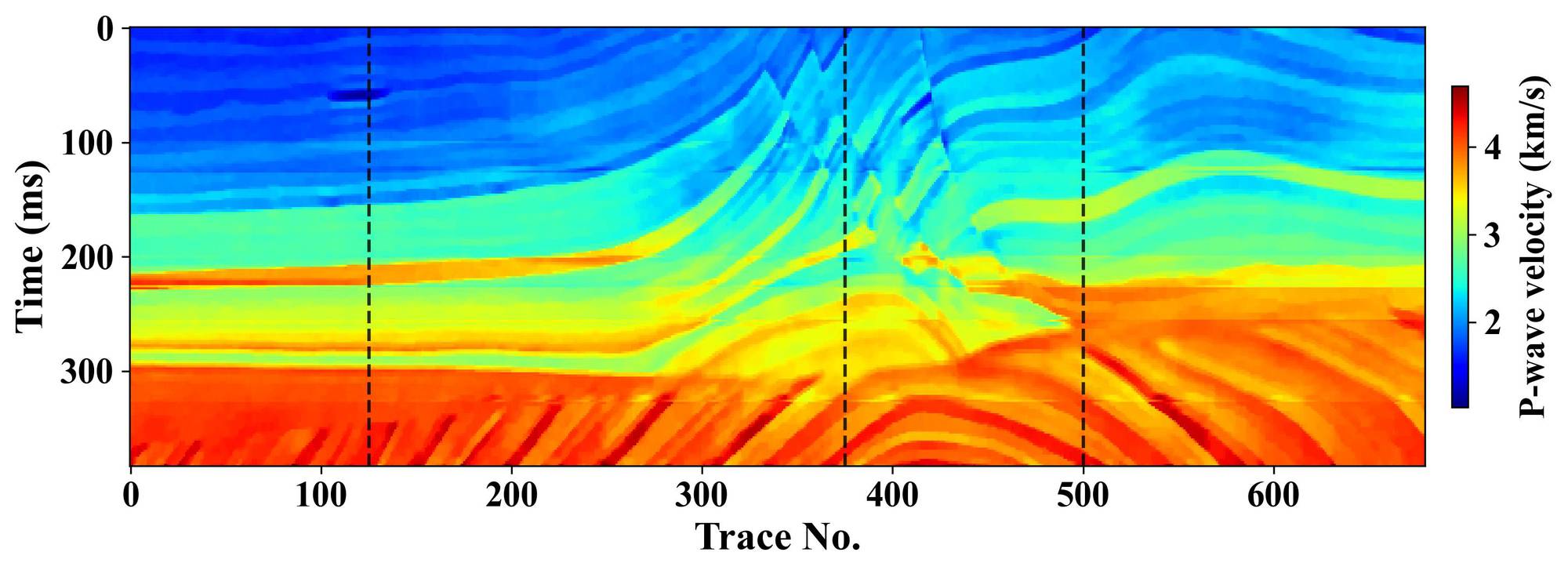}
 \label{fig:2DTVnolow-mar-vp}}
    \subfigure[]{\includegraphics[width=0.65\columnwidth]{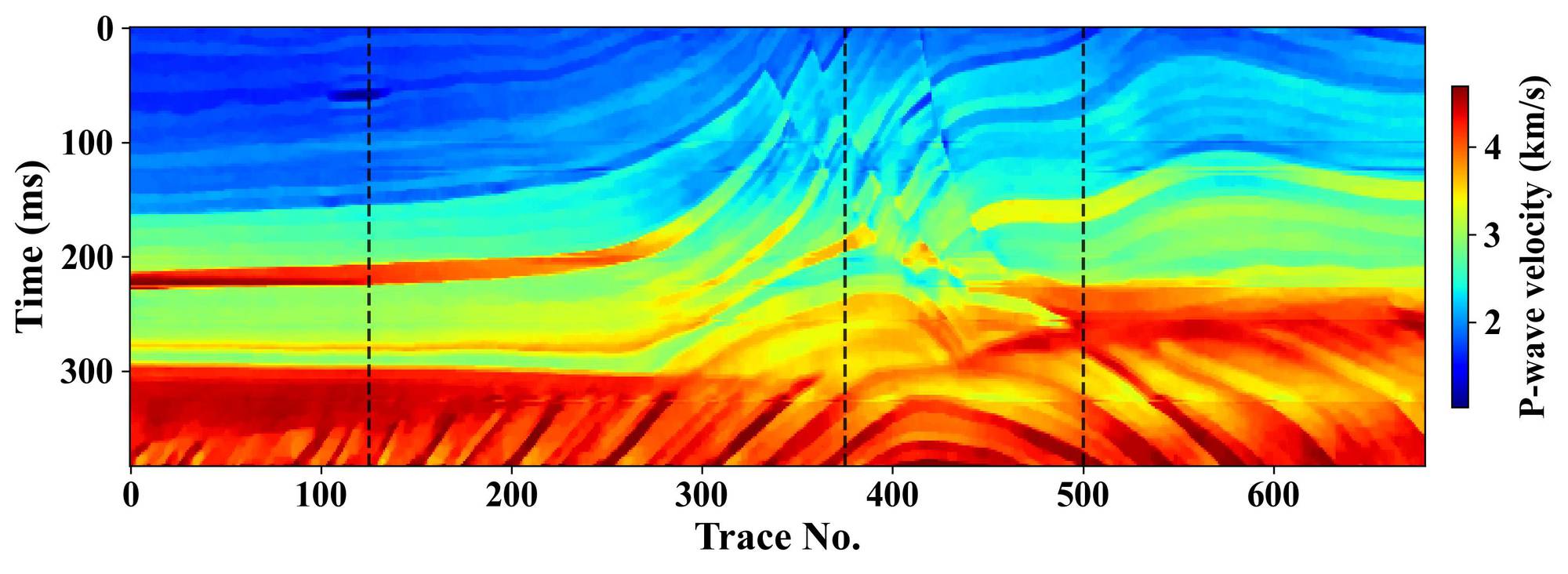}
 \label{fig:2DTV-mar-vp}}
  \subfigure[]{\includegraphics[width=0.65\columnwidth]{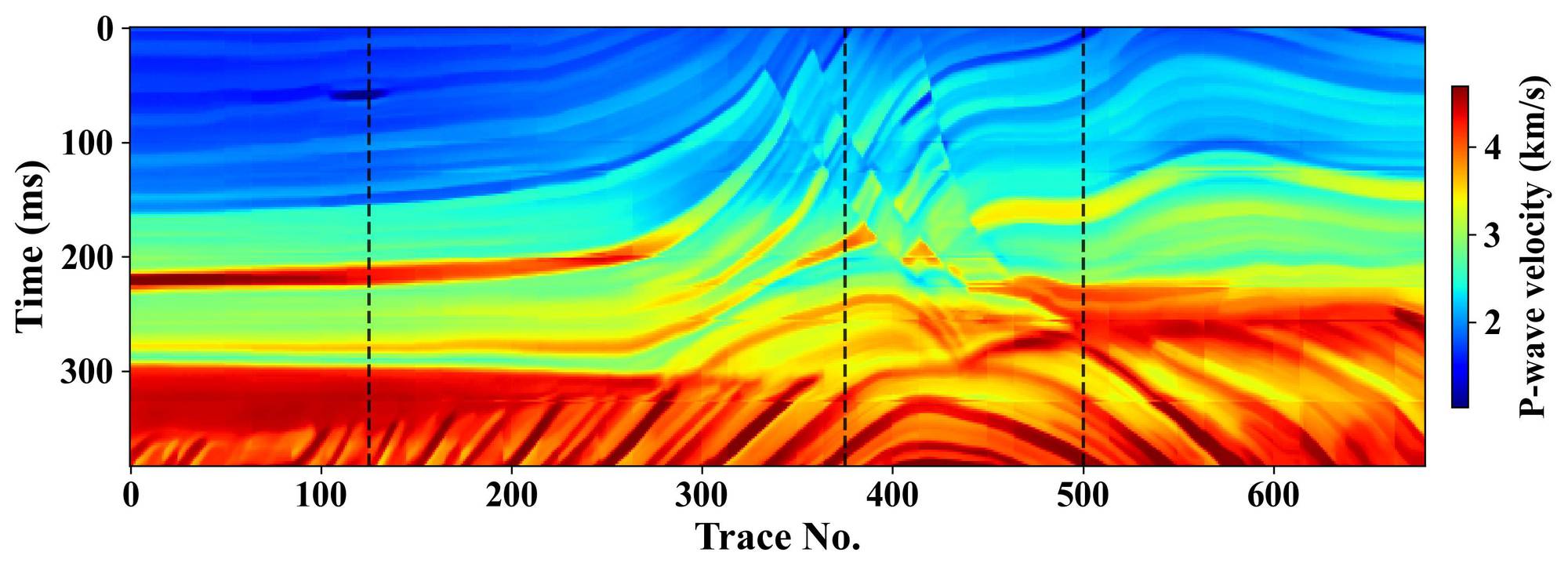}
 \label{fig:ddpm-condlowseis_mar-vp}}  
  \subfigure[]{\includegraphics[width=0.65\columnwidth]{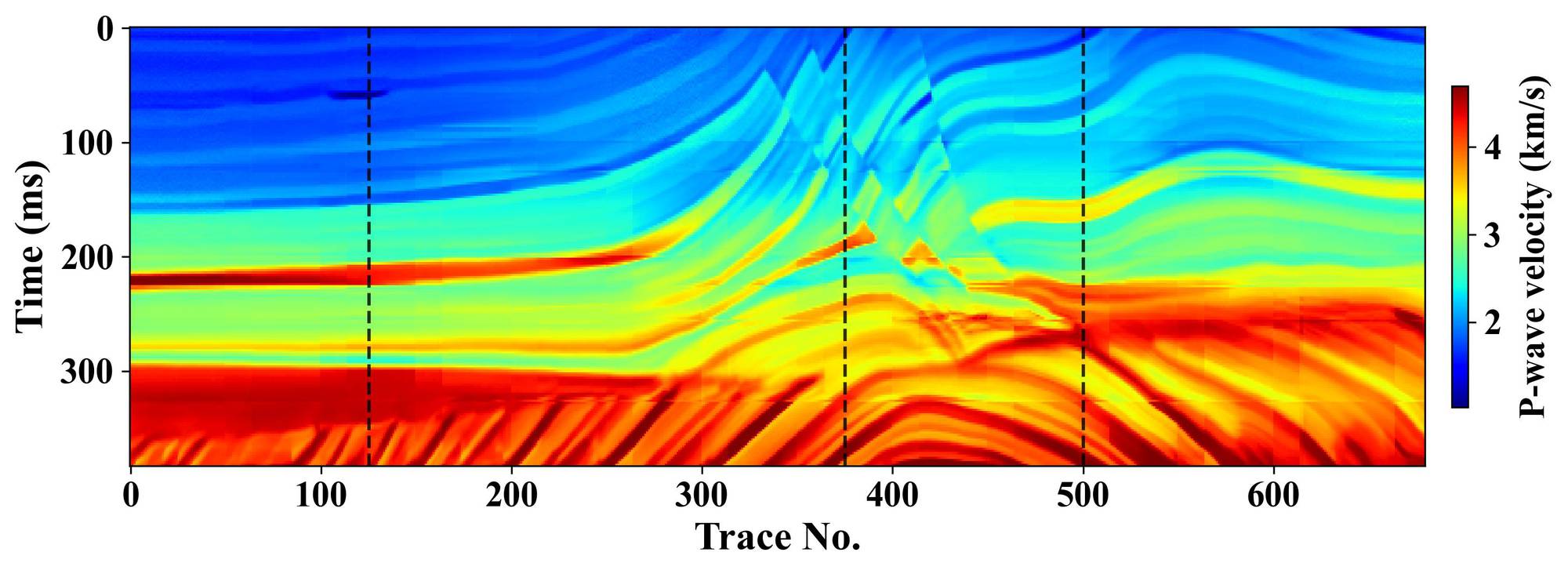}
 \label{fig:ddpm-condlowlogseis_mar-vp}}
  \subfigure[]{\includegraphics[width=0.65\columnwidth]{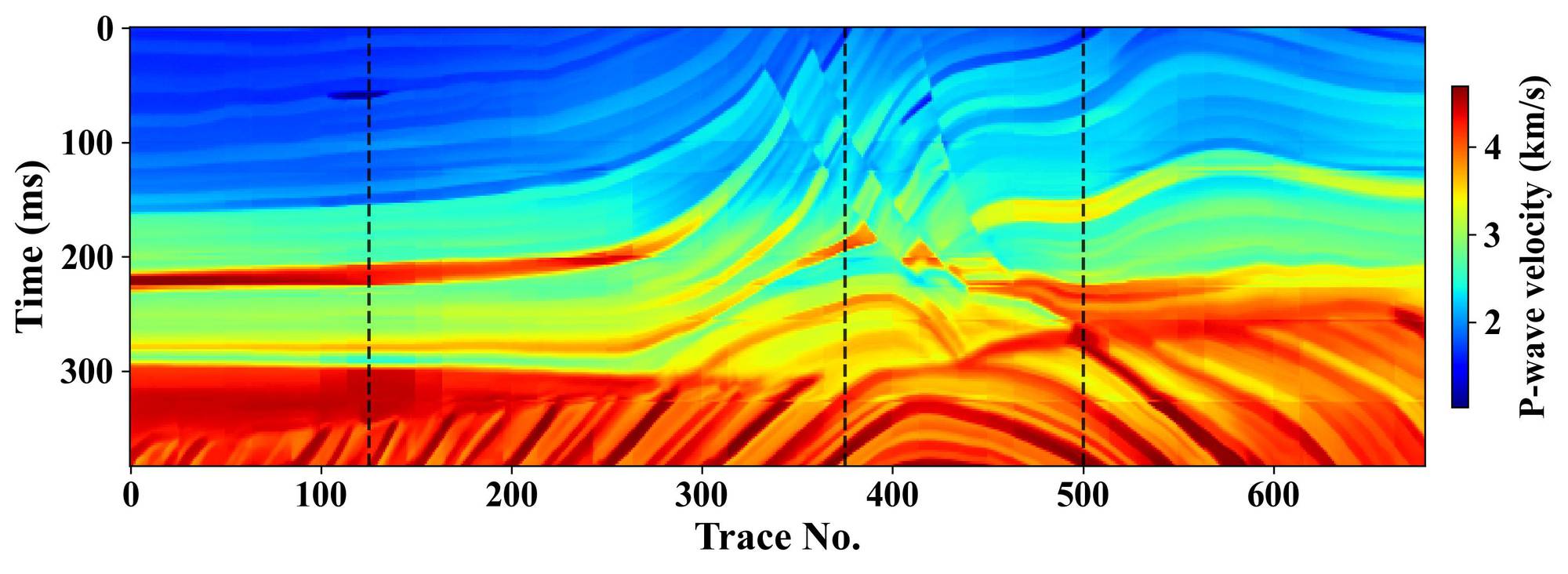}
 \label{fig:ddpm-condlogseis3216_mar-vp}}
     \subfigure[]{\includegraphics[width=0.65\columnwidth]{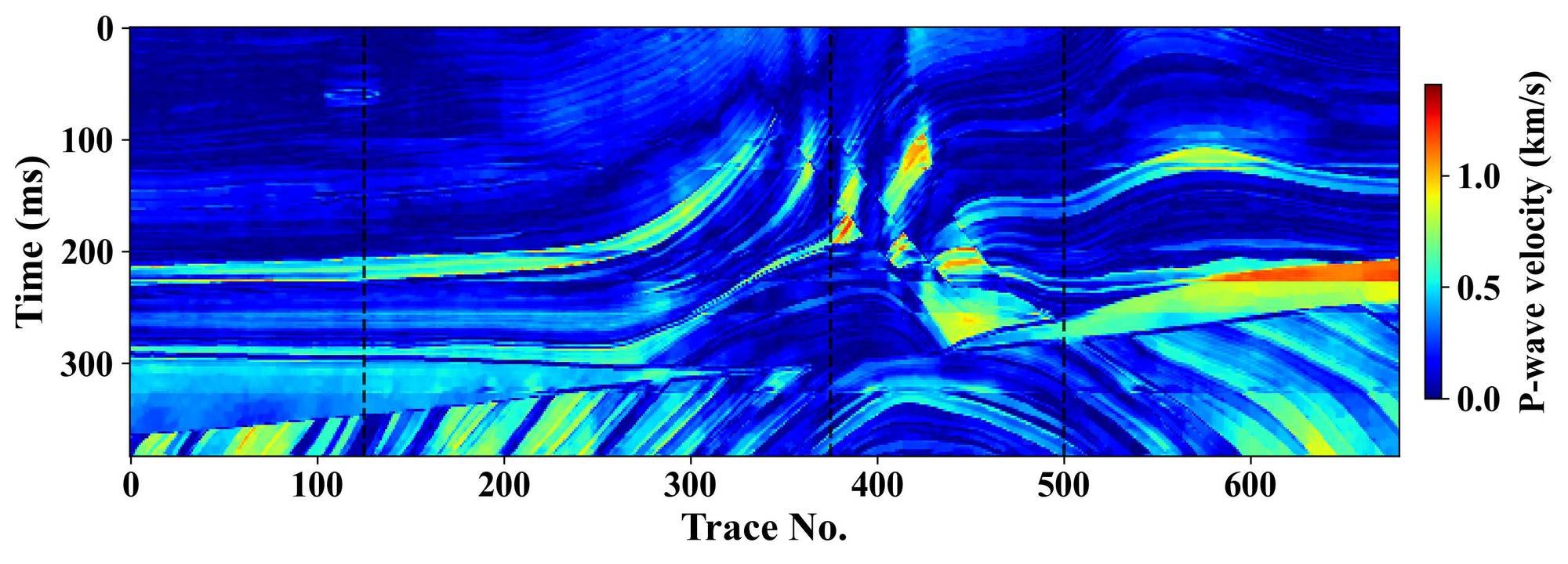}
 \label{fig:2DTVnolow-marerro-vp}} 
    \subfigure[]{\includegraphics[width=0.65\columnwidth]{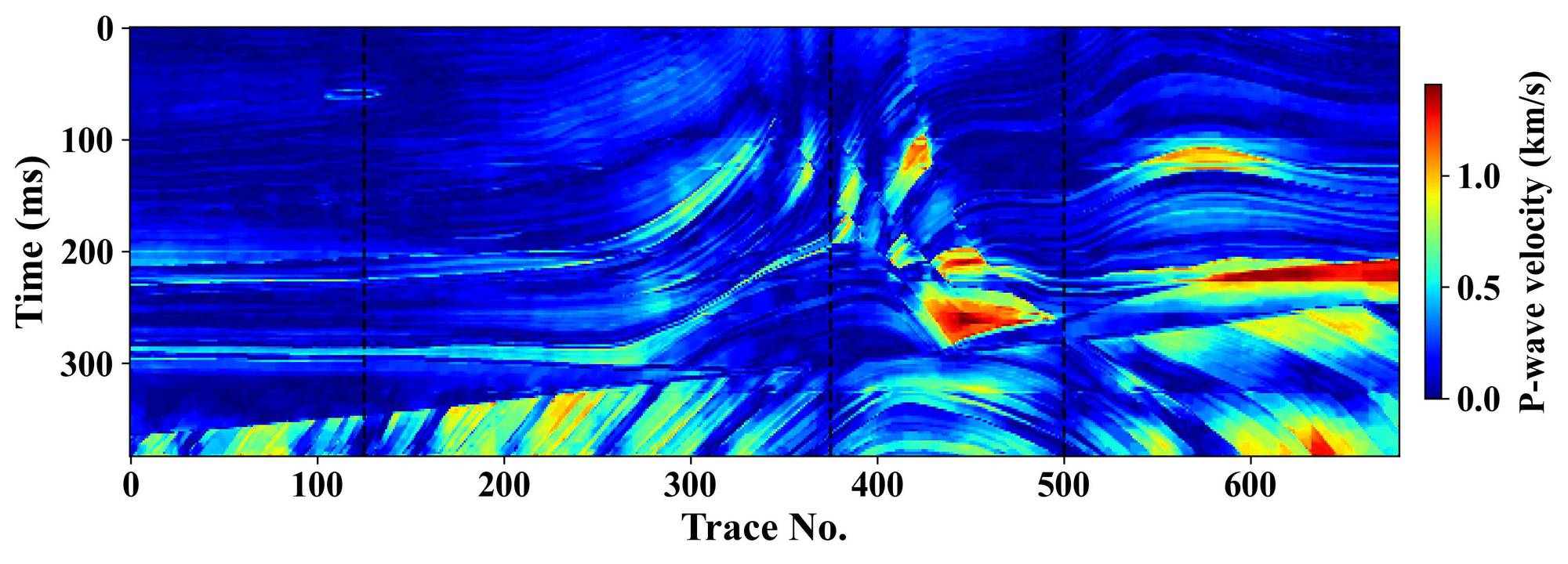}
 \label{fig:2DTV-marerro-vp}}
   \subfigure[]{\includegraphics[width=0.65\columnwidth]{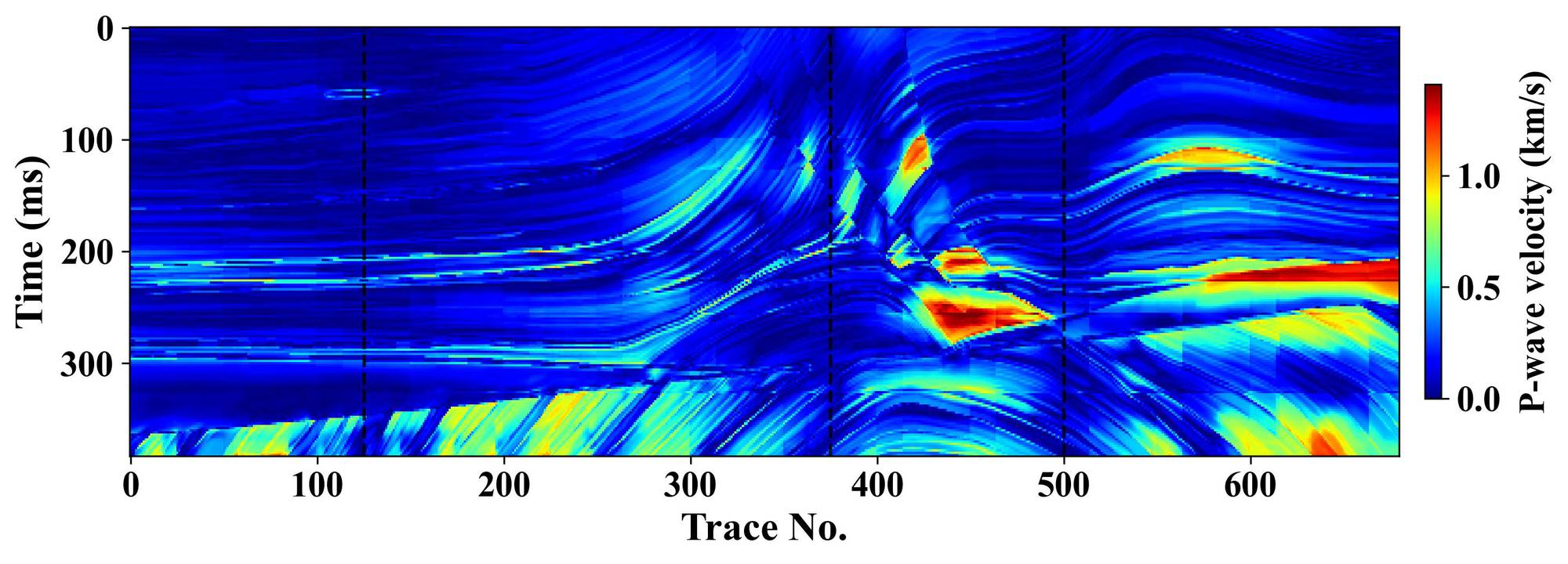}
 \label{fig:ddpm-condlowseis_marerro-vp}}
      \subfigure[]{\includegraphics[width=0.65\columnwidth]{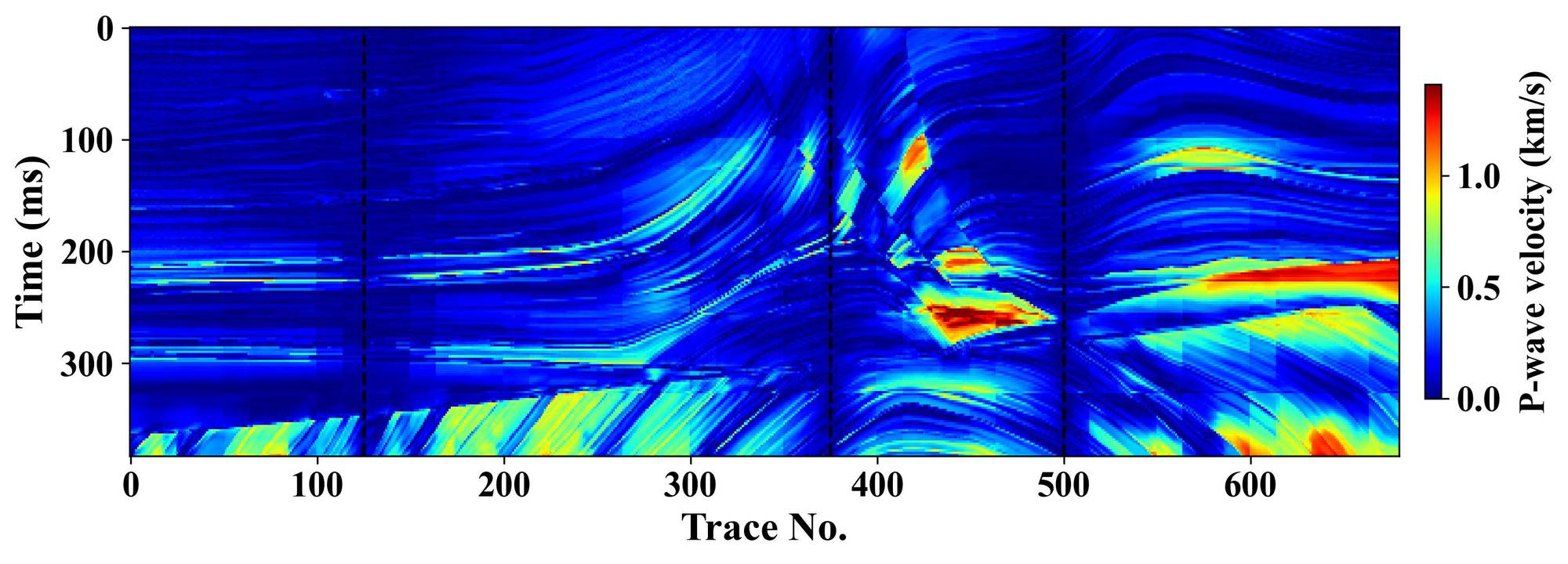}
 \label{fig:ddpm-condlowlogseis_marerro-vp}}
    \subfigure[]{\includegraphics[width=0.65\columnwidth]{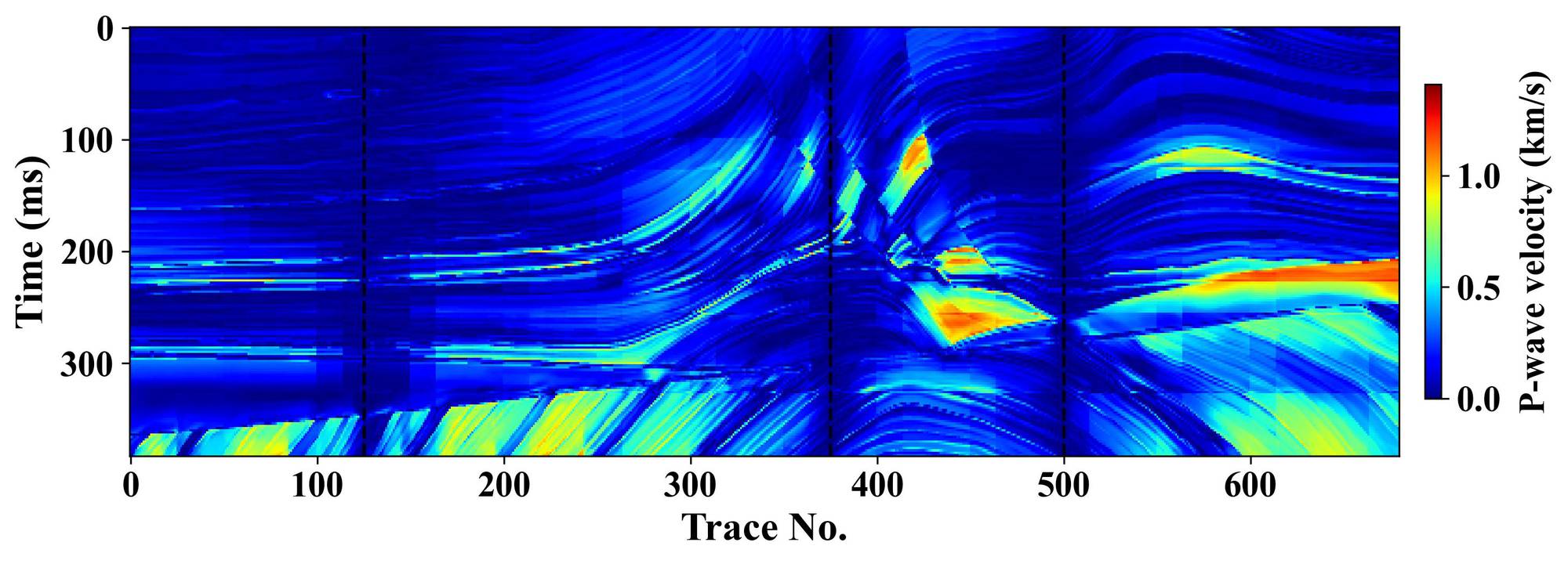}
 \label{fig:ddpm-condlogseis3216_marerro-vp}} 
 \caption{Predicted P-wave velocity models obtained using different methods. 
(a) True P-wave velocity model. 
(b)--(f) Predicted results obtained using 2D-TV, 2D-TVL, DM-SL, DM-SLW, and DM-SWI, respectively. 
(g)--(k) Errors between the predicted results in (b)--(f) and the true P-wave velocity model.
 }
\label{fig:marsamplesvp}
\end{figure*}

To validate the proposed method, we use two conventional regularization-based inversion methods as baselines: 2D total variation regularization (2D-TV) and 2D total variation regularization with an additional explicit low-frequency constraint (2D-TVL). These methods are widely used in seismic inversion and usually provide stable results by promoting spatial continuity or piecewise smoothness in the inverted models. For the proposed diffusion-based method, three condition combinations are tested: seismic data and low-frequency models (DM-SL), seismic data, low-frequency models, and pseudo-well logs (DM-SLW), and seismic data, pseudo-well logs, and interpolated well-log models (DM-SWI). Figs. \ref{fig:2DTVnolow-mar-vp}--\ref{fig:ddpm-condlogseis3216_mar-vp} show the predicted P-wave velocity models obtained using different methods. The 2D-TV result preserves the main structural features to some extent, but it shows noticeable deviations from the true model in the high-velocity regions. This suggests that 2D-TV regularization alone has a limited ability to recover the large-scale background trend and absolute amplitude level of the P-wave velocity model. By introducing an additional low-frequency constraint, 2D-TVL better recovers the amplitude and spatial distribution of the high-value components than 2D-TV. However, some local details and lateral continuity are still not well preserved. In comparison, the proposed DM-SL, DM-SLW, and DM-SWI methods produce more accurate P-wave velocity predictions than 2D-TV and 2D-TVL, indicating the advantage of using the diffusion model as a learned prior in the inversion process. Figs. \ref{fig:2DTVnolow-marerro-vp}--\ref{fig:ddpm-condlogseis3216_marerro-vp} show the errors between the predicted results and the true P-wave velocity model. Among the proposed methods, DM-SWI yields the smallest errors, indicating that interpolated well-log models provide an effective soft spatial constraint when combined with seismic data and well logs. This result differs from the $64 \times 64$ patch tests, where DM-SLW shows better performance, mainly because the smoothing operator used to impose the low-frequency constraint in DM-SLW only approximates the relationship between the elastic parameters and the low-frequency model. By contrast, the interpolated well-log model in DM-SWI is defined directly in the model domain and can be incorporated through ILVR as a soft reference constraint, which helps reduce inversion errors. In addition, the improvement introduced by well logs is mainly concentrated around the well locations. This is mainly because the patch-wise processing strategy used in the proposed method limits the long-range propagation of well log information across the entire model.
 
\begin{figure*}[htb!]
\setlength{\abovecaptionskip}{0.2cm}
 \centering
   \subfigure[]{\includegraphics[width=0.65\columnwidth]{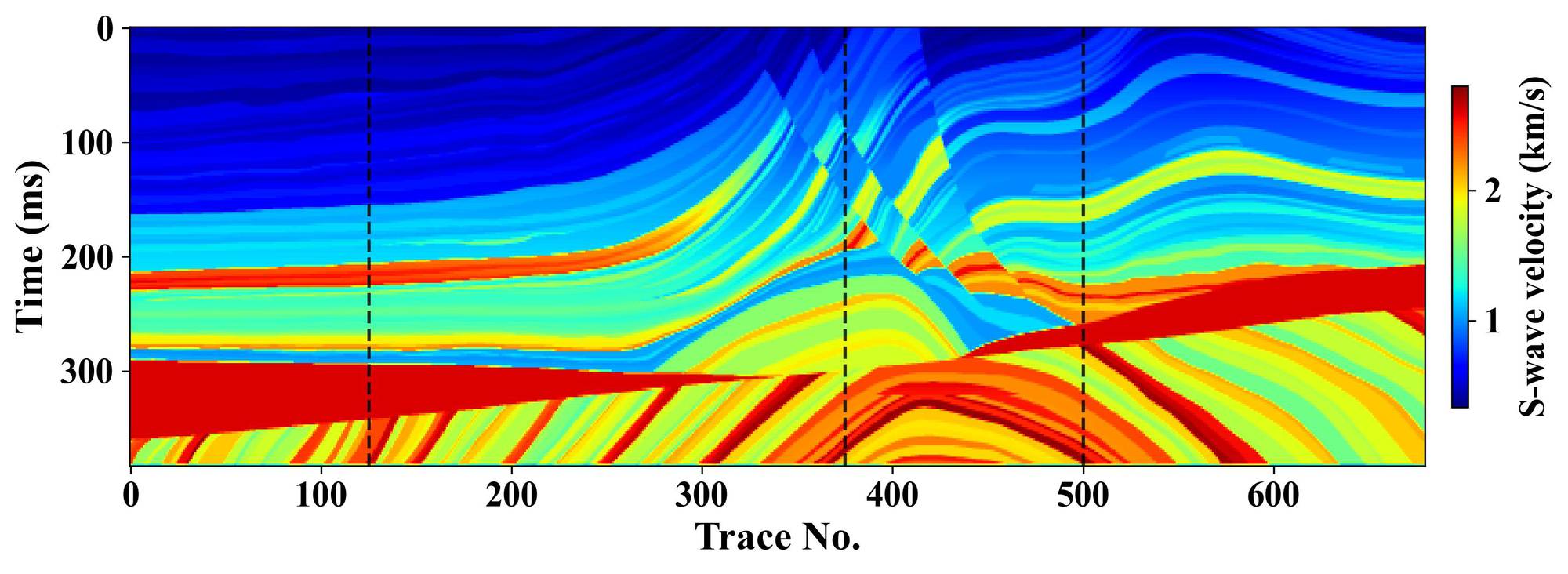}
 \label{fig:mar-vs}}
      \subfigure[]{\includegraphics[width=0.65\columnwidth]{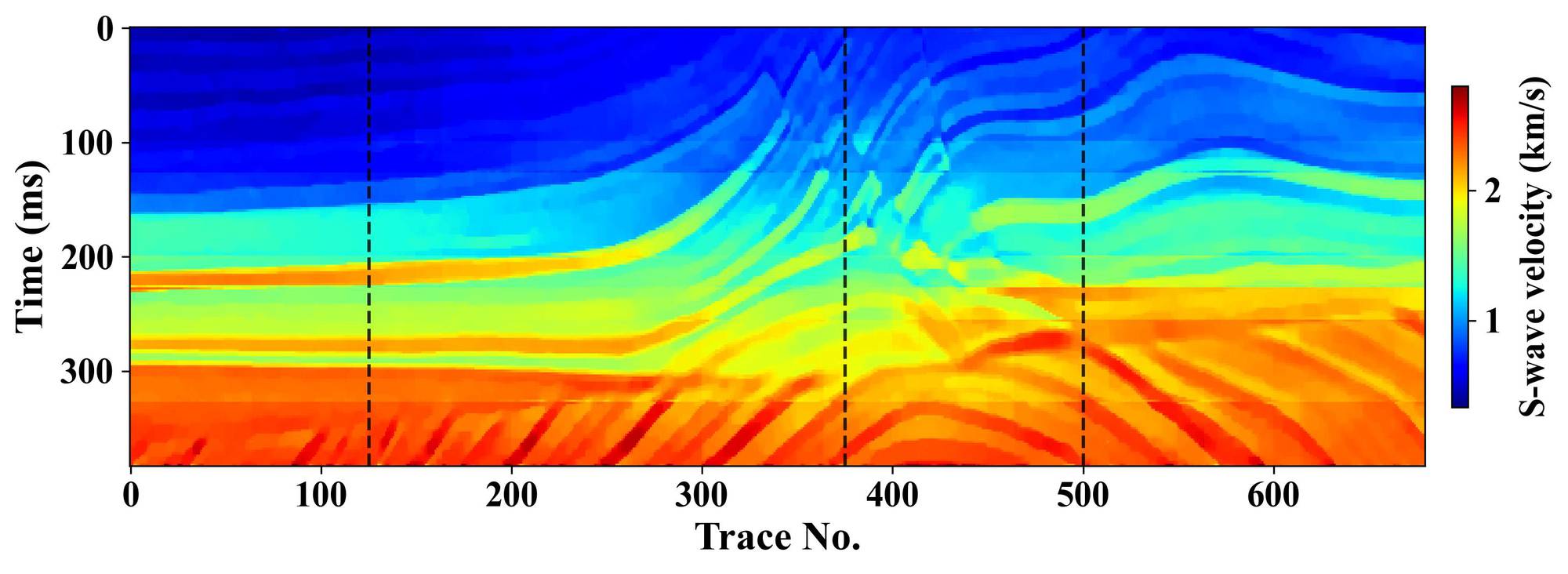}
 \label{fig:2DTVnolow-mar-vs}}
     \subfigure[]{\includegraphics[width=0.65\columnwidth]{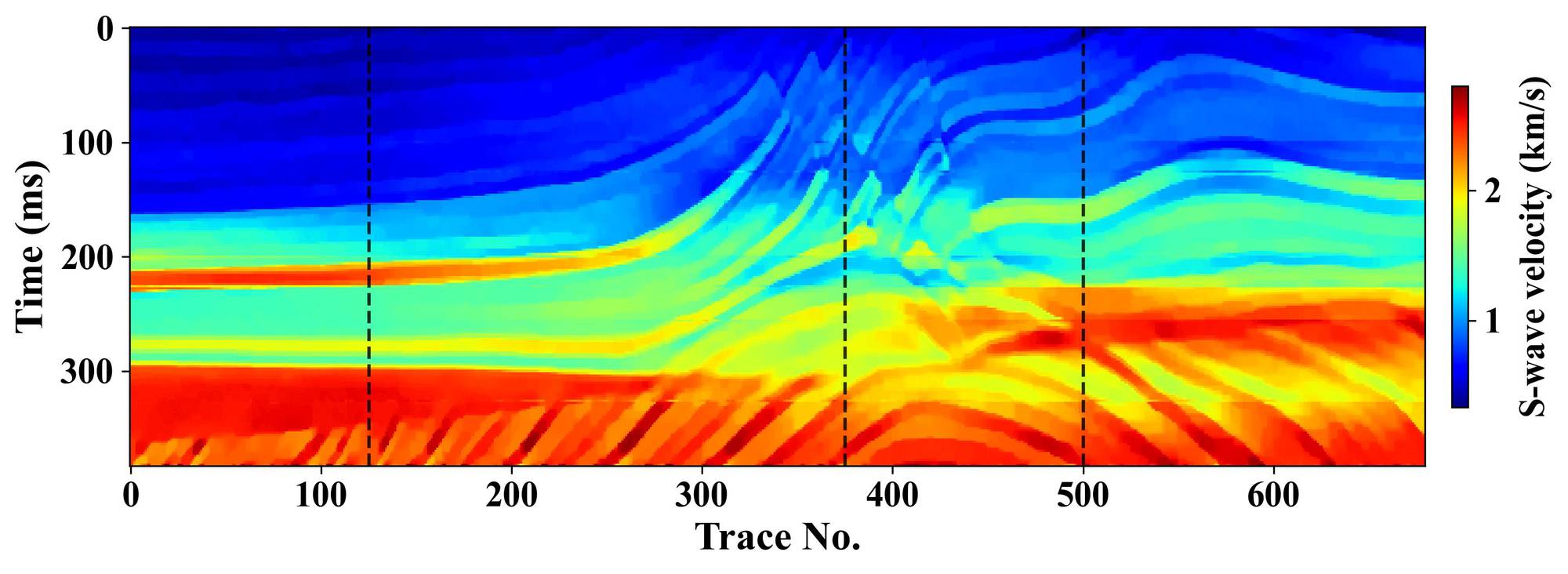}
 \label{fig:2DTV-mar-vs}}  
  \subfigure[]{\includegraphics[width=0.65\columnwidth]{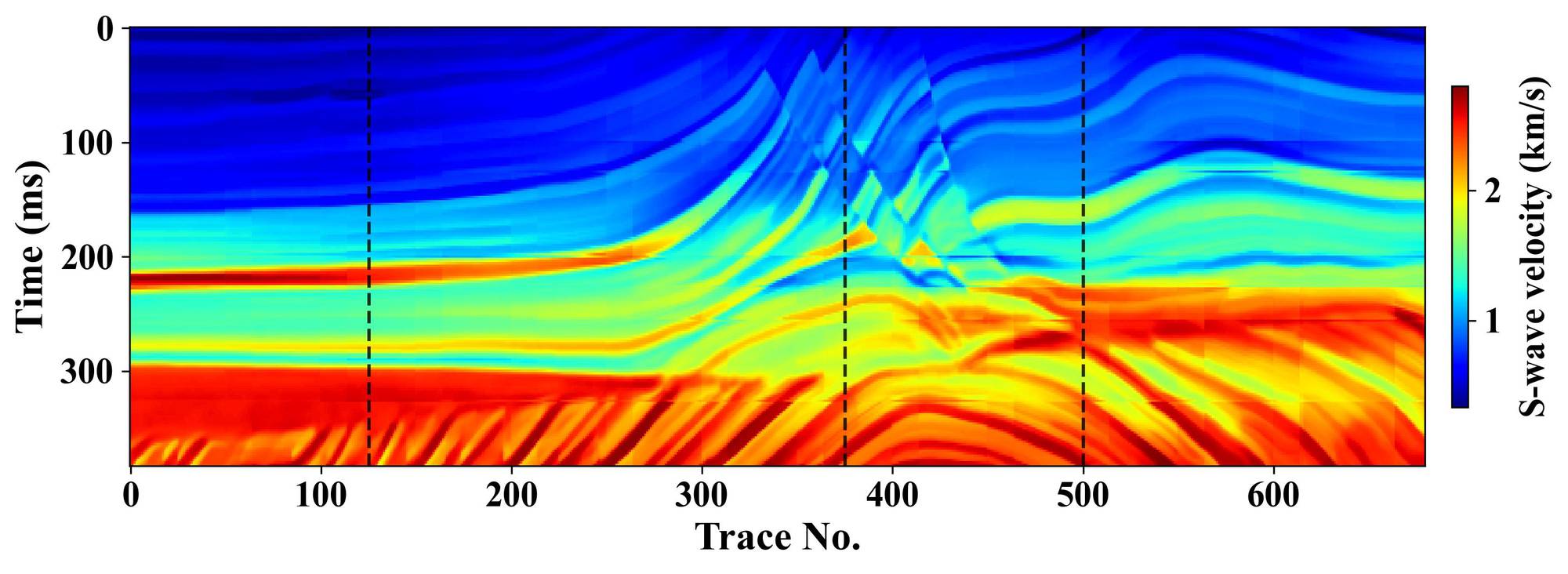}
 \label{fig:ddpm-condlowseis_mar-vs}}  
  \subfigure[]{\includegraphics[width=0.65\columnwidth]{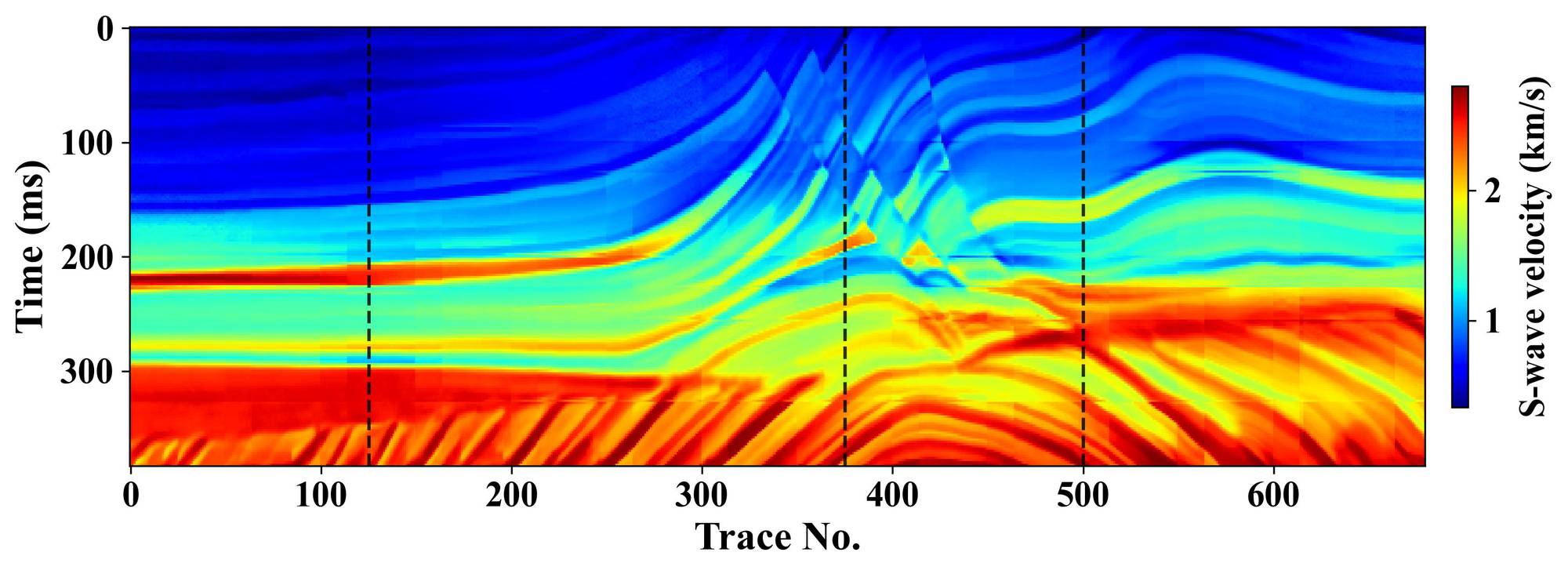}
 \label{fig:ddpm-condlowlogseis_mar-vs}}
  \subfigure[]{\includegraphics[width=0.65\columnwidth]{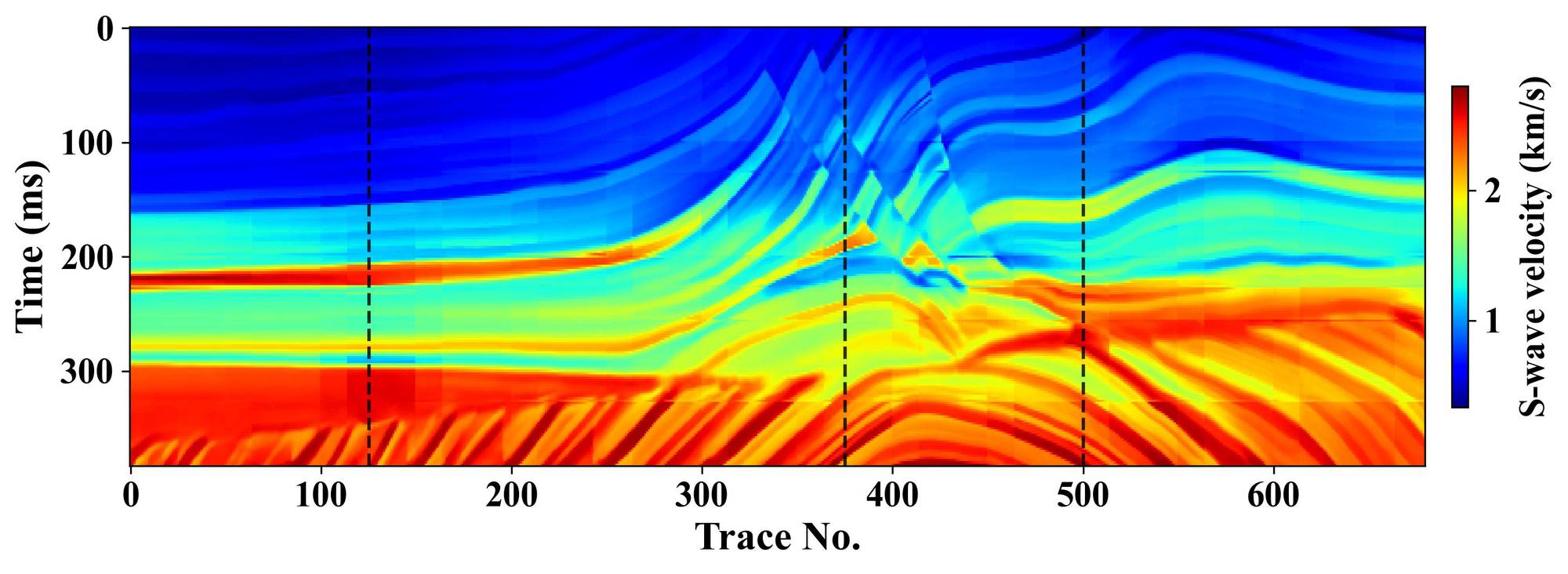}
 \label{fig:ddpm-condlogseis3216_mar-vs}}
      \subfigure[]{\includegraphics[width=0.65\columnwidth]{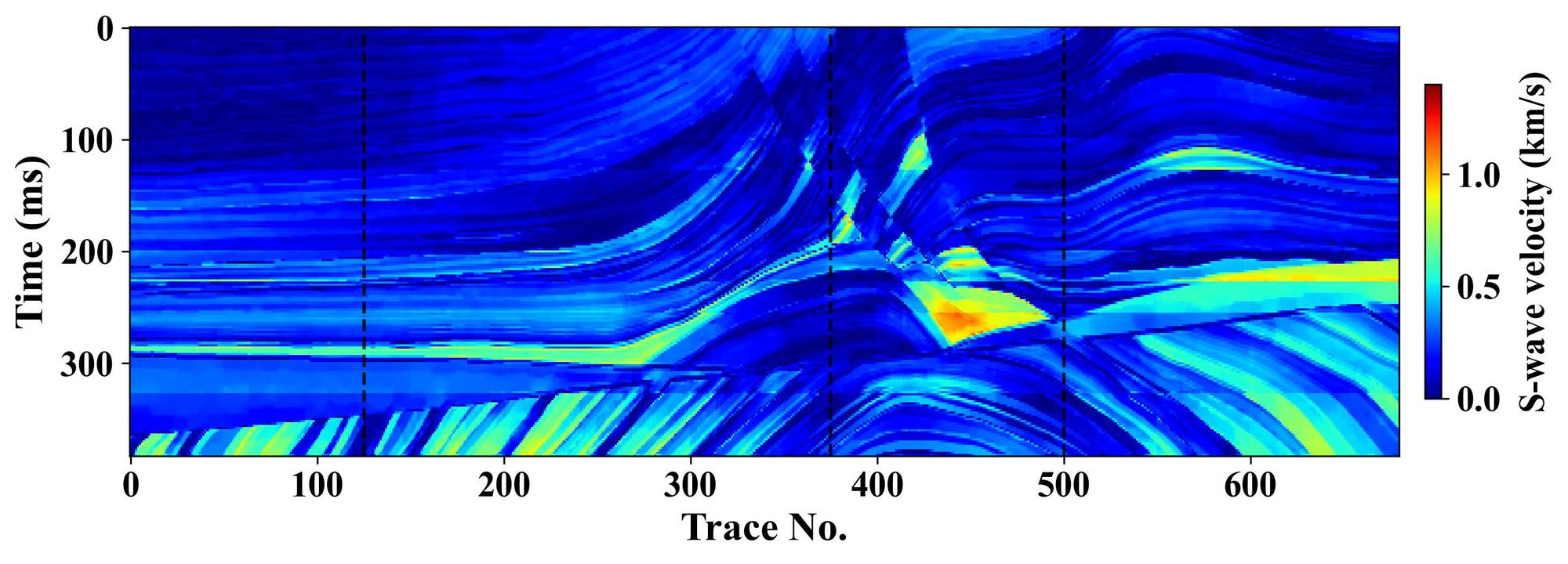}
 \label{fig:2DTVnolow-marerro-vs}}
     \subfigure[]{\includegraphics[width=0.65\columnwidth]{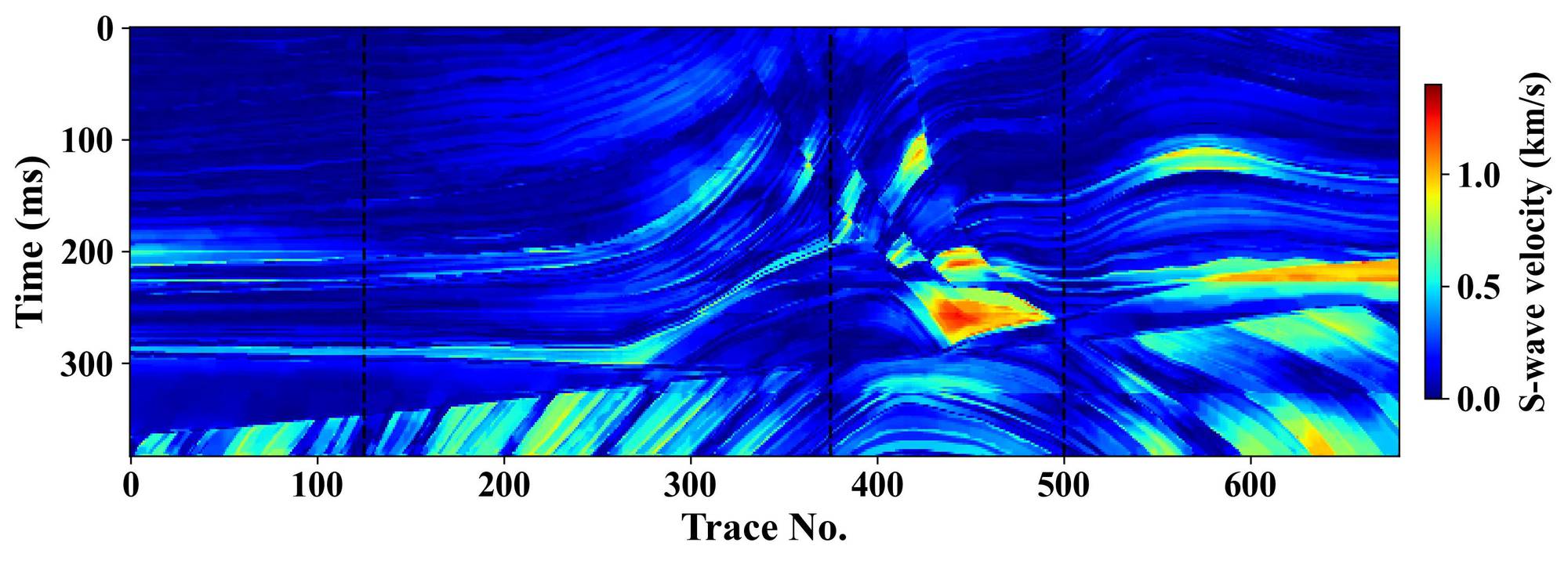}
 \label{fig:2DTV-marerro-vs}} 
    \subfigure[]{\includegraphics[width=0.65\columnwidth]{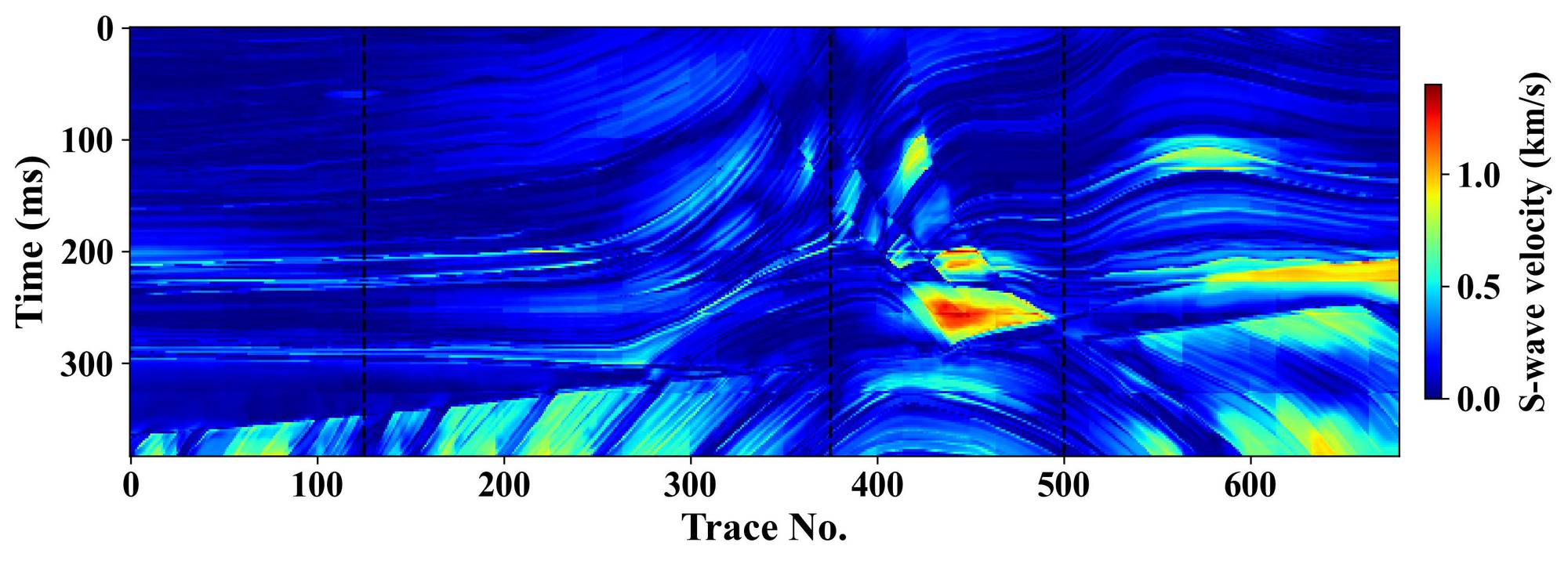}
 \label{fig:ddpm-condlowseis_marerro-vs}}
      \subfigure[]{\includegraphics[width=0.65\columnwidth]{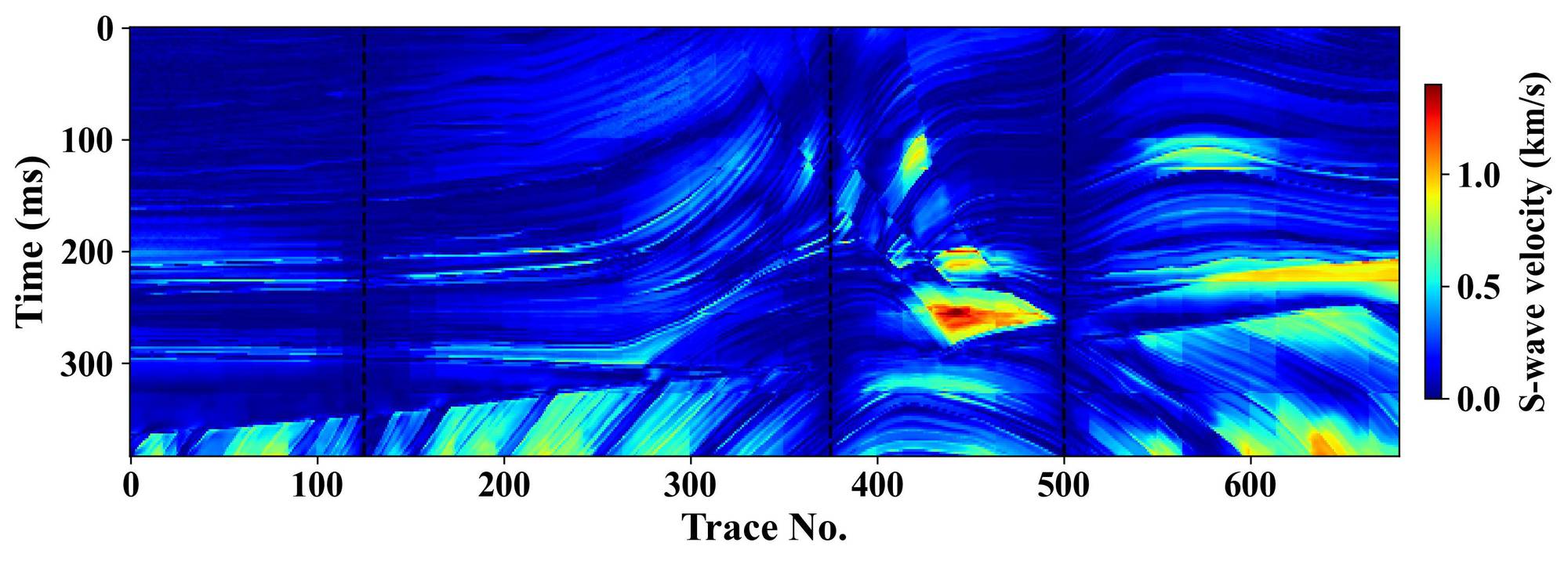}
 \label{fig:ddpm-condlowlogseis_marerro-vs}}
     \subfigure[]{\includegraphics[width=0.65\columnwidth]{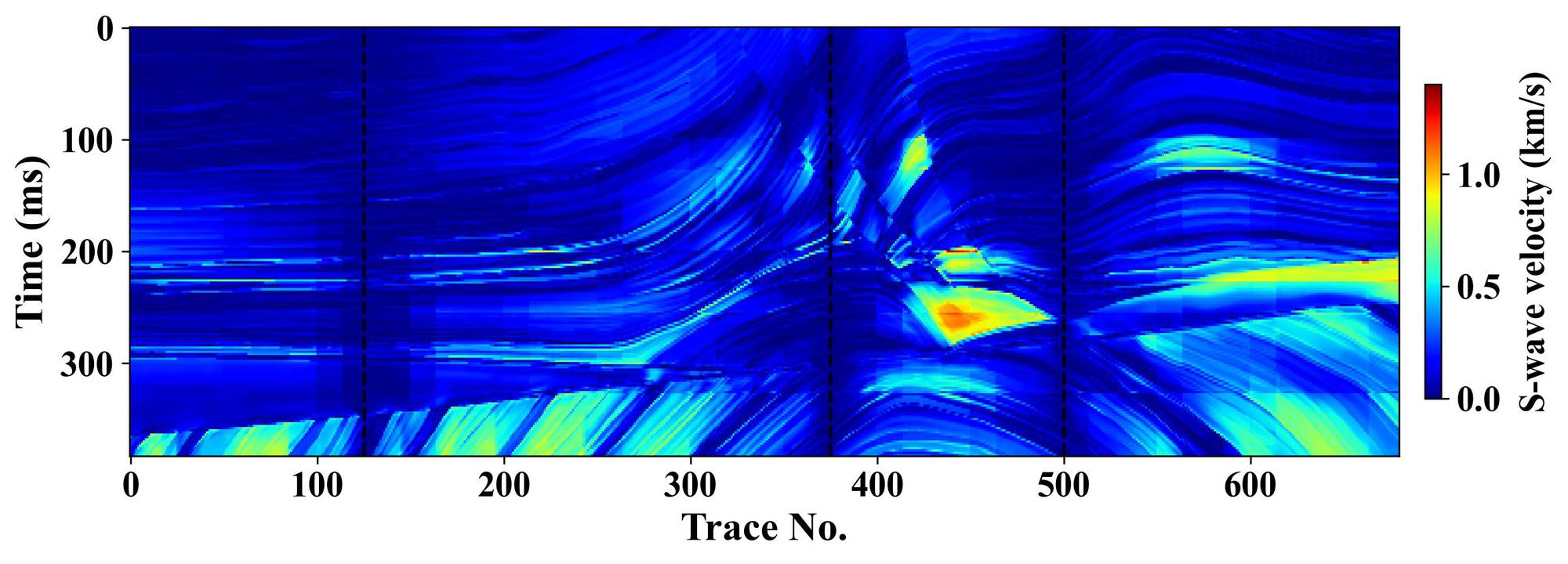}
 \label{fig:ddpm-condlogseis3216_marerro-vs}}
 \caption{Predicted S-wave velocity models obtained using different methods. 
(a) True S-wave velocity model. 
(b)--(f) Predicted results obtained using 2D-TV, 2D-TVL, DM-SL, DM-SLW, and DM-SWI, respectively. 
(g)--(k) Errors between the predicted results in (b)--(f) and the true S-wave velocity model.
 }
\label{fig:marsamplesvs}
\end{figure*}

\begin{figure*}[htb!]
\setlength{\abovecaptionskip}{0.2cm}
 \centering
   \subfigure[]{\includegraphics[width=0.65\columnwidth]{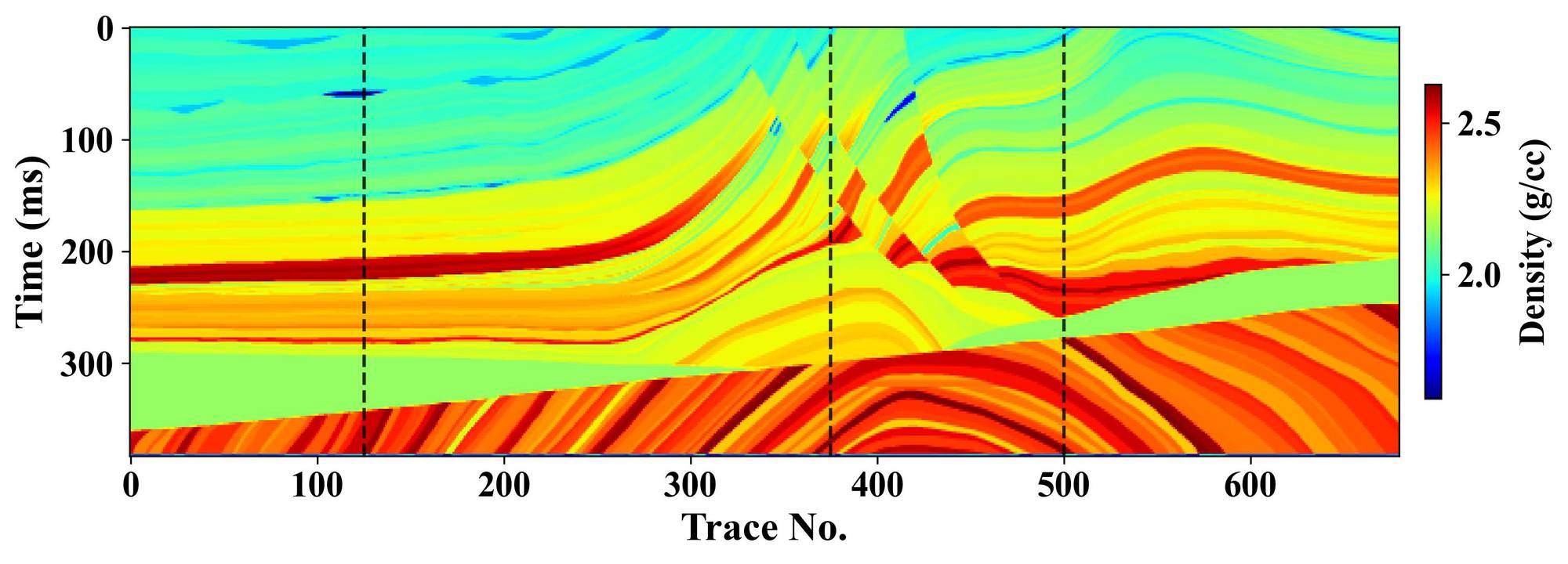}
 \label{fig:mar-rho}}
      \subfigure[]{\includegraphics[width=0.65\columnwidth]{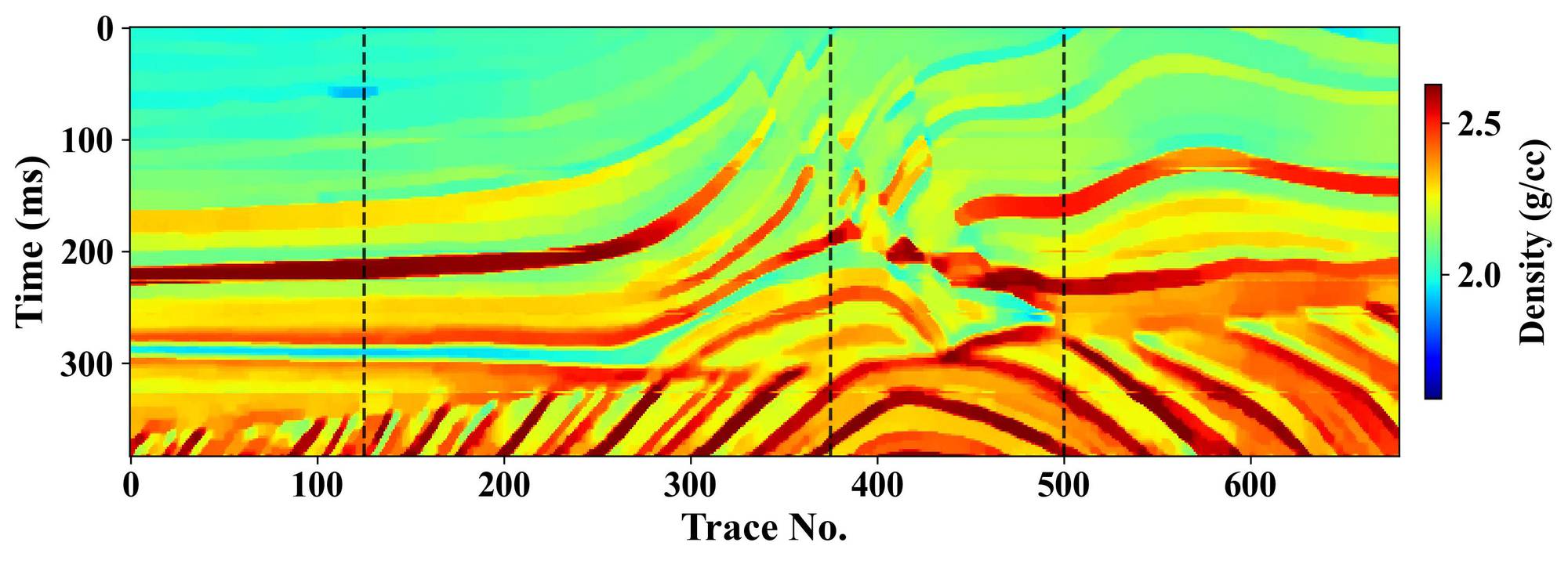}
 \label{fig:2DTVnolow-mar-rho}}
     \subfigure[]{\includegraphics[width=0.65\columnwidth]{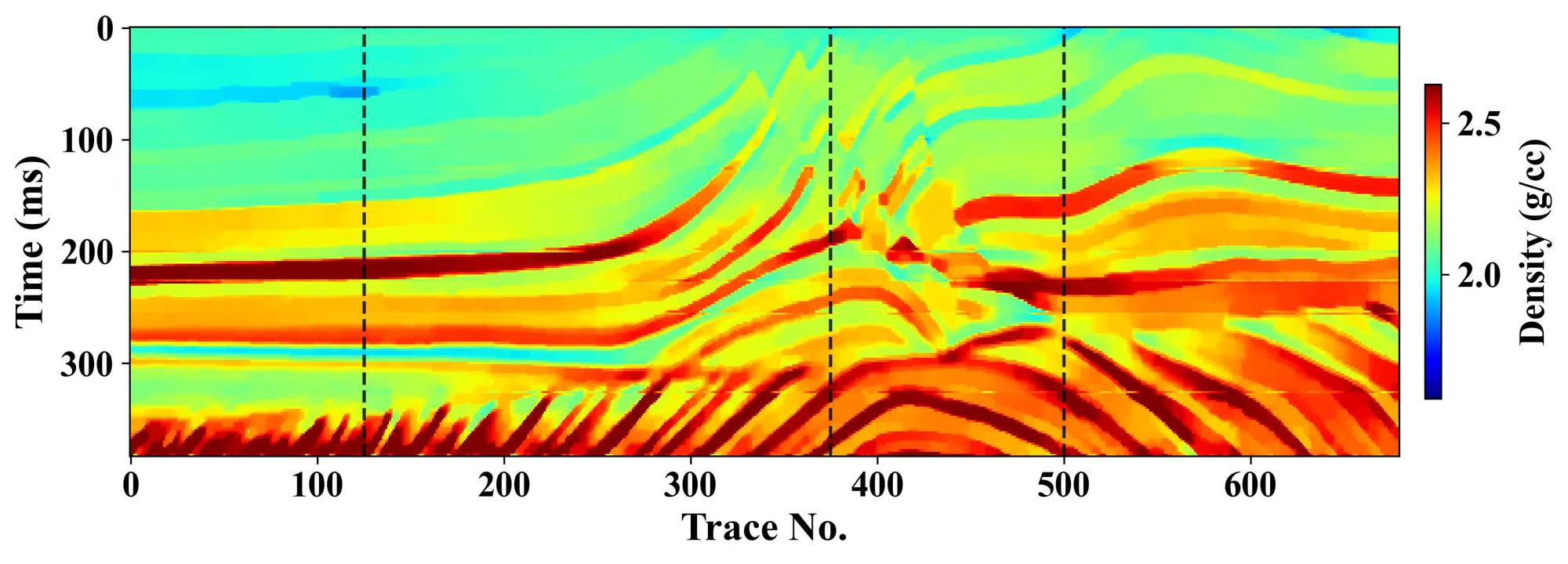}
 \label{fig:2DTV-mar-rho}}
  \subfigure[]{\includegraphics[width=0.65\columnwidth]{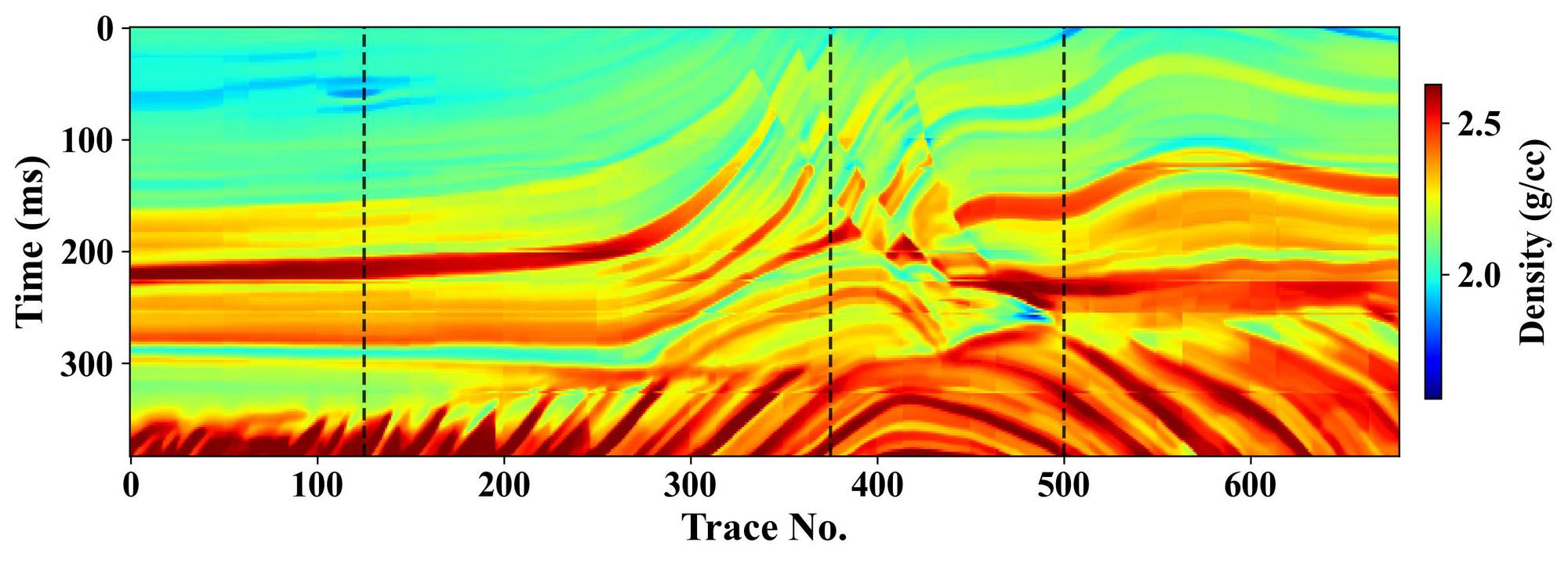}
 \label{fig:ddpm-condlowseis_mar-rho}} 
  \subfigure[]{\includegraphics[width=0.65\columnwidth]{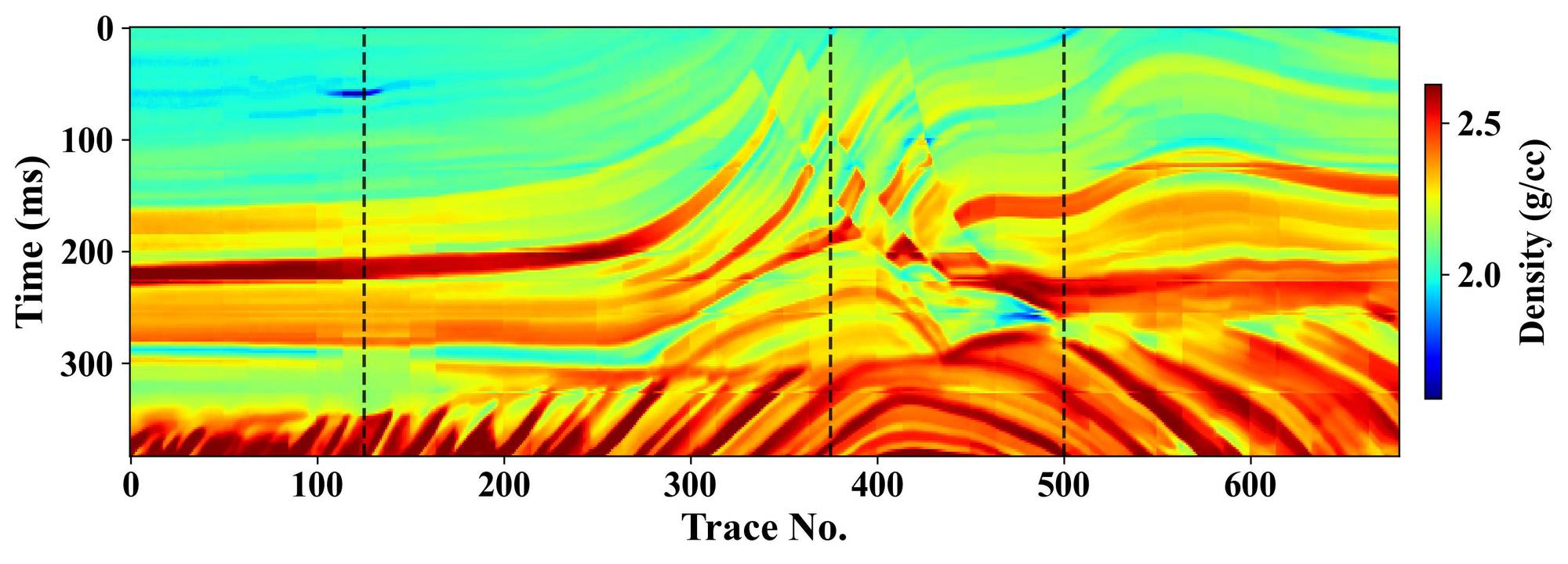}
 \label{fig:ddpm-condlowlogseis_mar-rho}}
  \subfigure[]{\includegraphics[width=0.65\columnwidth]{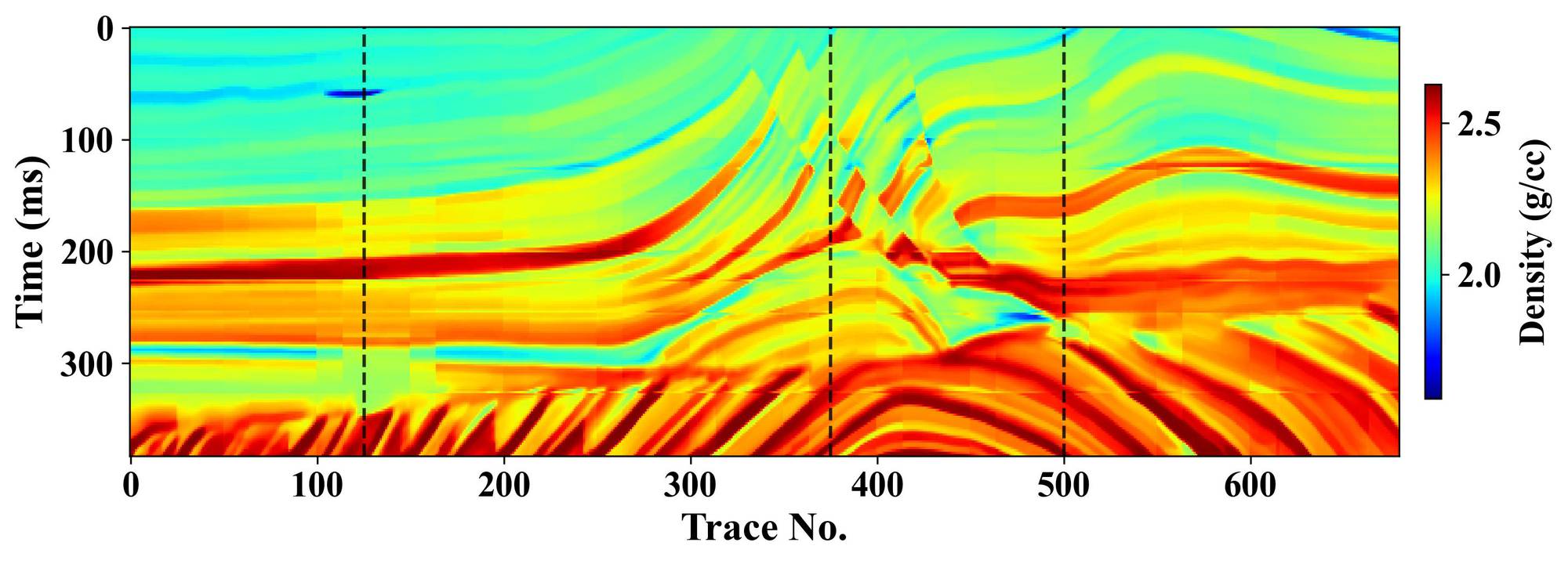}
 \label{fig:ddpm-condlogseis3216_mar-rho}} 
      \subfigure[]{\includegraphics[width=0.65\columnwidth]{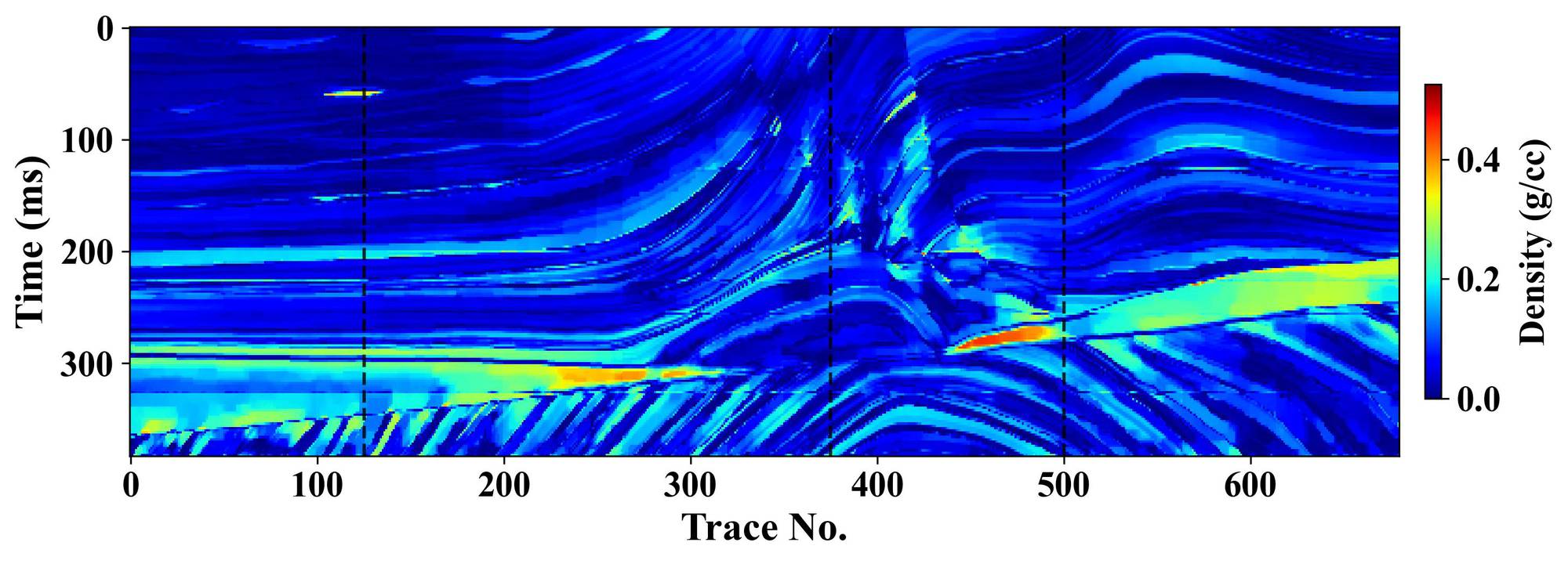}
 \label{fig:2DTVnolow-marerro-rho}}
     \subfigure[]{\includegraphics[width=0.65\columnwidth]{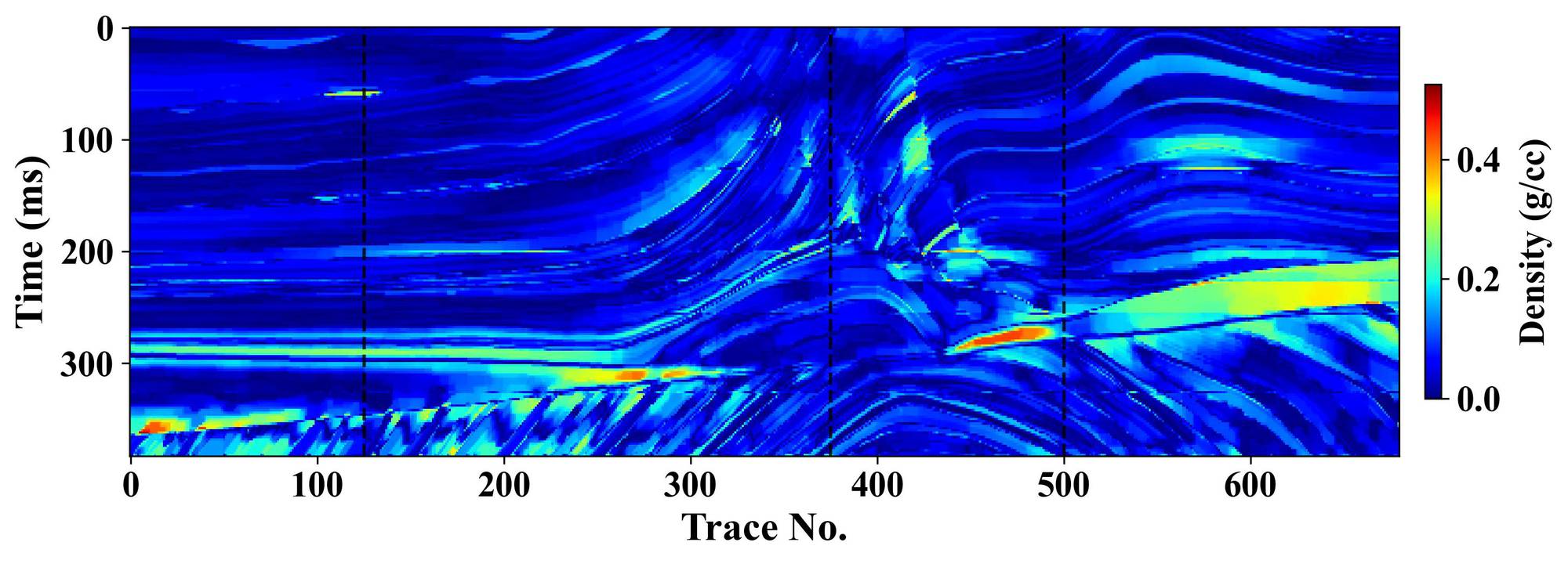}
 \label{fig:2DTV-marerro-rho}} 
    \subfigure[]{\includegraphics[width=0.65\columnwidth]{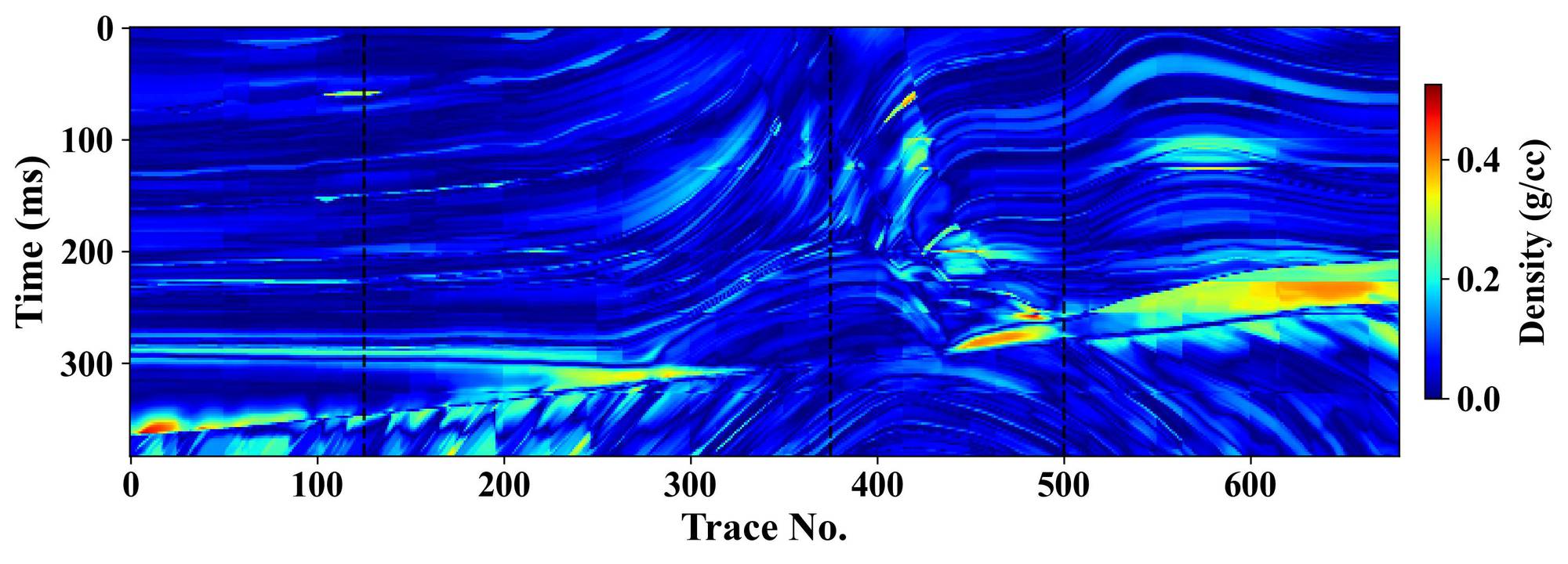}
 \label{fig:ddpm-condlowseis_marerro-rho}}
      \subfigure[]{\includegraphics[width=0.65\columnwidth]{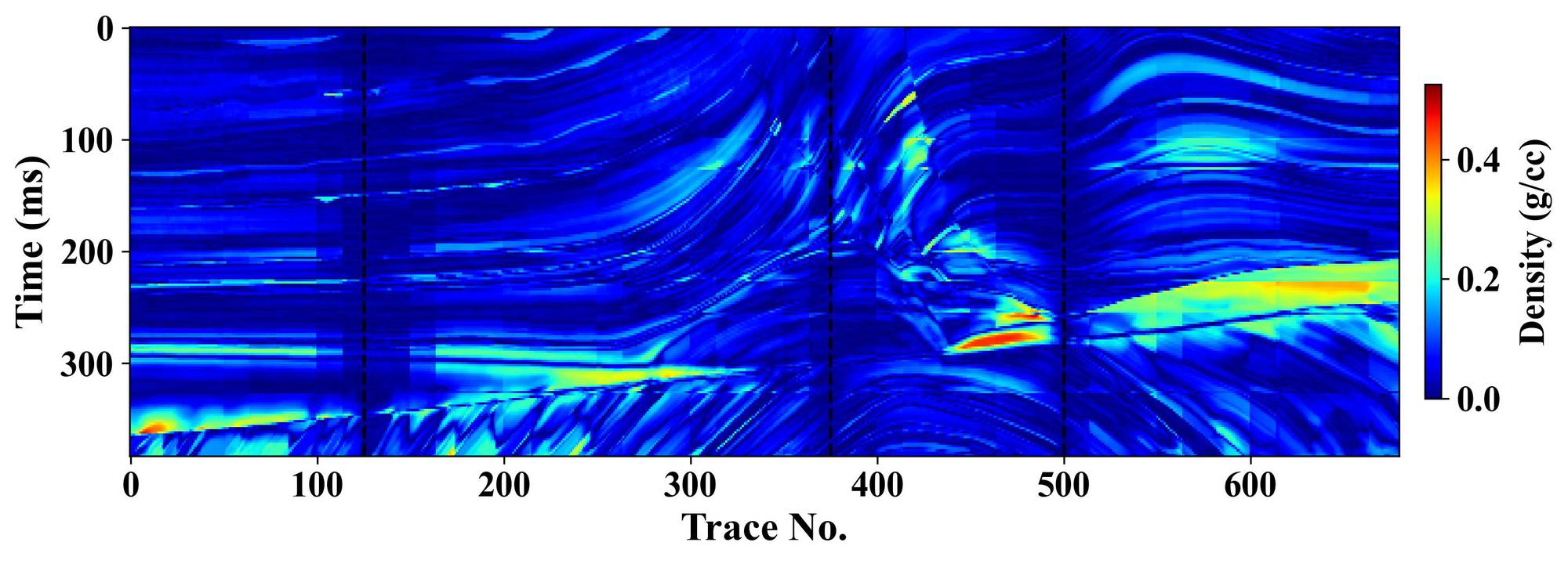}
 \label{fig:ddpm-condlowlogseis_marerro-rho}}
     \subfigure[]{\includegraphics[width=0.65\columnwidth]{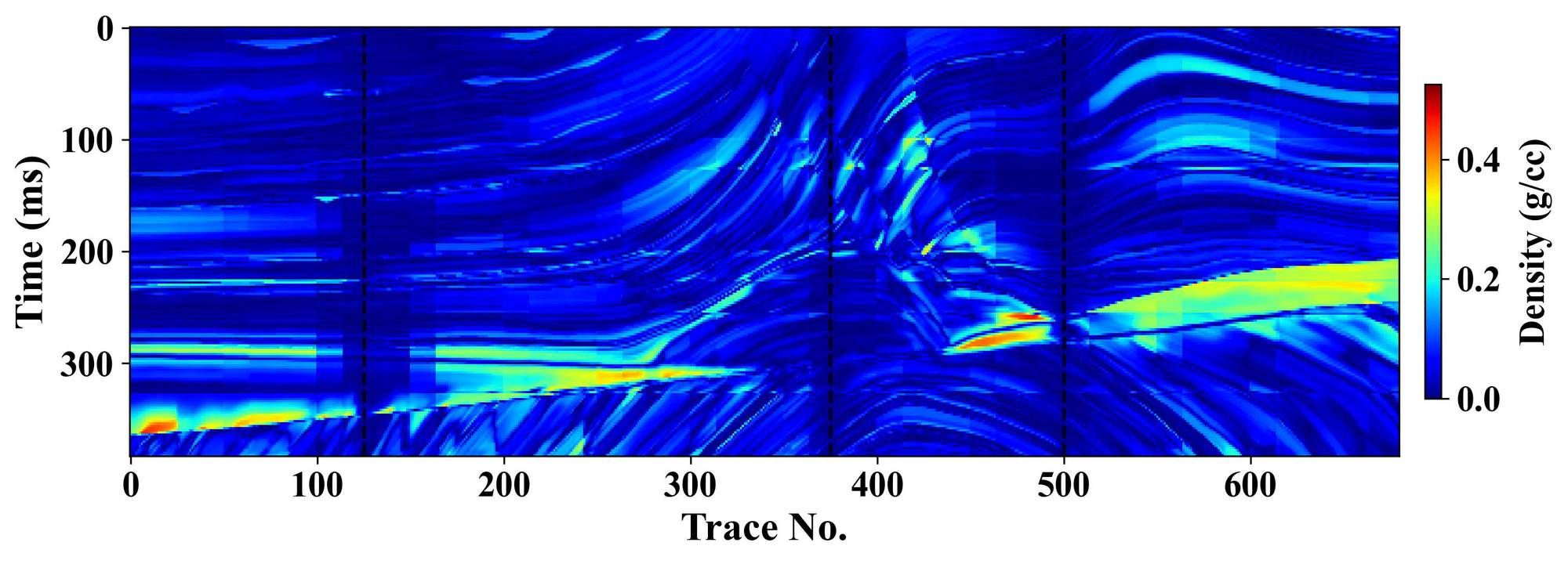}
 \label{fig:ddpm-condlogseis3216_marerro-rho}} 
 \caption{Predicted density models obtained using different methods. 
(a) True density model. 
(b)--(f) Predicted results obtained using 2D-TV, 2D-TVL, DM-SL, DM-SLW, and DM-SWI, respectively. 
(g)--(k) Errors between the predicted results in (b)--(f) and the true density model.
 }
\label{fig:marsamplesrho}
\end{figure*}

Fig. \ref{fig:marsamplesvs} shows the predicted S-wave velocity models obtained using different methods. Due to the coupling among elastic parameters, the 2D-TV and 2D-TVL methods fail to recover some weak structures and local parameter variations. In contrast, the proposed diffusion-based methods provide more accurate and higher-resolution S-wave velocity estimates. The corresponding error maps further show that the proposed methods yield smaller errors than 2D-TV and 2D-TVL, with DM-SWI producing the smallest error. Fig. \ref{fig:marsamplesrho} shows the estimated density models and the corresponding errors. Compared with 2D-TV and 2D-TVL, the proposed methods yield density estimates with better-resolved spatial details. To further evaluate the inversion results quantitatively, Table \ref{tab:comparsyn} reports the Pearson correlation coefficient (PCC), structural similarity index measure (SSIM), peak signal-to-noise ratio (PSNR), and mean squared error (MSE). The proposed methods consistently outperform 2D-TV and 2D-TVL in these metrics, and DM-SWI achieves the best overall performance.

The above results indicate that using the diffusion model as a learned prior can improve the inversion accuracy while maintaining a certain degree of robustness to noise. When seismic data and well logs are used as basic conditions, interpolated well-log models provide results comparable to, or even better than, those obtained using low-frequency models. This suggests that interpolated well-log models can act as effective soft spatial constraints for propagating sparse well log information. However, the recovery of deep-layer high values remains sensitive to the accuracy of the background or reference trend. If the given trend is inaccurate or insufficiently informative, the absolute values in deeper layers may not be accurately recovered. Future work will incorporate additional geological information, such as lithofacies constraints, to further improve elastic parameter recovery within the diffusion-based synthesis framework.

\begin{table*}[htbp]
\centering
\caption{Quantitative evaluation of elastic parameter inversion results obtained by different methods.}
\label{tab:comparsyn}
\small
\setlength{\tabcolsep}{4pt}
\begin{tabular}{p{1cm} p{1cm} p{2cm} p{2cm} p{2cm} p{2cm} p{2cm}}
\toprule
 &  & 2D-TV & 2D-TVL & DM-SL & DM-SLW & DM-SWI \\
\midrule
\multirow{3}{*}{PCC}
& $v_p$& 0.9348 & 0.9325 & 0.9363 & 0.9409 & \textbf{0.9481} \\
& $v_s$ & 0.9207& 0.9245 & 0.9334 & 0.9355 & \textbf{0.9408} \\
& $\rho$ &0.7998 & 0.8425 & 0.8504 & 0.8603 & \textbf{0.8605} \\
\midrule
\multirow{3}{*}{SSIM}
& $v_p$& 0.7951 & 0.7899 & 0.8045 & 0.8250 & \textbf{0.8447} \\
& $v_s$ & 0.7157& 0.7276 & 0.7670 & 0.7834 & \textbf{0.8035} \\
& $\rho$ &0.6289 & 0.6327 & 0.6260 & 0.6607 & \textbf{0.6751} \\
\midrule
\multirow{3}{*}{PSNR}
& $v_p$& 21.2534 & 20.8295 & 21.0790 & 21.4144 & \textbf{22.0037} \\
& $v_s$& 18.7995 & 19.1567 & 19.5668 & 19.7084 & \textbf{20.0924} \\
& $\rho$&19.9473 & 20.6651 & 21.0682 & \textbf{21.3689} & 21.3654 \\
\midrule
\multirow{3}{*}{MSE}
& $v_p$& 0.1010 & 0.1113 & 0.1051 & 0.0973 & \textbf{0.0850} \\
& $v_s$ &0.0802 & 0.0738 & 0.0672 & 0.0650 & \textbf{0.0595} \\
& $\rho$ &0.0108 & 0.0092 & 0.0084 & 0.0078 & \textbf{0.0078} \\
\bottomrule
\end{tabular}
\end{table*}

\subsection{Field data example} 
\begin{figure*}[htb!]
\setlength{\abovecaptionskip}{0.2cm}
 \centering
    \subfigure[]{\includegraphics[width=0.65\columnwidth]{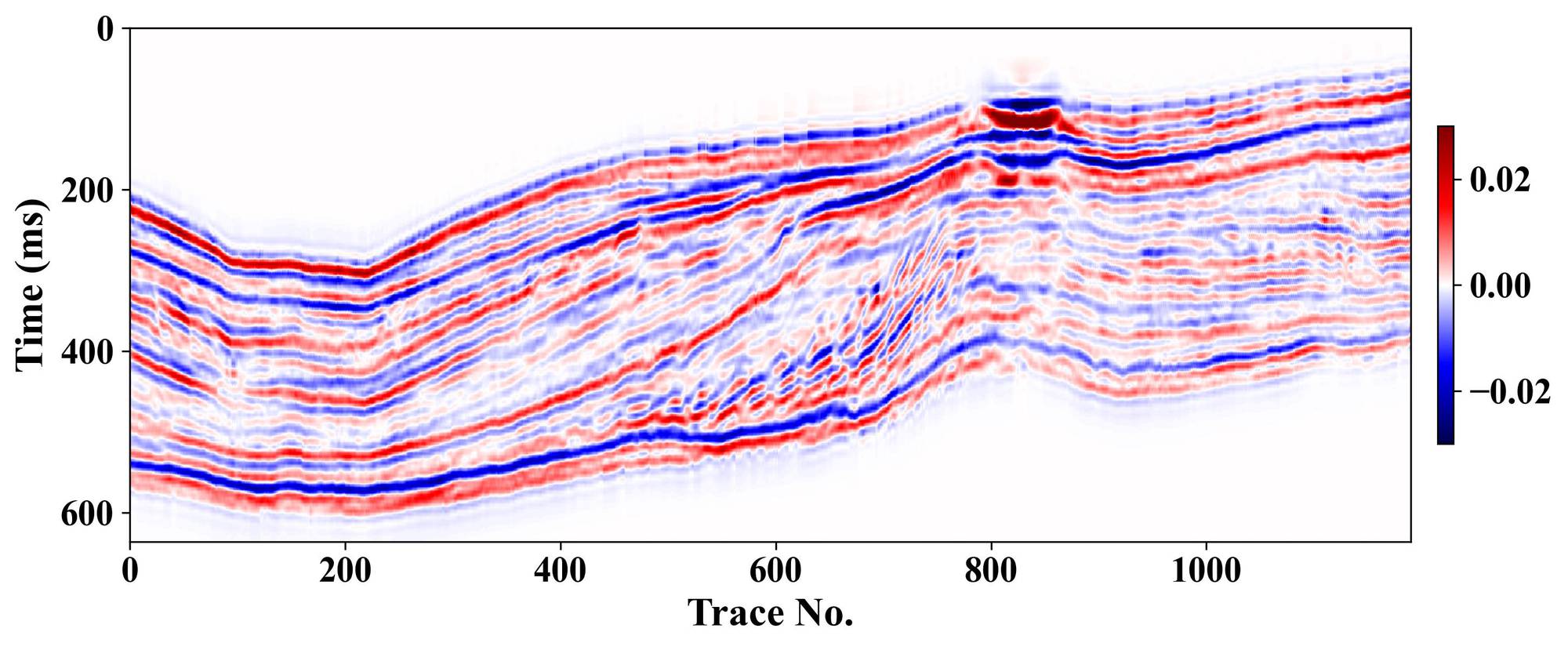}
 \label{fig:F3-seis0}}
   \subfigure[]{\includegraphics[width=0.65\columnwidth]{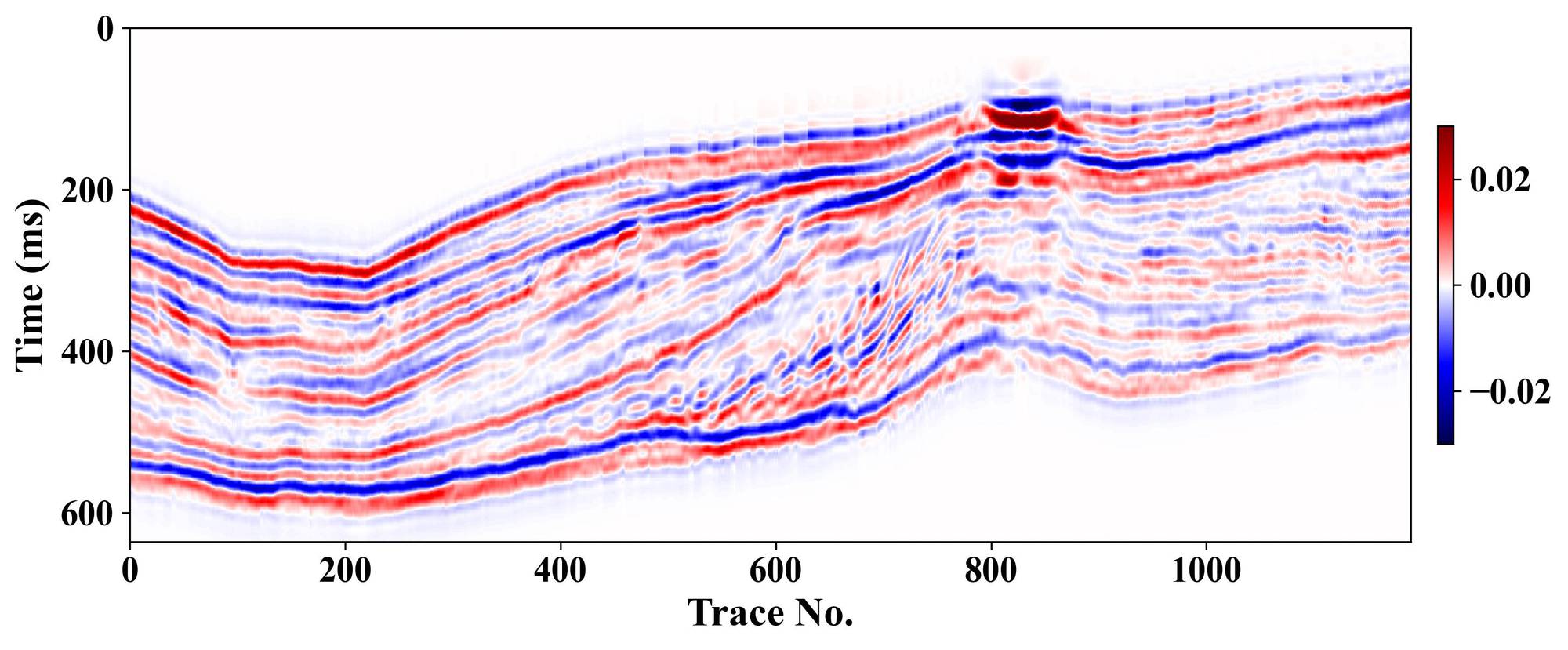}
 \label{fig:F3-seis1}}
   \subfigure[]{\includegraphics[width=0.65\columnwidth]{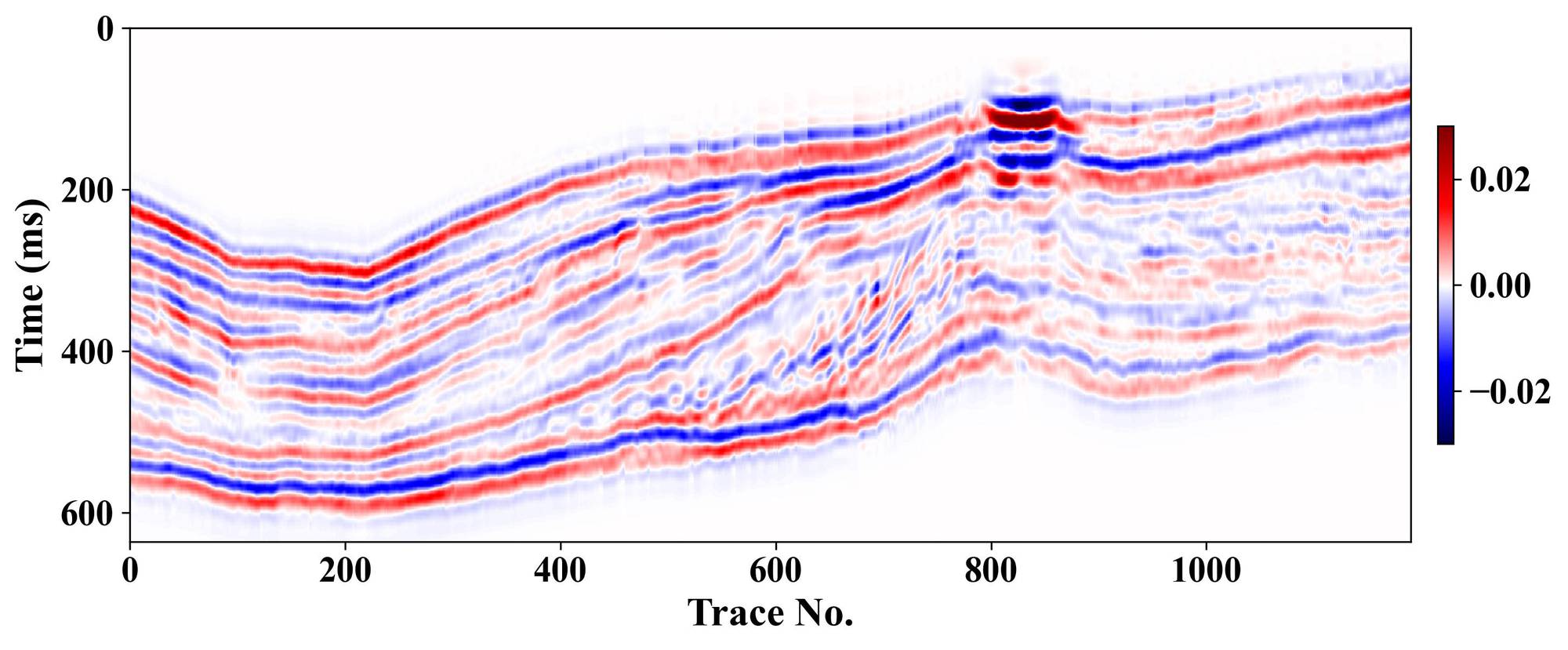}
 \label{fig:F3-seis2}}
   \subfigure[]{\includegraphics[width=0.65\columnwidth]{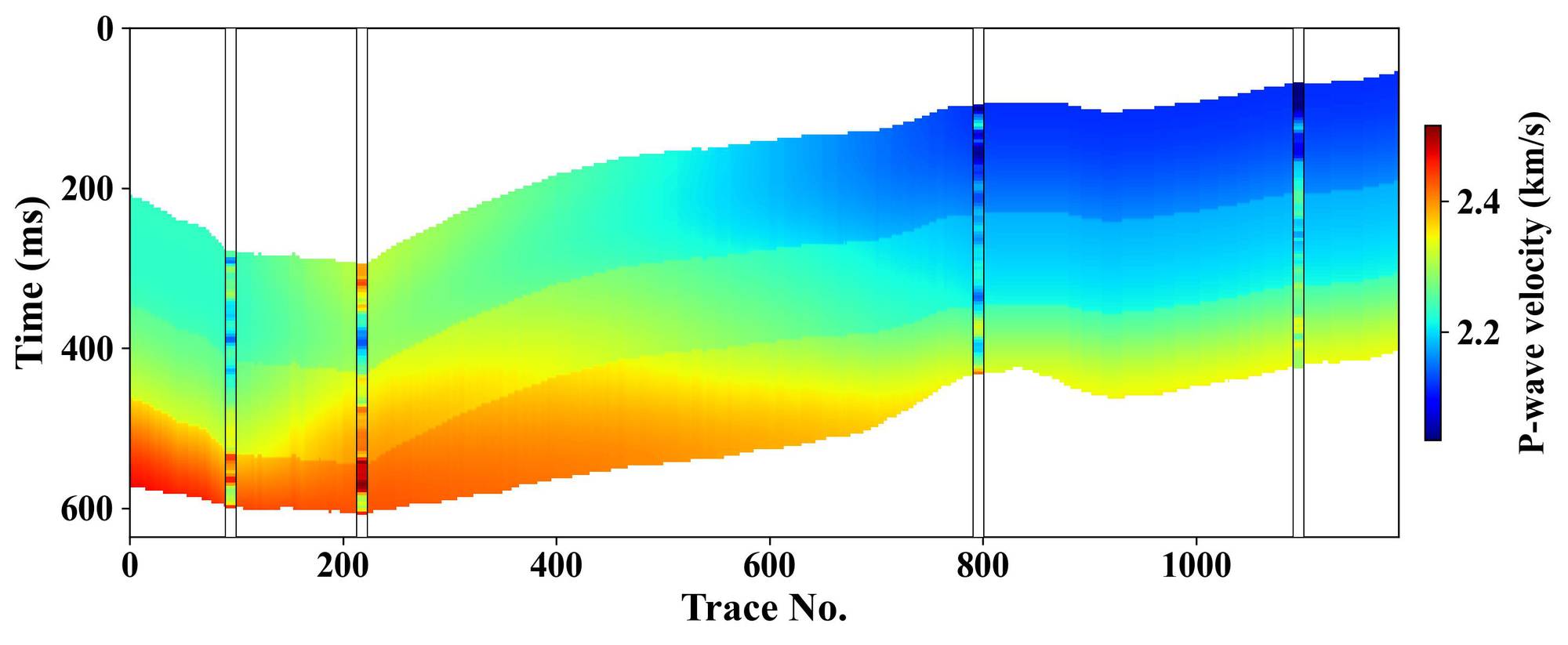}
 \label{fig:F3-lowvp}}
   \subfigure[]{\includegraphics[width=0.65\columnwidth]{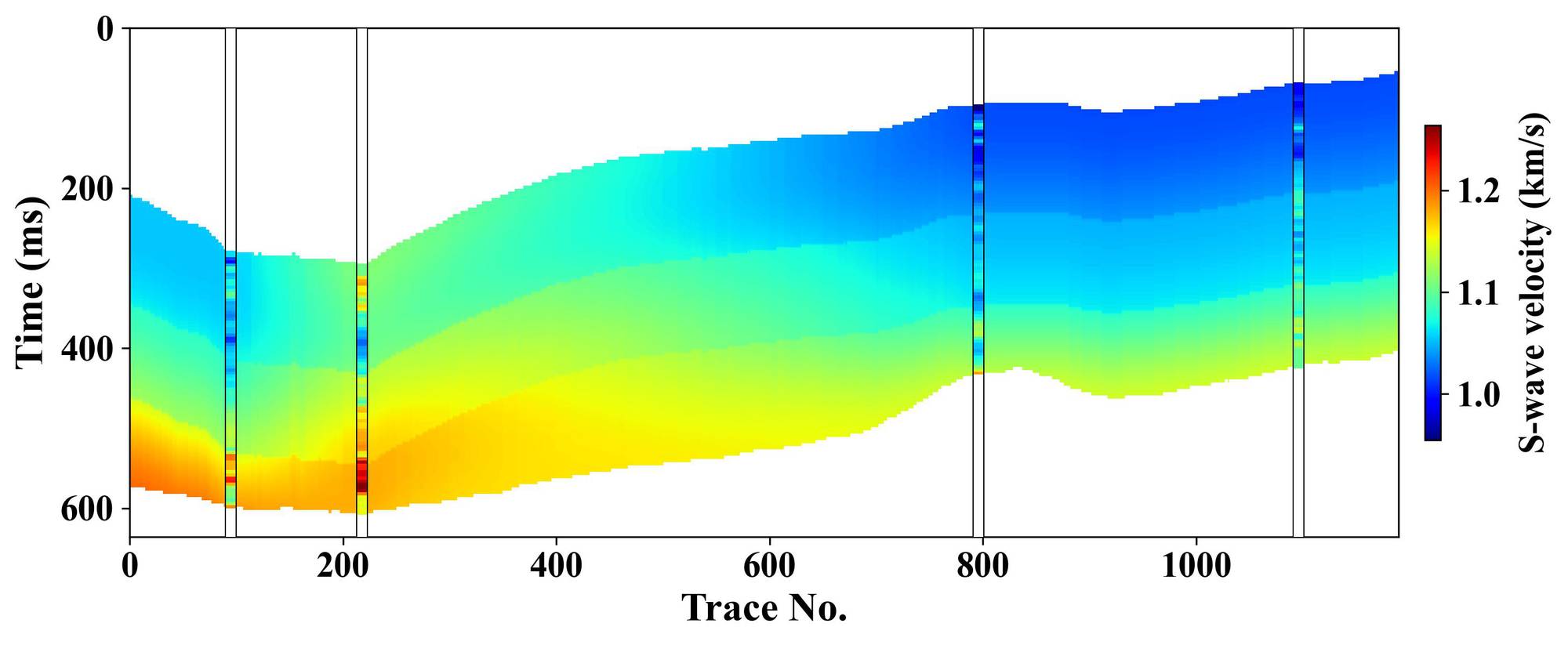}
 \label{fig:F3-lowvs}}
   \subfigure[]{\includegraphics[width=0.65\columnwidth]{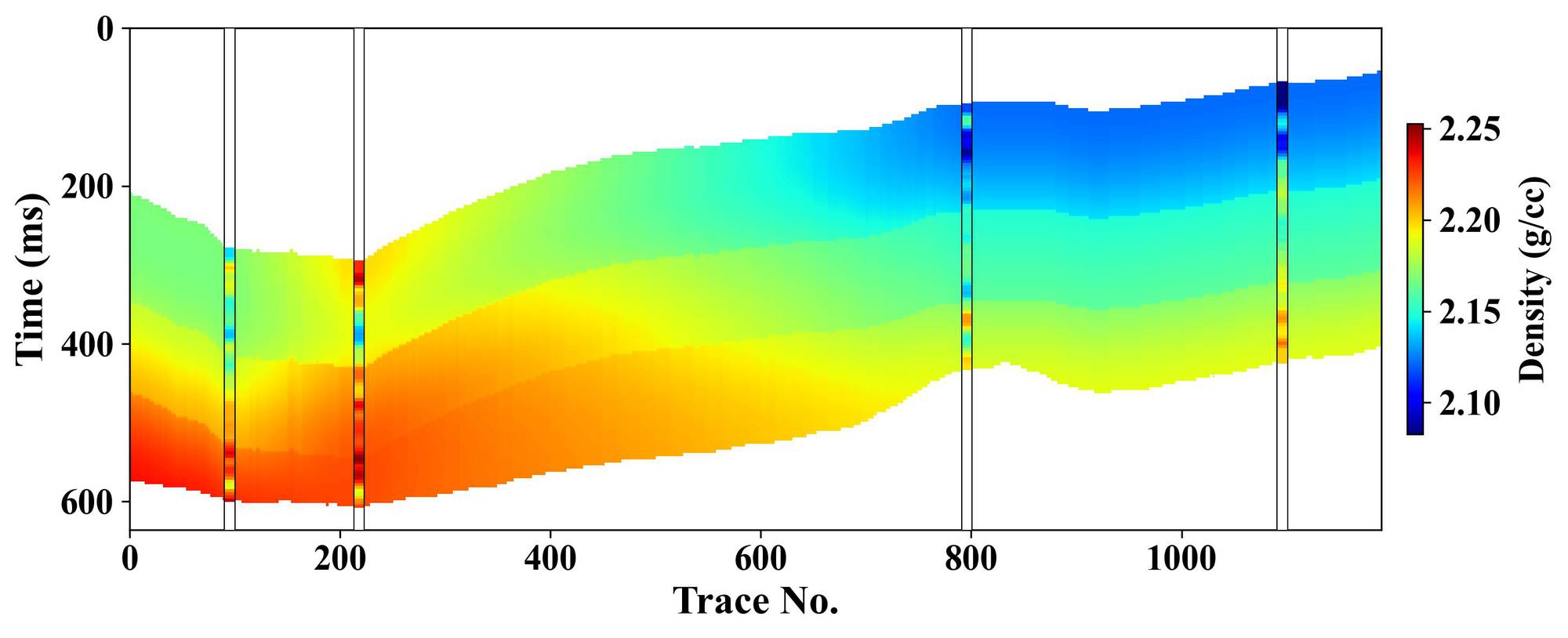}
 \label{fig:F3-lowrho}}
 \caption{F3 demo sample data used for elastic parameter synthesis. 
(a)--(c) Synthetic angle-stacked seismic data at incidence angles of $8^\circ$, $18^\circ$, and $28^\circ$. 
(d)--(f) Low-frequency models of P-wave velocity, S-wave velocity, and density, obtained by applying 1D spatial interpolation to the first three measured wells, followed by three successive applications of a $31 \times 31$ mean filter. }
\label{fig:logsamples}
\end{figure*}

To further validate the proposed method, we conduct an experiment on the F3 block to evaluate an inversion-oriented case of the proposed elastic parameter synthesis framework, where multiple types of conditioning information are incorporated. The F3 block is located in the Dutch sector of the North Sea and represents a more realistic geological setting than the Marmousi II model. Since the publicly available F3 project does not provide prestack seismic gathers, we use the F3 demo sample data provided by LTrace Geosciences, which consist of synthetic angle-stacked seismic data at incidence angles of $8^\circ$, $18^\circ$, and $28^\circ$. Although these data are not original prestack field seismic data, they provide a practical test case for evaluating the proposed method under more realistic subsurface conditions. The corresponding F3 survey area includes four measured wells, which are used to provide well log constraints in the experiment. Then, we extract a 2D profile passing through the wells as the test area, as shown in Figs. \ref{fig:F3-seis0}--\ref{fig:F3-seis2}. Figs. \ref{fig:F3-lowvp}--\ref{fig:F3-lowrho} show the low-frequency models of P-wave velocity, S-wave velocity, and density. These models are constructed by first applying 1D spatial interpolation to the first three well logs, followed by three successive applications of a $31 \times 31$ mean filter. Before applying the trained diffusion model to the F3 data, we examine whether the training datasets constructed in Section \ref{sec:IVA} contain the elastic-parameter statistics compatible with the measured F3 well logs. As shown in Fig. \ref{fig:scatter}, the generated training samples cover the main statistical trends of the measured well log data. Therefore, the trained diffusion model can be directly applied to this experiment without retraining.

To validate the proposed method, we also compare it with 2D-TV and 2D-TVL. For the proposed diffusion-based framework, we evaluate the same condition combinations as those used in the Marmousi II experiment: DM-SL, DM-SLW, and DM-SWI, where S, L, W, and I denote seismic data, low-frequency models, well logs, and interpolated well-log models, respectively. The diffusion-model settings are also kept consistent with that experiment, including the number of diffusion steps and the constraint-imposition strategy. In addition, we further investigate whether the synthesized samples can serve as training data to support downstream data-driven inversion under limited well log constraints. Specifically, well logs and interpolated well-log models are used as conditioning information to generate synthetic training datasets, which are then used to train the deep neural network in \cite{pang2025iterative} for learning the mapping from seismic data to elastic parameters. After the supervised training stage, semi-supervised learning is employed as a transfer learning strategy to fine-tune the network. The resulting supervised-to-semi-supervised workflow is denoted as SDL-ST.

\begin{figure*}[htb!]
\setlength{\abovecaptionskip}{0.2cm}
 \centering
    \subfigure[]{\includegraphics[width=0.65\columnwidth]{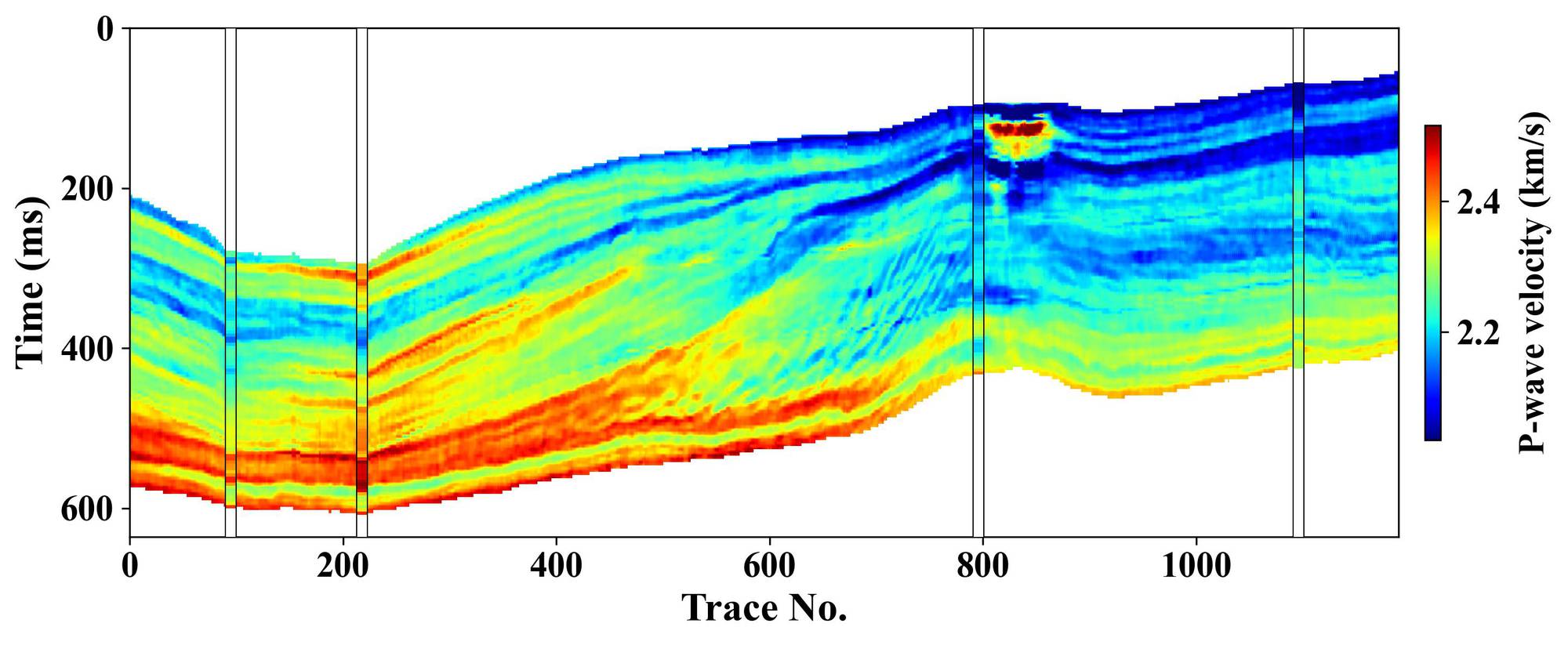}
 \label{fig:2DTVnolow-F3-vp}}
   \subfigure[]{\includegraphics[width=0.65\columnwidth]{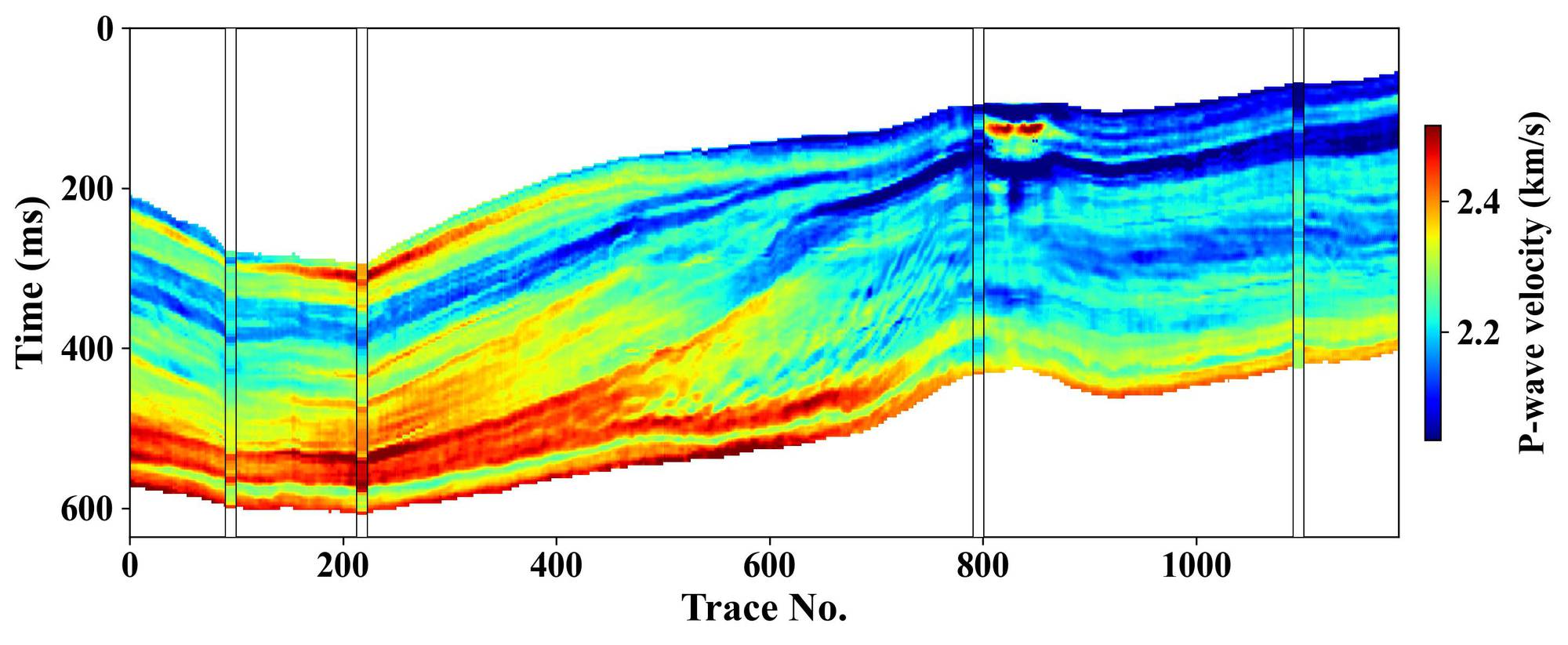}
 \label{fig:2DTV-F3-vp}}
    \subfigure[]{\includegraphics[width=0.65\columnwidth]{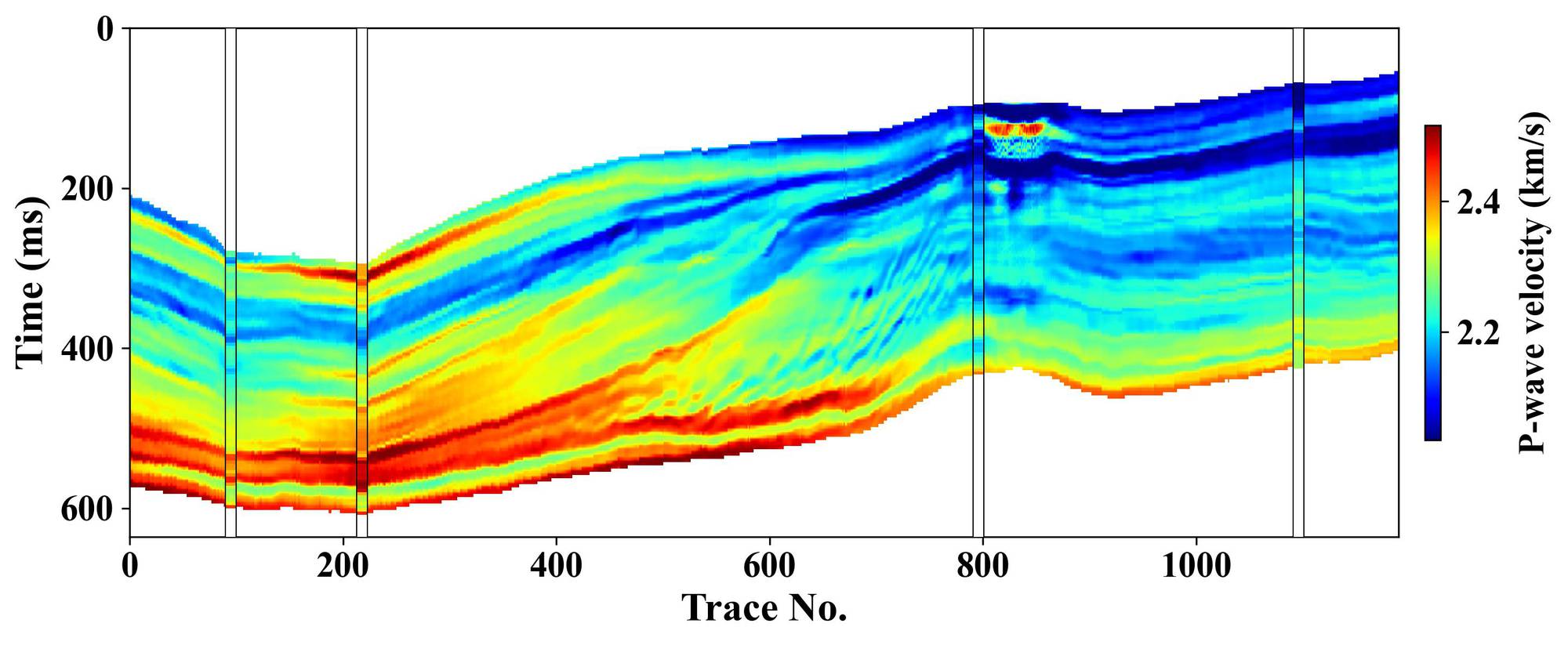}
 \label{fig:ddpm-condlowseis_F3-vp}}
   \subfigure[]{\includegraphics[width=0.65\columnwidth]{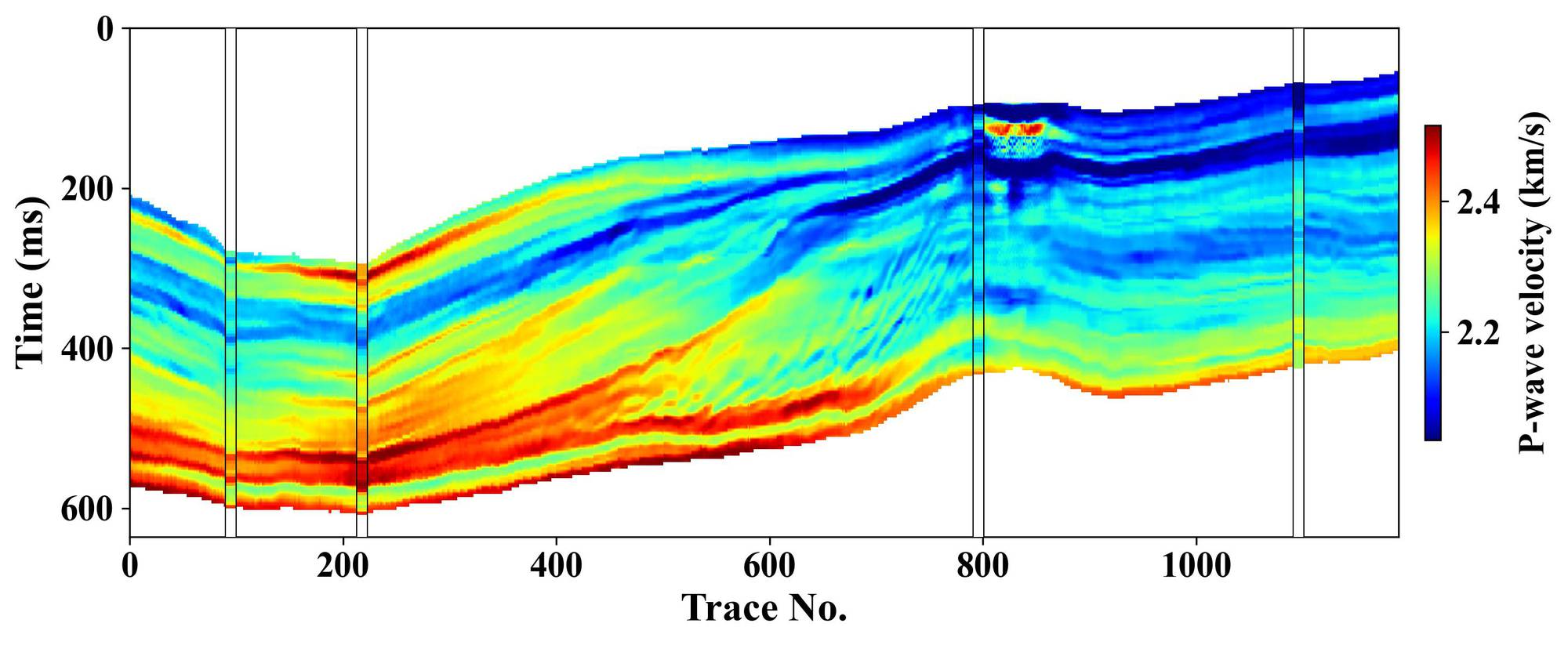}
 \label{fig:ddpm-condlowlogseis_F3-vp}} 
   \subfigure[]{\includegraphics[width=0.65\columnwidth]{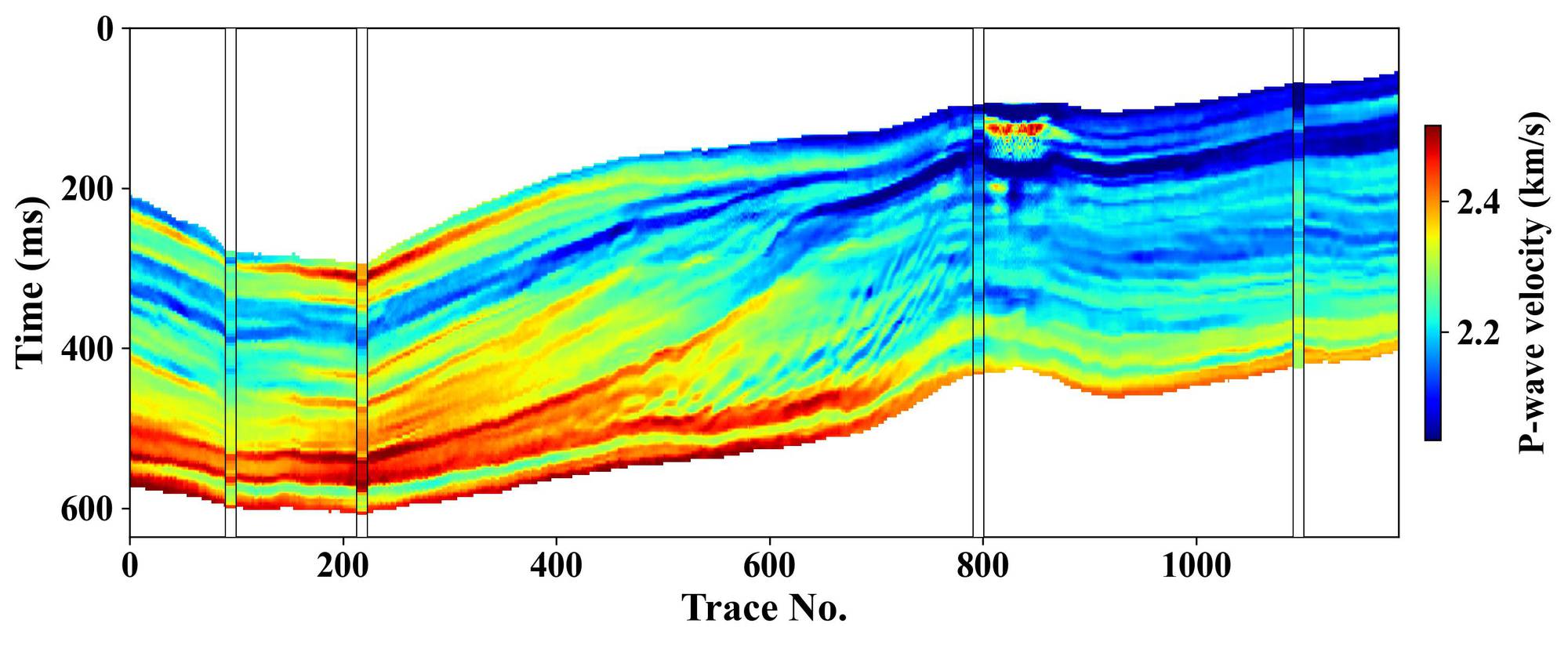}
 \label{fig:ddpm-condlogseis_F3-vp}}
    \subfigure[]{\includegraphics[width=0.65\columnwidth]{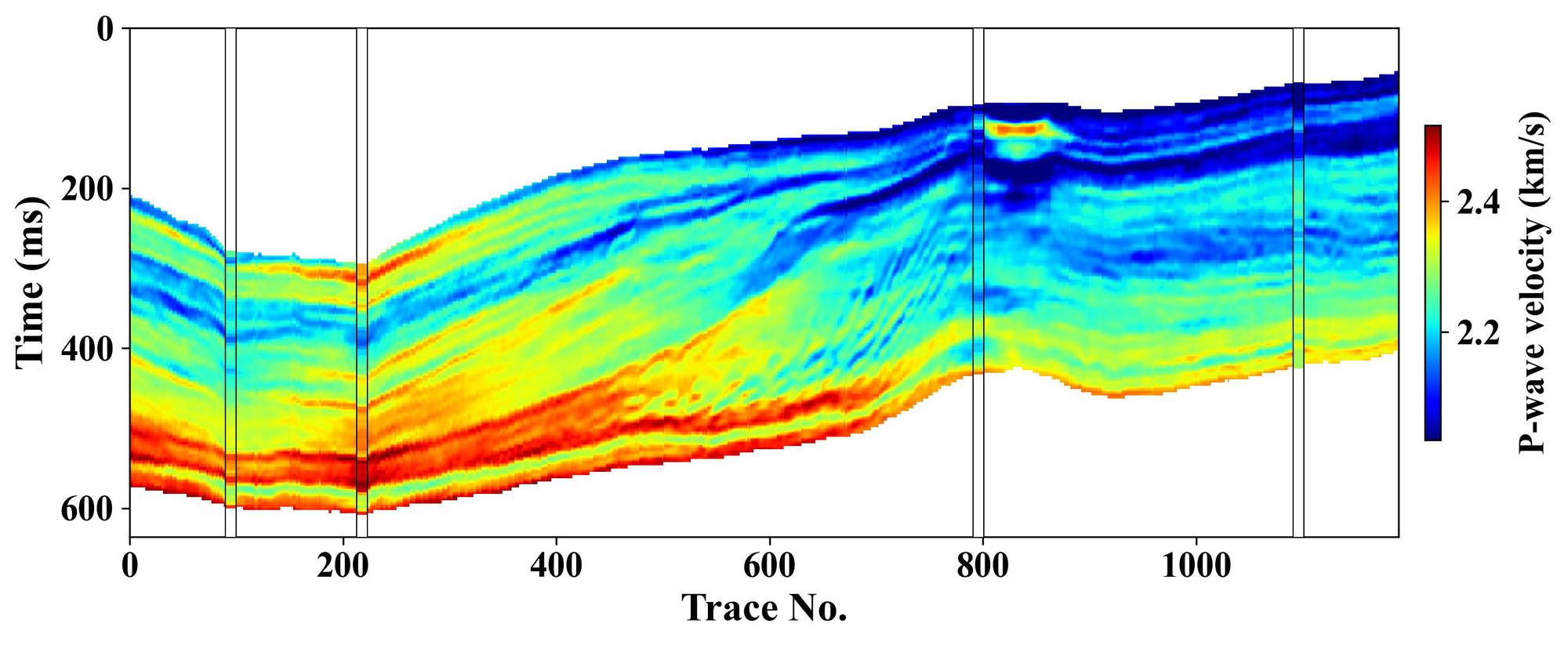}
 \label{fig:pre-semitune-vp}}
 \caption{Predicted P-wave velocity models obtained using different methods. 
(a)--(f) Predicted results obtained using 2D-TV, 2D-TVL, DM-SL, DM-SLW, DM-SWI, and SDL-ST, respectively. The four colored vertical profiles indicate the well logs, with the last well used as the test well.}
\label{fig:F3samplesvp}
\end{figure*}

\begin{figure*}[htb!]
\setlength{\abovecaptionskip}{0.2cm}
 \centering
    \subfigure[]{\includegraphics[width=0.65\columnwidth]{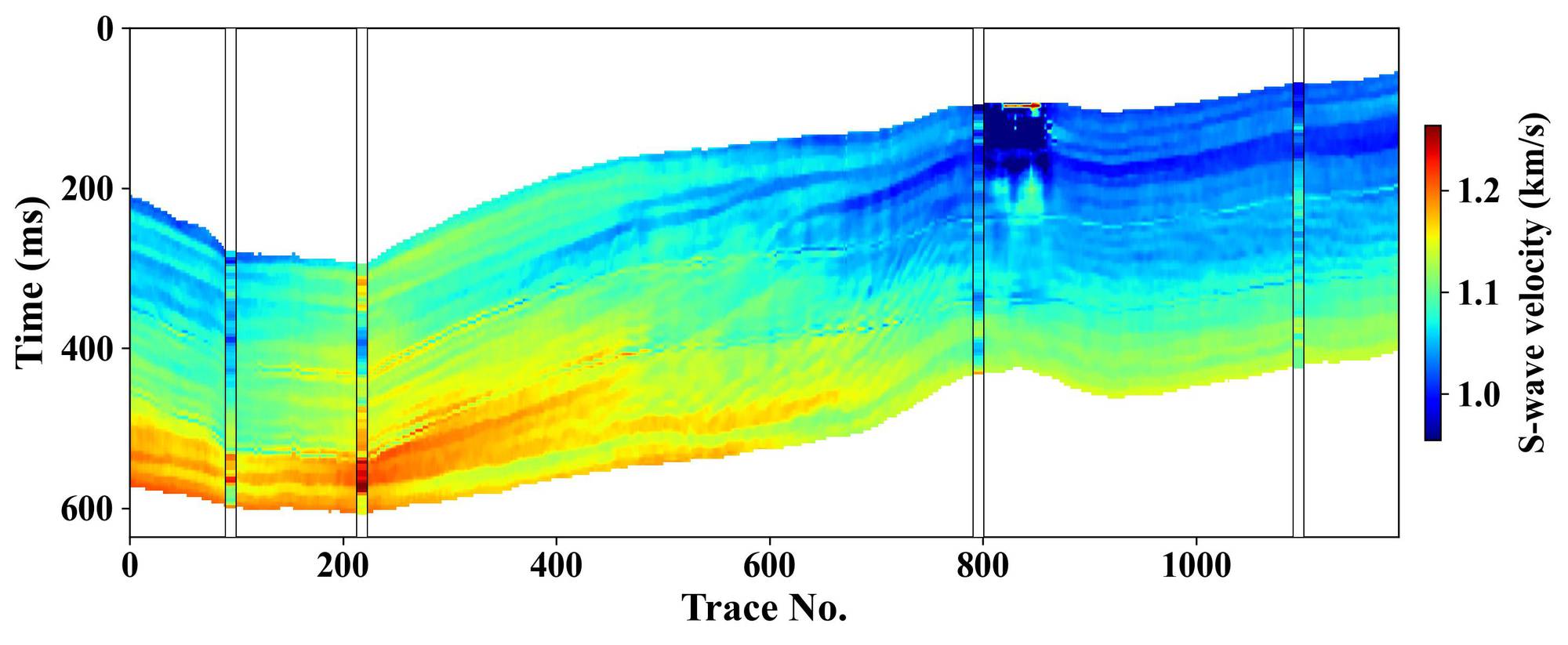}
 \label{fig:2DTVnolow-F3-vs}}
   \subfigure[]{\includegraphics[width=0.65\columnwidth]{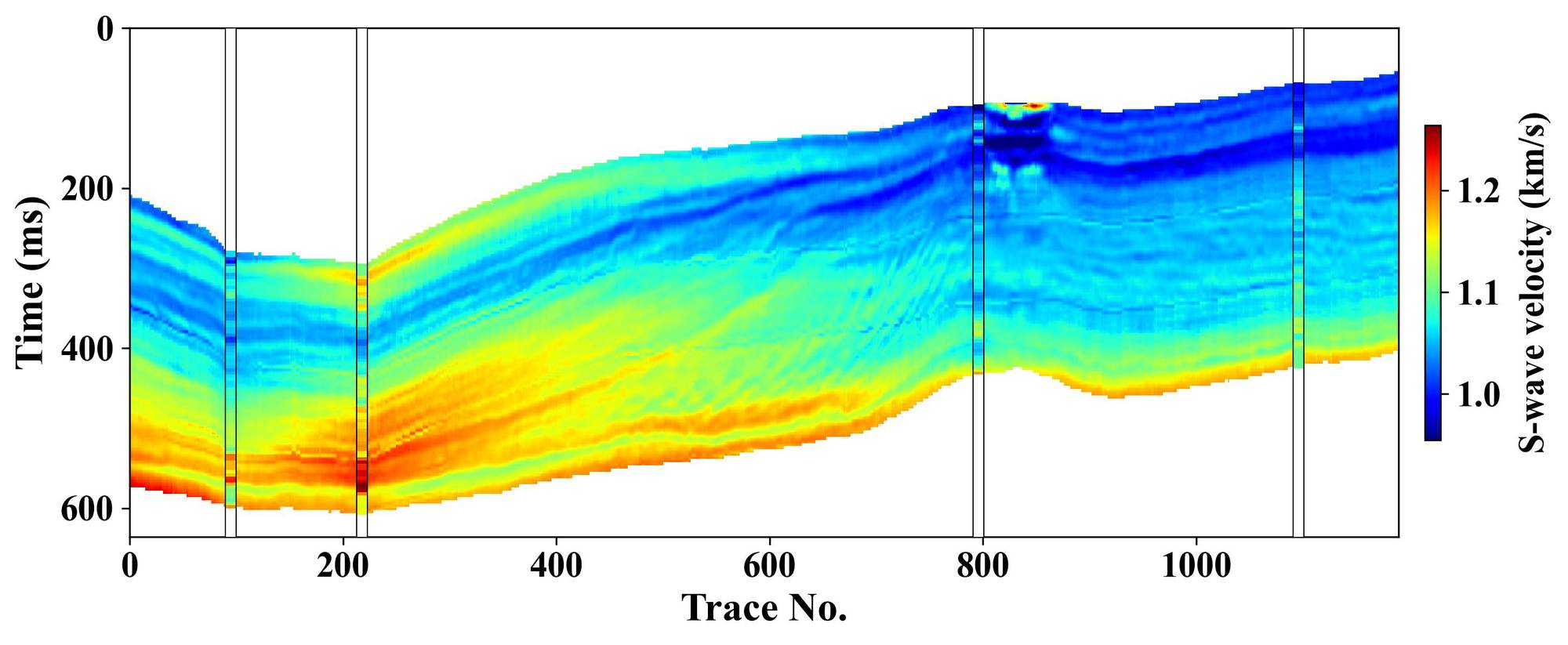}
 \label{fig:2DTV-F3-vs}}
    \subfigure[]{\includegraphics[width=0.65\columnwidth]{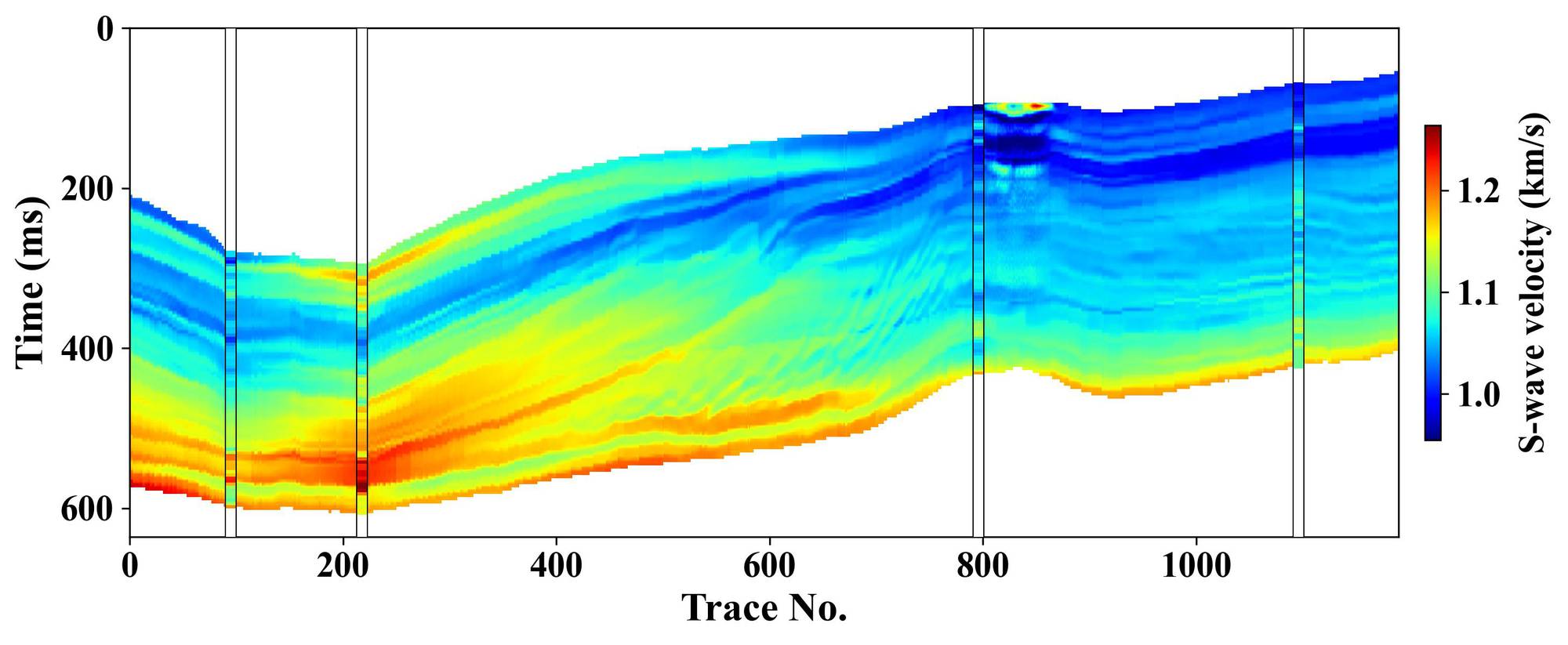}
 \label{fig:ddpm-condlowseis_F3-vs}}
   \subfigure[]{\includegraphics[width=0.65\columnwidth]{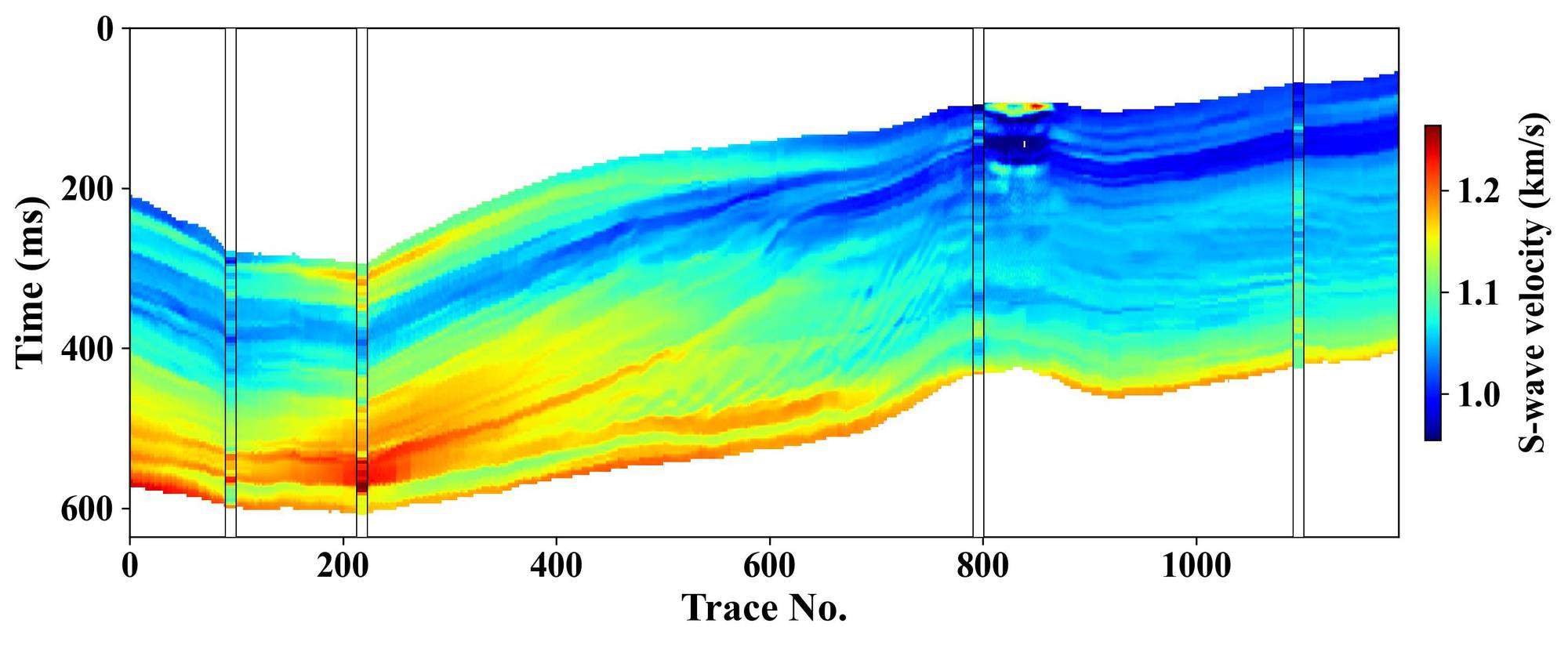}
 \label{fig:ddpm-condlowlogseis_F3-vp}} 
   \subfigure[]{\includegraphics[width=0.65\columnwidth]{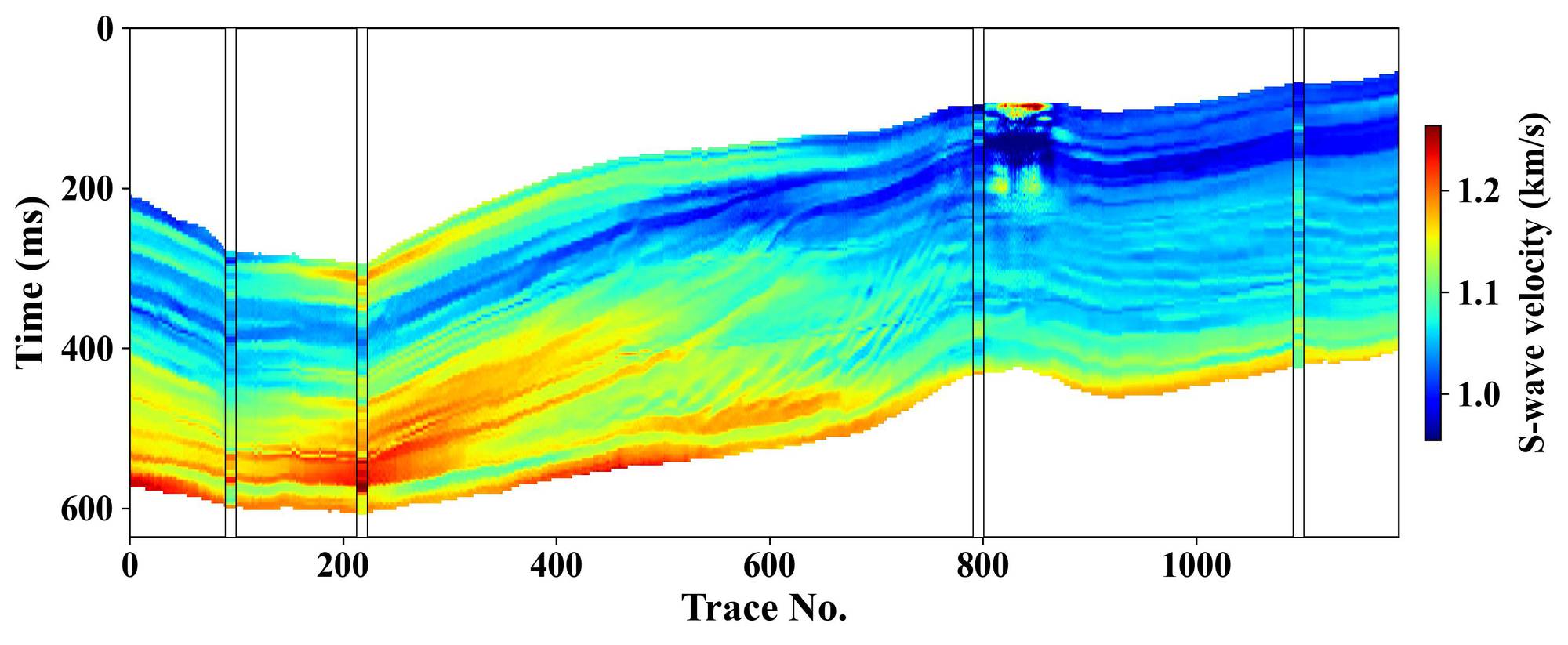}
 \label{fig:ddpm-condlogseis_F3-vs}}
    \subfigure[]{\includegraphics[width=0.65\columnwidth]{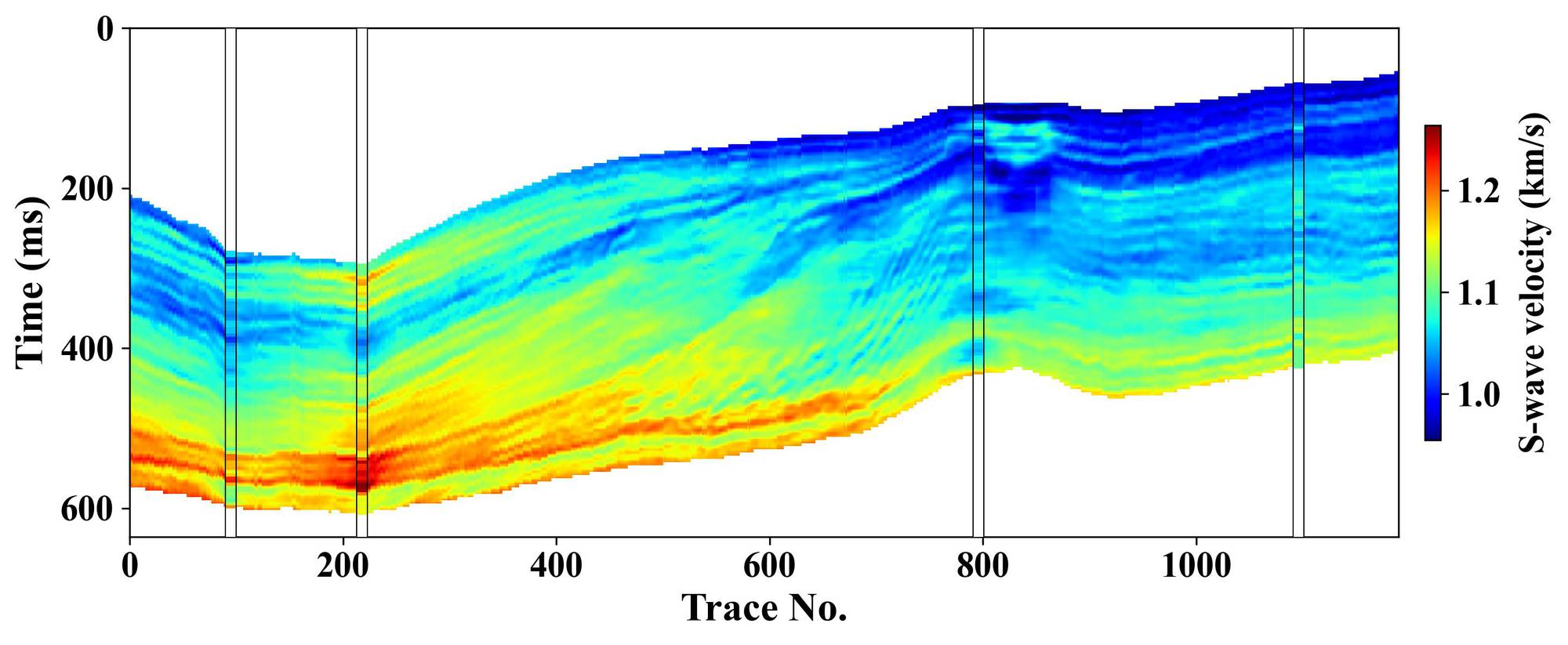}
 \label{fig:pre-semitune-vs}}
 \caption{Predicted S-wave velocity models obtained using different methods. 
(a)--(f) Predicted results obtained using 2D-TV, 2D-TVL, DM-SL, DM-SLW, DM-SWI, and SDL-ST, respectively. The four colored vertical profiles indicate the well logs, with the last well used as the test well.
 }
\label{fig:F3samplesvs}
\end{figure*}

\begin{figure*}[htb!]
\setlength{\abovecaptionskip}{0.2cm}
 \centering
    \subfigure[]{\includegraphics[width=0.65\columnwidth]{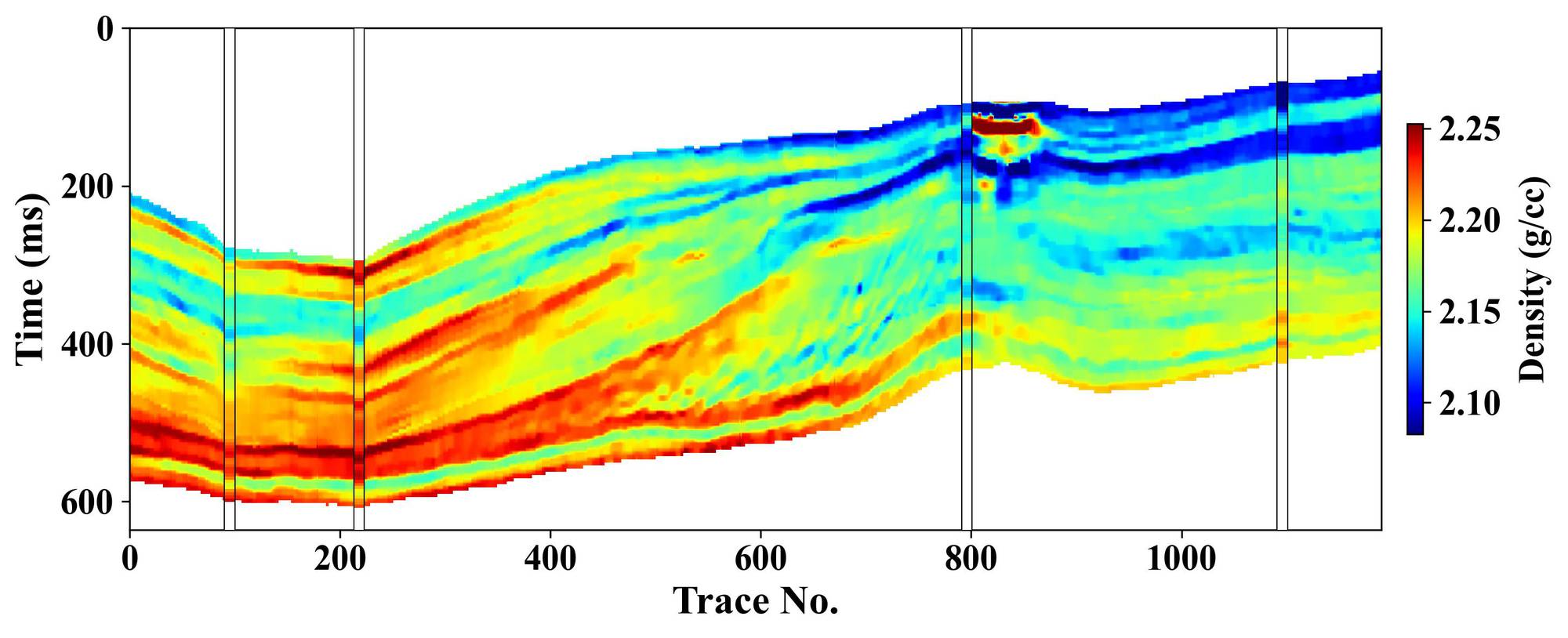}
 \label{fig:2DTVnolow-F3-rho}}
   \subfigure[]{\includegraphics[width=0.65\columnwidth]{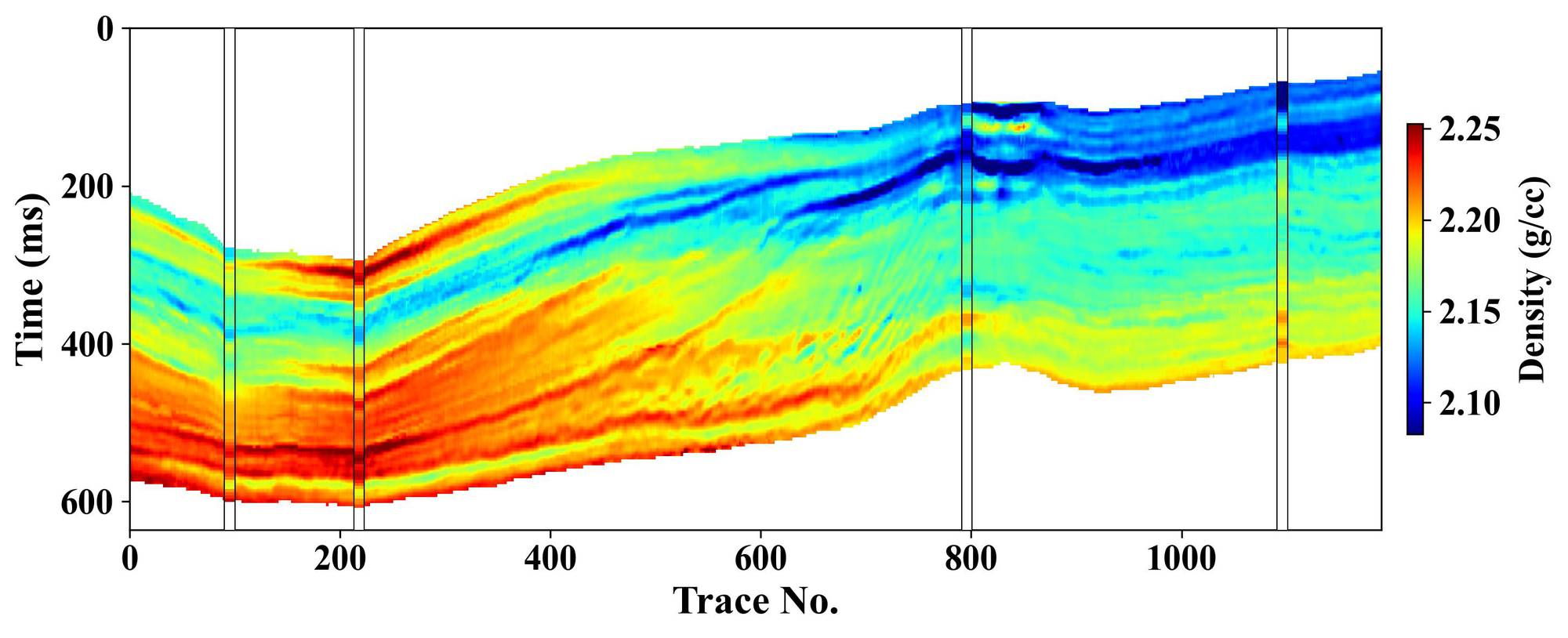}
 \label{fig:2DTV-F3-rho}}
    \subfigure[]{\includegraphics[width=0.65\columnwidth]{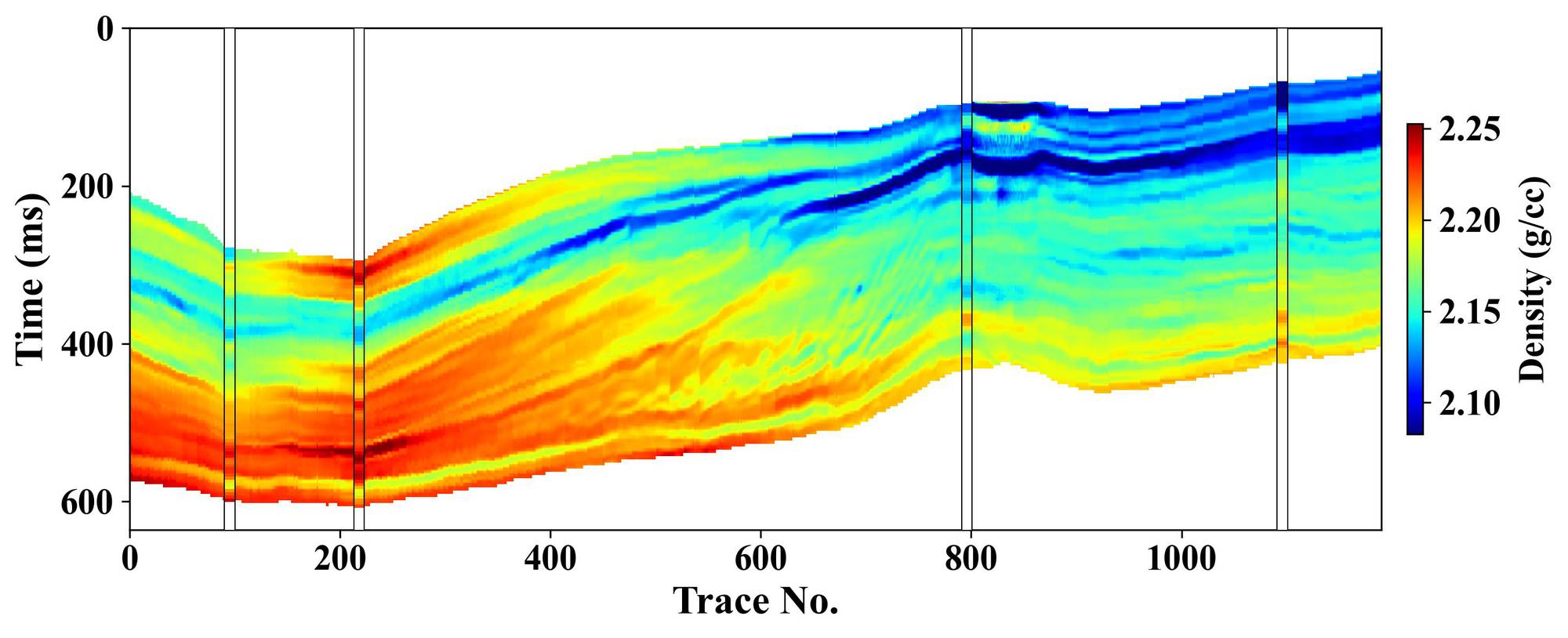}
 \label{fig:ddpm-condlowseis_F3-rho}}
   \subfigure[]{\includegraphics[width=0.65\columnwidth]{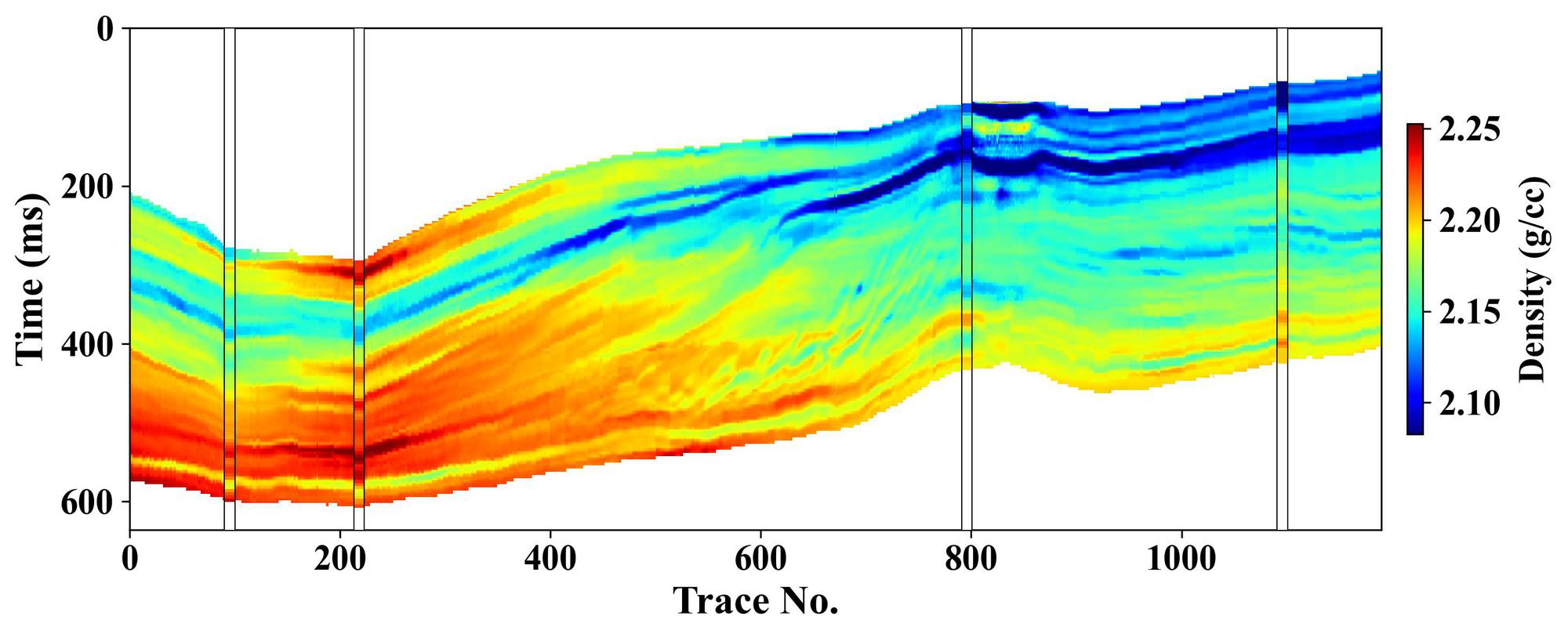}
 \label{fig:ddpm-condlowlogseis_F3-rho}} 
   \subfigure[]{\includegraphics[width=0.65\columnwidth]{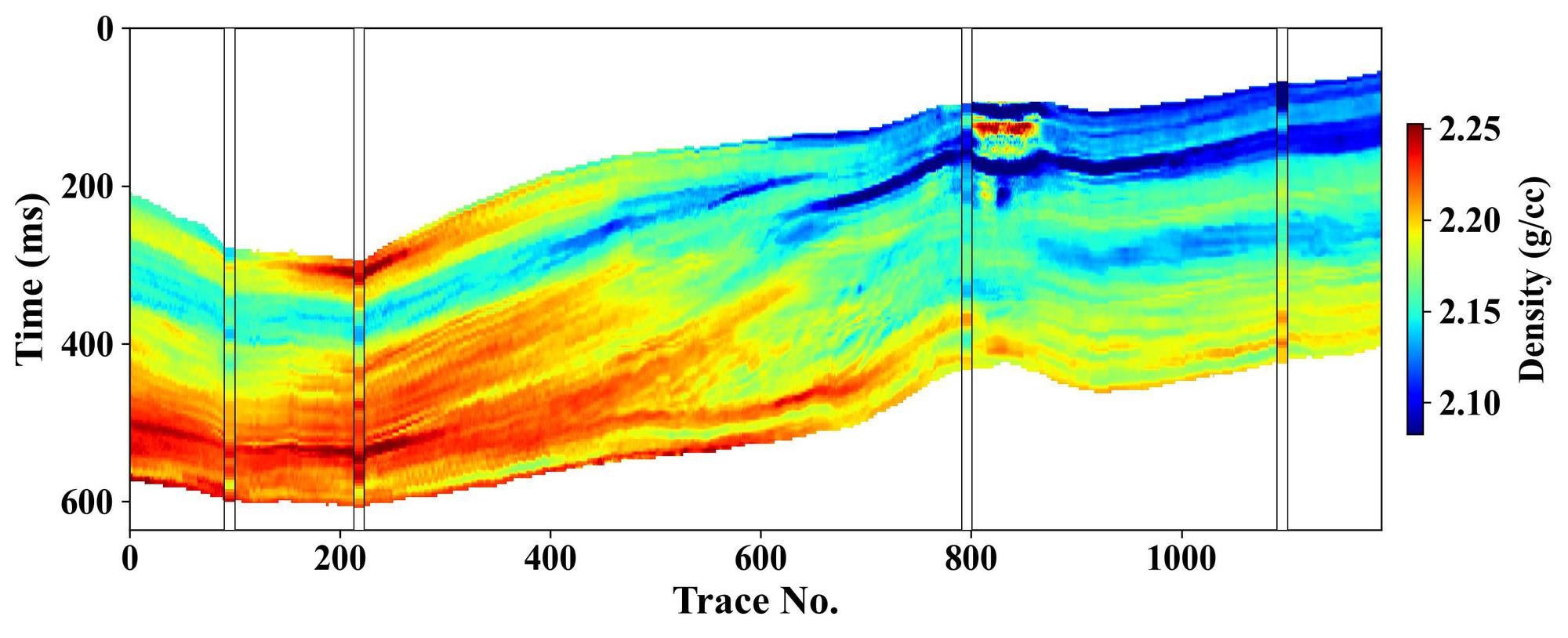}
 \label{fig:ddpm-condlogseis_F3-rho}}
    \subfigure[]{\includegraphics[width=0.65\columnwidth]{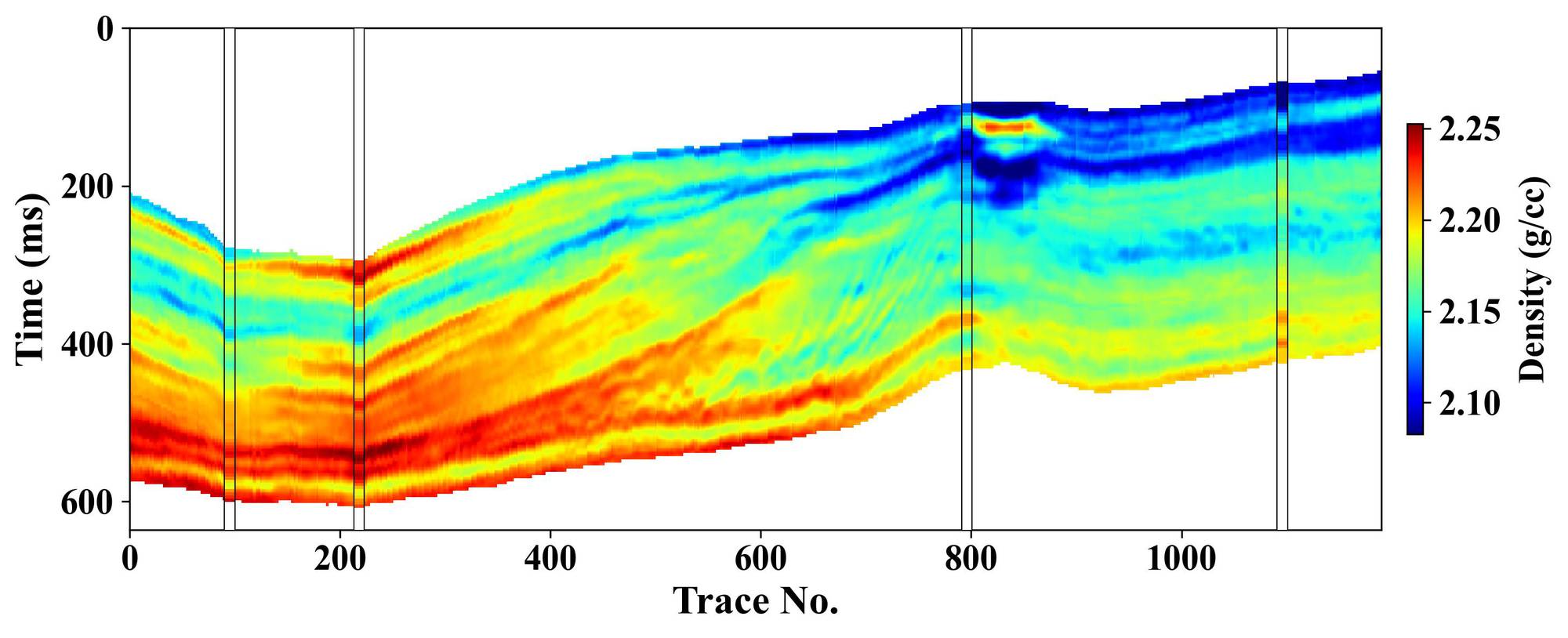}
 \label{fig:pre-semitune-rho}}
 \caption{Predicted density models obtained using different methods. 
(a)--(f) Predicted results obtained using 2D-TV, 2D-TVL, DM-SL, DM-SLW, DM-SWI, and SDL-ST, respectively. The four colored vertical profiles indicate the well logs, with the last well used as the test well.
 }
\label{fig:F3samplesrho}
\end{figure*}

Fig. \ref{fig:F3samplesvp} shows the predicted P-wave velocity models obtained using different methods. The results of 2D-TV and 2D-TVL are more susceptible to noise contamination, leading to degraded lateral continuity and poorly resolved local details. In comparison, the proposed diffusion-based methods produce P-wave velocity models with improved spatial continuity and better characterization of some local variations. Among them, DM-SWI recovers more detailed variations, which can be attributed to the model-domain spatial constraint provided by the interpolated well-log models. The SDL-ST result is also superior to those of 2D-TV and 2D-TVL and is generally comparable to the proposed diffusion-based methods. Fig. \ref{fig:F3samplesvs} shows the predicted S-wave velocity models obtained by different methods. Similar to the P-wave velocity results, the 2D-TV and 2D-TVL results show reduced structural continuity, especially in the right part of the profile, where local velocity variations are poorly resolved. Additionally, the 2D-TV result deviates from the second well log and exhibits an interwell variation between the second and third wells that is not observed in the other inversion results. In contrast, the proposed diffusion-based methods recover better-resolved structures and more coherent lateral variations, with DM-SWI preserving the richest local details among the proposed methods. The SDL-ST result also shows competitive performance, producing visually natural structures with well-preserved local details. Fig. \ref{fig:F3samplesrho} shows the predicted density models. For density prediction, all methods perform better than 2D-TV. The 2D-TVL result is comparable to those of the proposed diffusion-based methods, suggesting that the low-frequency constraint plays an important role in density estimation. The SDL-ST method also produces reasonable and visually plausible results, although its spatial resolution is slightly lower than that of the proposed methods. Overall, the proposed diffusion-based methods achieve the best overall performance across the three elastic parameters, particularly in terms of spatial resolution and local detail recovery, while SDL-ST provides competitive results with more natural spatial variations.

To quantitatively evaluate the inversion results, Table \ref{tab:F3compar} reports the PCCs between the predicted elastic parameters and the measured well logs at the well locations. Log-A denotes the average PCC calculated from the first three wells, which are used as available well constraints, whereas Log-V denotes the PCC calculated from the test well. For Log-A, DM-SLW, DM-SWI, and SDL-ST obtain relatively high PCCs because well log information is incorporated in these methods. Compared with 2D-TV, both 2D-TVL and DM-SL achieve higher PCCs, indicating the importance of explicitly incorporating low-frequency models in the inversion process. For the test well, SDL-ST achieves the highest PCC, while the remaining methods show similar PCCs. Notably, 2D-TV also gives a relatively high PCC, which may be attributed to its recovery of the main background trend. In addition, we calculate the PCCs between the reconstructed seismic data, generated from the inverted elastic parameters, and the input seismic data shown in Figs. \ref{fig:F3-seis0}--\ref{fig:F3-seis2}. All reconstructed seismic data achieve PCCs higher than 0.9, indicating that the inversion results satisfy seismic data consistency.

The above experiments demonstrate the effectiveness of the proposed method in elastic parameter inference from seismic data. Moreover, the proposed method can generate reliable synthetic training datasets, providing useful data support for subsequent deep-learning-based inversion frameworks.

\begin{table*}[htbp]
\centering
\caption{Quantitative comparison of prediction accuracy at well locations among different methods.}
\label{tab:F3compar}
\small
\setlength{\tabcolsep}{4pt}
\begin{tabular}{p{1cm} p{1cm} p{2cm} p{2cm} p{2cm} p{2cm} p{2cm} p{2cm}}
\toprule
 &  & 2D-TV & 2D-TVL & DM-SL & DM-SLW & DM-SWI & SDL-ST \\
\midrule
\multirow{3}{*}{Log-A}
& $v_p$& 0.8146 & 0.8376 & 0.8525 & 0.9922 & 0.9947 &0.9993 \\
& $v_s$ & 0.7018& 0.8594 & 0.8584 & 0.9920 & 0.9943 &0.9994 \\
& $\rho$ &0.8638 & 0.9244 & 0.9173 & 0.9971 & 0.9952 &0.9995\\
\midrule
\multirow{3}{*}{Log-V}
& $v_p$& 0.8714 & 0.8245 & 0.8345 & 0.8365 & 0.8263 & \textbf{0.8863}\\
& $v_s$ &0.8254 & 0.7676 & 0.7853 & 0.7836 & 0.7516 &\textbf{0.8592} \\
& $\rho$ &0.9063 & 0.8761 & 0.8655 & 0.8614 & 0.8997 &\textbf{0.9396}\\
\bottomrule
\end{tabular}
\end{table*}

\section{Discussion}
The proposed framework provides a flexible way to perform elastic parameter synthesis under multi-condition guidance. Nevertheless, several aspects still need to be further investigated. First, incorporating more types of conditions without explicit operators remains challenging. Although Adapter-based conditioning provides an effective solution for structural information, different types of geological or geophysical constraints may require specifically designed condition-injection modules. Developing a more general conditioning mechanism is therefore an important direction for future work. Second, the DPS-projection guidance strategy requires manually specified weights to balance different constraints. In practice, different conditions may have different levels of reliability, noise contamination, and spatial coverage. Therefore, adaptive weighting strategies should be developed to balance the contributions of different conditions according to their quality and credibility. Finally, the computational cost of diffusion sampling is still relatively high, especially for large-scale elastic parameter synthesis in field-data applications. Future work will investigate fast sampling strategies and more efficient latent-space diffusion models to improve the computational efficiency of the proposed framework.

\section{Conclusions}
In this work, we first construct synthetic elastic parameter datasets using well log statistics and representative geological characteristics. The constructed datasets are then used to train the diffusion model to learn the prior distribution of elastic parameter models. Based on the pretrained diffusion model, we propose a unified multi-condition guided diffusion framework for controllable elastic parameter synthesis. The proposed framework flexibly integrates different types of conditioning information according to their relationships with elastic parameters. ILVR and Adapter-based conditioning were used to incorporate model-domain reference information and non-model-domain structural information without explicit conditioning operators, whereas a DPS-projection hybrid guidance strategy was developed to enforce conditions with explicit conditioning operators. Synthetic experiments demonstrate that the proposed method can generate condition-consistent elastic parameter samples under both single-condition and multi-condition guidance. The inversion experiments further show that the proposed diffusion-based method achieves more accurate elastic parameter prediction than the conventional regularization-based methods, indicating the advantage of learned generative priors in solving ill-posed inverse problems. The synthesized samples can also support downstream end-to-end deep-learning-based inversion, demonstrating the potential of the proposed framework under limited labeled data.

\appendices

\section*{Acknowledgment}
%"自然基金  博后面上 以及基本科研业务费"
We thank dGB Earth Sciences for making the data available as an OpendTect project via their TerraNubis portal terranubis.com. We also gratefully acknowledge LTrace Geosciences for providing the F3 demo sample data used in this study.

\bibliographystyle{IEEEtran}
\bibliography{PCDL}

\end{document}